# PROJECT DYNAMICS AND EMERGENT COMPLEXITY

**Christopher M. Schlick**

Institute of Industrial Engineering and Ergonomics

RWTH Aachen University

**Bruno Demissie**

Fraunhofer Institute for Communication, Information Processing and Ergonomics FKIE

Fraunhofer Society

## ABSTRACT

This paper presents theoretical and empirical analyses of dynamics and emergent complexity in new product development (NPD) projects that are organized according to the management concept of concurrent engineering. Concurrent engineering is a systematic approach to the integrated, concurrent design of products and their related processes. It is intended to cause the developers, from the outset, to consider all elements of the product life cycle and therefore requires intensive cooperation between and within development teams. To analyze cooperative work in such an open organizational system, a model-driven approach is taken and different mathematical models are formulated.

Two classes of models are considered. The basic class are vector autoregression (VAR) models of finite order which can capture the typical cooperative processing of the development tasks with short iteration length. Through an augmented state-space formulation it is also possible to represent more complex autoregressive processes with periodically correlated components. These processes incorporate a hierarchical coordination structure and therefore are also able to simulate long-scale effects of intentionally withholding the release of information. Beyond VAR models, the theoretically interesting class of linear dynamical systems (LDS) with additive Gaussian noise are considered. In LDS models the state of the project cannot be directly observed but is rather inferred through a causal model from other variables that are directly measured. In that sense a “hidden” state process of cooperative product development is distinguished from the observation process. The internal state information is not completely accessible by the project manager but must be estimated on the basis of repeated readings from dedicated performance measurement instruments. This fundamental uncertainty in the project state and its evolution can generate a non-negligible fraction of long-term correlations between development tasks and therefore significantly increase emergent complexity. In addition to the mathematical models, least squares and maximum likelihood estimation methods are introduced to demonstrate how the independent parameters can be efficiently estimated from time series of task processing. To validate the models with field data, a case study was carried out in a German industrial company. The validation results are presented and discussed in detail in separate sections.

Furthermore, the complexity framework, theories and measures that have been developed in organizational theory, systematic engineering design and basic scientific research on complex systems are reviewed and applied to project management. To evaluate emergent complexity in NPD projects, an information-theory measure from basic scientific research—termed “effective measure complexity” (EMC)—is chosen because it can be derived from first principles and therefore has high construct validity. In addition, EMC can be calculated efficiently not only from generative dynamic models but also from purely historical data, without intervening models. EMC measures the mutual information between the infinite past and future histories of a stochastic process. According to this principle, EMC is of particular interest for evaluating time-dependent complexity of NPD projects and identifying the relevant interactions between development tasks.

The formulated VAR models provide the mathematical foundation for the calculation of several closed-form solutions of EMC in the original state space, solutions that allow an explicit complexity evaluation based on the model´s independent parameters. A transformation into the spectral basis is carried out to obtain additional, more expressive solutions in matrix form. In the spectral basis, the essential parameters driving emergent complexity, which are surprisingly few in number, can be identified and the effects of cooperative relationships can be interpreted directly. The essential parameters include the eigenvalues of the work transformation matrix as a dynamical operator of the VAR model and the correlation coefficients between components of unpredictable performance fluctuations. Furthermore, a closed-form solution of EMC in the original state-space coordinates is presented for an arbitrary LDS. This solution is not only interesting for the evaluation of emergent complexity in NPD projects, it can also be used to analyze, design and control technical systems.

Finally, the theoretical complexity analyses are elucidated in practical terms through two applied studies. The first study deals with optimizing project organization design. The objective is to minimize emergent complexity by selecting the optimal staffing of three CE teams with developers who have different productivities in a simulated NPD project. The aim of the second study is to optimize the period for minimal emergent complexity, in which information about integration and tests of geometric/topological entities is deliberately withheld by system-level teams and not released to component-level teams. This kind of noncooperative behavior is justified by the aim to improve solution maturity and reduce coordination efforts. In both studies we formulate and solve unconstrained optimization problems and consider the total effort that taken in the project as an additional constraint. This constraint has significant consequences for the solutions. The applied studies shows that EMC is not only a highly satisfactory quantity in theory, it has also led to useful results in organizational optimization regarding key performance indicators, such as the lead time of the project and its empirical variance.

# Contents

# 1 INTRODUCTION

In times of economic volatility and uncertainty of future returns, successful development of innovative products and effective management of the associated new product development (NPD) projects are particularly important for gaining competitive advantage. To shorten time-to-market and lower development/production costs, NPD projects are often subjected to concurrent engineering (CE). In their landmark report, Winner et al. (1988) define CE as “a systematic approach to the integrated, concurrent design of products and their related processes, including manufacture and support. This approach is intended to cause the developers, from the outset, to consider all elements of the product life cycle from conception through disposal, including quality, cost, schedule, and user requirements”. A good example is a large-scale vehicle development project in the automotive industry. In the late development stage, such a project involves hundreds of engineers collaborating in dozens of CE teams. The CE teams are usually structured according to the subsystems of the product to be developed (e.g. car body, powertrain, chassis frame, etc.) and are coordinated by system-integration and management teams. The members are recruited to the CE teams from different branches of the company, such as engineering design, manufacturing or marketing, as well as from suppliers and engineering service providers. Teams vary in number and size depending on the project phase. The needs, requirements, functions and objectives are discussed and “orchestrated” by the subject-matter experts in so-called “CE team meetings” and mapped onto design parameters in highly cooperative work processes. The CE team meetings are scheduled at regular intervals, typically weeks apart. Additional ad-hoc team meetings are sometimes also carried out to solve time-or-quality critical design problems.

NPD projects—not only in the automotive industry—are often characterized as “creative” and can show very informative (but also complex and difficult-to-manage) patterns of organizational dynamics. These patterns emerge from cooperative work processes that are not only fundamentally cyclic, with analysis, synthesis and decision-making stages, but also tightly coupled through the product structure with many interfaces between mechanical, electronic and software components. The workflow is characterized by frequent and sometimes irregular iterations because of the availability of new or updated information about geometric/topological entities, parameters, etc. As a consequence, the development tasks are both highly variable and strongly dependent on each other and on elements of “surprise” in the form of seemingly erratic evolutionary events that occur. The coexistence of variability and dependency is typical of complex systems in different domains, as stressed by Shalizi (2006) and Nicolis and Nicolis (2007). In that sense, NPD projects with long-range spatiotemporal coordination can be regarded as one of the most authentic prototypes of complex sociotechnical systems. They serve as inspiration for raising new issues and stimulate research in complex systems.

One important source of variability is the human factor. The manifold project states usually create a demand for knowledge resources that exceeds the deliberative capacity of any project manager, team leader or team member, and many assumptions about design ranges or physical functions of the product under development have to be made for successful execution. During the course of the development project, some assumptions turn out to be wrong and must be reexamined, while others turn out to be too vague and must be refined, and a few may turn out to be too rigid and must be relaxed. Although most changes and revisions are processed rather quickly and successfully, there is an inevitable level of “ignorance”, which continually generates unpredictable performance fluctuations. These fluctuations are present in each level of the organization and are an irreducible property of the participating units. Furthermore, on the level of the individual, human errors must be taken into account in such an “open organizational system”. The random influences in conjunction with tight coordination structures often render the project difficult to predict and control, because a large body of knowledge of prior history is necessary, and the evolution toward a stable design solution can differ significantly from the expected (unperturbed) process (Huberman and Wilkinson 2005). Depending on the kind and intensity of cooperative relationships, some of the development teams can enter “vicious cycles” of multiple revisions, which demand significant unplanned and unwanted effort as well as long delays (Huberman and Wilkinson 2005). Moreover, the simultaneous revision cycles can be reinforced, and a fatal pattern of organizational dynamics termed “design churns” (Yassine et al. 2003) or “problem-solving oscillations” (Mihm et al. 2003, Mihm and Loch 2006) can emerge. In this case, the progress of the project irregularly oscillates between being on, ahead of, or behind schedule. Ultimately, the project must be abandoned to break the cycle. The design churn effect was analyzed by Terwiesch et al. (2002) and Yassine et al. (2003) in NPD projects in the automotive industry. An anonymous product development manager at an automobile manufacturer commented, “We just churn and chase our tails until someone says they won’t be able to make the launch date” (Yassine et al. 2003). According to the literature review of Mihm and Loch (2006), design churns occur not only in the automotive industry but also in large development projects across different domains.

Design churns are an intriguing example of emergent (also termed “self-generated”) complexity in NPD projects, which can lead to disastrous results and painful financial losses. The emergence is strong in the sense that patterns of organizational dynamics can only be reliably forecast from the observation of the past of each particular instance of task processing and with significant knowledge of prior history (Chalmers 2002). A deeper understanding of the interrelationships between performance variability and project dynamics is needed to cope with this kind of emergent complexity, together with new methods for quantitative complexity analysis and evaluation.

The goal of this paper is therefore twofold: first, to present a model of cooperative task processing in an open organizational system on the basis of the theory of vector-autoregressive (VAR) processes; second, to introduce an information-theory complexity measure that is underpinned by a theory of basic research (Grassberger 1986; Bialek et al. 2001) and enables quantifying strong emergence in terms of mutual information between past and future history. The challenge is that even if the composition of the development tasks, their rate of processing and the laws of interaction are given, it is difficult to anticipate the performance of the whole project. Multiple interactions can affect performance variability and generate effects that cannot be simply reduced to properties of the constituent parts. Instead, these phenomena emerge from higher-order interactions and can be considered properties of the organization as a whole (Huberman and Wilkinson 2005). The information-theory approach to project management builds on our previous work on project modeling and simulation (Schlick et al. 2007, 2008, 2009, Tackenberg et al. 2009, 2010) and differs from the abovementioned studies in the following way. First, a VAR model of cooperative task processing in NPD was formulated and validated against field data acquired in a small industrial company in Germany; second, the approach aimed to analyze explicitly emergent complexity and to formulate closed-form solutions of different strengths, which can be used for identifying the essential complexity-driving variables and to optimize project organization design.

The paper is organized as follows. In section 2, the foundations for deterministic and stochastic modeling of cooperative task processing in NPD projects are laid, and the corresponding state equations are formulated. The self-developed stochastic state equation is explained in detail and validated on the basis of traces of work remaining that were acquired in industry. Moreover, a transformation into the spectral basis is performed to uncover the essential dynamic mechanisms in open organizational systems. In section 3, concepts and measures of complex systems science are reviewed and applied to project management. An information-theory quantity—termed the “effective measure complexity”—is analyzed in detail because of its outstanding construct validity and computational merits for the evaluation of emergent complexity in the application area. The stochastic state-space model developed in section 2 explicitly allows calculation of the EMC and presentation of closed-form solutions with varying numbers of independent parameters. These solutions are derived and discussed in detail in section 4. Several closed-form solutions in the original state-space coordinates and the spectral basis are presented. A solution with a minimal number of independent parameters obtained through a canonical correlation analysis is introduced as well. Full, simplified polynomial-based solutions for projects with two and three tasks are given. Upper and lower bounds are put on the EMC at the end of this section. Section 5 elucidates the theoretical complexity considerations in practical terms by an applied example in project organization design optimization. It is sought to minimize project complexity by systematically choosing the "best" CE team design from within a predefined group of individuals with different productivities, in a simulated project. Finally, section 6, covers the main conclusions of the paper and a brief outlook for future research.

*Notation.* Throughout this paper we will use the following mathematical notation: $A_{:i}$ denotes the *i*th column of the matrix $A$. $A^{\mathrm{T}}$ is the transpose, and $A^*$ the conjugate of $A$. The conjugate matrix is obtained by taking the complex conjugate of each element of $A$. The inverse of $A$ is denoted by $A^{-1}$. The elements of a matrix are either written as sub-scripted, unbolded lower-case letters, e.g. $a_{ij}$, or are indexed by $i$ and $j$ as $A_{[\![i,j]\!]}$. The index form stems from the notation of the Mathematica® modeling and simulation environment. Additional operations on matrices and vectors, although quite unusual, begin with the capital letter of the unary or binary operation, and the argument is written in square brackets, e.g. $E[.]$, $\mathrm{Var}[.]$, $\mathrm{Det}[.]$, $\mathrm{Exp}[.]$, $H[.]$ or $I[.;.]$. This representation is also derived from the Mathematica® modeling and simulation environment. Similarly, the linear algebraic product of matrices, vectors or vector/matrices is written explicitly, that is for instance as $A \cdot A$ and not as $AA$. This rule is only violated if the terms get too long and their meaning is clear from the context, e.g. in some paragraphs in sections 2.8, 2.9, 4.2 and the appendix. The multiplication of a scalar $a$ with a matrix $A$ is written as $\{a\} \cdot A$. An identity matrix of size $n$ is denoted by the symbol $I_n$. A zero column vector with $n$ components is denoted by $0_n$. A continuous-type or discrete-type random state variable is denoted by a Latin capital letter, e.g. $X$. An observed value (realization) of a random state variable is indicated by a lower-case letter symbol, e.g. $x$. A random variable that represents unpredictable fluctuations or noise is denoted by a lower-case Greek letter, e.g. $\varepsilon$. The symbol ~ means that a random variable is distributed according to a certain probability distribution, e.g. $\varepsilon \sim \mathcal{N}(\mu, \Sigma)$. A multivariate Gaussian distribution with location (mean) $\mu$ and covariance matrix $\Sigma$ is written as $\mathcal{N}(\mu, \Sigma)$. The corresponding probability density function with parameter vector $\theta = (\mu, \Sigma)$ is denoted by $f_\theta[x] = \mathcal{N}(x; \mu, \Sigma)$. Equations that use or generate a time series include a time index of state and noise variables, e.g. $x_t$ or $\varepsilon_t$. The complete stochastic state process is written as $\{X_t\}$. Finite process fragments $\left(X_{t_1}, X_{t_1+1}, \ldots, X_{t_2}\right)$ from time step $t_1 \in \mathbb{Z}$ to $t_2 \in \mathbb{Z}$ are written as $\{X_t\}_{t_1}^{t_2}$. Similarly, the term $\{x_t\}_{t_1}^{t_2} = \left(x_{t_1}, x_{t_1+1}, \ldots, x_{t_2}\right)$ denotes the sequence of states that was observed across the same interval of time steps.

# 2 MATHEMATICAL MODELS OF COOPERATIVE WORK IN NEW PRODUCT DEVELOPMENT PROJECTS

## 2.1 Deterministic Formulation

To analyze the interrelationships between project dynamics and emergent complexity explicitly and through a special theoretical perspective, a dynamic model of cooperative work in NPD has to be formulated and the independent parameters have to be defined. We begin with the deterministic formulation of a continuous-state, discrete time model based on the seminal work of Smith and Eppinger (1997), according to whom a distinct phase of an NPD project with $p$ concurrent and interacting tasks can be modeled by a linear first-order difference equation as

$$x_t = A_0 \cdot x_{t-1} \qquad t = 1, \dots, T. \tag{1}$$

The above state equation is also termed a linear homogeneous recurrence relation. The $p$-dimensional state vector $x_t \in \mathbb{R}^p$ represents the work remaining for all $p$ tasks at time step $t$. It is assumed that the state vector is observed (or estimated) by the project manager at equally spaced time instants and therefore time can be indexed by the discrete variable $t$. The amount of work remaining can be measured by the time left to finalize a specific design, the number of engineering drawings requiring completion before the design is released, the number of engineering design studies required before design release, or the number of open issues that need to be addressed/resolved before design release (Yassine et al. 2003). The matrix $A_0 = (a_{ij})$ is a dynamical operator for the iteration over all $p$ tasks, also called the "work transformation matrix" (WTM). The WTM is a square real matrix of dimension $p \times p$, i.e. $A_0 \in \mathbb{R}^{p \times p}$. We assume that it is has full rank. The WTM can be regarded as a task-oriented variant of the popular design structure matrix (Steward 1981), which is often used in industry and academia to analyze and optimize complex products. It enables the project manager to model, visualize and evaluate the dependencies among the development tasks and to derive suggestions for improvement or reorganization. It is clear that not only the tasks to be processed but also the structure of the product (in terms of an envisioned physical solution) and the formalized design problem are important in meeting the project goals and satisfying the functional and nonfunctional requirements. However, for the sake of simplicity, in the following we focus on the tasks and their interactions and assume that additional dependencies from the product or problem domain were integrated into a joint work transformation model. This approach is supported by the management-oriented complexity definition of Tatikonda and Rosenthal (2000): "We define project complexity as the nature, quantity and magnitude of organizational subtasks and subtask interactions posed by the project". Given a distinct phase of an NPD project, it is assumed that the WTM does not vary with time, and that the state equation is autonomous.

In this paper, we use the improved WTM concept of Yassine et al. (2003) and Huberman and Wilkinson (2005). Hence, the diagonal elements $a_{ii}$ $(i = 1 \dots p)$ account for different productivity levels of developers when processing tasks. This is in contrast to the original WTM model of Smith and Eppinger (1997) in which tasks are processed at the same rate. The diagonal elements $a_{ii}$ are defined as autonomous task-processing rates. They indicate the part of the work left incomplete after an iteration over task $i$ and therefore must be nonnegative real numbers ($a_{ii} \in \mathbb{R}^+$). The off-diagonal elements $a_{ij} \in \mathbb{R}$ $(i \neq j)$, however, model the informational coupling among tasks and indicate the intensity and nature of cooperative relationships between developers. Depending on their value, they have different meanings: 1) if $a_{ij} = 0$, work carried out on task $j$ has no direct effect on task $i$; 2) if $a_{ij} > 0$, work on task $j$ slows down the processing of task $i$, and one unit of work on task $j$ at time step $t$ generates $a_{ij}$ units of extra work on task $i$ at time step $t + 1$; 3) if $a_{ij} < 0$, work on task $j$ accelerates the processing of task $i$, and one unit of work on task $j$ reduces the work on task $i$ by $a_{ij}$ units at time step $t+1$. The only limitation on the use of negative entries is that negative values of the work remaining in the state vector $x_t$ are not permitted at any time instant. In practice, many off-diagonal elements must be expected to be nonnegative, because NPD projects usually require intensive cooperation, leading to additional work. For instance, Klein et al. (2003) analyzed the design of the Boeing 767 and found that half of the engineering labor budget was spent on redoing work because the original work did not lead to satisfactory results. About 25%–30% of design decisions required reworking, and in some instances up to 15 iterations had to be done to reach a stable design state.

This paper only consider a distinct phase of an NPD project in which a subsets of tasks must be processed in parallel and no task in the subset is theoretically processed independently of the others, because input about components under development by other tasks is required regularly. This assumption does not limit the generality of the approach. In NPD projects in which work is broken down into multiple phases that are performed in series, the analysis simply holds for each phase in the sequence.

The initial time step $t=0$ usually indicates the beginning of a project phase. If the project initialization phase is modeled by the above linear homogeneous recurrence relation, the initial time step represents the beginning of the whole project. The end of the project phase of interest is indicated by time instant $T \in \mathbb{N}$. For small development projects the time index $T$ can also cover the complete duration. It is often assumed that all parallel tasks are initially 100% to be completed, and so the initial state $x_0$ is

$$x_0 = \begin{pmatrix} 1 \\ \vdots \\ 1 \end{pmatrix}. \tag{2}$$

However, the project manager can also assign nonnegative values to vector components of $x_0$. By doing so, it is possible to model overlapping tasks (see eq. 45). Moreover, if an NPD project undergoes major reorganization, one can define separate initial states and WTMs. The analysis would then apply separately to each reorganized phase of the project.

From the theory of linear dynamic systems it is known that the rate and nature of convergence of the modeled cooperative work processes within the project phase are determined by the eigenmodes of the dynamical operator $A_0$. Following Smith and Eppinger (1997), we use the term "design mode" $\phi_i = \big(\lambda_i(A_0), \vartheta_i(A_0)\big)$ to refer to an eigenvalue $\lambda_i(A_0)$ inherent to $A_0$ associated to its eigenvector $\vartheta_i(A_0)$ $(1 \le i \le p)$. Strictly speaking, there is an infinitive number of eigenvectors associated to each eigenvalue of a dynamical operator. Because any scalar multiple of an eigenvector is still an eigenvector, an infinite family of eigenvectors exists for each eigenvalue. However, these vectors are all proportional to each other. In that sense, each design mode $\phi_i$ has both temporal (eigenvalue) and structure-organizational (scalar multiple of eigenvector) characteristics. Every dynamical operator $A_0$ has exactly $p$ eigenvalues, which are not necessarily distinct. Eigenvectors corresponding to distinct eigenvalues are linearly independent.

According to Luenberger (1979), if all solutions of the linear system from eq. 1 that start out near an equilibrium state $x_e$ of work remaining stay near or converge to $x_e$, the state is called stable or asymptotically stable. The origin $\bar{x} = 0$ is always a singular point of the vector field $x \longmapsto A_0 \cdot x$ on $\mathbb{R}^p$ and therefore an equilibrium point of the linear homogenous recurrence relation given by eq. (1). A linear homogeneous recurrence relation is internally stable if its dynamical operator is stable in the sense of Lyapunov (Hinrichsen and Pritchard 2005). A square real matrix is said to be asymptotically stable in the sense of Lyapunov if and only if for an operator $A_0$ and any positive semi-definite matrix $D$ there exists a positive-definite symmetric matrix $P$ satisfying the following Lyapunov criterion (see e.g. Halanay and Rasvan 2000, Hinrichsen and Pritchard 2005 or Siddiqi 2010):

$$P - A_0 \cdot P \cdot A_0^{\mathrm{T}} = D. \tag{3}$$

For the first-order linear autoregressive model that will be introduced in the next section, $A_0$ is the cited WTM, $P$ is the steady-state state covariance matrix (eq. 173) and $D$ is the covariance matrix of the random performance fluctuations (eq. 5). According to Siddiqi (2009), the Lyapunov criterion can be interpreted as satisfied for a linear autoregressive model if, for a given observation covariance, there exists a legitimate belief distribution in which the predicted belief over project state is equivalent to the previous belief over project state, that is, if there exists an equilibrium point of the distribution.

In the following, we follow the convention of listing the eigenvalues of the dynamical operator $A_0$ in order of decreasing magnitude ($|\lambda_1(A_0)| \geq |\lambda_2(A_0)| \geq \cdots$). For the matrix $A_0$ with these eigenvalues, we define the spectral radius as the greatest-magnitude eigenvalue and denote it by $\rho(A_0) = \max|\lambda_i|$. An eigenvalue corresponding to $\max|\lambda_i|$ (that is, $\lambda_1$) is called the dominant eigenvalue (Gentle 2007).

The Lyapunov criterion (eq. 3) holds for the linear homogenous recurrence relation given by eq. (1) if and only if the spectral radius is less than or equal to one, i.e. $\rho(A_0) \leq 1$ (Hinrichsen and Pritchard 2005). Please recall that a matrix $M$ is positive semidefinite if and only if it holds that $v^{\mathrm{T}} \cdot M \cdot v \geq 0$ for all non-zero column vectors $v$ of $m = \mathrm{Dim}[M]$ real numbers. Let $\lambda$ be a left eigenvalue of $A_0$ and $\vartheta_l$ a corresponding normalized eigenvector satisfying $\vartheta_l^{\mathrm{T}} \cdot A_0 = \{\lambda\} \cdot \vartheta_l^{\mathrm{T}}$, then the Lyapunov equation can be written as

$$\begin{aligned} v^{\mathrm{T}} \cdot D \cdot v &= v^{\mathrm{T}} \cdot \left(P - A_0 \cdot P \cdot A_0^{\mathrm{T}}\right) \cdot v \\ &= v^{\mathrm{T}} \cdot P \cdot v - v^{\mathrm{T}} \cdot \{\lambda\} \cdot P \cdot \{\lambda\} \cdot \vartheta \\ &= v^{\mathrm{T}} \cdot P \cdot v \cdot \{1 - |\lambda|^2\}. \end{aligned}$$

Since the matrix $P$ is a positive-definite symmetric matrix, it holds that $v^{\mathrm{T}} \cdot P \cdot v \geq 0$ and it follows that $|\lambda| \leq 1$ is equivalent to $v^{\mathrm{T}} \cdot M \cdot v \geq 0$ (Siddiqi 2009). Therefore, the Lyapunov criterion from eq. 3 is satisfied if $\rho(A_0) \leq 1$.

A modelled NPD project is said to be asymptotically stable, if and only if the spectral radius is less than 1: that is, $\rho(A_0) < 1$. In this case, irrespective of the initial state $x_0$ the work remaining converges to the zero vector, meaning that all tasks are fully completed. If it holds that $\rho(A_0) = 1$, the project is stable but not asymptotically stable, and the work remaining $x_t$ would eventually oscillate around $x_0$ indefinitely. Therefore, the notion of asymptotic stability is stronger than stability. For the first design mode $\phi_1$ with the dominant eigenvalue $\lambda_1$ the equation $|\lambda_1| = 1$ determines the bound of asymptotic stability of the project. If the project is neither stable nor asymptotically stable and $|\lambda_1(A_0)| > 1$, it is said to be unstable. If the project is unstable, a redesign of tasks and their interactions is necessary, because the work remaining then exceeds all given limits.

Unfortunately, even if the modeled project is asymptotically stable, theoretically an infinite number of iterations are necessary to reach the final state where zero work remains for all tasks. Therefore, project managers have to specify an additional stopping criterion. In the following we use a simple one-dimensional parametric criterion $\delta \in ]0;1[$ indicating that the goal is reached, if the work remaining is at most $100\delta$ percent for all $p$ tasks. According to Huberman and Wilkinson (2005), the zero vector represents a theoretical optimal solution, and the values of the state vector are an abstract measure of the amount of work left to be done before a task's solution is optimal.

## 2.2 Stochastic Formulation in Original State Space

In their seminal paper on performance variability and project dynamics, Huberman and Wilkinson (2005) showed how to model NPD projects based on stochastic processes theory, and how to apply formal methods of statistics to analyze, predict and evaluate the dynamics of open organizational systems. An open organizational system is a sociotechnical system in which humans continuously interact with each other and with their work environment. These interactions usually take the form of goal-directed information exchange within and through the system boundary and lead to a kind of self-organization, since patterns of coordination can emerge that convey new properties, such as oscillations or pace-setting. Furthermore, there is a regular supply of energy and matter from the environment. In the work presented here, we follow the basic ideas of Huberman and Wilkinson and formulate a stochastic model of cooperative task processing based on the theory of Gauss–Markov processes (Cover and Thomas 1991, Papoulis and Pillai 2002). However, we do not incorporate "multiplicative noise" to represent performance variability in NPD projects as Huberman and Wilkinson do, but rather assume that the effects of performance fluctuations on work remaining are cumulative. Clearly, there are subtle conceptual differences between the both approaches, but they are beyond the scope of this paper, which is to validate the model with field data from an industrial NPD project (section 2.5) and to analyze explicitly the interrelationships between projects dynamics and emergent complexity (section 5).

Our model generalizes the first-order difference equation (eq. 1) according to Smith and Eppinger (1997) to a deterministic random process $\{X_t\}$ (Puri 2010) with the state equation

$$X_t = A_0 \cdot X_{t-1} + \varepsilon_t \qquad t = 1, \dots, T. \tag{4}$$

In this first-order linear autoregressive model, the multivariate random variable $X_t$ represents the measured (or estimated) work remaining at time step $t$ of the project phase under consideration. $A_0$ is the cited WTM. The random vector $\varepsilon_t$ is used to model unpredictable performance fluctuations (noise).

In NPD projects there are many performance-shaping factors. Although we do not know their exact number or distribution, the central limit theorem tells us that, to a large degree, the sum of independently and identically distributed factors can be represented by a Gaussian distribution $\mathcal{N}(x; \mu, C)$ with location $\mu = E[\varepsilon_t]$ and covariance $C = E[(\varepsilon_t - \mu)(\varepsilon_t - \mu)^{\mathrm{T}}]$. The location is also often simply termed "mean". We assume that the performance fluctuations are independent of the work remaining and therefore that the location and covariance do not depend on the time index. Hence, we can also write $\varepsilon_\diamond$ in place of $\varepsilon_t$ in the following definitions of the entries of the covariance matrix.

The covariance matrix $C$ is a square matrix of size $p$, whose entry $C_{[[i,j]]}$ in the $i, j$ position is the covariance between the $i$-th element $\varepsilon_\diamond^{(i)}$ and the $j$-th element $\varepsilon_\diamond^{(j)}$ of the random vector $\varepsilon_\diamond$, i.e.

$$\begin{aligned} C_{[[i,j]]} &= \mathrm{Cov}\left[\varepsilon_\diamond^{(i)}, \varepsilon_\diamond^{(j)}\right] \\ &= E[\left(\varepsilon_\diamond^{(i)} - \mu^{(i)}\right)\left(\varepsilon_\diamond^{(j)} - \mu^{(j)}\right)]. \end{aligned} \tag{5}$$

$C$ is symmetric by definition and also positive-semidefinite (Lancaster and Tismenetsky 1985). We assume that $C$ has full rank. The diagonal elements $C_{[[i,i]]}$ represent the scalar-valued variances $c_{ii}^2$ of vector components $\varepsilon_\diamond^{(i)}$ (i.e. performance fluctuations in work tasks $i$):

$$\begin{aligned} c_{ii}^2 &= \mathrm{Var}\left[\varepsilon_\diamond^{(i)}\right] \\ &= E[\left(\varepsilon_\diamond^{(i)} - \mu^{(i)}\right)^2]. \end{aligned} \tag{6}$$

The square root of the scalar-valued variance $c_{ii}^2$ is the well-known standard deviation $c_{ii}$. The off-diagonal elements $C_{[[i,j]]}$ $(i \neq j)$ represent the scalar-valued covariances and can be factorized as

$$\rho_{ij} c_{ii} c_{jj} = \rho_{ij} \sqrt{\mathrm{Var}\left[\varepsilon_\diamond^{(i)}\right] \mathrm{Var}\left[\varepsilon_\diamond^{(j)}\right]} \qquad (i \neq j), \tag{7}$$

where the first factor is Pearson´s famous product-moment coefficient

$$\rho_{ij} := \mathrm{Corr}\left[\varepsilon_\diamond^{(i)}, \varepsilon_\diamond^{(j)}\right] = \frac{\mathrm{Cov}\left[\varepsilon_\diamond^{(i)}, \varepsilon_\diamond^{(j)}\right]}{\sqrt{\mathrm{Var}\left[\varepsilon_\diamond^{(i)}\right] \mathrm{Var}\left[\varepsilon_\diamond^{(j)}\right]}}. \tag{8}$$

The Pearson correlation $\rho_{ij}$ is $+1$ in the case of a perfect positive linear relationship (correlation) and $-1$ in the case of a perfect negative linear relationship (anticorrelation). It has values between $-1$ and 1 in all other cases, indicating the degree of linear dependence between the variables.

In the developed autoregression model of cooperative work in NPD projects it is assumed that the performance fluctuations have no systematic component and that $\mu = 0_p = (\,0 \;\; 0 \;\; \ldots \;\; 0\,)^{\mathrm{T}}$. We imposed no additional a priori constraints on the covariance matrix $C$. Hence, the noise term

$$\varepsilon_t \sim \mathcal{N}\left(0_p, C\right)$$

in the state equation can be expressed explicitly by its Gaussian probability density function $f[x] = \mathcal{N}\left(x; 0_p, C\right)$ (*pdf*, see e.g. Puri 2010) as

$$\mathcal{N}\left(x; 0_p, C\right) = \frac{1}{(2\pi)^{p/2} (\mathrm{Det}[C])^{1/2}} \mathrm{Exp}\left[-\frac{1}{2} x^{\mathrm{T}} \cdot C^{-1} \cdot x\right]. \tag{9}$$

The covariance matrix $C$ can be written in vector form as

$$C = \mathrm{Cov}[\varepsilon_t, \varepsilon_t] = E[\varepsilon_t \varepsilon_t^{\mathrm{T}}].$$

We assume that the performance fluctuations are uncorrelated from time step to time step and that it holds for all time steps $\{\mu, \nu\} \in \mathbb{Z}$ that

$$E\left[\varepsilon_\mu \varepsilon_\nu^{\mathrm{T}}\right] = \{\delta_{\mu\nu}\} \cdot C.$$

$\delta_{\mu\nu}$ is the Kronecker delta which is defined as

$$\delta_{\mu\nu} = \begin{cases} 1 & \mu = \nu \\ 0 & \mu \neq \mu \end{cases}. \tag{10}$$

If the covariance matrix is a nonzero scalar multiple of the identity matrix $I_p$, that is $C = \{\sigma^2\} \cdot I_p$, we speak of isotropic noise, and the variance $\sigma^2$ represents the overall noise strength ($\sigma^2 \in \mathbb{R}^+$). In spite of the stochastic task processing, it is assumed in the following that perfect initial conditions exist, and that the components of the initial state vector according to state eq. 4 are positive real numbers and not random variables. This assumption is justified by the fact that in most projects the initial state $x_0$ represents the planned amount of work at the beginning of a given project phase (cf. eq. 2), which is predefined by the project manager. In this case a real valued parameter vector $\theta = [x_0 \quad A_0 \quad C]$ is sufficient to parameterize the model. Alternatively, the initial state vector $X_0$ can be assumed to be a Gaussian random vector with location $\mu_0$ and covariance $C_0$. In section 2.8 we will present a stochastic model formulation with hidden state variables that can cover this case under a more general theoretical framework. When this alternative formulation is used, the parameter vector must be extended, and becomes $\theta = [\mu_0 \quad C_0 \quad A_0 \quad C]$. A graphical representation of the first-order autoregression model is shown in Figure 1 in the form of a dynamic Bayesian network (see e.g. Pearl 1988). In a dynamic Bayesian network the random state variables are related to each other over adjacent time steps and are drawn as nodes of the graph. At any point in time *t*, the value of a state variable can be calculated from the internal regressors and the immediate prior value (time step *t*-1). The directed arcs represent conditional dependencies between the variables. Exogenous inputs to the model are not considered in the following.

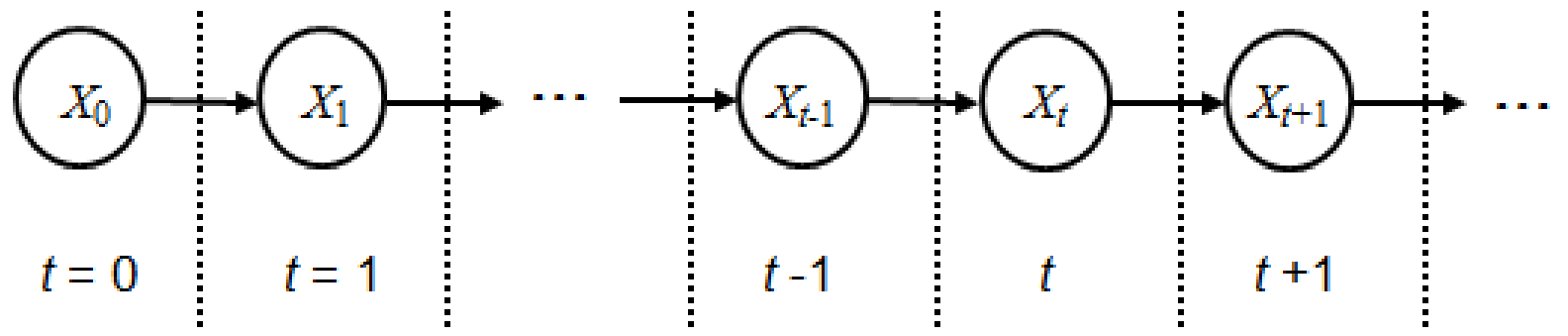

**Figure 1**

*Figure 1*. Graphical representation of the first-order autoregression model in the form of a dynamic Bayesian network. The nodes in the graph represent the random state variables of the stochastic process. The directed arcs encode conditional dependencies between the variables.

It is not difficult to see that the process $\{X_t\}$ can be decomposed into a deterministic and stochastic part as

$$X_t = A_0^t \cdot x_0 + \sum_{v=1}^{t} A_0^{t-v} \cdot \varepsilon_v \quad (t \geq 1).$$

The deterministic part represents the mean vectors

$$\begin{aligned} E[X_t, \mathcal{N}(x; 0_p, C)] &= E[X_t] \\ &= A_0^t \cdot x_0 \end{aligned}$$

of work remaining, which evolve unperturbed. For an arbitrary project with predefined initial state $x_0$ and WTM $A_0$, a closed-form solution to the sum of the mean work remaining can be calculated across a time interval $T$ as

$$\begin{aligned} E\left[\sum_{t=0}^{T} X_t\right] &= \sum_{t=0}^{T} E[X_t] \\ &= \sum_{t=0}^{T} (A_0^t \cdot x_0) \end{aligned}$$

$$= \left(\sum_{t=0}^{T} A_0^t\right) \cdot x_0$$

$$= \left(I_p - A_0\right)^{-1} \cdot \left(I_p - A_0^T\right) \cdot x_0 \,. \tag{11}$$

The above solution is based on the Neumann series generated by $A_0$; $I_p$ denotes the $p \times p$ identity matrix. In the limit $T \to \infty$ we have for an asymptotically stable project the expected cumulated work remaining:

$$\lim_{T\to\infty} E\left[\sum_{t=0}^{T} X_t\right] = \left(I_p - A_0\right)^{-1} \cdot x_0 \,.$$

The expected total work $x_{tot}$ in the project, as an indicator of the total effort involved in completing the deliverables can be estimated by summing over the vector components of the expected cumulated work remaining:

$$x_{tot} = \text{Total}\left[\left(I_p - A_0\right)^{-1} \cdot x_0\right]. \tag{12}$$

The function Total[…] computes the sum of the components of the argument vector. By formulating the problem as a recurrence relation it is possible to to consider rework easily and to accurately estimate the total effort involved. The effort-centered approach can also cope with deliverables that do not meet all the original requirements or that have a quality problem and therefore require rework. This is not to be confused with changes of scope, where separate state variables must be defined and a dedicated scope change management system should be utilized.

In addition to the deterministic evolution of the mean work remaining the stochastic part of the process $\{X_t\}$ represents the accumulated unpredictable performance fluctuations. The formulation of the linear model means that the variances and covariances of the vector components of the fluctuations are independent of the work remaining. In view of an information processing system, the process $\{X_t\}$ satisfies the Markov property. The Markov property describes a special kind of "memorylessness" in the sense that conditional on the present state $x_t$ of the modeled project, its future $\{X_{t+1}, \ldots\}$ and past $\{X_1, \ldots, X_{t-1}\}$ are rendered independent:

$$f_\theta[x_{t+1}|x_0, \ldots, x_t] = f_\theta[x_{t+1}|x_t] \quad \forall t \geq 0. \tag{13}$$

$f_\theta[x_{t+1}|x_0, \ldots, x_t]$ denotes the conditional probability density function of vector $X_{t+1}$, given the sequence of vectors $X_0, \ldots, X_t$ (Papoulis and Pillai 2002).

According to the decomposition of the process into a deterministic and stochastic part it is obvious that the probability density function of the project state $X_t$ at time step $t$ is Gaussian with location $A_0^t x_0$ and covariance $\sum_{v=1}^{t} A_0^{t-v}\, C\left(A_0^{\mathrm{T}}\right)^{t-v}$, that is

$$X_t \sim \mathcal{N}\left(A_0^t \cdot x_0, \sum_{v=1}^{t} A_0^{t-v} \cdot\, C \cdot \left(A_0^{\mathrm{T}}\right)^{t-v}\right).$$

The density function $f_\theta[x_t]$ of state $X_t$ can be written explicitly as (Puri 2010)

$$f_\theta[x_t] = \frac{1}{(2\pi)^{p/2}(\text{Det}[\Sigma_t])^{1/2}} \text{Exp}\left[-\frac{1}{2}(x_t - A_0^t \cdot x_0)^{\mathrm{T}} \cdot \Sigma_t^{-1} \cdot (x_t - A_0^t \cdot x_0)\right], \tag{14}$$

where

$$\Sigma_t = \sum_{v=1}^{t} A_0^{t-v} \cdot C \cdot \left(A_0^{\mathrm{T}}\right)^{t-v}.$$

The conditional density of state $X_{t+1}$ given state $X_t = x_t$ (eq. 13) is

$$f_\theta[x_{t+1}|x_t] = \frac{1}{(2\pi)^{p/2}(\mathrm{Det}[C])^{1/2}} \mathrm{Exp}\left[-\frac{1}{2}(x_{t+1}-A_0 \cdot x_t)^{\mathrm{T}} C^{-1}(x_{t+1}-A_0 \cdot x_t)\right]. \tag{15}$$

At first glance, the chosen memoryless perturbation mechanism may appear to over-simplify the problem. However, the correlations $\rho_{ij}$ between performance fluctuations between tasks $i$ and $j$ can strongly influence the course of the project not only at single time steps but also on long time scales and therefore lead to unexpected stateful behavior. This is the case if the correlations are reinforced through the informational coupling between the development tasks. To reinforce the correlations, the covariance matrix $C$ must have nonzero off-diagonal elements: in other words, the noise must be nonisotropic. Depending on the structure of the dynamical operator $A_0$, the correlations $\rho_{ij}$ can significantly excite the design modes and lead to unexpected effects of emergent complexity, such as the cited problem-solving oscillations in the preasymptotic range of development projects (Mihm and Loch 2006, Schlick et al. 2008). We will return to the interesting phenomenon of excitation of design modes in section 4.2, where the interrelationships between project dynamics and emergent complexity are analyzed in detail in the spectral basis.

Following the theoretical considerations of system stability from the previous section, the first-order linear autoregressive model defined in eq. 4 is asymptotically stable in the sense of Lyapunov (eq. 3) if and only if the spectral radius of the dynamical operator $A_0$ is strictly less than one, i.e. $\rho(A_0) < 1$, and the matrix in eq. 3 is positive definite. In contrast to the deterministic model formulation, an autoregression model with Gaussian performance fluctuations without drift and unit spectral radius $\rho(A_0) = 1$ would steadily move away from the equilibrium state $x_e$ and therefore not be stable (Papoulis and Pillai 2002, Siddiqi 2010). If $\rho(A_0) = 1$, the autoregressive process is said to be marginally stable (Halanay and Rasvan 2000).

In extension of state eq. 4, we can formulate a dynamic project model on the basis of a forcing matrix $K$ in conjunction with a noise variable $\eta_t$ whose covariance matrix does not indicate correlations among vector components (i.e. work tasks) and is therefore diagonal. To do so, the covariance matrix $C$ is decomposed into eigenvectors and eigenvalues through an eigendecomposition:

$$C = K \cdot \Lambda_K \cdot K^{-1}, \tag{16}$$

where

$$K \cdot K^{\mathrm{T}} = I_p \text{ and } K^{-1} = K^{\mathrm{T}}.$$

Because the covariance matrix $C$ is symmetric by definition, the forcing matrix $K$ resulting from the eigendecomposition has mutually orthogonal column vectors $k_i(C) = K_{:i}$ and is therefore orthogonal. These vectors are the eigenvectors of $C$. $\Lambda_K$ is simply a diagonal matrix with the eigenvalues $\lambda_i(C)$ along the principal diagonal. The associated state equation is

$$X_t = A_0 \cdot X_{t-1} + K \cdot \eta_t, \tag{17}$$

with

$$\eta_t \sim \mathcal{N}(0_p, \Lambda_K) \tag{18}$$

and

$$\Lambda_K = \mathrm{Diag}[\lambda_i(C)] \quad 1 \le i \le p. \tag{19}$$

According to the above equation, the eigenvalues $\lambda_i(C)$ of the decomposed covariance matrix $C$ can be interpreted as the variances of the performance fluctuations along the rotated axes of the identified eigenvectors $k_i(C)$. Following our terminology we will use the term “fluctuation mode”, $\Psi_i = \big(\lambda_i(C), k_i(C)\big)$, to refer to an eigenvalue $\lambda_i(C)$ of $C$ along with its eigenvector $k_i(C)$ $(1 \le i \le p)$.

The introduced stochastic models of cooperative task processing are quite closely related to the dynamical model of product development on complex directed networks that was introduced by Braha and Bar-Yam (2007). However, there are some important differences: 1) the autoregression models are defined over a continuous range of state values and can therefore represent different kinds of cooperation relationships as well as precedence relations (e.g. overlapping); 2) each task is nonequally influenced by other tasks; 3) correlations $\rho_{ij}$ between performance fluctuations among tasks i and j can be captured.

## 2.3 Stochastic Formulation in Spectral Basis

In order to analyze explicitly the intricate interrelationships between project dynamics and emergent complexity (section 3 and 4), we work in the spectral basis (Neumaier and Schneider 2001). To carry out the transformation of the state-space coordinates, the WTM $A_0$ as dynamical operator is diagonalized through an eigendecomposition (cf. eq. 16) as

$$A_0 = S \cdot \Lambda_S \cdot S^{-1}, \tag{20}$$

with

$$\Lambda_S = \mathrm{Diag}[\lambda_i(A_0)] \qquad 1 \le i \le p. \tag{21}$$

The eigenvectors $\vartheta_i(A_0) = S_{:i}$ of the design modes $\phi_i$ of $A_0$ are the column vectors of $S$ $(i = 1 \dots p)$. However, because $A_0$ must not be symmetric, the eigenvectors are in general not mutually orthogonal, and their elements can be complex numbers. The diagonal matrix $\Lambda_s$ stores the ordered eigenvalues $\lambda_i(A_0)$ along the principal diagonal.

It is very easy to analyze the stability of the modelled NPD project in the spectral basis. According to the previous section, the autoregressive model defined in eq. 4 is asymptotically stable in the sense of Lyapunov (eq. 3) if and only if it holds for the spectral radius of the dynamical operator $A_0$ that $\rho(A_0) < 1$. Based on eq. 20 we can conclude that the limit

$$\begin{aligned}\lim_{k\to\infty} A_0^k &= \lim_{k\to\infty} S \cdot \Lambda_S^k \cdot S^{-1} \\ &= S \cdot \left(\lim_{k\to\infty} \Lambda_S^k\right) \cdot S^{-1}\end{aligned}$$

is a null matrix, since it is obvious that the entries of $\Lambda_S^k$ along the principal diagonal are just the eigenvalues raised to power $k$, which converge to zero when $\rho(A_0) < 1$. Hence, we have

$$\lim_{k\to\infty} \Lambda_S^k = \begin{pmatrix} \lambda_1 & 0 & \dots \\ 0 & \lambda_2 & \dots \\ \vdots & \vdots & \ddots \end{pmatrix} = 0.$$

If $\Lambda_S = I_p$, the project is said to be marginally stable but not asymptotically stable, because the work remaining would steadily move away from the equilibrium state $x_e$.

In the spectral basis, the dynamic model from eq. 4 can be represented by the state vector $X_t$ and the vector $\varepsilon_t$ of unpredictable performance fluctuations as simple linear combinations, as follows:

$$\begin{aligned} X_t &= S \cdot X'_t \\ &= \sum_{i=1}^{p} X_t^{(i)} \cdot \vartheta_i(A_0) \end{aligned} \tag{22}$$

and

$$\begin{aligned} \varepsilon_t &= S \cdot \varepsilon'_t \\ &= \sum_{i=1}^{p} \varepsilon_t^{(i)} \cdot \vartheta_i(A_0), \end{aligned} \tag{23}$$

with coefficient vectors

$$X'_t = \begin{pmatrix} X_t'^{(1)} \\ \vdots \\ X_t'^{(p)} \end{pmatrix}$$

and

$$\varepsilon'_t = \begin{pmatrix} \varepsilon_t'^{(1)} \\ \vdots \\ \varepsilon_t'^{(p)} \end{pmatrix}.$$

For the initial state,

$$x_0 = S \cdot x'_0.$$

We obtain the transformed stochastic process $\{X'_t\}$ that is generated by the coefficient vectors on the basis of the state equation

$$X'_t = \Lambda_S \cdot X'_{t-1} + \varepsilon'_t \qquad t = 1, \dots, T, \tag{24}$$

with

$$\varepsilon'_t \sim \mathcal{N}(0_p, C') \tag{25}$$

and

$$C' = S^{-1} \cdot C \cdot ([S^{\mathrm{T}}]^*)^{-1}. \tag{26}$$

The transformed covariance matrix $C' = E\left[\varepsilon'_t \left[\varepsilon_t'^{\mathrm{T}}\right]^*\right]$ is also positive-semidefinite.

When we substitute the eigendecomposition of $C$ according to eq. 16 in eq. 26 we have

$$\begin{aligned} C' &= S^{-1} \cdot K \cdot \Lambda_K \cdot K^{-1} \cdot \left([S^{\mathrm{T}}]^*\right)^{-1} \\ &= S^{-1} \cdot K \cdot \Lambda_K \cdot \left([S^{\mathrm{T}}]^* \cdot K\right)^{-1} \\ &= S^{-1} \cdot \left(K^{\mathrm{T}}\right)^{\mathrm{T}} \cdot \Lambda_K \cdot \left([S^{\mathrm{T}}]^* \cdot K\right)^{-1} \\ &= S^{-1} \cdot \left(K^{\mathrm{T}}\right)^{-1} \cdot \Lambda_K \cdot \left([S^{\mathrm{T}}]^* \cdot K\right)^{-1} \end{aligned}$$

$$= (K^{\mathrm{T}} \cdot S)^{-1} \cdot \Lambda_K \cdot ([S^{\mathrm{T}}]^* \cdot K)^{-1}. \tag{27}$$

Let $(K^{\mathrm{T}} \cdot S)^{-1} = (d_{ij})$ and $([S^{\mathrm{T}}]^* \cdot K)^{-1} = \left(e_{ij}\right),\ 1 \le i, j \le p$. On the basis of the matrix elements, we can derive simple formulas for the diagonal and off-diagonal elements of $C'$, which are needed in section 3 to calculate the EMC in an expressive closed form (see eq. 190 in conjunction with eqs. 188 and 189). The diagonal elements are

$$C'_{[[i,i]]} = \sum_{n=1}^{p} d_{in} \lambda_n(C) e_{ni}$$

and the off-diagonal elements are

$$C'_{[[i,j]]} = \sum_{n=1}^{p} d_{in} \lambda_n(C) e_{nj}.$$

Hence, the correlations $\rho'_{ij}$ in the spectral basis are

$$\begin{aligned} \rho'_{ij} &:= \frac{C'_{[[i,j]]}}{\sqrt{C'_{[[i,i]]} C'_{[[j,j]]}}} \\ &= \frac{c'^{2}_{ii}}{c'_{ii} c'_{jj}} \\ &= \frac{\sum_{n=1}^{p} d_{in} \lambda_n(C) e_{nj}}{\sqrt{\left(\sum_{n=1}^{p} d_{in} \lambda_n(C) e_{ni}\right)\left(\sum_{n=1}^{p} d_{jn} \lambda_n(C) e_{nj}\right)}}. \end{aligned} \tag{28}$$

Interestingly, the correlations $\rho'_{ij}$ can be interpreted in a geometrical framework (de Cock 2002). Let $\epsilon'^{(i)}_t$ be the row vector of the normalized (by $1/\sqrt{t}$) sequence of samples that were drawn from the *i*-th vector component of the multivariate distribution of the random variable $\varepsilon'_t$ representing the performance fluctuations in the spectral basis, that is

$$\epsilon'^{(i)}_t := \frac{1}{\sqrt{t}} \left(\varepsilon'^{(i)}_0,\ \dots, \varepsilon'^{(i)}_{t-1}\right).$$

The correlations $\rho'_{ij}$ between the *i*-th and *j*-th components of $\varepsilon'_t$ are defined as the angle between the vectors $\epsilon'^{(i)}_t$ and $\epsilon'^{(j)}_t$ for $t \to \infty$:

$$\rho'_{ij} = \lim_{t \to \infty} \cos\left(\epsilon'^{(i)}_t \angle\, \epsilon'^{(j)}_t\right). \tag{29}$$

The independence of the *i*-th and *j*-th components of $\varepsilon'_t$ implies that $\rho'_{ij} = 0$, and hence that the vectors of the normalized sequence of samples $\epsilon'^{(i)}_t$ and $\epsilon'^{(j)}_t$ are orthogonal for $t \to \infty$. In an analogous way, let $x'^{(i)}_t$ be the row vector of the normalized sequence of samples that were drawn from the *i*-th vector component of the multivariate distribution of the transformed state variable $X'_t$, that is

$$x'^{(i)}_t := \frac{1}{\sqrt{t}} \left(x'^{(i)}_0,\ \dots, x'^{(i)}_{t-1}\right).$$

The correlations between the $i$-th and $j$-th components of $X'_t$ in the steady state ($t \to \infty$) of the stochastic process are the reinforced correlations $\rho'_{ij}$ between noise components. The reinforcement factor is $1/\left(1-\lambda_i(A_0)\overline{\lambda_j(A_0)}\right)$, and there holds (Neumaier and Schneider 2001)

$$\frac{1}{1-\lambda_i(A_0)\overline{\lambda_j(A_0)}}\rho'_{ij} = \lim_{t\to\infty} \cos\left(x_t'^{(i)} \angle\, x_t'^{(j)}\right). \tag{30}$$

In the above equation the terms $\overline{\lambda_j(A_0)}$ denote the complex conjugates of the eigenvalues. As mentioned earlier, this interesting reinforcement phenomenon will be discussed again in section 4.

If $A_0$ is a symmetric matrix, we can also obtain expressive vector calculus forms of both the diagonal (representing the variances along the rotated coordinate axes in the spectral basis) and off-diagonal (representing the correlations $\rho'_{ij}$ among them) elements of $C'$, as shown in the following steps. This is because the eigenvectors $\vartheta_i(A_0)$ are mutually orthogonal and have only real components. However, if $A_0$ is not symmetric, the eigenvectors are not orthogonal, and the following simplifications are impossible.

If $A_0$ is symmetric, all $p$ components of the eigenvectors are real, and eq. 26 can be rewritten as

$$C' = S^{-1} \cdot C \cdot (S^{\mathrm{T}})^{-1}.$$

The eigenvectors $\vartheta_i(A_0) = S_{:i}$ are mutually orthogonal. The normalized eigenvectors are denoted by $\tilde{\vartheta}_i$ ($\|\tilde{\vartheta}_i\| = 1$), the corresponding orthonormal matrix by $S_\perp$, and the mutually orthogonal (but not normalized) column vectors of the forcing matrix $K$ by $k_i = K_{:i}$. We can write

$$C' = (S_\perp \cdot N)^{-1} \cdot C \cdot \left((S_\perp \cdot N)^{\mathrm{T}}\right)^{-1}$$

with

$$N = \mathrm{Diag}[||\vartheta_i(A_0)||]\,,\ 1 \le i \le p.$$

Because $N$ is diagonal, the transpose can be written as

$$C' = (S_\perp \cdot N)^{-1} \cdot C \cdot (N \cdot S_\perp^{\mathrm{T}})^{-1},$$

and the inverse can be factorized:

$$C' = N^{-1} \cdot S_\perp^{-1} \cdot C \cdot (S_\perp^{\mathrm{T}})^{-1} \cdot N^{-1}.$$

For orthonormal matrices, $S_\perp^{-1} = S_\perp^{\mathrm{T}}$, and we have

$$\begin{aligned} C' &= N^{-1} \cdot S_\perp^{\mathrm{T}} \cdot C \cdot S_\perp \cdot N^{-1} \\ &= N^{-1} \cdot S_\perp^{\mathrm{T}} \cdot K \cdot \Lambda_K \cdot K^{\mathrm{T}} \cdot S_\perp \cdot N^{-1} \\ &= N^{-1} \cdot (S_\perp^{\mathrm{T}} \cdot K) \cdot \Lambda_K \cdot (S_\perp^{\mathrm{T}} \cdot K)^{\mathrm{T}} \cdot N^{-1} \end{aligned} \tag{31}$$

with

$$S_\perp^{\mathrm{T}} \cdot K = \begin{pmatrix} \tilde{\vartheta}_1 \cdot k_1 & \tilde{\vartheta}_1 \cdot k_2 & \dots \\ \tilde{\vartheta}_2 \cdot k_1 & \tilde{\vartheta}_2 \cdot k_2 & \dots \\ \dots & \dots & \ddots \end{pmatrix}$$

$$(S_\perp^{\mathrm{T}} \cdot K)^{\mathrm{T}} = \begin{pmatrix} \tilde{\vartheta}_1 \cdot k_1 & \tilde{\vartheta}_2 \cdot k_1 & \dots \\ \tilde{\vartheta}_1 \cdot k_2 & \tilde{\vartheta}_2 \cdot k_2 & \dots \\ \dots & \dots & \ddots \end{pmatrix}$$

$$
(S_\perp^{\mathrm{T}} \cdot K) \cdot \Lambda_K = \begin{pmatrix} \tilde{\vartheta}_1 \cdot k_1\{\lambda_1(C)\} & \tilde{\vartheta}_1 \cdot k_2\{\lambda_2(C)\} & \dots \\ \tilde{\vartheta}_2 \cdot k_1\{\lambda_1(C)\} & \tilde{\vartheta}_2 \cdot k_2\{\lambda_2(C)\} & \dots \\ \dots & \dots & \ddots \end{pmatrix}.
$$

Now we can formulate a simple geometric relationship for the diagonal elements of $C'$

$$
C'_{[[i,i]]} = \frac{1}{||\vartheta_i||^2} \sum_{n=1}^{p} \lambda_n(C)\left(\tilde{\vartheta}_i \cdot k_n\right)^2,
$$

as well as for the off-diagonal elements

$$
C'_{[[i,j]]} = \frac{1}{||\vartheta_i||\,||\vartheta_j||} \sum_{n=1}^{p} \lambda_n(C)\left(\tilde{\vartheta}_i \cdot k_n\right)\left(\tilde{\vartheta}_j \cdot k_n\right).
$$

The closed-form solution for the correlations $\rho'_{ij}$ embedded in $C'$ is:

$$
\begin{aligned}
\rho'_{ij} &= \frac{||\vartheta_i||\,||\vartheta_j||}{||\vartheta_i||\,||\vartheta_j||} \cdot \frac{\sum_{n=1}^{p} \lambda_n(C)\left(\tilde{\vartheta}_i \cdot k_n\right)\left(\tilde{\vartheta}_j \cdot k_n\right)}{\sqrt{\left(\sum_{n=1}^{p} \lambda_n(C)\left(\tilde{\vartheta}_i \cdot k_n\right)^2\right)\left(\sum_{n=1}^{p} \lambda_n(C)\left(\tilde{\vartheta}_j \cdot k_n\right)^2\right)}} \\
&= \frac{\sum_{n=1}^{p} \lambda_n(C)\left(\tilde{\vartheta}_i \cdot k_n\right)\left(\tilde{\vartheta}_j \cdot k_n\right)}{\sqrt{\left(\sum_{n=1}^{p} \lambda_n(C)\left(\tilde{\vartheta}_i \cdot k_n\right)^2\right)\left(\sum_{n=1}^{p} \lambda_n(C)\left(\tilde{\vartheta}_j \cdot k_n\right)^2\right)}}.
\end{aligned} \tag{32}
$$

When we analyze eqs. 31 and 32, it is not difficult to see that the correlations $\rho'_{ij}$ in the spectral basis are zero, in either of the following cases.

i. The column vectors of the forcing matrix $K$ and the column vectors of the transformation matrix $S$ are pairwise collinear, and $S = K \cdot \Lambda_{SK}$ ($\Lambda_{SK} = \mathrm{Diag}[c_i], c_i \in \mathbb{R}$).
ii. The forcing matrix $K$ is equal to the identity matrix $I_p$, and the noise is isotropic with overall strength $\sigma^2 \in \mathbb{R}^+$, that is

$$
\Lambda_K = \mathrm{Diag}[\lambda(C)] = \{\sigma^2\} \cdot I_p.
$$

In case (i) the off-diagonal elements $C'_{[[i,j]]}$ are zero, because $\tilde{\vartheta}_i \cdot k_{n'} \neq 0$ for only one column vector $k_{n'}$ with index $n'$ that is aligned with $\tilde{\vartheta}_{n'}$, while for this index $\tilde{\vartheta}_j \cdot k_{n'} = 0$ for all $j \neq i$, because the vectors are mutually orthogonal (keeping in mind that $A_0$ is supposed to be symmetric). Therefore,

$$
\sum_{n=1}^{p} \lambda_n(C)\left(\tilde{\vartheta}_i \cdot k_n\right)(\tilde{\vartheta}_j \cdot k_n) = 0,
$$

from which $C'_{[[i,j]]} = 0$ and $\rho'_{ij} = 0$ follow.

In case (ii) we can substitute $K = I_p$, as well as $\Lambda_K = \{\sigma^2\} \cdot I_p$ in eq. 31:

$$
\begin{aligned}
C' &= N^{-1} \cdot \left(S_\perp^{\mathrm{T}} \cdot I_p\right) \cdot \{\sigma^2\} \cdot I_p \cdot \left(S_\perp^{\mathrm{T}} \cdot I_p\right)^{\mathrm{T}} \cdot N^{-1} \\
&= \{\sigma^2\} \cdot N^{-1} \cdot S_\perp^{\mathrm{T}} \cdot S_\perp \cdot N^{-1} \\
&= \{\sigma^2\} \cdot (N^{-1})^2.
\end{aligned}
$$

Since $N$ is a diagonal matrix, again ${C'}_{[[i,j]]} = 0$ and $\rho'_{ij} = 0$ follow.

Interestingly, under the assumption of isotropic noise, that is $C = \{\sigma^2\} \cdot I_p$ or $K = I_p$ and $\Lambda_K = \{\sigma^2\} \cdot I_p$, only the dynamic and not the pure noise part of emergent complexity are relevant (see eqs. 178 and 179). This also holds for a dynamical operator $A_0$ that is not symmetric.

We can also write the state equation (24) in the spectral basis component-wise (Neumaier and Schneider 2001). For each vector component the eigenvalue $\lambda_i = \lambda_i(A_0)$ is the dynamical operator of the scalar state equation with coefficient $X_t'^{(i)}$:

$$X_t'^{(i)} = \lambda_i X_{t-1}'^{(i)} + \varepsilon_t'^{(i)} \qquad t = 1, \dots, T. \tag{33}$$

It is important to note that in the scalar state equations the noise is correlated and that for the expectations it holds for all time steps $\mu \in \mathbb{Z}$ that

$$\begin{aligned} E\left[\varepsilon_\mu'^{(i)} \left[\varepsilon_t'^{(j)}\right]^*\right] &= \delta_{\mu t} C'_{[[i,j]]} \\ &= \delta_{\mu t} \rho'_{ij} C'_{[[i,i]]} C'_{[[j,j]]}, \end{aligned} \tag{34}$$

where $\rho'_{ij}$ $(k, l = 1, \dots, p)$ is the correlation coefficient defined in eq. 28 and $\delta_{\mu t}$ is the Kronecker delta (eq. 10). From eq. 33 it follows that for $\mathrm{Arg}[\lambda_i] \neq 0$ the expectations

$$E\left[X_t'^{(i)}\right] = \lambda_i E\left[X_{t-1}'^{(i)}\right]$$

describe spirals in the Gaussian plane represented by

$$\begin{aligned} E\left[X_{t+\tau}'^{(i)}\right] &= \lambda_i^\tau E\left[X_t'^{(i)}\right] \\ &= e^{-\frac{\tau}{\tau_i}} e^{(\mathrm{Arg}[\lambda_i]) i\tau} E\left[X_t'^{(i)}\right] \end{aligned} \tag{35}$$

with damping time scales (Neumaier and Schneider 2001)

$$\tau_i := -\frac{1}{\log|\lambda_i|} \tag{36}$$

and periods

$$T_i := \frac{2\pi}{|\mathrm{Arg}[\lambda_i]|}. \tag{37}$$

The function $\mathrm{Arg}[\lambda_\mathrm{i}]$ denotes the argument of the possibly complex eigenvalue $\lambda_i = a_i + ib_i$, which can be computed as

$$\mathrm{Arg}[\lambda_i] = \tan^{-1}\left(\frac{a_i}{b_i}\right).$$

We use the convention that $-\pi \leq \mathrm{Arg}[\lambda_i] \leq \pi$ to ensure that a pair of complex conjugate eigenvalues of the dynamical operator is associated with a single period of the described spiral. For a stable stochastic process $\{X_t\}$ all eigenvalues must be less than one in magnitude and therefore the damping time scale $\tau_i$ is positive and bounded. If the eigenvalue $\lambda_i$ of interest has a nonzero imaginary part or is real but negative, the period $T_i$ is also bounded. In this case we can consider the scalar stochastic process $\left\{X_t'^{(i)}\right\}$ as a stochastically driven damped oscillator (Neumaier and Schneider 2001). The period of the damped oscillator is minimal if $\lambda_i$ is real and negative and we have $T_i = 2$ (or $|\mathrm{Arg}[\lambda_i]| = \pi$). This period is equivalent to the famous Nyquist frequency which plays an important role in Fourier analysis and system theory (see e.g. Puri 2010). In contrast, if the eigenvalue $\lambda_i$ is real and positive, then the period $T_i$ grows over all limits ($T_i \to \infty$) and the scalar-valued system spirals indefinitely around the expectation value. In this case, the scalar stochastic process $\left\{X_t'^{(i)}\right\}$ can be regarded as a stochastically driven relaxator (Neumaier and Schneider 2001). Therefore, the linear combinations according to eqs. 22 and 23 decompose the vector autoregression process $\{X_t\}$ generated by state eq. 4 into linear combinations of damped oscillators and relaxators with oscillation and relaxation modes $\vartheta_i(A_0)$ that operate on damping time scales $\tau_i$ and have periods $T_i$.

Since the work transformation matrix $A_0$ is real, its eigenvalues and eigenvectors can be written as complex conjugate pairs. Based on these pairs we can also decompose the introduced linear autoregression model (eq. 4) into real rather than complex design modes (see e.g. Neumaier and Schneider 2001 or Hinrichsen and Pritchard 2005). In fact, for any complex eigenvalue $\lambda_i = a_i + ib_i$ the scalar process $\left\{X_t'^{(i)}\right\}$ defined by state eq. 33 can be expressed as a real bivariate autoregressive process of first order

$$\begin{pmatrix} \mathrm{Re}\left[X_t'^{(i)}\right] \\ \mathrm{Im}\left[X_t'^{(i)}\right] \end{pmatrix} = \begin{pmatrix} a_i & -b_i \\ b_i & a_i \end{pmatrix} \begin{pmatrix} \mathrm{Re}\left[X_{t-1}'^{(i)}\right] \\ \mathrm{Im}\left[X_{t-1}'^{(i)}\right] \end{pmatrix} + \varepsilon_t'^{(i)}$$

with the bivariate noise vectors

$$\varepsilon_t'^{(i)} = \begin{pmatrix} \mathrm{Re}\left[\varepsilon_t^{(i)}\right] \\ \mathrm{Im}\left[\varepsilon_t^{(i)}\right] \end{pmatrix} = \frac{1}{2} \begin{pmatrix} \varepsilon_t^{(i')} + \varepsilon_t^{(i)} \\ i\left(\varepsilon_t^{(i')} - \varepsilon_t^{(i)}\right) \end{pmatrix}.$$

In the above equation the noise components indexed by $i'$ are conjugates of the noise components indexed by $i$. From the definition of the correlated noise in eq. 34, the noise covariance matrix can be written as (Neumaier and Schneider 2001):

$$\begin{aligned} &E\left[\varepsilon_\mu'^{(k)} \left[\varepsilon_\nu'^{(l)}\right]\right] \\ &= \frac{\delta_{\mu\nu}}{4} \begin{pmatrix} C'_{[\![k,l]\!]} + C'_{[\![k',l']\!]} + C'_{[\![k',l]\!]} + C'_{[\![k,l']\!]} & -i\left(-C'_{[\![k,l]\!]} + C'_{[\![k',l']\!]} - C'_{[\![k',l]\!]} + C'_{[\![k,l']\!]}\right) \\ -i\left(-C'_{[\![k,l]\!]} + C'_{[\![k',l']\!]} - C'_{[\![k',l]\!]} + C'_{[\![k,l']\!]}\right) & C'_{[\![k,l]\!]} + C'_{[\![k',l']\!]} - C'_{[\![k',l]\!]} - C'_{[\![k,l']\!]} \end{pmatrix} \\ &= \frac{\delta_{\mu\nu}}{2} \begin{pmatrix} \mathrm{Re}\left[C'_{[\![k,l]\!]}\right] + \mathrm{Re}\left[C'_{[\![k',l]\!]}\right] & \mathrm{Im}\left[C'_{[\![l,k]\!]}\right] + \mathrm{Im}\left[C'_{[\![l,k']\!]}\right] \\ \mathrm{Im}\left[C'_{[\![k,l]\!]}\right] + \mathrm{Im}\left[C'_{[\![k',l]\!]}\right] & \mathrm{Re}\left[C'_{[\![k,l]\!]}\right] - \mathrm{Re}\left[C'_{[\![k',l]\!]}\right] \end{pmatrix} \\ &= \frac{\delta_{\mu\nu}}{2} \begin{pmatrix} \mathrm{Re}\left[\rho'_{kl} C'_{[\![k,k]\!]} C'_{[\![l,l]\!]}\right] + \mathrm{Re}\left[\rho'_{k'l} C'_{[\![k',k']\!]} C'_{[\![l,l]\!]}\right] & \mathrm{Im}\left[\rho'_{lk} C'_{[\![l,l]\!]} C'_{[\![k,k]\!]}\right] + \mathrm{Im}\left[\rho'_{lk'} C'_{[\![l,l]\!]} C'_{[\![k',k']\!]}\right] \\ \mathrm{Im}\left[\rho'_{kl} C'_{[\![k,k]\!]} C'_{[\![l,l]\!]}\right] + \mathrm{Im}\left[\rho'_{k'l} C'_{[\![k',k']\!]} C'_{[\![l,l]\!]}\right] & \mathrm{Re}\left[\rho'_{kl} C'_{[\![k,k]\!]} C'_{[\![l,l]\!]}\right] - \mathrm{Re}\left[\rho'_{k'l} C'_{[\![k',k']\!]} C'_{[\![l,l]\!]}\right] \end{pmatrix} \end{aligned}$$

The eigenmodes of the decomposed process are given by the real and imaginary parts of the eigenvector $\vartheta_i(A_0) = S_{:i}$.

For small NPD projects with only $p = 2$ tasks, we can obtain simple analytical solutions for the eigenvalues, eigenvectors and correlation coefficient in the spectral basis. To do so, we use the following parametric representation of the WTM and the noise covariance matrix in the original state space coordinates:

$$A_0 = \begin{pmatrix} a_{11} & a_{12} \\ a_{21} & a_{11} + \Delta a \end{pmatrix}$$

$$C = \begin{pmatrix} \sigma_{11}^2 & \rho\sigma_{11}\sigma_{22} \\ \rho\sigma_{11}\sigma_{22} & \sigma_{22}^2 \end{pmatrix},$$

where $\{a_{11}, \sigma_{11}, \sigma_{22}\} \in \mathbb{R}^+$, $\{\Delta a, a_{12}, a_{21}\} \in \mathbb{R}$ and $\rho \epsilon [-1\,;1]$. Please note that the parametric representation of the autonomous task processing rates $a_{11}$ and $a_{22} = a_{11} + \Delta a$ through the rate difference Δa is slightly different than in the applied example which will be presented section 2.5. The rate difference is introduced in order to obtain solutions that are easier to interpret.

The above parametric representation leads to the eigenvalues

$$\lambda_1 = \frac{1}{2}\left(2a_{11} + \Delta a - \sqrt{\Delta a^2 + 4a_{12}a_{21}}\right)$$

$$\lambda_2 = \frac{1}{2}\left(2a_{11} + \Delta a + \sqrt{\Delta a^2 + 4a_{12}a_{21}}\right),$$

the infinite families of eigenvectors $\{c_1, c_2\} \in \mathbb{R}$

$$\vartheta_1 = \{c_1\} \cdot \begin{pmatrix} -\dfrac{\Delta a + \sqrt{\Delta a^2 + 4a_{12}a_{21}}}{2a_{21}} \\ 1 \end{pmatrix}$$

$$\vartheta_2 = \{c_2\} \cdot \begin{pmatrix} -\dfrac{\Delta a - \sqrt{\Delta a^2 + 4a_{12}a_{21}}}{2a_{21}} \\ 1 \end{pmatrix},$$

and therefore to the matrix

$$S = \begin{pmatrix} -\dfrac{\Delta a + \sqrt{\Delta a^2 + 4a_{12}a_{21}}}{2a_{21}} & -\dfrac{\Delta a - \sqrt{\Delta a^2 + 4a_{12}a_{21}}}{2a_{21}} \\ 1 & 1 \end{pmatrix}$$

for the basis transformation. In order to obtain an instance of an autoregressive model that is asymptotically stable in the sense of Lyapunov (eq. 3), the spectral radius $\rho(A_0)$ must be less than one and therefore it must hold that $1/2\left(2a_{11} + \Delta a + \sqrt{\Delta a^2 + 4a_{12}a_{21}}\right) < 1$.

In the spectral basis, the transformed variances ${\sigma'_{11}}^2 = C'_{[\![1,1]\!]}$ and ${\sigma'_{22}}^2 = C'_{[\![2,2]\!]}$ are given by

$${\sigma'_{11}}^2 = \frac{4a_{21}^2\sigma_{11}^2 + 4\rho\left(\Delta a - \mathrm{Re}\left[\sqrt{g}\right]a_{12}\sigma_{11}\sigma_{22}\right) + \left(\mathrm{Abs}[g] + \Delta a\left(\Delta a - 2\mathrm{Re}\left[\sqrt{g}\right]\right)\right)\sigma_{22}^2}{4\mathrm{Abs}[g]}$$

$${\sigma'_{22}}^2 = \frac{4a_{21}^2\sigma_{11}^2 + 4\rho\left(\Delta a + \mathrm{Re}\left[\sqrt{g}\right]a_{12}\sigma_{11}\sigma_{22}\right) + \mathrm{Abs}[g] + \Delta a\left(\Delta a + 2\mathrm{Re}\left[\sqrt{g}\right]\right)\sigma_{22}^2}{4\mathrm{Abs}[g]}.$$

The correlation coefficient $\rho'$ can be expressed in the spectral basis as

$$\rho' = \frac{\left(-4a_{21}^2\sigma_{11}^2 - 4\rho a_{21}\left(\Delta a - \mathrm{Re}\left[\sqrt{g}\right] + \sqrt{g}\right)\right)\sigma_{11}\sigma_{22} + \left(\mathrm{Abs}[g] - \Delta a\left(\Delta a + 2i\mathrm{Im}\left[\sqrt{g}\right]\right)\right)\sigma_{22}^2}{\sigma'_{11}\sigma'_{22}}.$$

In the three equations above the coupling constant $g$ determines the strength of the interaction between both tasks. It is defined as

$$g \coloneqq \Delta a^2 + 4a_{12}a_{21}.$$

If all entries of the WTM $A_0$ are nonnegative, the correlation coefficient $\rho'$ can be further simplified and we have

$$\rho' = \frac{\sigma_{22}(a_{12}\sigma_{22} - \Delta a\rho\sigma_{11}) - a_{21}\sigma_{11}^2}{\sqrt{(a_{21}^2\sigma_{11}^4 + 2\Delta a\rho a_{21}\sigma_{11}^3\sigma_{22} + (\Delta a^2 + 2(1-2\rho^2)a_{12}a_{21})\sigma_{11}^2\sigma_{22}^2 - 2\Delta a\rho a_{12}\sigma_{11}\sigma_{22}^3 + a_{12}^2a_{22}^4)}}.$$

Furthermore, if not only all the entries of the WTM are nonnegative, but also the correlation coefficient of the noise in the original state space coordinates is zero, i.e. $\rho = 0$, we arrive at the most simple parametric form:

$$\rho' = \frac{a_{12}\sigma_{22}^2 - a_{21}\sigma_{11}^2}{\sqrt{a_{21}^2\sigma_{11}^4 + (\Delta a^2 + 2a_{12}a_{21})\sigma_{11}^2\sigma_{22}^2 + (a_{12}\sigma_{22}^2)^2}}.$$

It is obvious that the transformed correlation coefficient $\rho'$ is zero if the noise is isotropic and the WTM is symmetric.

### 2.4 Least Squares Parameter Estimation and Model Selection

Neumaier and Schneider (2001) term the stochastic process that is generated by the introduced linear autoregression models (eqs. 4 and 24) a vector autoregressive process of order 1, abbreviated as VAR(1) process, with zero mean of the time series. A VAR(1) process with zero mean is the least complex stochastic process in its class. In the original state-space coordinates, a VAR(1) model can be easily generalized by increasing the regression order and therefore the correlation length of the process. When the regression order is increased not only the present state of the development project is considered to make accurate predictions of its future course but also states that lay in the finite past. The corresponding vector autoregression model of order $n$, abbreviated as VAR($n$) model, is defined by the extended state equation:

$$X_t = \sum_{i=0}^{n-1} A_i \cdot X_{t-i-1} + \varepsilon_t \qquad t = 1, \dots, T. \tag{38}$$

The definition of the noise vector $\varepsilon_t$ was already given in eq. 9. One could also include a $p$-dimensional parameter vector $\omega$ of intercept terms to allow for a nonzero mean of the time series (see Neumaier and Schneider 2001 or Lütkepohl 2005). From a theoretical point of view, this systematic shift or drift of work remaining does not necessarily have to be considered in the model formulation and is consequently ignored in the following analysis.

The VAR($n$) model is one of the most flexible models for the analysis of multivariate time series. This model has proven especially useful for describing the dynamic behavior of economic time series and for forecasting. Neumaier and Schneider (2001), Franses and Paap (2004), Lütkepohl (2005) and others developed efficient and robust methods to estimate the order of the model, the values of its parameters, spectral information and confidence regions based on empirically acquired time series. We will describe the parameter estimation for vector autoregression models based on the material from Neumaier and Schneider (2001). The description of the criteria for model selection is based on the textbook by Lütkepohl (2005).

An interesting finding from the theory of stochastic dynamical systems is that every linear autoregression model of finite order can be rewritten as a first-order model based on the state equation

$$\tilde{X}_t = \tilde{A} \cdot \tilde{X}_{t-1} + \tilde{\varepsilon}_t \qquad t = 1, \dots, T,$$

where $\tilde{X}_t$ is the augmented state vector

$$\tilde{X}_t = \begin{pmatrix} X_t \\ X_{t-1} \\ \vdots \\ X_{t-n+1} \end{pmatrix},$$

$\tilde{\varepsilon}_t$ is the augmented noise vector

$$\tilde{\varepsilon}_t = \begin{pmatrix} \varepsilon_t \\ 0 \\ \vdots \\ 0 \end{pmatrix}$$

and $\tilde{A}$ is the extended dynamical operator

$$\tilde{A} = \begin{pmatrix} A_0 & A_1 & \cdots & A_{n-2} & A_{n-1} \\ I_p & 0 & \cdots & 0 & 0 \\ 0 & I_p & \cdots & 0 & 0 \\ & & & & \\ 0 & 0 & \ddots & 0 & 0 \\ 0 & 0 & \dots & I_p & 0 \end{pmatrix}.$$

Through this order reduction by state-space augmentation, the complexity measures that will be presented in section 4 in different closed-forms do not have to be reworked but rather can be applied directly to analyze and evaluate emergent complexity in NPD projects.

If the process order $n$ is known in advance, various techniques have been developed to estimate the work transformation matrices $A_0, \dots, A_{n-1}$ and the covariance matrix $C$ from a time series of empirically acquired state vectors $x_t$ (see literature review in Neumaier and Schneider, 2001). The most prominent techniques are based on the well-known Maximum Likelihood Estimation (MLE, see for instance Brockwell and Davis 1991). For a fixed series of data and an underlying parameterized model, MLE picks the value of parameters that maximize the probability of generating the observations or minimize the prediction error. It is usually assumed that all states consist only of correlated noise of which the covariance matrix is determined by state eq. 38. Since the state covariance matrix depends on the regression parameters, the likelihood must maximized in several iterations. The iterative estimates are not only asymptotically efficient as the sample size grows to infinity, they also show good small-sample performance (Brockwell and Davis 1991). An alternative iterative MLE technique with good small-sample performance that can also deal with “hidden” (not directly observable) state variables will be introduced in section 2.8. Although these are advantageous properties for estimation techniques, the iterative procedure is computationally quite demanding and slows the parameter estimation for large projects significantly. A much simpler technique is based on the classic least squares estimation (Neumaier and Schneider 2001). Our analyses have shown that in the application area of project management least-square estimation is comparatively accurate and robust. The least-square estimates can also be used as a bootstrap for subsequent maximum likelihood iterations. Furthermore, least-square estimation can be intuitively extended by information-theoretic or Bayesian model selection techniques. This reveals interesting theoretical connections between selecting a predictive model based on universal criteria and evaluating its statistical or information-theoretical complexity. We will return to this important point in sections 3.2.2 and 3.2.3.

According to Neumaier and Schneider (2001) the least squares estimates for a VAR($n$) process are most conveniently derived when the process generated by state eq. 38 is cast in the classic linear regression form

$$X_t = A \cdot U_t + \varepsilon_t \qquad t = 1, \dots, T \tag{39}$$
$$\varepsilon_t \sim \mathcal{N}(0_p, C),$$

with the coefficient matrix

$$A = (A_0 \quad \cdots \quad A_{n-1})$$

and predictors

$$U_t = \begin{pmatrix} X_{t-1} \\ \vdots \\ X_{t-n} \end{pmatrix}.$$

The key insight in the derivation of the least squares estimates of the independent parameters is to consider the casted regression model as a model with fixed predictors $U_t$. Clearly, this is only an approximation, because for the multivariate time series the $U_t$ are realizations of a random variable. However, the above definition of the predictors implies that the assumption of fixed predictors leads to treating

$$U_1 = \begin{pmatrix} X_0 \\ \vdots \\ X_{1-n} \end{pmatrix}$$

as a vector of fixed initial states. Since the relative effect of the initial state vanishes as the length $T$ of the time series approaches infinity, using corresponding parameter estimates for the regression model in the autoregressive model formulation can be expected to be asymptotically correct (Neumaier and Schneider 2001). In fact, when using a linear regression model, the least squares principle leads to the best unbiased parameter estimates. We define the following data-related matrices:

$$U = \sum_{t=1}^{T} u_t u_t^{\mathrm{T}}$$

$$V = \sum_{t=1}^{T} x_t x_t^{\mathrm{T}}$$

$$W = \sum_{t=1}^{T} x_t u_t^{\mathrm{T}}.$$

The column vector $x_t$ represents the work remaining that had been measured for all $p$ development tasks at time step $t \geq 0$. The least square estimate for the coefficient matrix $A$ can be written as the matrix product

$$\hat{A} = W \cdot U^{-1}. \tag{40}$$

The corresponding estimate for the covariance matrix is given by

$$\hat{C} = \frac{1}{T - pn} \sum_{t=1}^{T} (x_t - \hat{A} \cdot u_t)(x_t - \hat{A} \cdot u_t)^{\mathrm{T}}. \tag{41}$$

The leading factor $1/(T - pn)$ is used to adjust the degrees of freedom of the covariance matrix. In terms of a multivariate regression, the estimated noise covariance matrix can be interpreted as a point estimate of the inherent prediction error.

Alternatively, the estimate of the covariance matrix can be expressed as

$$\hat{C} = \frac{1}{T - pn} \left(V - W \cdot U^{-1} \cdot W^{\mathrm{T}}\right).$$

This estimate is proportional to the Schur complement of the composed matrix

$$\begin{pmatrix} U & W^{\mathrm{T}} \\ W & V \end{pmatrix} = \sum_{t=1}^{T} \begin{pmatrix} u_t \\ x_t \end{pmatrix} \begin{pmatrix} u_t^{\mathrm{T}} & u_t^{\mathrm{T}} \end{pmatrix} = K \cdot K^{\mathrm{T}},$$

where the aggregated data matrix $K$ is defined as

$$K = \begin{pmatrix} u_1^{\mathrm{T}} & x_1^{\mathrm{T}} \\ \vdots & \vdots \\ u_T^{\mathrm{T}} & x_T^{\mathrm{T}} \end{pmatrix},$$

and therefore assured to be positive semidefinite. Neumair and Schneider (2001) have shown that a QR factorization of the data matrix $K$ as

$$K = Q \cdot R$$

with an orthogonal matrix $Q$ and an upper triangular matrix

$$R = \begin{pmatrix} R_{11} & R_{12} \\ 0 & R_{22} \end{pmatrix}$$

allows the development of an efficient procedure to compute the parameter estimates numerically. Therefore, the above Schur complement is rewritten as

$$\begin{pmatrix} U & W^{\mathrm{T}} \\ W & V \end{pmatrix} = R^{\mathrm{T}} \cdot R = \begin{pmatrix} R_{11}^{\mathrm{T}} \cdot R_{11} & R_{11}^{\mathrm{T}} \cdot R_{12} \\ R_{12}^{\mathrm{T}} \cdot R_{11} & R_{12}^{\mathrm{T}} \cdot R_{12} + R_{22}^{\mathrm{T}} \cdot R_{22} \end{pmatrix}.$$

Based on the rewritten Schur complement, the following least squares estimates are obtained:

$$\hat{A} = (R_{11}^{-1} \cdot R_{12})^{\mathrm{T}}$$

$$\hat{C} = \frac{1}{T - pn} \left(R_{22}^{\mathrm{T}} \cdot R_{22}\right).$$

The estimate for the initial state is simply

$$\hat{x}_0 = x_0.$$

As an alternative to the QR factorization, the matrix $R$ can be obtained from a Cholesky decomposition. Furthermore, regularization schemes can be used to reduce rounding error. More details are available in Neumair and Schneider (2001).

If not only a single time series of empirically acquired state vectors $x_t$ is given but multiple realizations of the random process had been acquired in $N$ independent measurement trials, the additional time series are simply appended as additional $N-1$ blocks of rows in the regression eq. 39. Similarly, the predictors are extended by additional row blocks. The initial state is determined by averaging over all initial data points.

If not only the coefficient matrices and the covariance matrix of the VAR($n$) process have to be estimated from data but also the model order, a good trade-off between the predictive accuracy gained by increasing the number of independent parameters and the danger of overfitting the model has to be found. Overfitting means in our context of project management that the model is fitted to random performance fluctuations instead of the implicit or explicit rules of cooperative work that are necessary for the functioning of the project. In order to find an optimal solution to the trade-off in terms of a universal principle, Rissanen´s (1989, 2007) minimum description length principle aims at selecting the model with the briefest recording of all attribute information − not only the likelihood of a fixed series of data and an underlying parameterized model. This integrative view provides a natural safeguard against overfitting as it defines a method to reduce the part of the data that looks like noise by using a more elaborate − but in the sense of Occam´s Razor not unnecessary complex − model. We will come back to this important model selection principle in section 3.2.2 when we discuss the stochastic complexity of a generative model within a simple parametric model class comprising of distributions indexed by a specific parameter set. In the following, we take a more pragmatic view and select the order of an autoregression model on the basis of the standard selection criteria of Akaike (Final Prediction Error Criterion as well as Information Criterion, see Akaike 1971, 1974a and 1974b) and Schwarz (Schwarz-Bayes Criterion, see Schwarz 1978, also termed Bayesian Information Criterion). These criteria had been validated in many scientific studies and can also be calculated efficiently. For regular parametric distribution families of dimension $p$ the simplified two-stage minimum description length criterion takes the form of the Schwarz-Bayes Criterion (eq. 43) and therefore has within the scope of this paper the same consequences for model selection (Hansen and Yu 2001). By the simplified two-stage minimum description length criterion, we mean an encoding scheme in which the description length for the best-fitting member of the model class is calculated in the first stage and then the description length of data based on the parameterized probability distribution is determined (see section 3.2.2). However, it is important to point out that the Schwarz´s Bayesian approximation only holds if the number of parameters is kept fixed and the number of observations goes to infinity. If the number of parameters is either infinite or grows with the number of observations, then model selection based on minimum description length can lead to quite different results (see Grünwald 2007). A detailed comparison of these (and other) criteria can be found in chapter 4.3 of Lütkepohl (2005) and Lütkepohl (1985).

Akaike´s (1971) Final Prediction Error (FPE) is one of the historically most significant criteria and provides a generalized measure of model quality by simulating the situation where a parameterized model is tested on a different dataset. Clearly, the candidate model with smallest FPE is favored. The FPE is defined for a VAR($n$) model according to eq. 38 as

$$FPE(n) = \ln\left(\left(\frac{T+pn}{T-pn}\right)^{p} \mathrm{Det}\left[\hat{\Sigma}_{(n)}\right]\right),$$

where

$$\hat{\Sigma}_{(n)} := \mathrm{Det}\left[\frac{\hat{\Delta}_{(n)}}{T}\right]$$

is a scalar measure for the not biased corrected variance of the estimator. The scalar measure is simply a multiple of the least square estimate of the noise covariance matrix for the *n*th order model that had been fitted to the time series of task processing in the project and we have

$$\begin{aligned}\hat{\Delta}_{(n)} &= (T - pn)\hat{C} \\ &= R_{22}^{\mathrm{T}} \cdot R_{22}.\end{aligned}$$

In the chosen model formulation with Gaussian random variables the least square estimation is equivalent to the maximum likelihood estimation. $\hat{\Sigma}_{(n)}$ can also be interpreted as the one-step prediction error of order $n$. Please note that the above definition of the FPE is only valid for an autoregression model without a parameter vector of intercept terms. If intercept terms are included, the model-related product $pn$ in the denominator and numerator of the first factor has to be corrected by $pn + 1$ (Lütkepohl 2005). It is obvious that the FPE can also be expressed as the sum

$$FPE(n) = \ln \mathrm{Det}\left[\hat{\Sigma}_{(n)}\right] + p\ln\frac{T + pn}{T - pn},$$

which is the form of representation most frequently found in textbooks. Based on the FPE, Akaike (1974a, 1974 b) later developed an information-theoretic criterion that the literature calls the Akaike Information Criterion (AIC). The AIC is the most widely known and used quantity for model selection among scientists and practitioners. To develop the AIC, Akaike proposed an information-theoretic framework wherein the estimation of the model parameters and model selection could be simultaneously accomplished. In fact, the AIC is an asymptotically unbiased estimator of the expected relative Kullback-Leibler distance, which represents the amount of information lost when we use candidate models $g_i$ within a candidate family $G$ to approximate another model $f$. The AIC chooses the candidate model with the smallest expected Kullback-Leibler distance. Model $f$ may be the "true" model or not. If $f$ itself is only a poor approximation of the true generative mechanisms, the AIC selects the relatively best among the poor models. Details about the Kullback-Leibler divergence and its underlying information-theory principles can be found in Cover and Thomas (1991). For a VAR($n$) model, the AIC is defined as:

$$AIC(n) = \ln \mathrm{Det}\left[\hat{\Sigma}_{(n)}\right] + \frac{2}{T}np^2. \qquad (42)$$

The term $np^2$ in the above definition of the criterion represents the number of freely estimated parameters of the candidate model. According to Lütkepohl (2005) the FPE and AIC criteria are very similar and it holds that

$$\ln FPE(n) = AIC(n) + 2\frac{p}{T} + o(T^{-2})..$$

The term $o(T^{-2})$ denotes an arbitrary sequence indexed by $T$ that remains bounded when multiplied by $T^{-2}$. Due to this property the difference between $\ln FPE(n)$ and $AIC(n)$ tends rapidly towards zero for $T \to \infty$ and $FPE(n)$ and $AIC(n)$ are asymptotically equivalent.

Finally, the Schwarz-Bayes Criterion (SBC, also termed Bayesian information criterion − BIC) is introduced and evaluated. Schwarz derived the SBC to serve as an asymptotic approximation to a transformation of the Bayesian posterior probability of a candidate model. In settings with a large sample size, the model favored by this criterion ideally corresponds to the candidate model, which is most probable a posteriori and therefore rendered most plausible by the observed data. This is indicated by minimum scores. The calculation of the SBC is based on the empirical logarithmic likelihood function and does not require the specification of prior distributions. To briefly explain Schwarz´s deductive approach, we can refer to Bayes factors. They are the Bayesian analogues of the famous likelihood ratio tests (e.g. Honerkamp 2002). Under the assumption that two candidate models $g_1$ and $g_2$ are regarded as equally probable a priori, the Bayes factor $l(g_1, g_2)$ represents the ratio of the posterior probabilities of the models. Which model is most probable a posteriori is determined by whether the Bayes factor $l(g_1, g_2)$ is greater or is less than one. Under certain circumstances, model selection based on BIC is very similar to model selection through Bayes factors. For a VAR($n$) model the SBC is defined as

$$SBC(n) = \ln \mathrm{Det}\big[\hat{\Sigma}_{(n)}\big] + \frac{\ln T}{T} np^2. \tag{43}$$

In a similar manner to the Akaike Information Criterion, the term $np^2$ represents the number of freely estimated parameters of the candidate model. The principle of counting free parameters can be easily generalized to other model classes, such as linear dynamical systems. This more complex model class with latent state variables will be introduced and discussed in section 2.8.

Under all three selection criteria, the order $n_{opt}$ of the VAR($n$) model is considered to be the optimal one if it is assigned minimum scores, that is

$$n_{opt} = \arg\min_n \begin{cases} FPE(n) \\ AIC(n) \\ SBC(n) \end{cases}. \tag{44}$$

A substantial advantage in using information-theoretic or Bayesian criteria is that they are also valid for nonnested models and can therefore also be used to evaluate sinusoidal performance curves with given amplitudes. Traditional likelihood ratio tests are defined only for nested models such as vector autoregression models of finite order, and this represents another substantial limitation on the use of hypothesis testing in model selection (sensu Burnham and Anderson 2002). Vector autoregression models are nested in the sense that a VAR($n_1$) model can be considered as a special case of a VAR($n_2$) model if it holds that $n_1 < n_2$. This is a direct consequence of the formulation of the state equation (eq. 38). It is important to note that for $FPE$ and $AIC$ the estimate $\hat{n}(.) = n_{opt}$ is inconsistent under the assumption that the maximum order is larger than the maximum evaluated order $n_{max}$. Furthermore, it can be shown that, under quite general conditions, the limiting probability for underestimating the model order is zero for both criteria and the probability of overfitting is a nonnegative constant.

Hence, for FPE and AIC it holds that

$$probability\ of\ overfitting \to 0$$
$$probability\ of\ overfitting \to constant > 0.$$

as $T \to \infty$ (Stoica and Selen 2004, Lütkepohl 2005). In other words, $FPE$ and $AIC$ tend to overestimate the true autoregression order.

In contrast, Schwarz´s $SBC$ is strongly consistent for any dimension of the state space. For this criterion it can be shown that, under the assumption that the data-generating process belongs to the considered model class, the autoregression order is consistently selected and the

$$probability\ of\ correct\ selection \rightarrow 1$$

as $T \rightarrow \infty$ (Stoica and Selen 2004, Lütkepohl 2005).

## 2.5 Project Management Example from Industry and Model Validation

To demonstrate the state-space concept introduced and validate the developed dynamic project models with field data, a detailed analysis of the course of NPD projects was carried out at a small industrial company in Germany (Schlick et al. 2008, Schlick et al. 2012). The company develops mechanical and electronic sensor components for the automotive industry. We investigated task processing by a team of three engineers in a multiproject setting comprising projects A, B and C. Project A was the research focus, comprising 10 partially overlapping development tasks covering all project phases—from conceptual design of the particular sensor to product documentation for the customer. In addition, the workloads of two concurrent smaller projects, B and C, were acquired. The acquired time data of all three projects were very accurate, because the company used a barcode-based labor time system: an engineer in this company who starts processing a development task has to indicate this with a manual barcode scan on a predefined task identification sheet. When the task was finished, the task identifier was scanned again. The recorded "time-on-development-task" had a resolution of 1 min and was used as an estimator for the values of the components of the introduced state variable $X_t$ representing the work remaining. Let $y_t^{(i)}$ be the recorded time-on-task for task $i$ at time step $t$. The estimated mean work remaining for the $i$-th component of the state variable is $x_t^{(i)} = 1 - y_t^{(i)}/y_{t_{max}}^{(i)}$, where $t_{max}$ represents the time step in which the processing of task *i* was completed. The time scale is weeks.

Figure 2 shows the obtained time series for the initial five tasks of project A and the first two tasks of project C. Figure 2 also shows the complete work remaining in project B, a "fast-track project". Its detailed work breakdown structure was not considered and only the accumulated time data was analyzed (Schlick et al. 2012).

The validation study included interviews and informal discussions with management and engineers to understand the development projects in detail. We found that the development of sensor technologies is a good subject for NPD project modeling because tasks are largely processed in parallel and frequent iterations occur.

For simplicity, we focus on the first two overlapping development tasks of project A, "conceptual sensor design" (task 1) and "design of circuit diagrams" (task 2), and model only their overlapping range (cf. Figure 2). Concerning the left bound of this range, the conceptual sensor design had reached a completion level of 39.84% when the design of the circuit diagram began. Therefore, the estimated work remaining at the initial time step is therefore:

$$\hat{x}_0 = \begin{pmatrix} x_0^{(1)} = 0.6016 \\ x_0^{(2)} = 1.0000 \end{pmatrix}. \quad (45)$$

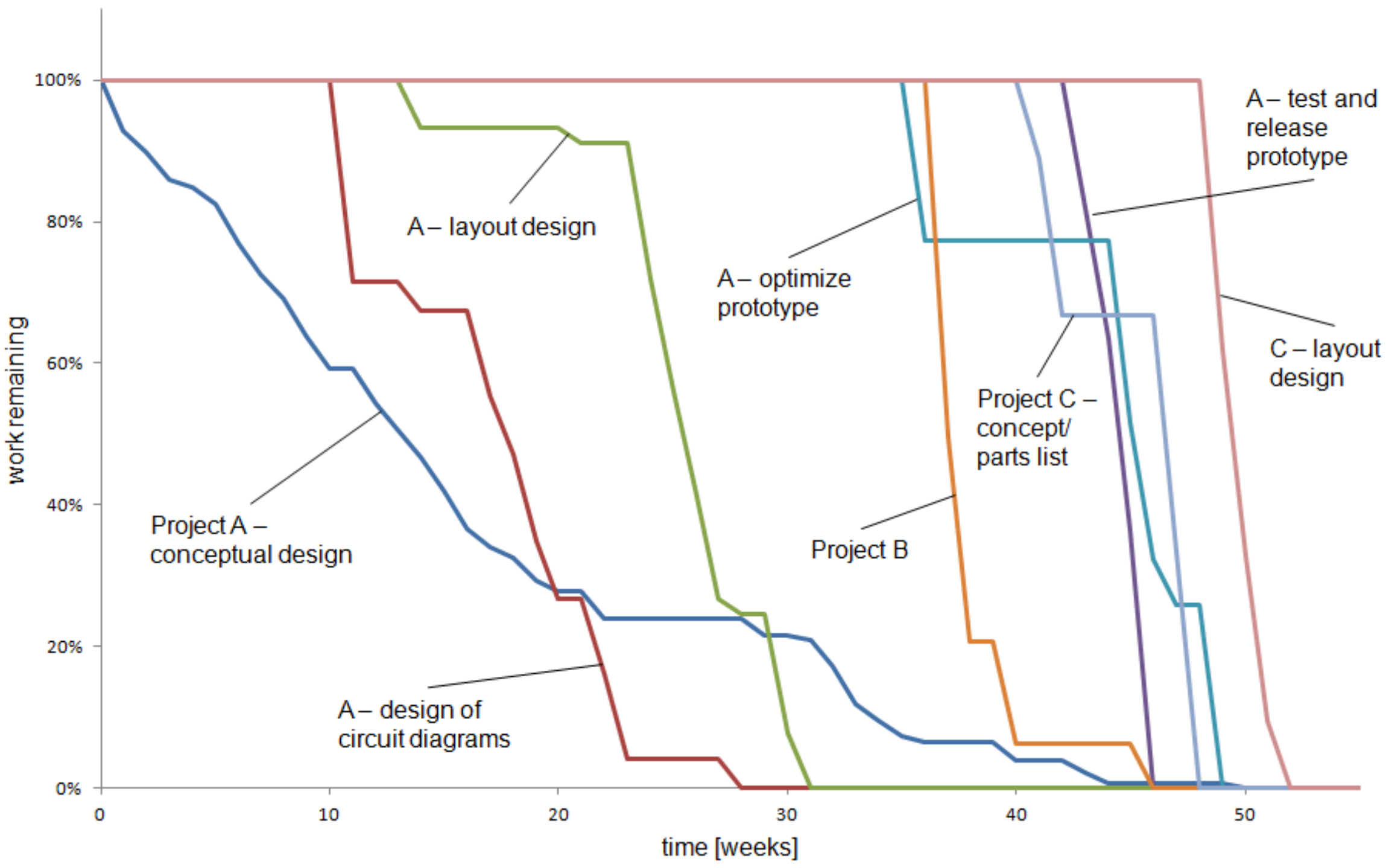

**Figure 2**

*Figure 2*. Time series of work remaining in three real NPD projects. The data were acquired in a small industrial company in Germany. Only the first five tasks of project A and the first two tasks of project C are shown. Project B was a "fast-track project". Its work breakdown structure is not considered and only the complete work remaining is shown. The time scale is weeks.

The least square method developed by Neumaier and Schneider (2001) was used to estimate the additional parameters of the VAR$(n)$ process (see section 2.4). The maximum model order to be considered in model selection was set to $n_{max} = 6$. The estimation algorithm and model selection procedures were implemented in Mathematica® based on the Matlab® source code of the ARfit toolbox. The ARfit toolbox was published by Schneider and Neumaier (2001) in a paper accompanying their theoretical considerations. The estimation results were also verified through the ARfit toolbox in the original Matlab® simulation environment. Please note that the fast algorithm introduced by Neumaier and Schneider (2001) in section 4.2 of their paper was used to improve computational efficiency. The fast algorithm does not require separate QR factorizations (see section 2.4 ) for each candidate autoregression model, but instead only one factorization for the most complex model of order $n_{max} = 6$. Due to this simplification, the least square estimates are slightly less accurate and therefore only approximate order selection criteria for lower order models can be obtained.

We start by presenting and discussing the parameter estimates for a vector autoregression model of first order. This model with least independent parameters was selected due to Schwarz´ Bayesian Criterion (section 2.4). Afterwards, the parameters of a second order model that is favored by Akaike´s criteria will be presented.

The model selection procedure showed that Schwarz´ Bayesian Criterion is minimal for a VAR model of first order and we have $SBC(n_{opt} = 1) = -15.74$ for this model. Due to the cited design of the selection algorithm for fast processing, the first $n_{max} - 1$ data points of the mean work remaining are ignored in the implementation of the ARfit toolbox. Although this approach does not affect the accuracy of results asymptotically, for short time series such as in our industrial case study it can constitute a significant loss of information. Therefore, the least square fitting was repeated with $n_{max} = 1$. The estimated WTM $A_0$ for this model is given by

$$\hat{A}_0 = \begin{pmatrix} a_{11} = 0.9406 & a_{12} = -0.0017 \\ a_{21} = 0.0085 & a_{22} = 0.8720 \end{pmatrix}. \tag{46}$$

The estimated covariance matrix $\hat{C}$ of the normally distributed random variable $\varepsilon_t$ is given the representation

$$\varepsilon_t \sim \mathcal{N}\left( \begin{pmatrix} 0 \\ 0 \end{pmatrix}, \begin{pmatrix} \sigma_{11}^2 = (0.0135)^2 & \rho\sigma_{11}\sigma_{22} = -0.38 \cdot 0.0135 \cdot 0.0416 \\ \rho\sigma_{11}\sigma_{22} & \sigma_{22}^2 = (0.0416)^2 \end{pmatrix} \right). \tag{47}$$

In the above probability distribution, the variable $\rho \epsilon [-1\,;1]$ represents Pearson´s correlation coefficient in the original state-space coordinates. We can also rewrite the formulation of the covariance matrix and assume that the standard deviation $\sigma_{ii}$ of the fluctuations is proportional with the proportionality constant $s_i$ to the work remaining after each time step $a_{ii}$:

$$\varepsilon_t \sim \mathcal{N}\left( \begin{pmatrix} 0 \\ 0 \end{pmatrix}, \begin{pmatrix} (s_1 a_{11})^2 = (0.0144 \cdot 0.9406)^2 & \rho(s_1 a_{11})(s_2 a_{22}) \\ \rho(s_1 a_{11})(s_2 a_{22}) & (s_2 a_{22})^2 = (0.0477 \cdot 0.8720)^2 \end{pmatrix} \right).$$

Figure 3 shows the list plots of work remaining in the overlapping range of tasks 1 and 2 over 50 weeks. The figure not only presents the empirical time series of task processing which had been acquired in the industrial company but also the results of a Monte Carlo simulation based on the parameterized project model (see state eq. 4 in conjunction with parameter estimates from eqs. 45, 46 and 47). A total of 1,000 separate and independent project simulations were performed.

Concerning the field data, the fact that the conceptual design of the sensor was processed entirely in parallel with the design of the circuit diagram and continued after the circuit diagram was finished is of particular interest (Figure 3). More than 10 iterations are necessary to reach a stable conceptual design state after week 35. In the list plot of Figure 3, the simulated task processing is represented by the means and 95% confidence intervals of work remaining. The estimated stopping criterion of $\delta = 0.02$ is plotted as a dashed line at the bottom of the chart. According to Figure 3, 49 out of 50 confidence intervals of the simulated work remaining of task 1 include the empirical data points from the real project before the stopping criterion is met. Only the confidence interval computed for week 1 is a little too small. In contrast, when the processing of task 2 is compared against the field data, the goodness-of-fit is significantly lower, and only 47 out of the 50 confidence intervals cover the empirical data points before the stopping criterion is met. Moreover, the quite abrupt completion of task 2 in week 18 is poorly predicted by the smoothly decaying means of the autoregression model of first order (Schlick et al. 2008, Schlick et al. 2012). However, the prediction in singular cases can be much better because of the large variance of the fluctuation variable $\varepsilon_t$ in the second dimension covering task 2.

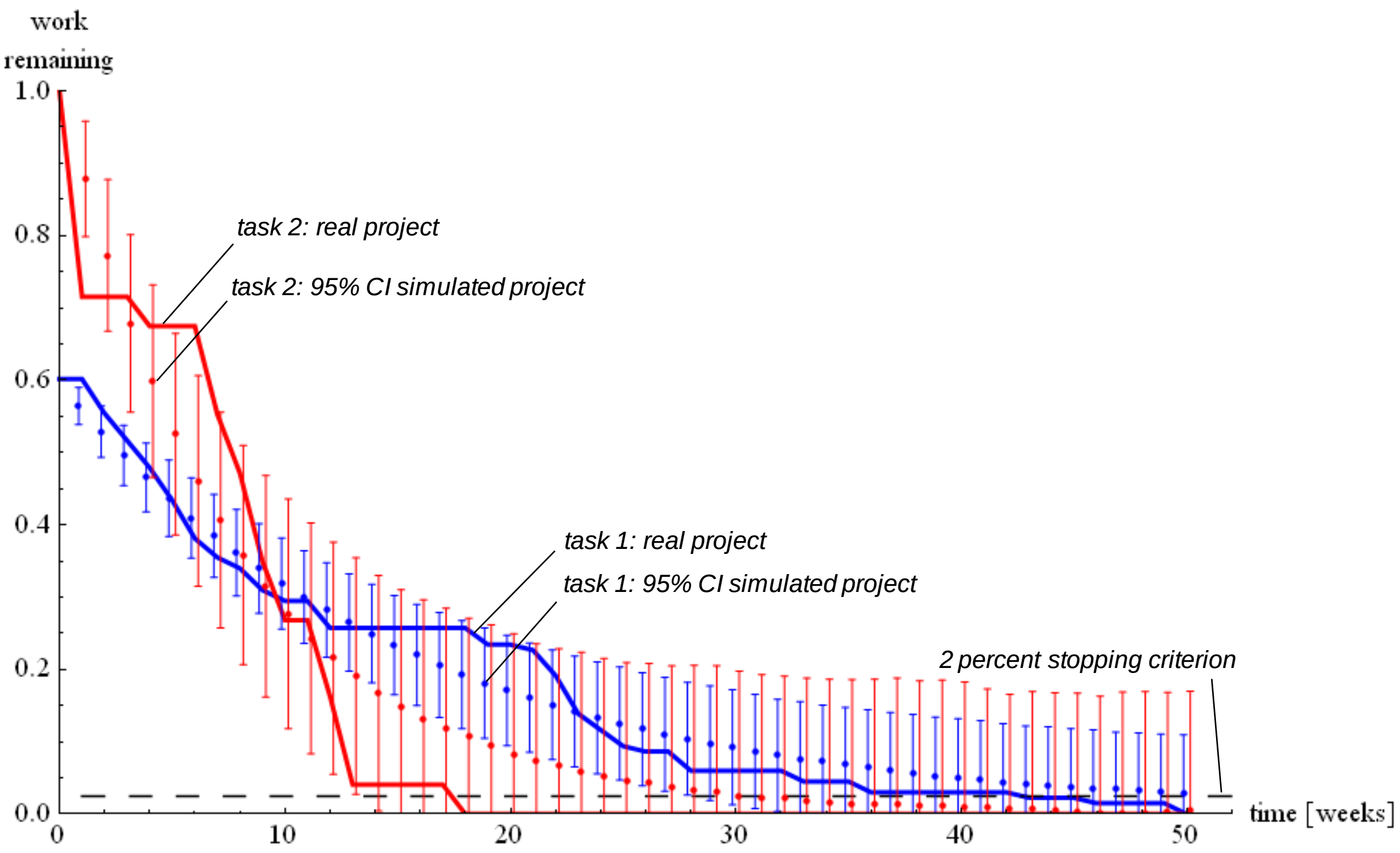

**Figure 3**

*Figure 3*. List plot of work remaining in the real NPD project. The data were acquired in a small industrial company in Germany. Only the overlapping range of the first two tasks, "conceptual sensor design" (task 1) and "design of circuit diagram" (task 2), is shown. The plot also shows means of simulated time series of task processing as note points and 95% confidence intervals as error bars. The Monte Carlo simulation was based on state eq. 3 in conjunction with the least square estimates of the independent parameters according to eqs. 45, 46 and 47. A total of 1000 separate and independent runs were calculated. Note points have been offset to distinguish the error bars. The stopping criterion of 2% is marked by a dashed line at the bottom of the plot.

The root-mean-square deviation (RMSD) between the work remaining in task 1 predicted by the 1000 simulation runs and the empirical data is $\mathrm{RMSD}_{\mathrm{task1}} = 0.046$. For task 2, the deviation is more than twice that value: $\mathrm{RMSD}_{\mathrm{task2}} = 0.107$. Regarding the established scientific standards of organizational simulation (see Rouse and Boff 2005), the total deviation is low, and this confirms the validity of the model.

In total, a parameter vector $\theta_1 = \left[x_0^{(1)} \quad x_0^{(2)} \quad a_{11} \quad a_{12} \quad a_{21} \quad a_{22} \quad c_{11} \quad c_{22} \quad \rho \quad \delta\right]$ with 10 components is necessary for modeling task 1 and 2 in this phase of the project based on a VAR(1) model.

If the alternative formulation with a forcing matrix $K$ in the original state-space coordinates is used for project modeling (eq. 17), we have in addition to the initial state $x_0$ (eq. 45) and $A_0$ (eq. 46) the estimated independent parameters

$$\hat{K} = \begin{pmatrix} -0.1355 & 0.9908 \\ 0.9908 & 0.1355 \end{pmatrix}$$

and

$$\eta_t \sim \mathcal{N}\left(\begin{pmatrix}0\\0\end{pmatrix}, \begin{pmatrix}\lambda_1(C) = (0.00176)^2 & 0\\ 0 & \lambda_2(C) = (0.00015)^2\end{pmatrix}\right).$$

The parameter vector is $\theta_2 = \begin{bmatrix}x_0^{(1)} & x_0^{(2)} & a_{11} & a_{12} & a_{21} & a_{22} & k_{11} & k_{12} & \lambda_1(C) & \lambda_2(C) & \delta\end{bmatrix}$.

For the sake of completeness, the estimated independent parameters in the spectral basis (eqs. 24 and 25) are presented as well. The eigendecomposition of the dynamical operator $A_0$ according to eq. 16 leads to the matrix of eigenvectors:

$$S = \begin{pmatrix}0.9924 & 0.0247\\ 0.1228 & 0.9997\end{pmatrix}.$$

The estimate of the initial state $\hat{x}_0' = S^{-1} \cdot \hat{x}_0$ is given by

$$\hat{x}_0' = \begin{pmatrix}x_0'^{(1)} = 0.5830\\ x_0'^{(2)} = 0.9287\end{pmatrix},$$

the transformed dynamical operator is given by

$$\hat{\Lambda}_S = \begin{pmatrix}\lambda_1(A_0) = 0.9404 & 0\\ 0 & \lambda_2(A_0) = 0.8722\end{pmatrix},$$

and the estimated covariance $\hat{C}'$ of the normally distributed performance fluctuations is represented by

$$\varepsilon_t' \sim \mathcal{N}\left(\begin{pmatrix}0\\0\end{pmatrix}, \begin{pmatrix}\sigma_{11}'^{\,2} = (\,0.0141)^2 & \rho'\sigma_{11}'\sigma_{22}' = -\,0.48 \cdot 0.0141 \cdot 0.0424\\ \rho'\sigma_{11}'\sigma_{22}' & \sigma_{22}'^{\,2} = (0.0424)^2\end{pmatrix}\right).$$

Interestingly, the basis transformation slightly reinforces variances $c_{11}'$ and $c_{22}'$ and correlation coefficient $\rho_{12}'$. The corresponding parameter vector is $\theta_3 = \begin{bmatrix}x_0'^{(1)} & x_0'^{(2)} & \lambda_1(A_0) & \lambda_2(A_0) & \sigma_{11}' & \sigma_{22}' & \rho' & \delta\end{bmatrix}$.

It is important to note that in accordance with Akaike´s two model selection criteria (FPE and AIC), minimum scores were assigned to a model of second order, $n_{opt} = 2$, not to a first order autoregression model. In this case we have $FPE(n_{opt} = 2) = -16.06$ and $AIC(n_{opt} = 2) = -12.45$. The estimated WTMs $\hat{A}_0$ and $\hat{A}_1$ for this extended model are given by

$$\hat{A}_0 = \begin{pmatrix}1.1884 & -0.1476\\ 0.0470 & 1.1496\end{pmatrix} \tag{48}$$

$$\hat{A}_1 = \begin{pmatrix}-0.2418 & 0.1344\\ -0.0554 & -0.2622\end{pmatrix}. \tag{49}$$

The estimated covariance matrix $\hat{C}$ is given by the representation

$$\varepsilon_t \sim \mathcal{N}\left(\begin{pmatrix}0\\0\end{pmatrix}, \begin{pmatrix}(0.0116)^2 & -0.013 \cdot 0.0116 \cdot 0.0257\\ -0.013 \cdot 0.0116 \cdot 0.0257 & (0.0257)^2\end{pmatrix}\right). \tag{50}$$

Figure 4 shows the corresponding list plots of real and simulated work remaining in the overlapping range of tasks 1 and 2 over 50 weeks.

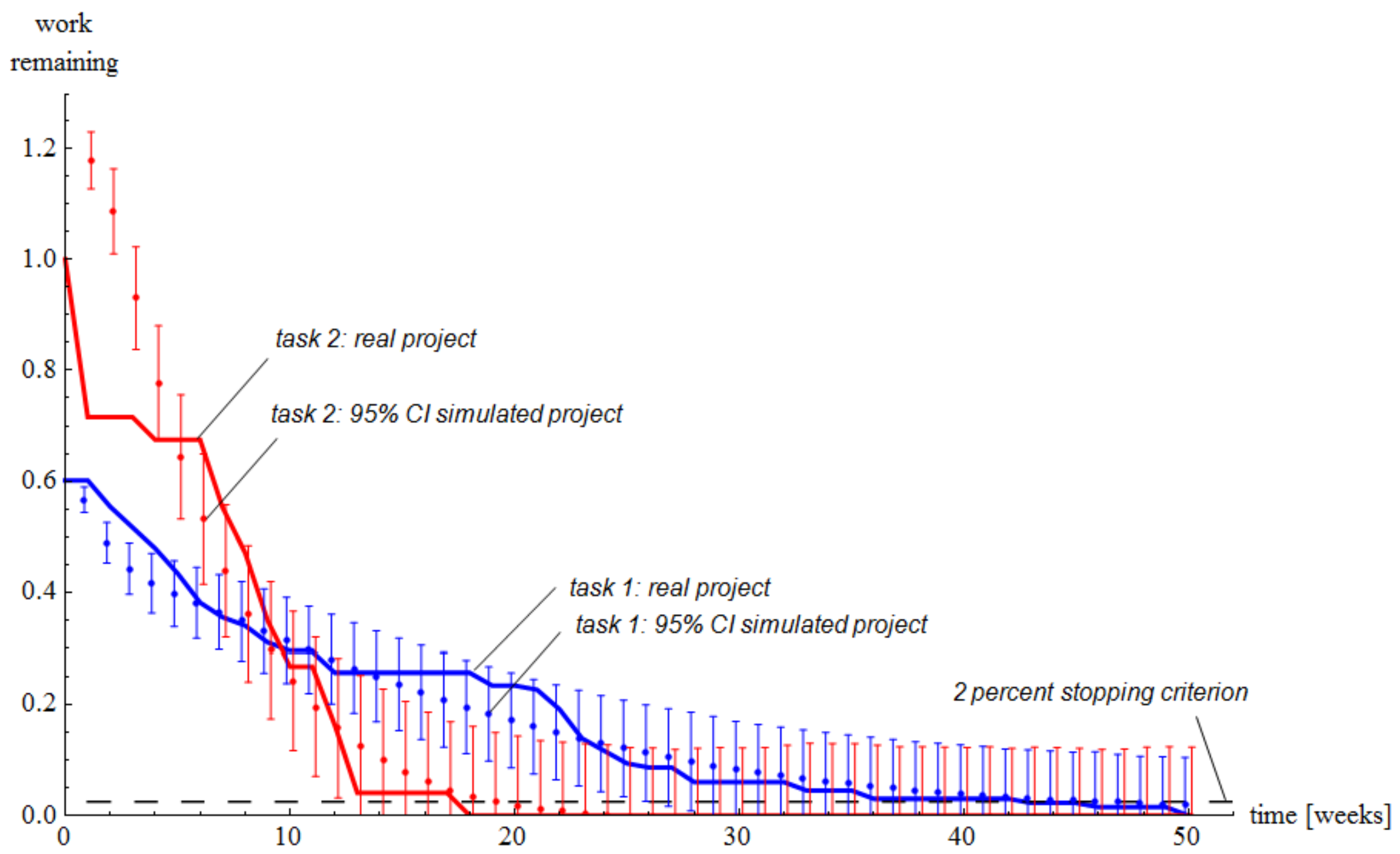

**Figure 4**

*Figure 4*. List plot of work remaining in the real and simulated NPD projects. As in Figure 3, only the overlapping range of the first two tasks is shown. The means of simulated time series of task processing are shown as note points and 95% confidence intervals appear as error bars. The Monte Carlo simulation was based on state eq. 38. The selected order of the autoregression model was $n_{opt} = 2$. The least square estimates of the independent parameters are given in eqs. 45, 48, 49 and 50. A total of 1000 separate and independent runs were calculated. Note points have been offset to distinguish the error bars. The stopping criterion of 2% is marked by a dashed line at the bottom of the plot.

The simulation model is based on state eq. 38 in conjunction with parameter estimates from eqs. 45, 48, 49 and 50. As before, 1,000 separate and independent simulation runs were calculated. The RMSD between the work remaining in task 1 predicted by the project simulations and the field data is $\mathrm{RMSD_{task1}} = 0.050$. For task 2, the deviation is more than twice that value: $\mathrm{RMSD_{task2}} = 0.115$. For both tasks the deviations for the second order model are slightly larger than for the first order model. Although the prediction error of the second-order model is only slightly larger, significant qualitative differences between both models exist and can be clearly seen in Figure 4 between the first and third week of both tasks in which the confidence intervals do not include the data points from the real project. In particular the confidence intervals computed for weeks 1, 2 and 3 of task 1 demonstrate low goodness-of-fit indices and the mean values as centers of the confidence intervals are obviously far too high. However, Figure 4 also shows that these are the most important deviations and in total 47 out of 50 confidence intervals of the simulated work remaining in task 1 include the data points from the real project before the stopping criterion is met. For task 2 the goodness-of-fit is comparable, and 46 out of the 50 confidence intervals cover the empirical data points before the stopping criterion is satisfied. In contrast to the first-order autoregression model, the relatively abrupt completion of task 2 in week 18 is accurately predicted by the smoothly decaying means of the second-order model. According to the center of the confidence interval computed for week 19 the simulated processing of task 2 will on average be finished at week 19.

## 2.6 Stochastic Formulation with Periodically Correlated Processes

An extension of the introduced autoregressive approach to dynamic modeling of CE projects that is especially interesting from theoretical and practical perspectives is to formulate a so-called “periodic vector autoregressive" (PVAR) stochastic process (Franses and Paap 2004, Ursu and Duchesne 2009). In principle, a PVAR model can capture not only the dynamic processing of the development tasks with short iteration length but also the long-scale effects of withholding the release of information on purpose. According to the previous validation study, short iterations for a given amount of work are necessary to process and disseminate component-level design information within development teams. Frequent iterations between teams carrying out component level- and system-level design are also necessary if the scope of predictability for the development project is small and only a few stable assumptions can be made about the design ranges or physical functions of the product under development. A corresponding simulation study is presented in section 5.1. In fact, the organization of the problem-solving processes is such that, by definition, the tasks are cooperatively processed, in the sense that information about their individual progress and problems cannot be hidden. An additional long-scale effect, however, often occurs in large-scale CE projects with a hierarchical coordination structure (Yassine et al. 2003), because system-level teams may withhold the release of a certain fraction of information about integration and tests of geometric/topological entities for a limited period. Between the releases, new information is “hidden” (kept secret), and work in the subordinate teams is based on the old state of knowledge. Such a hold-and-release policy is typical for CE projects in the automotive or aerospace industry. This kind of “noncooperative behavior” is justified by the desire to improve solution maturity and reduce the burden of coordination. A deterministic model capable of capturing both cooperative and noncooperative task processing was developed by Yassine et al. (2003). In their seminal paper, a time-variant (nonautonomous) state equation was formulated and validated based on simulation runs. We built directly upon their results in the following. However, the PVAR approach can also account for unpredictable fluctuations in performance and can be the basis for analytical complexity evaluations (see section 3).

To formulate a PVAR model of a CE project with a hierarchical coordination structure, it is assumed that a certain amount of finished system-related work is released by the development teams responsible for system integration and testing to component-level teams only at time steps $ns$ ($n \in \mathbb{N}$) with fixed, predetermined period $s \geq 2$. At all other time steps, $ns + v$ ($n = 0,1,\ldots; v = 1,\ldots,s-1$), the tasks are processed by short iterations without withholding information. Under these assumptions, the stochastic difference equation 4 can be generalized to a process $\{X_{ns+v}\}$ with periodically correlated components. The state equation is

$$X_{ns+v} = \Phi_1(v) \cdot X_{ns+v-1} + \varepsilon_{ns+v}, \tag{51}$$

where the index $n$ indicates the long time scale with period $s$, and $v$ the short time scale. $X_t = \left(X_t(1),\ldots,X_t(d)\right)^{\mathrm{T}}$ is a $d \times 1$ single random vector encoding the state of the project at time step $t = ns + v$ ($n = 0,1,\ldots; v = 1,\ldots,s$). The leading components of the state vector $X_t$ represent the work remaining of the $p^{\mathrm{C}} \in \mathbb{N}$ component-level and $p^{\mathrm{S}} \in \mathbb{N}$ system-level tasks that are processed on the short time scale. For these tasks, the work transformation can be captured by a combined WTM $A_0$ as

$$A_0 = \begin{pmatrix} A_0^{\mathrm{C}} & A_0^{\mathrm{SC}} \\ A_0^{\mathrm{CS}} & A_0^{\mathrm{S}} \end{pmatrix}.$$

In the combined WTM $A_0$, the submatrix $A_0^{\mathrm{C}}$ of size $p^{\mathrm{C}} \times p^{\mathrm{C}}$ is the dynamic operator for the cooperative processing of component-level tasks. The $p^{\mathrm{S}} \times p^{\mathrm{S}}$ submatrix $A_0^{\mathrm{S}}$ refers to system-level tasks in an analogous manner. The $p^{\mathrm{C}} \times p^{\mathrm{S}}$ submatrix $A_0^{\mathrm{SC}}$ determines the rework fraction created by system-level tasks for the corresponding component-level tasks in each short iteration, whereas the $p^{\mathrm{S}} \times p^{\mathrm{C}}$ submatrix $A_0^{\mathrm{CS}}$ determines the rework fraction created by component-level tasks for the system-level tasks. Moreover, the substates $\left(X_t(1), \dots, X_t\left(p^{\mathrm{C}} + p^{\mathrm{S}}\right)\right)$ have to be augmented by other $p^{\mathrm{S}}$ substates to account for the periodic hold-and-release policy of system-level design information. The augmented $p^{\mathrm{S}}$ substates do not represent the work remaining as the leading states do, but represent the amount of finished work on the system level that has accumulated over the short iterations. The finished work remains hidden for the component-level teams until it is released at time step $ns$. Through the PVAR model formulation, the finished work can be placed in a hold state. The associated $p^{\mathrm{S}} \times p^{\mathrm{S}}$ submatrix $A_0^{\mathrm{SH}}$ captures the rework fraction created by the system-level tasks in each iteration at time step $v = 1, \dots, s-1$. After release, additional rework is generated for the component-level tasks. This rework is calculated based on the WTM $A_0^{\mathrm{HC}}$. This WTM is of size $p^{\mathrm{C}} \times p^{\mathrm{S}}$. There, $d = p^{\mathrm{C}} + 2p^{\mathrm{S}}$ holds.

The periodically correlated work processes in the project are represented by the time evolution of the state vector $X_{ns+v}$ based on the autoregressive coefficients $\Phi_1(v)$. The two time scales correspond to indices $n$ and $v$. The long time scale is indexed by $n$. In seasonal macroeconomic models, for instance, $n$ indicates the year that the acquired samples of the time series refer to (e.g. income and consumption). However, in large-scale CE projects the release period is much shorter and covers typically intervals of four to eight weeks. On the other hand, the short time scale is indexed by $v$. On this scale, the iterations usually occur on a daily or weekly basis (see section 2.5). In the terminology of macroeconomic models, $v$ indicates a specific "season" of the "year". Furthermore, the length $s$ of the period between releases of hidden information ("number of seasons within year") has to be specified by the project manager. For a predetermined period length $s$, the random vector $X_{ns+v}$ contains the realization of work remaining during the $v$th iteration over all component-level and system-level tasks at the release period $n$ and the amount of finished work on the system level that is ready to be released to component-level tasks in period $n+1$. According to the previous analysis the state vector can be separated into three substate vectors and defined as

$$X_t = \begin{pmatrix} X_t^{\mathrm{C}} \\ X_t^{\mathrm{S}} \\ X_t^{\mathrm{H}} \end{pmatrix}$$

with

$$X_t^{\mathrm{C}} = \begin{pmatrix} X_t(1) \\ \vdots \\ X_t\left(p^{\mathrm{C}}\right) \end{pmatrix}$$

$$X_t^{\mathrm{S}} = \begin{pmatrix} X_t\left(p^{\mathrm{C}} + 1\right) \\ \vdots \\ X_t\left(p^{\mathrm{C}} + p^{\mathrm{S}}\right) \end{pmatrix}$$

$$X_t^{\mathrm{H}} = \begin{pmatrix} X_t\left(p^{\mathrm{C}} + p^{\mathrm{S}} + 1\right) \\ \vdots \\ X_t\left(p^{\mathrm{C}} + 2p^{\mathrm{S}}\right) \end{pmatrix}.$$

Furthermore, the task processing on the long and short time scales has to be modeled, and for this aim, two independent dynamical operators are introduced (Yassine et al. 2003). These operators correspond to the autoregressive coefficients $\Phi_1(v)$ in state equation 51. The release of hidden information over $s$ time steps is modeled by the first dynamical operator $\Phi_1(s)$. It is assumed that the release occurs at the end of the period, that is at relative time step $v = s$. The operator $\Phi_1(s)$ can be composed of the previously defined submatrices as

$$\Phi_1(s) = \begin{pmatrix} A_0^{\mathrm{C}} & A_0^{\mathrm{SC}} & A_0^{\mathrm{HC}} \\ A_0^{\mathrm{CS}} & A_0^{\mathrm{S}} & 0 \\ 0 & 0 & \{\varepsilon\} I_{p^{\mathrm{S}}} \end{pmatrix}. \tag{52}$$

In the above equation the $\varepsilon$-symbol denotes an arbitrarily small positive quantity. The definition of positive interactions between the augmented $p^{\mathrm{S}}$ substates is necessary for explicitly evaluating the emergent complexity of the modeled large-scale CE project on the basis of the chosen information-theoretic measure EMC, which is presented and discussed in the next section. EMC simply scales linearly with $\varepsilon$. For practical purposes, it is recommended to calculate with $\varepsilon = 10^{-3}$. By doing so, the finished work after release is set back to a nonzero but negligible amount in terms of productivity.

The task processing in the $v = 1, \dots, s-1$ iterations before release is modeled on the basis of a second dynamical operator $\Phi_1(1)$. In contrast to macroeconomic models, it is assumed that the cooperative processing of the development tasks by short iterations without withholding information follows a regime in which the intensity of interactions does not change before the end of period $s$ and therefore the autoregressive coefficients are constant, that is $\Phi_1(1) = \cdots = \Phi_1(s-1)$. No other dynamical operators are needed to capture project dynamics within the period. $\Phi_1(1)$ can be composed of the previously defined submatrices in an analogous manner as

$$\Phi_1(1) = \begin{pmatrix} A_0^{\mathrm{C}} & A_0^{\mathrm{SC}} & 0 \\ A_0^{\mathrm{CS}} & A_0^{\mathrm{S}} & 0 \\ 0 & A_0^{\mathrm{SH}} & \{1-\varepsilon\} I_{p^{\mathrm{S}}} \end{pmatrix}. \tag{53}$$

By the same reasoning, it is assumed that the covariance of the noise vector $\varepsilon_{ns+v}$ representing unpredictable performance fluctuations when processing the development tasks cooperatively by short iterations is constant for all $v = 1, \dots, s-1$ iterations before the end of period $s$ and no other random effects influence the cooperative work within the period.

The evolution of the process state $X_t$ over $t$ time steps can therefore be expressed by two dynamical operators $\Phi_1(1)$ and $\Phi_1(s)$ in conjunction with two noise vectors $\varepsilon_1$ and $\varepsilon_s$ as

$$X_t = \begin{cases} \Phi_1(1) \cdot X_{t-1} + \varepsilon_1 & \text{for } t = ns + v \text{ with } n = 0,1,\dots \text{ and } v = 1,\dots,s-1 \\ \Phi_1(s) \cdot X_{t-1} + \varepsilon_s & \text{for } t = ns \text{ with } n = 1,2,\dots . \end{cases}$$

The noise vectors $\varepsilon_1$ and $\varepsilon_s$ correspond to zero-mean white noise and have zero means and covariances $C_1$ and $C_s$, respectively:

$$\varepsilon_1 \sim \mathcal{N}(0_d, C_1)$$

$$\varepsilon_s \sim \mathcal{N}(0_d, C_s).$$

Hence, the combined error process $\{\varepsilon_{ns+v}\}$ can also be expressed by a zero-mean periodic white noise process. $\{\varepsilon_{ns+v}\}$ is composed of $d \times 1$ random vectors, such that $E[\varepsilon_{ns+v}] = 0_d$ and $E[\varepsilon_{ns+v}\, {\varepsilon_{ns+v}}^{\mathrm{T}}] =$ $C_1$ for $v = 1, \ldots, s-1$, and $E[\varepsilon_{ns+v}] = 0_d$ and $E[\varepsilon_{ns+v}\, \varepsilon_{ns+v}{}^{\mathrm{T}}] = C_s$ for $v = s$. It is assumed that the covariance matrices $C_1$ and $C_s$ are not singular.

If all parallel tasks are initially to be completed in full and no finished work exists in hold state, the initial state is simply

$$x_0 = \begin{pmatrix} \begin{pmatrix} 1 \\ \vdots \\ 1 \end{pmatrix} \\ \begin{pmatrix} 1 \\ \vdots \\ 1 \end{pmatrix} \\ \begin{pmatrix} 0 \\ \vdots \\ 0 \end{pmatrix} \end{pmatrix},$$

where the first row vector of ones has size $|A_0^{\mathrm{C}}|$, the second row vector of ones has size $|A_0^{\mathrm{S}}|$ and the third vector of zeros has size $|A_0^{\mathrm{SH}}|$.

An important result from the theory of stochastic processes (Franses and Paap 2004, Ursu and Duchesne 2009) is that the PVAR process in eq. 51 also offers a compact representation as a VAR(1) model:

$$\Phi_0^* \cdot X_n^* = \Phi_1^* \cdot X_{n-1}^* + \varepsilon_n^*, \tag{54}$$

where $X_n^* = \left(X_{ns+s}^{\mathrm{T}}, X_{ns+s-1}^{\mathrm{T}}, \ldots, X_{ns+1}^{\mathrm{T}}\right)^{\mathrm{T}}$ and $\varepsilon_n^* = \left(\varepsilon_{ns+s}^{\mathrm{T}}, \varepsilon_{ns+s-1}^{\mathrm{T}}, \ldots, \varepsilon_{ns+1}^{\mathrm{T}}\right)^{\mathrm{T}}$ are $ds \times 1$ state and error vectors, respectively. The matrix $\Phi_0^*$ and the autoregressive coefficient $\Phi_1^*$ are given by the nonsingular matrices

$$\begin{aligned} \Phi_0^* &= \begin{pmatrix} I_d & -\Phi_1(s) & 0 & \cdots & 0 & 0 \\ 0 & I_d & -\Phi_1(s-1) & \cdots & 0 & 0 \\ \vdots & & & \ddots & & \vdots \\ 0 & 0 & 0 & \cdots & I_d & -\Phi_1(2) \\ 0 & 0 & 0 & \cdots & 0 & I_d \end{pmatrix} \\ &= \begin{pmatrix} I_d & -\Phi_1(s) & 0 & \cdots & 0 & 0 \\ 0 & I_d & -\Phi_1(1) & \cdots & 0 & 0 \\ \vdots & & & \ddots & & \vdots \\ 0 & 0 & 0 & \cdots & I_d & -\Phi_1(1) \\ 0 & 0 & 0 & \cdots & 0 & I_d \end{pmatrix} \end{aligned} \tag{55}$$

and

$$\Phi_1^* = \begin{pmatrix} 0 & 0 & \cdots & 0 \\ 0 & 0 & & 0 \\ \vdots & & \ddots & \vdots \\ \Phi_1(1) & 0 & \cdots & 0 \end{pmatrix}. \tag{56}$$

The matrices $\Phi_0^*$ and $\Phi_1^*$ are both of size $ds \times ds$.

The error process $\{\varepsilon_n^*\}$ also corresponds to a zero-mean periodic white noise process with $E[\varepsilon_n^*] = 0_{ds}$ and $E\left[\varepsilon_n^*\, \varepsilon_n^*{}^{\mathrm{T}}\right] = C^*$. The covariance matrix $C^*$ is not singular. We assume that the vectors $\varepsilon_{ns+s}^{\mathrm{T}}, \varepsilon_{ns+s-1}^{\mathrm{T}}, \ldots$ are uncorrelated and that $C^*$ can be expressed by

$$C^* = \begin{pmatrix} C_s & 0 & 0 & 0 \\ 0 & C_1 & 0 & 0 \\ 0 & 0 & \ddots & 0 \\ 0 & 0 & 0 & C_1 \end{pmatrix}. \tag{57}$$

If all parallel tasks are initially to be completed in full, the initial state in the VAR(1) representation is

$$x_0^* = \begin{pmatrix} x_0 \\ \begin{pmatrix} 0 \\ \vdots \\ 0 \end{pmatrix} \\ \vdots \\ \begin{pmatrix} 0 \\ \vdots \\ 0 \end{pmatrix} \end{pmatrix}, \tag{58}$$

where the row vectors of zeros following the original initial state $x_0$ have size $|x_0|$. A total of $s-1$ row vectors of zeros must be appended to the original initial state for a complete state representation.

It is clear that the VAR(1) model according to eq. 54 can be rewritten as

$$X_n^* = (\Phi_0^*)^{-1} \cdot \Phi_1^* \cdot X_{n-1}^* + (\Phi_0^*)^{-1} \cdot \varepsilon_n^*. \tag{59}$$

The matrix $\Phi_0^*$ can be easily inverted:

$$\begin{aligned} &(\Phi_0^*)^{-1} \\ &= \begin{pmatrix} I_d & \Phi_1(s) & \Phi_1(s-1)\Phi_1(s-2) & \cdots & \Phi_1(s-1)\Phi_1(s-2)\ldots\Phi_1(3) & \Phi_1(s)\Phi_1(s-1)\ldots\Phi_1(2) \\ 0 & I_d & \Phi_1(s-1) & \cdots & \Phi_1(s-2)\ldots\Phi_1(3) & \Phi_1(s-1)\ldots\Phi_1(2) \\ \vdots & & & \ddots & & \vdots \\ 0 & 0 & 0 & \cdots & I_d & \Phi_1(2) \\ 0 & 0 & 0 & \cdots & 0 & I_d \end{pmatrix} \\ &= \begin{pmatrix} I_d & \Phi_1(s) & \Phi_1(1)^2 & \cdots & \Phi_1(1)^{s-3} & \Phi_1(s)\Phi_1(1)^{s-2} \\ 0 & I_d & \Phi_1(1) & \cdots & \Phi_1(1)^{s-4} & \Phi_1(1)^{s-2} \\ \vdots & & & \ddots & & \vdots \\ 0 & 0 & 0 & \cdots & I_d & \Phi_1(1) \\ 0 & 0 & 0 & \cdots & 0 & I_d \end{pmatrix}. \end{aligned} \tag{60}$$

The most convenient VAR(1) representation is therefore

$$X_n^* = A_0^* \cdot X_{n-1}^* + \varepsilon_n^\star,$$

with the combined dynamical operator

$$A_0^* = (\Phi_0^*)^{-1} \cdot \Phi_1^* = \begin{pmatrix} \Phi_1(s)(\Phi_1(1))^{s-1} & 0 & 0 & \cdots & 0 \\ (\Phi_1(1))^{s-1} & 0 & 0 & \cdots & 0 \\ (\Phi_1(1))^{s-2} & 0 & 0 & \cdots & 0 \\ \vdots & \vdots & 0 & \ddots & \vdots \\ \Phi_1(1) & 0 & 0 & \cdots & 0 \end{pmatrix}$$

and the transformed error vector $\varepsilon_n^\star = (\Phi_0^*)^{-1} \cdot \varepsilon_n^*$ with covariance

$$C^\star = (\Phi_0^*)^{-1} \cdot C^* \cdot (\Phi_0^*)^{-\mathrm{T}}.$$

In an analogous manner to section 2.2, a closed-form solution to the sum of the mean work remaining can be calculated across $T = ns$ time steps for an arbitrary project with predefined initial state $x_0^*$ and combined dynamical operator $A_0^*$ as

$$
\begin{aligned}
E\left[\sum_{i=0}^{n} X_n^*\right] &= \sum_{i=0}^{n} E[X_n^*] \\
&= \sum_{i=0}^{n} \left((A_0^*)^i \cdot x_0\right) \\
&= \left(\sum_{i=0}^{n} (A_0^*)^i\right) \cdot x_0^* \\
&= \left(I_p - A_0^*\right)^{-1} \cdot \left(I_p - (A_0^*)^n\right) \cdot x_0^*.
\end{aligned}
$$

In the limit $T \to \infty$, we have for an asymptotically stable project with periodically correlated processes the expected cumulated work remaining:

$$
\lim_{n\to\infty} E\left[\sum_{i=0}^{n} X_n^*\right] = \left(I_p - A_0^*\right)^{-1} \cdot x_0^*.
$$

The expected total work $x_{tot}$ in the project can be calculated by summing over the vector components of the expected cumulated work remaining:

$$
x_{tot} = \text{Total}\left[\left(I_p - A_0^*\right)^{-1} \cdot x_0^*\right]. \tag{61}
$$

To evaluate explicitly the intricate interrelationships between project dynamics and emergent complexity, the stochastic process must satisfy the criterion of strict stationarity. We will return to this point in section 4.1. A strictly stationary process has a joint probability density that is invariant with shifting the origin, and therefore the locus and variance do not change over time. It is clear that the periodic autoregression model in eq. 51 is a non-stationary model as the variance and autocovariances take different values in different time steps ("seasons"). In order to facilitate the analysis of stationarity, the introduced time-invariant representations as VAR(1) models (cf. eq. 54) have to be considered. These models are stationary if the roots of the characteristic equation

$$
\text{Det}[\Phi_0^* - \Phi_1^* z] = 0
$$

are outside the unit circle ($z \in \mathbb{C}$). The characteristic equation can be simplified to

$$
\begin{aligned}
\text{Det}[\Phi_0^* - \Phi_1^* z] &= \text{Det}[I_{ds} - ((\Phi_0^*)^{-1} \cdot \Phi_1^*) z] \\
&= \text{Det}\left[I_d - \left(\prod_{k=0}^{s-1} \Phi_1(s-k)\right) z\right] \\
&= \text{Det}\left[I_d - \Phi_1(s) \cdot \left(\Phi_1(1)\right)^{s-1} z\right] = 0.
\end{aligned}
$$

Hence, the series of task processing are stationary if the roots of the matrix $\Phi_1(s) \cdot \left(\Phi_1(1)\right)^{s-1}$ are within the unit circle (Franses and Paap 2004, Ursu and Duchesne 2009).

## 2.7 Extended Least Squares Parameter Estimation and Model Selection

The autoregressive coefficients and the error covariances of a PVAR model of arbitrary order (not necessarily limited to a first-order autoregression order for each season as previously formulated) without linear constraints on the independent parameters can be calculated efficiently on the basis of standard least-squares or maximum-likelihood estimation techniques from textbooks (see, e.g. Brockwell and Davis 1991, Franses and Paap 2004 and Lütkepohl 2005). To apply these techniques, one only has to bring the introduced linear recurrent relations into a standard regression form and then execute the (usually iterative) estimation procedures. However, in the developed model formulation we had to pose the constraint that some entries of the dynamical operators $\Phi_1(1)$ (eq. 52) and $\Phi_1(s)$ (eq. 53) must be zero in order to be able to model the typical hold-and-release policy of design information in a CE project with periodically correlated work processes or must be equal to $\varepsilon$ or $(1-\varepsilon)$ for an analytical evaluation of emergent complexity in later sections. Furthermore, some coefficients of the work transformation submatrices are linear dependent. Consequently, we cannot use the standard estimation techniques but instead have to use an extended algorithm developed by Ursu and Duchesne (2009) that is able to carry out least-squares estimation with linear constraints on the regression parameters. For the given model formulation with zero-mean periodic white noise the least squares estimators are equivalent to maximum likelihood estimators. In the following sections we will present closed-form solutions of the estimators of different strength based on the original work of Ursu and Duchesne (2009). To allow the interested reader to follow the accompanying proofs in the original material, we will use a similar notation to the developers of the algorithm. We start by presenting a convenient regression form and then proceed with the specification of the least square estimators for the full unconstraint case and the more complex restricted case.

In principle, the asymptotic properties of the least square estimators could be derived from generalized results for time series based on the multivariate representation from eq. 54 (see, e.g. Brockwell and Davis 1991 and Lütkepohl 2005). However, to estimate the statistical properties of the autoregressive coefficient matrices for each release period $v$ of the formulated model of a CE project with periodically correlated work processes, the multivariate stochastic process generated by state eq. 54 needs to be inverted. This operation seems to be unnecessarily complex in the multivariate setting. Instead, it is more efficient to use the individual PVAR components directly in parameter estimation. Consider the sequence of aggregated random variables $\{X_{ns+v}, 0 \leq ns+v < Ts\}$ representing the task processing in the CE project over $T$ time steps, $t = 0, 1, \ldots, T-1$. At each time step the $v = 1, \ldots, s$ short iterations of the development teams without purposefully withholding design information are aggregated and therefore the states in between the $n$ long iterations in which the release of hidden information occurs are combined. Hence, we have a total sample size equal to $n' = Ts$. For a convenient state representation let

$$Z(v) = \left(X_v, X_{s+v}, \ldots, X_{(T-1)s+v}\right) \tag{62}$$

$$E(v) = \left(\varepsilon_v, \varepsilon_{s+v}, \ldots, \varepsilon_{(T-1)s+v}\right) \tag{63}$$

$$\boldsymbol{X}(v) = \left(\boldsymbol{X}_0(v), \ldots, \boldsymbol{X}_{T-1}(v)\right) \tag{64}$$

all be $d \times T$ random matrices, where

$$\boldsymbol{X}_t(v) = X_{ts+v-1},$$

denotes the $d \times 1$ random vectors of work remaining of the component-level and system-level tasks at time steps $t = 0, 1, \ldots, T-1$. Utilizing these aggregated variables, the PVAR model from eq. 51 can be recast and written in the following convenient regression form:

$$Z(v) = B(v) \cdot \boldsymbol{X}(v) + E(v), \qquad v = 1, \ldots, s. \tag{65}$$

The independent parameters of the regression model are collected in the $d \times d$ parameter matrix $B(v)$. Using the definitions of the dynamical operators $\Phi_1(1)$ and $\Phi_1(s)$, the parameter matrix can be defined as:

$$B(v) = \begin{cases} \Phi_1(1) & \text{for } v = 1, \ldots, s-1 \\ \Phi_1(s) & \text{for } v = s \,. \end{cases}$$

Since for all $v = 1, \ldots, s-1$ the regression equation 65 contains the same unknown regression parameters, it is convenient to concatenate them into one equation

$$Z_c = \Phi_1(1) \cdot X_c + E_c \,, \tag{66}$$

where

$$Z_c = \big(Z(1), \ldots, Z(s-1)\big) \tag{67}$$

$$\boldsymbol{X}_c = \big(\boldsymbol{X}(1), \ldots, \boldsymbol{X}(s-1)\big) \tag{68}$$

$$E_c = \big(E(1), \ldots, E(s-1)\big). \tag{69}$$

Convenient vector representations of the regression equations 65 and 66 can be obtained by using the Kronecker product $\otimes$. A vectorization of the dependent variables based on the Kronecker product leads to

$$z_1 = vec[Z_c] = \big(\boldsymbol{X}_c^{\mathrm{T}} \otimes I_d\big) \cdot \beta(1) + vec[E_c] \tag{70}$$

and

$$z_2 = vec[Z(s)] = \big(\boldsymbol{X}^{\mathrm{T}}(s) \otimes I_d\big) \cdot \beta(s) + vec[E(s)]. \tag{71}$$

The vectors of the regression coefficients are given by

$$\beta(v) = vec[\Phi_1(v)] \qquad v = 1, \ldots, s.$$

In general, the vector operator $vec[A]$ represents the vector obtained by stacking the columns of the matrix $A$ onto each other. We can also combine both regression equations in one large equation:

$$z = \begin{bmatrix} z_1 \\ z_2 \end{bmatrix} = \begin{bmatrix} \boldsymbol{X}_c^{\mathrm{T}} \otimes I_d & 0 \\ 0 & \boldsymbol{X}^{\mathrm{T}}(s) \otimes I_d \end{bmatrix} \cdot \begin{bmatrix} \beta(1) \\ \beta(s) \end{bmatrix} + e \,. \tag{72}$$

The combined noise vector is given by

$$e = \begin{bmatrix} vec[E_c] \\ vec[E(s)] \end{bmatrix}.$$

The vector of regression coefficients

$$\beta = \begin{bmatrix} \beta(1) \\ \beta(s) \end{bmatrix}$$

contains by parts the same elements (see the definitions of the matrices in eqs. 52 and 53), i.e. they are linear dependent. Furthermore, many elements of $\beta$ are known to be zero. This can be expressed by the following linear relation:

$$\beta = R \cdot \xi + b. \tag{73}$$

The vector $\xi$ represents a $K \times 1$ vector of unknown regression parameters in the restricted case. It is obvious that the parameter setting

$$R = I_{2d^2} \quad \text{and} \quad b = 0$$

represents the full unconstraint case, in which no constraints are imposed on the entries of the dynamical operators $\Phi_1(1)$ and $\Phi_1(s)$. Through the specification of the entries in $R$ and $b$, additional linear constraints can be imposed on the parameters for each release period $v$. If, for instance, a matrix entry in $\Phi_1(1)$ must be zero, the corresponding row vector of $R$ and vector component of $b$ are set to zero. Through this encoding the null entries in the dynamical operators are ignored in least squares estimation and only the non-zero informational couplings among tasks are determined. Secondly, if the $n$-th component of $\beta$ is related to the $m$-th component of the irreducible regression vector $\xi$ then we set the element $R_{[[n,m]]} = 1$.

When we convert the linear relation from eq. 73 into a vector representation that is similar to eq. 71, we arrive at the following expression for $z$:

$$\begin{aligned} z &= \left(\boldsymbol{X}^{\mathrm{T}} \otimes I_d\right) \cdot \beta + e \\ &= \left(\boldsymbol{X}^{\mathrm{T}} \otimes I_d\right) \cdot (R \cdot \xi + b) + e, \end{aligned} \tag{74}$$

where

$$\boldsymbol{X}^{\mathrm{T}} = \begin{bmatrix} \boldsymbol{X}_c^{\mathrm{T}} & 0 \\ 0 & \boldsymbol{X}^{\mathrm{T}}(s) \end{bmatrix}.$$

The least squares estimators of the parameter vectors $\xi$ are calculated by minimizing the generalized least squares criterion:

$$\Im_g(\xi) = e^{\mathrm{T}} (I_T \otimes C_e)^{-1} e. \tag{75}$$

The matrix $C_e$ represents the covariance matrix of the combined noise vector $e$, that is $C_e = E[e\, e^{\mathrm{T}}]$. It can be easily composed of the individual covariance matrices $C_1$ and $C_s$.

In the unrestricted case, an equivalent representation of the least squares estimators based on the above generalized least squares criterion $\Im_g(\xi)$ can be obtained by minimizing the ordinary least squares:

$$\Im(\beta) = e^{\mathrm{T}} \cdot e\ .$$

A similar result holds for VAR models, see Schneider and Neumaier (2001) and Lütkepohl (2005). To obtain the ordinary least squares estimators, the function $\Im(\beta)$ is differentiated with respect to each "vectorized" dynamical operator $\Phi_1(v)$:

$$\frac{\delta \Im(\beta)}{\delta vec[\Phi_1(v)]} = -2 \sum_{t=0}^{T-1} (X_{ts+v-1} \otimes \varepsilon_{ts+v}), \quad v = 1, \dots, s.$$

Setting the derivatives to zero yields the following system of equations for a given release period $v$:

$$\sum_{t=0}^{T-1} (\boldsymbol{X}_t(v) \otimes \varepsilon_{ts+v}) = 0_{d^2}.$$

In the above equation $0_{d^2}$ is the $d^2 \times 1$ null vector. Since the noise can be expressed as $\varepsilon_{ns+v} = X_{ns+v} - \left(\boldsymbol{X}_n^{\mathrm{T}}(v) \otimes I_d\right)\beta(v)$, the normal equations for each short iteration $v$ are given by

$$\sum_{t=0}^{T-1} (\boldsymbol{X}_t(v) \otimes X_{ts+v}) = \left( \sum_{t=0}^{T-1} \left( \boldsymbol{X}_t(v) \cdot \boldsymbol{X}_t^{\mathrm{T}}(v) \otimes I_d \right) \right) \beta(v).$$

Hence, the desired least squares estimators $\hat{\beta}(v)$ satisfy the relation:

$$\hat{\beta}(v) = \left(\left(\boldsymbol{X}(v) \cdot \boldsymbol{X}^{\mathrm{T}}(v)\right)^{-1} \boldsymbol{X}(v) \otimes I_d\right) z(v). \tag{76}$$

The estimated residuals are given by the difference:

$$\hat{\varepsilon}_{ns+v} = X_{ns+v} - \left(\boldsymbol{X}_t^{\mathrm{T}}(v) \otimes I_d\right)\hat{\beta}(v). \tag{77}$$

In the above estimators the independent variables $\boldsymbol{X}(v)$ and $X_{ns+v}$, respectively denote the time series of empirically acquired state vectors as single realizations of the periodically correlated work processes in the CE project. If multiple realizations of the PVAR process had been acquired in $N$ independent trials as opposed to merely a single time series is being given, the additional time series are simply appended as additional $N-1$ blocks of rows in the regression eq. 65. Similarly, the predictors are extended by additional row blocks. The initial state is determined by averaging over all initial state vectors.

Solving the least squares problem directly, Ursu and Duchesne (2009) give the following alternative equation for the least squares estimators:

$$\hat{B}(v) = Z(v) \cdot \boldsymbol{X}^{\mathrm{T}}(v) \left(\boldsymbol{X}(v) \cdot \boldsymbol{X}^{\mathrm{T}}(v)\right)^{-1}. \tag{78}$$

Based on the above relation one can also express the difference between estimator $\hat{B}(v)$ and $B(v)$ as:

$$\hat{B}(v) - B(v) = \left\{\frac{1}{T}\right\} \cdot E(v) \cdot \boldsymbol{X}^{\mathrm{T}}(v) \left(\left\{\frac{1}{T}\right\} \cdot \boldsymbol{X}(v) \cdot \boldsymbol{X}^{\mathrm{T}}(v)\right)^{-1}.$$

Noting that for the sum over $T$ time steps it holds that

$$\sum_{t=0}^{T-1} vec[\varepsilon_{ns+v} \cdot X_n^{\mathrm{T}}(v)] = vec[E(v) \cdot X^{\mathrm{T}}(v)],$$

it follows for the convergence in distribution (symbol "$\xrightarrow{d}$") that

$$\left\{\frac{1}{\sqrt{T}}\right\} \cdot vec[E(v) \cdot \boldsymbol{X}^{\mathrm{T}}(v)] \xrightarrow{d} \mathcal{N}\left(0_{d^2}, \Omega(v) \otimes C(v)\right)$$

and for the convergence in probability (symbol "$\xrightarrow{p}$") that

$$\left\{\frac{1}{T}\right\} \cdot vec[E(v) \cdot \boldsymbol{X}^{\mathrm{T}}(v)] \xrightarrow{p} 0_{d^2}.$$

The function $\mathcal{N}\left(0_{d^2}, \Omega(v) \otimes C(v)\right)$ denotes the $d^2$-variate Gaussian distribution with location $0_{d^2}$ and covariance $\Omega(v) \otimes C(v)$. The *pdf* of this distribution is given in eq. 9. $\Omega(v)$ denotes the $d \times d$ covariance matrix of the aggregated random vector $\boldsymbol{X}_t(v)$, see Ursu and Duchesne (2009).

After the derivation of the least square estimators for the full unconstraint case, we proceed with the restricted case, i.e. the case in which additional linear constraints must be satisfied. If the parameters satisfy the linear constraint in eq. 65, the least squares estimators of $\xi(v)$ minimize the generalized criterion $\Im_g(\xi)$ (eq. 75). It is obvious that the generalized criterion is not equivalent to the ordinary least squares criterion $\Im(\beta)$ (see e.g. Lütkepohl 2005). Rearranging the regression eq. 74 leads to the following relation for the combined noise vector:

$$e = z - \left(\boldsymbol{X}^{\mathrm{T}} \otimes I_d\right)(R \cdot \xi + b).$$

This relation is sufficient to derive the asymptotic properties of the least squares estimator of $\xi$ under linear constraints. Owing to limited space we will not present the stepwise derivation of $\xi$ in this paper but only cite the result from the original work by Ursu and Duchesne (2009):

$$\hat{\xi} = \left(R^{\mathrm{T}}\left(\boldsymbol{X} \cdot \boldsymbol{X}^{\mathrm{T}} \otimes C_e^{-1}\right)R\right)^{-1} R^{\mathrm{T}}\left(\boldsymbol{X} \otimes C_e^{-1}\right)\left(z - \left(\boldsymbol{X}^{\mathrm{T}} \otimes I_d\right)b\right).$$

Ursu and Duchesne (2009) show that the estimator $\hat{\xi}$ is consistent for $\xi$ and asymptotically follows a Gaussian distribution:

$$\{\sqrt{T}\} \cdot (\hat{\xi} - \xi) \xrightarrow{d} \mathcal{N}\left(0_K, \left(R^{\mathrm{T}}(\Omega \otimes C_e^{-1})R\right)^{-1}\right).$$

However, the estimator $\hat{\xi}$ is unfeasible in almost all practical applications in project management because it relies on the (usually) unknown covariance matrix $C_e$. Instead, a consistent estimator $\tilde{C}_e$ of the covariance matrix $C_e$ can be used and we have the alternative representation:

$$\hat{\hat{\xi}} = \left(R^{\mathrm{T}}\left(\boldsymbol{X} \cdot \boldsymbol{X}^{\mathrm{T}} \otimes \tilde{C}_e^{-1}\right)R\right)^{-1} R^{\mathrm{T}}\left(\boldsymbol{X} \otimes \tilde{C}_e^{-1}\right)\left(z - \left(\boldsymbol{X}^{\mathrm{T}} \otimes I_d\right)b\right). \tag{79}$$

According to Ursu and Duchesne (2009) good candidate consistent estimators are given by the unconstrained least squares estimators:

$$\tilde{C}_e = \left\{\frac{1}{T-d}\right\} \cdot \left(Z - \hat{B} \cdot \boldsymbol{X}\right)\left(Z - \hat{B} \cdot \boldsymbol{X}\right)^{\mathrm{T}}. \tag{80}$$

In the above equation $\hat{B}$ denotes the least squares estimators from eq. 78, which were obtained for the full unconstraint case. The resulting estimator of $\beta$ is given by

$$\hat{\hat{\beta}} = R \cdot \hat{\hat{\xi}} + b.$$

Its asymptotic distribution is Gaussian:

$$\{\sqrt{T}\} \cdot \left(\hat{\hat{\beta}} - \beta\right) \xrightarrow{d} \mathcal{N}\left(0_{2d^2}, R\left(R^{\mathrm{T}}(\Omega \otimes \tilde{C}_e^{-1})R\right)^{-1} R^{\mathrm{T}}\right).$$

The detailed proof of the above results can be found in Ursu and Duchesne (2009). It follows similar lines of reasoning to the proof in Lütkepohl (2005). However, Lütkepohl (2005) established the asymptotic properties of least squares estimators only for VAR models in which the model parameters satisfy linear constraints according to eq. 73 and he did not generalize his results to PVAR models.

For applied studies in project management and schedule management/control control, it can also be of interest not only to estimate the coefficient matrices $\Phi_1(v)$ and the error covariance matrices $C(v)$ of the PVAR process from time series data based on the introduced model formulation, but to follow a fully data-driven approach and also include the regression order for each iteration $v$ and the corresponding multiple dynamical operators in a combined estimation procedure. Similar to model selection in the class of VAR$(n)$ models from the previous section, in a fully data-driven approach a good trade-off must be found between the predictive accuracy gained by increasing the number of independent parameters and the danger of overfitting the model to random performance fluctuations and not inherent patterns of cooperation. We start by formulating an extended model of periodically correlated work processes with iteration-dependent correlation lengths and then proceed with solving the more subtle problem of selecting the "right" regression order for each iteration. To incorporate iteration-dependent correlation lengths into a PVAR process, the state eq. 51 has to be extended towards multiple interacting autoregression models (Ursu and Duchesne 2009):

$$X_{ns+v} = \sum_{k=1}^{n(v)} \Phi_k(v) \cdot X_{ns+v-1} + \varepsilon_{ns+v}, \tag{81}$$

The variable $n(v)$ denotes the autoregressive model order at iteration $v$ of the work process and $\Phi_k(v)$ represents the multiple dynamical operators holding for that period of time. $n(v)$ must be smaller than the the length $s$ of the period between releases of hidden information. Both the autoregressive model order $n(v)$ and the dynamical operators $\Phi_k(v)$, $k = 1, \dots, n(v)$, are the model coefficients during iteration $v = 1, \dots, s$. Therefore, the regression order of the extended PVAR model is not just a non-negative integer as for the VAR($n$) model, but an $s$-tuple $\big(n(1), \dots, n(s)\big)$ of multiple regression orders in which the vector components determine the regression order for the individual iteration $v = 1, \dots, s$.

Similar to the previous model formulation, the combined error process $\{\varepsilon_{ns+v}\}$ corresponds to a zero-mean periodic white noise process. $\{\varepsilon_{ns+v}\}$ is composed of $d \times 1$ random vectors, such that $E[\varepsilon_{ns+v}] = 0_d$ and $E[\varepsilon_{ns+v}\, \varepsilon_{ns+v}{}^{\mathrm{T}}] = C(v)$ for $v = 1, \dots, s$. It is assumed that the covariance matrices $C(v)$ for the iterations are not singular.

Following the same procedure as before, we can develop a generalized state representation for the extended PVAR model:

$$Z(v) = \big(X_v, X_{s+v}, \dots, X_{(T-1)s+v}\big) \tag{82}$$

$$E(v) = \big(\varepsilon_v, \varepsilon_{s+v}, \dots, \varepsilon_{(T-1)s+v}\big) \tag{83}$$

$$\boldsymbol{X}(v) = \big(\boldsymbol{x}_0(v), \dots, \boldsymbol{x}_{T-1}(v)\big). \tag{84}$$

In the generalized state representation $Z(v)$ and $E(v)$ are the same $d \times T$ random matrices as before. By contrast, $\boldsymbol{X}(v)$ is a $(d^2 n(v)) \times T$ matrix, where the entries

$$\boldsymbol{x}_t(v) = \big(X^{\mathrm{T}}_{ts+v-1}, \dots, X^{\mathrm{T}}_{ts+v-n(v)}\big)^{\mathrm{T}},$$

denote the $(d^2 n(v)) \times 1$ random vectors of work remaining of the component-level and system-level tasks at time steps $t = 0, 1, \dots, T-1$. It is obvious that the full PVAR model can be rewritten as $Z(v) = B(v) \cdot \boldsymbol{X}(v) + E(v)$, $v = 1, \dots, s$. This regression form was already introduced in eq. 65.

The dynamical operators of the full PVAR model are collected for each short iteration in the extended $d \times (dn(v))$ parameter matrix $B(v)$. The parameter matrix is defined as

$$B(v) = \Big(\Phi_1(v), \dots, \Phi_{n(v)}(v)\Big). \tag{85}$$

It is important to note that the generalized state representation according to eqs. 82–85 is in principle sufficient to estimate the independent parameters in the full unconstraint case, and the least squares estimators for the parameter matrix (eq. 76) and the error covariance (eq. 77) can be directly applied. To make the estimation procedure fully operational, the parameters just have to be stacked into the parameter vector

$$\beta(v) = \big(vec^{\mathrm{T}}[\Phi_1(1)], \dots, vec^{\mathrm{T}}\big[\Phi_{n(v)}(v)\big]\big),$$

of dimension $d^2 n(v) \times 1$.

If a least square estimation with linear constraints on the parameters of the dynamical operators needs to be carried out, we have to define an extended $(d^2 n(v)) \times K(v)$ matrix $R(v)$ of rank $K(v)$ and an extended $(d^2n(v)) \times 1$ vector $b(v)$ to satisfy the linear relation given by eq. 73. Similar as before, the vector $\xi(v)$ represents a $K(v) \times 1$ vector of unknown regression parameters. The parameter setting $R(v) = I_{d^2n(v)}$ and $b(v) = 0$ reflects the full unconstraint case. If certain matrix entries in $\Phi_k(v)$ must be zero, the corresponding row vectors of $R(v)$ and vector components of $b(v)$ have to be set to zero. Such a coherent theoretical framework for constraint satisfaction also allows us to use the feasible least squares estimator from eq. 79 directly. A more complicated estimation relation is not necessary. According to Ursu and Duchesne (2009) good candidate consistent estimators for the error covariance matrix $\tilde{C}(v)$ are also given by the unconstrained least squares estimators from eq. 80.

If the $s$-tuple $\big(n(1), \dots, n(s)\big)$ of regression orders holding for the individual short iterations $v = 1, \dots, s$ also has to be estimated from time series data for an unconstraint model in a fully data-driven approach, the cited trade-off between the predictive accuracy gained by increasing the regression order and the danger of overfitting the model can be resolved in a similar fashion as in the previous section by using the standard selection criteria of Akaike (1971, 1974a and 1974b) and Schwarz (1978). This is due to the fact that PVAR processes do not constitute a model class in their own right, but can be expressed as basic vector autoregressive processes (see eq. 59). In this section we focus on the Schwarz-Bayes Criterion (cf. eq. 43) because within the scope of this paper it has the same consequences for regression order selection as the (simplified two-stage) minimum description length criterion (Hansen and Yu 2001), which in turn is well grounded in Rissanen´s theory of minimum description length that will be presented and discussed in the complexity-theoretic section 3.2.2. Generalizing the fundamental ideas of McLeod (1994) on diagnostic checking periodic autoregression models, Ursu and Duchesne (2009) introduce a heuristic approach in which Schwarz´s SBC criterion is decomposed to obtain separate selection criteria for each short iteration $v = 1, \dots, s$. They define the cumulative criterion

$$SBC = \sum_{v=1}^{s} SBC_{n(v)}(v) \tag{86}$$

and the iteration dependent criteria

$$SBC_{n(v)}(v) = \ln \mathrm{Det}\big[\hat{\Sigma}_{(n(v))}(v)\big] + \frac{\ln T}{T} n(v) d^2. \tag{87}$$

For each separate criterion, the variable $\hat{\Sigma}_{(n(v))}(v)$ denotes the not bias corrected least squares estimate of $C(v)$ (cf eq. 77) for the candidate autoregression model of order $n(v)$. $\hat{\Sigma}_{(n(v))}(v)$ can also be interpreted as the one-step prediction error resulting from the separate autoregression model with parameter matrix $\big(\Phi_1(v), \dots, \Phi_{n(v)}(v)\big)$. In a similar manner to the Akaike Information Criterion, for a given iteration $v$ the last factor $n(v)d^2$ represents the number of freely estimated parameters of the candidate autoregression model. As already stated in section 2.5, the principle of counting free parameters can be easily generalized to other model classes, such as linear dynamical systems, which will be introduced and discussed in the next section. For practical purposes, the regression order can be varied systematically in the range of $n(v) \in \{1, \dots, s\}$ at each iteration $v$ and the one-step-ahead prediction error is evaluated using the criterion from eq. 87.

The regression order $n_{opt}(v)$ of the generative model holding at a given short iteration $v$ is considered to be the optimal one if it is assigned minimum scores, that is

$$n_{opt}(v) = \arg\min_{n(v)} SBC_{n(v)}(v). \tag{88}$$

The optimum tuple $n_{opt}$ of regression orders for the extended PVAR model is given by:

$$n_{opt} = \left(n_{opt}(1), \dots, n_{opt}(s)\right). \tag{89}$$

Ursu and Duchesne (2009) used the separate model selection criteria to fit a PVAR model to quarterly seasonally unadjusted West German income and consumption data for the years 1960-1987 and found the autoregressive orders $(2,1,3,1)$ to be the optimal ones. The same data were also analyzed by Lütkepohl (2005) based on the classic PVAR model formulation from section 2.4. Using the SBC selection criterion according to eq. 43, Lütkepohl (2005) obtained a minimum score for a VAR$(5)$ model.

After a model for the full unconstraint case has been fitted to data from a CE project on the basis of the above two-step procedure, it is sometimes also possible to reduce the number of independent parameters by setting selected entries in the dynamical operators $\Phi_1(v), \dots, \Phi_{n(v)}(v)$ to zero. Ursu and Duchesne (2009) use a straightforward selection heuristic in which the standard errors of the individual regression coefficients are evaluated: If the absolute value of the $t$-statistic of the given autoregressive parameter is less than one, the corresponding parameter is set to zero. The $t$-statistic is computed as the value of the least squares estimator divided by its standard error. In a third step these additionally identified constraints on the parameters are defined in the form of the linear relationship (eq. 73) and the parameters are re-estimated using the feasible estimators from eq. 79 in conjunction with the consistent estimators from eq. 80. The effectiveness of this kind of heuristic parameter reduction was also demonstrated by the authors on the basis of the quarterly seasonally unadjusted West German income and consumption data. They were able to reduce the number of independent parameters from 28 for the full unconstraint case to only 22 for a PVAR model with 6 null regression coefficients.

## 2.8 Stochastic Formulation with Hidden State Variables

A significant theoretical extension of the previously introduced approaches to modeling CE projects through autoregressive modeling techniques with periodically correlated or non-correlated processes is the formulation of a stochastic state-space model with “hidden” (latent) variables (Gharahmani 2001). In statistics, an independent variable is termed a latent variable (as opposed to an observable variable) if it cannot be directly observed but is rather inferred through a causal model from other variables that are directly measured. In our context, the hidden state variable $X_t \in \mathbb{R}^q$ represents the comprehensive internal state of the project at a specific time instant $t$ in vector form. We assume the state vector $x_t$ not only captures the essential dynamics of the work remaining from the $p$ predefined component-level and system-level tasks (which can be measured, for instance, by the time left to finalize a specific design or the number of engineering drawings requiring completion before the design is released; see Yassine et al. 2003 and sections 2.1 and 2.5) but also the efforts that must be made to communicate design decisions. The communication conveys the design information from one person or team to another and contributes to the common understanding of the design problem, product and processes. Communication is initiated more or less spontaneously and can also occur across the organizational hierarchy (Gibson and Hodgetts 1991). If communication occurs between hierarchically positioned persons, we speak of vertical communication. Horizontal communication occurs on the same hierarchical level. Diagonal communication refers to communication between managers and working persons located in different functional divisions. Due to these multiple channels, the internal state information is not completely known to the project manager as an intelligent external observer but must be estimated through the mental model of the possible state evolution in conjunction with readings from dedicated performance measurement instruments (earned value analysis, etc.). It is assumed that the estimates are obtained periodically in the form of observation vectors $Y_t \in \mathbb{R}^p$. These vectors directly refer to the work remaining of the predefined work breakdown structure and can be associated to the corresponding internal state $X_t$ without sequencing errors. Furthermore, it is assumed that the state process can be decomposed into $q$ interacting subtasks. These subtasks either represent concurrent development activities on individual or team levels or vertical, horizontal, or diagonal communication processes based on speech acts. It is important to note that the dimensions of the state space can differ from the observation space. In most cases of practical interest, the internal state vectors have significantly more components than the observation vectors ($\mathrm{Dim}[X_t] > \mathrm{Dim}[Y_t]$). Because we are aiming at a predictive model of a complex sociotechnical system, the inherent random performance fluctuations must also be taken into account for the representation of the hidden process. We represent the performance fluctuations by the random variable $\varepsilon_t$ and assume that they have no systematic component, that is $E[\varepsilon_t] = 0_q$. Furthermore, we develop the model under the assumption that the reliability of the measurement instruments is limited and non-negligible fluctuations $\nu_t$ of the readings around the true means occur. However, the instruments are not biased and there is $E[\nu_t] = 0_p$.

Formally, we define the state process $\{X_t\}$ to be linear and influenced by Gaussian noise $\varepsilon_t$ $(t = 0, \ldots, T)$. It is assumed that the observation process $\{Y_t\}$ directly depends on the state process in the sense that each vector $Y_t$ being acquired through observation at time instant $t$ is linearly dependent on the state vector $X_t$ and not on other instances of the state process. The observation process itself is perturbed by another Gaussian variable $\nu_t$. Hence, we have the simultaneous system of equations

$$X_{t+1} = A_0 \cdot X_t + \varepsilon_t \tag{90}$$

$$Y_t = H \cdot X_t + \nu_t. \tag{91}$$

The Gaussian vectors $\varepsilon_t$ and $\nu_t$ have zero means and covariances $C$ and $V$, respectively:

$$\varepsilon_t \sim \mathcal{N}(0_q, C)$$
$$\nu_t \sim \mathcal{N}(0_p, V).$$

In contrast to the vector autoregression models, we assume a Gaussian initial state density with location $\pi_0$ and covariance $\Pi_0$ in the above state-space model:

$$X_0 \sim \mathcal{N}(\pi_0, \Pi_0). \tag{92}$$

Like the stochastic model formulation without hidden variables (section 2.2), we assume that the performance and measurements fluctuations are uncorrelated from time step to time step and it holds for all time steps $\{\mu, \nu\} \in \mathbb{Z}$ that:

$$E\left[\begin{pmatrix}\varepsilon_\mu \\ \nu_\mu\end{pmatrix} \quad (\varepsilon_\nu^{\mathrm{T}} \quad \nu_\nu^{\mathrm{T}})\right] = \begin{pmatrix} C & S_{\varepsilon\nu} \\ S_{\varepsilon\nu}^{\mathrm{T}} & V \end{pmatrix} \cdot \{\delta_{\mu\nu}\}.$$

$\delta_{\mu\nu}$ is the Kronecker delta which was defined in eq. (10). To simplify the parameter estimation and complexity analysis, we assume that the partial covariances $S_{\varepsilon\nu}$ and $S_{\varepsilon\nu}^{\mathrm{T}}$ are zero:

$$E\left[\begin{pmatrix}\varepsilon_\mu \\ \nu_\nu\end{pmatrix} \quad (\varepsilon_\mu^{\mathrm{T}} \quad \nu_\nu^{\mathrm{T}})\right] = \begin{pmatrix} C & 0 \\ 0 & V \end{pmatrix} \cdot \{\delta_{\mu\nu}\}.$$

In the literature, the state-space model formulation introduced above is termed the “forward form” (e.g. van Overschee and de Moor 1996, de Cock 2002). Eq. 90 is termed the state equation and eq. 91 the output equation. Additional input (predictor) variables are not considered in the following. The linear state and output processes correspond to matrix operations, which are denoted by the operators $A_0$ and $H$, respectively. The dynamical operator $A_0$ is a square matrix of size $q \times q$. The output operator $H$ is a rectangular matrix of size $p \times q$. The literature often calls $A_0$ the state transition matrix, and $H$ the measurement, observation or generative matrix. We assume that $A_0$ is of rank $q$, $H$ of rank $p$ and that $C$, $V$ and $\Pi_0$ are always of full rank. The complete parameter vector is $\theta = [A_0 \quad H \quad C \quad V \quad \pi_0 \quad \Pi_0]$.

In the engineering literature, the complete state-space model of a CE project is termed a linear dynamical system (LDS) with additive Gaussian noise (de Cock 2002) or − using a more historical terminology − discrete time Kalman Filter (Puri 2010). In this model, only the vector $Y_t$ can be observed in equidistant time steps, whilst the true state vector $X_t$ and its past history $\{X_t\}_1^{t-1} = (X_1, \dots, X_{t-1})$ must be inferred through the stochastic linear model from the observable variables. A graphical representation in the form of a dynamic Bayesian network of the LDS without exogenous inputs is shown in Figure 5 (cf. Figure 1).

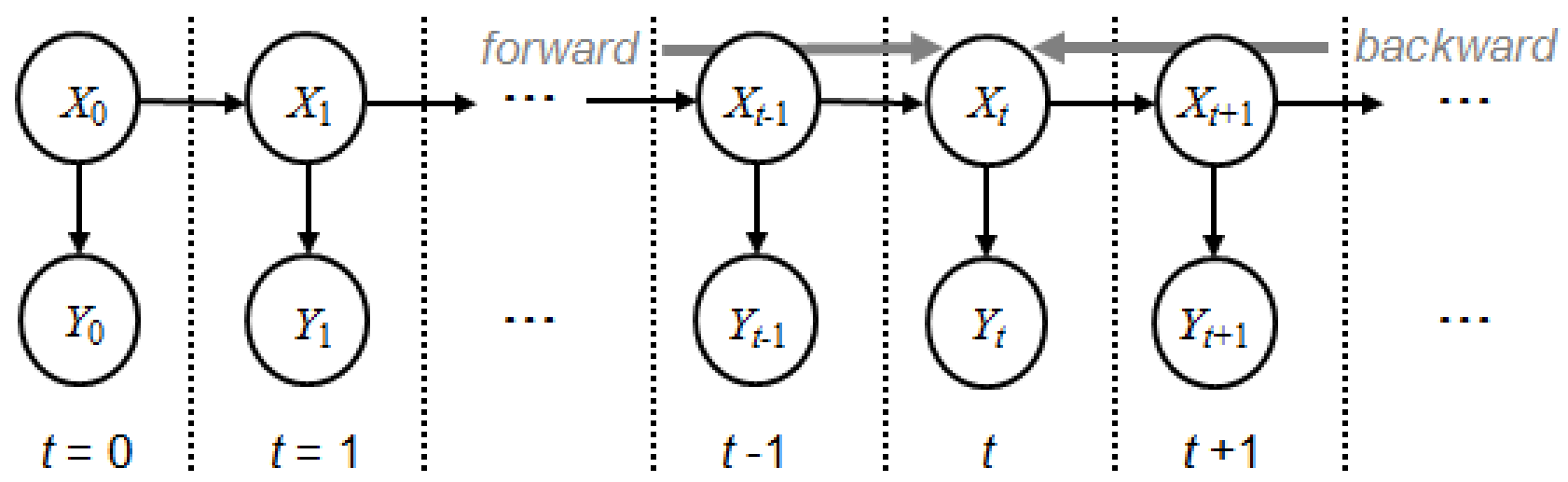

**Figure 5**

*Figure 5*. Graphical representation of the stochastic linear dynamical system in the form of a dynamic Bayesian network. The nodes in the graph represent the random state and observation variables of the stochastic process. The directed arcs encode conditional dependencies between the variables. The grey arrows indicate the forward and backward messages passed during the re-estimation of the parameters using the expectation-maximization algorithm (see section 2.9 for a detailed explanation of the algorithm).

An LDS model is one of the most prominent models in statistical signal processing. The model has also proven very useful for sensor data fusion and target tracking (see e.g. Koch 2010). Ghahramani and Hinton (1996), Yamaguchi et al. (2007) and others developed numerical methods to estimate the independent parameters based on multivariate time series of the observable variables.

In view of the theory of LDS it is important to point out that the generated stochastic process $\{Y_t\}$ can have a large memory depth in the sense that the past must be observed across a long time interval in order to make good predictions of the future. It is obvious that the Markov property (cf. eq. 13) holds for the hidden state process $\{X_t\}$ and the conditional *pdf* can be expressed as

$$f_\theta[x_{t+1}|x_t, \dots, x_0] = f_\theta[x_{t+1}|x_t] \quad \forall t \geq 0.$$

Therefore, the state evolution seems to be memorylessness in the sense that properties of random variables related to the future depend only on information about the present state and not on information from past instances of the process (see section 2.2). However, for the sequence of observations $\{Y_t\}$ the Markov property does not necessarily need to be satisfied. This is due to the fact that the information which is communicated from the past to the future by the hidden process must not completely flow through the observation $Y_t$. A portion of the total predictive information (see section 3.2.3) can be “kept secret” from the external observer even over long time intervals. Therefore, the stochastic process generated by an LDS can have a certain “crypticity” (Ellison et al. 2009) and not reveal all internal correlations and structures during the observation time. Formally speaking, for the conditional *pdf* of the output process it usually holds that

$$f_\theta[y_{t+1}|y_t,\dots,y_0] \neq f_\theta[y_{t+1}|y_t]$$

or equivalently expressed based on Bayes theorem as

$$\frac{f_\theta[y_0,\dots,y_{t+1}]}{f_\theta[y_0,\dots,y_t]} \neq \frac{f_\theta[y_t,y_{t+1}]}{f_\theta[y_t]}.$$

In fact, for the linear-Gaussian state-space model we can factorize the joint *pdf* of the system over time steps $t = 0$ to $t = T$ as

$$f_\theta[x_0,\dots,x_T,y_0,\dots,y_T] = f_\theta[x_0]\prod_{t=0}^{T-1} f_\theta[x_{t+1}|x_t]\prod_{t=0}^{T} f_\theta[y_t|x_t] \tag{93}$$

and calculate the marginal *pdf* by integrating out the hidden states of the process:

$$\begin{aligned} f_\theta[y_0,\dots,y_T] &= \int_{\mathbb{X}^p}\dots\int_{\mathbb{X}^p} f_\theta[x_0,\dots,x_T,y_0,\dots,y_T]\,dx_0\dots dx_T \\ &= \int_{\mathbb{X}^p}\dots\int_{\mathbb{X}^p} f_\theta[x_0]\prod_{t=0}^{T-1} f_\theta[x_{t+1}|x_t]\prod_{t=0}^{T} f_\theta[y_t|x_t]\,dx_0\dots dx_T. \end{aligned} \tag{94}$$

It is obvious that in the case of an arbitrary LDS, not only the time evolution of observables between two consecutive time steps is relevant for making good predictions, but also all possible transitions between hidden states of the dynamics process in the past that could give rise to the sequence of observations. For two consecutive time steps, we find:

$$f_\theta[y_{t+1}|y_t] = \frac{1}{f_\theta(y_t)}\int_{\mathbb{X}^p}\int_{\mathbb{X}^p} f_\theta(x_t)f_\theta(y_t|x_t)f_\theta(x_{t+1}|x_t)f_\theta(y_{t+1}|x_{t+1})dx_t dx_{t+1}.$$

For Gaussian noise vectors $\varepsilon_t$ and $\nu_t$, the joint *pdf* of the system can be written as

$$f_\theta[x_0,\dots,x_T,y_0,\dots,y_T] = \mathcal{N}(x_0;\pi_0,\Pi_0)\prod_{t=0}^{T-1}\mathcal{N}(x_{t+1};A_0x_t,C)\prod_{t=0}^{T}\mathcal{N}(y_t;Hx_t,V), \tag{95}$$

where the Gaussian density $\mathcal{N}(.)$ with location $\mu_x$ and covariance $\Sigma_x$ is defined as (cf. eq. 9)

$$\mathcal{N}(x;\mu_x,\Sigma_x) = \frac{1}{(2\pi)^{p/2}(\mathrm{Det}[\Sigma_x^{-1}])^{1/2}}\mathrm{Exp}\left[-\frac{1}{2}(x-\mu_x)^{\mathrm{T}}\cdot\Sigma_x^{-1}\cdot(x-\mu_x)\right]. \tag{96}$$

The density function $f_\theta[y_t]$ of state $Y_t$ given the initial location $\pi_0$, and the system and covariance matrices can be written explicitly as (cf. eq. 14)

$$f_\theta[y_t] = \frac{1}{(2\pi)^{p/2}\left(\mathrm{Det}\left[\Sigma_{y;t}\right]\right)^{1/2}}\cdot\mathrm{Exp}\left[-\frac{1}{2}(y_t - H\cdot A_0^t\cdot\pi_0)^{\mathrm{T}}\cdot\Sigma_{y;t}^{-1}\cdot(y_t - H\cdot A_0^t\cdot\pi_0)\right], \tag{97}$$

where

$$\Sigma_{y;t} = H \cdot \left( \Pi_0 + \sum_{v=1}^{t} A_0^v \cdot C \cdot \left(A_0^{\mathrm{T}}\right)^v \right) \cdot H + V.$$

The conditional density $f_\theta[y_{t+1}|y_t]$ between two consecutive observations is given by (cf. eq. 15)

$$f_\theta[y_{t+1}|y_t]$$
$$= \frac{1}{f_\theta(y_t)} \int_{\mathbb{R}^p} \int_{\mathbb{R}^p} \mathcal{N}\left(x_t; A_0^t \pi_0, \Sigma_{x;t}\right) \mathcal{N}(y_t; Hx_t, V) \mathcal{N}(x_{t+1}; A_0 x_t, C) \mathcal{N}(y_{t+1}; Hx_{t+1}, V) dx_t dx_{t+1},$$

where

$$\Sigma_{x;t} = \Pi_0 + \sum_{v=1}^{t} A_0^v \cdot C \cdot \left(A_0^{\mathrm{T}}\right)^v.$$

The complete explicit form of this density will be derived in section 4.2, based on the assumption that the state process is in steady state.

Following the procedure from section 2.2, a closed-form solution to the sum of the mean work remaining can be calculated across $T$ time steps for an arbitrary project with hidden initial state $x_0$ and operators $A_0$ and $H$ as

$$\begin{aligned} E\left[\sum_{t=0}^{T} Y_t\right] &= \sum_{t=0}^{T} E[Y_t] \\ &= \sum_{t=0}^{T} H \cdot (A_0^t \cdot x_0) \\ &= H \cdot \left(\sum_{t=0}^{T} A_0^t\right) \cdot x_0 \\ &= H \cdot \left(I_p - A_0\right)^{-1} \cdot \left(I_p - (A_0)^T\right) \cdot x_0 . \end{aligned}$$

If all subtasks of the state process are initially 100% to be completed, the initial state is simply

$$x_0 = \begin{pmatrix} 1 \\ \vdots \\ 1 \end{pmatrix}.$$

In the limit $T \to \infty$ we have for an asymptotically stable project with hidden variables the sum of the mean work remaining

$$\lim_{T\to\infty} E\left[\sum_{t=0}^{T} Y_t\right] = H \cdot \left(I_p - A_0\right)^{-1} \cdot x_0 .$$

As with the vector autoregression models, the expected total work $y_{tot}$ in the project can be estimated by:

$$y_{tot} = \mathrm{Total}\left[H \cdot \left(I_p - A_0\right)^{-1} \cdot x_0 \right]. \tag{98}$$

Following the concept from section 2.3, we can also transform the LDS into the spectral basis and therefore decompose the state process into a system with uncoupled processes with correlated noise. To transform the state-space coordinates, the state transitions matrix $A_0$ is diagonalized through an eigendecomposition as shown in eq. 20. The eigenvectors $\vartheta_i(A_0) = S_{:i}$ $(i = 1 \ldots q)$ of the state transition matrix are the column vectors of the linear transformation represented by the transformation matrix $S$. In the spectral basis transformed stochastic processes $\{X_t'\}$ and $\{Y_t'\}$ are obtained which are generated by the coefficient vectors on the basis of the system of equations

$$X_{t+1}' = \Lambda_S \cdot X_t' + \varepsilon_t'$$

$$Y_t' = H' \cdot X_t' + \nu_t'$$

with the additional representation

$$H' = S^{-1} \cdot H$$

$$\varepsilon_t' \sim \mathcal{N}(0_q, C')$$

$$C' = S^{-1} \cdot C \cdot ([S^{\mathrm{T}}]^*)^{-1}$$

$$\nu_t' \sim \mathcal{N}(0_p, V')$$

$$V' = S^{-1} \cdot V \cdot ([S^{\mathrm{T}}]^*)^{-1}$$

$$X_0' \sim \mathcal{N}(\pi_0', \Pi_0').$$

$$\pi_0' = S^{-1} \cdot \pi_0$$

$$\Pi_0' = S^{-1} \cdot \Pi_0 \cdot ([S^{\mathrm{T}}]^*)^{-1}.$$

The transformed covariance matrices $C'$ and $V'$ are also positive-semidefinite. The transformed LDS can be helpful in evaluating emergent complexity of the modeled NPD project, because the steady-state covariance matrix $\Sigma'$ of the state process can be expressed in a simple and expressive matrix form (see eq. 184).

Moreover, in specific application contexts it can also be interesting to use the inherent "degeneracy" in the LDS model (see e.g. Roweis and Gharahmani 1999). Degeneracy means that the complete informational structure contained in the covariance matrix $C$ of the performance fluctuations can be shifted into the state transition matrix $A_0$ and the observation matrix $H$. The informational structure can be shifted by decomposing $C$ into independent covariance components through the same eigendecomposition that was used to transform the state-space coordinates:

$$C = U \cdot \Lambda_U \cdot U^{-1},$$

with

$$\Lambda_U = \mathrm{Diag}[\lambda_i(C)] \qquad 1 \le i \le q.$$

Because $C$ is symmetric by definition, the eigenvectors $\vartheta_i(C) = U_{:i}$ are mutually orthogonal and $U^{-1} = U^{\mathrm{T}}$ holds. Therefore, for any LDS that is not driven by performance fluctuations represented by a normal distribution with identity covariance matrix $I_q$, we can build an equivalent model based on the following rescaling:

$$X' = \Lambda_U^{-1/2} \cdot U^{\mathrm{T}} \cdot X$$

$$A_0' = \Lambda_U^{-1/2} \cdot U^{\mathrm{T}} \cdot A_0 \cdot U \cdot \Lambda_U^{1/2} \tag{99}$$

$$H' = U \cdot \Lambda_U^{1/2} \cdot H \tag{100}$$

The rescaling can be interpreted as a “whitening” of the hidden random vectors (DelSole and Tippett 2007, cf eqs. 181 and 182). A random vector in a sample space is said to be white if the vector components are statistically independent of each other. If the independent vector components are also identically distributed, as in our case, then the random vector is said to be an i.i.d random vector. In practice, variables can be transformed to whitened space by projecting them onto their principal components and then normalizing the principal components to unit variance. It is obvious that the informational structure of the covariance matrix $V$ cannot be changed in the same way since the realizations $Y_t$ are definitely observed and we are not free to rescale them.

To support the time-dependent statistical analysis of the work processes and to simplify the analytical complexity evaluation in section 4.2, we introduce the autocovariance function

$$\begin{aligned} C_{YY}(t,s) &:= E[(Y_t - \mu_t)(Y_s - \mu_s)^{\mathrm{T}}] \\ &= E[Y_t Y_s^{\mathrm{T}}] - \mu_t \mu_s^{\mathrm{T}}, \end{aligned}$$

of the observation process $\{Y_t\}$, based on the assumption that the state process $\{X_t\}$ is in steady state. If the modeled project is asymptotically stable and therefore the modulus of the largest eigenvalue of the dynamical operator $A_0$ is less than 1, the mean of the state process in steady state is equal to the zero vector, indicating that there is no remaining work and we have $\mu_X = 0$. A detailed analysis of steady-state process dynamics will be carried out in section 4.1, and we only present some basic results from system theory. If $\{X_t\}$ is a stationary process, the autocovariance of the observation process becomes

$$\begin{aligned} C_{YY}(k,l) &= C_{YY}(l-k) \\ &= C_{YY}(\tau), \end{aligned}$$

where $\tau = l - k$ is the number of time steps by which the observation has been shifted. As a result, the autocovariance can be expressed as

$$\begin{aligned} C_{YY}(\tau) &= E[(Y_t - \mu_Y)(Y_{t+\tau} - \mu_Y)^{\mathrm{T}}] \\ &= E\left[Y_t Y_{t+\tau}^{\mathrm{T}}\right] - \mu_Y \mu_Y^{\mathrm{T}} \\ &= E\left[Y_t Y_{t+\tau}^{\mathrm{T}}\right] - (H \cdot \mu_X)(H \cdot \mu_X)^{\mathrm{T}} \\ &= E\left[Y_t Y_{t+\tau}^{\mathrm{T}}\right]. \end{aligned} \qquad (101)$$

Hence, in a signal processing sense, the autocovariance $C_{YY}(\tau)$ and the autocorrelation

$$R_{YY}(\tau) := E\left[Y_t Y_{t+\tau}^{\mathrm{T}}\right]$$

are equal in steady state and we have

$$C_{YY}(\tau) = R_{YY}(\tau). \qquad (102)$$

According to the work of van Overschee and de Moor (1996) on subspace identification of purely stochastic systems the autocovariance function of the observation process is given by

$$C_{YY}(\tau) = \begin{cases} H \cdot \Sigma \cdot H^{\mathrm{T}} + V & \tau = 0 \\ H \cdot A_0^{\tau-1} \cdot G & \tau > 0 \end{cases}, \qquad (103)$$

where the coefficient $G$ can be expressed as the expected value

$$G = E\left[X_{t+1} Y_t^{\mathrm{T}}\right]$$

$$= A_0 \cdot \Sigma \cdot H^{\mathrm{T}}.$$

In that sense $G$ describes the cross-covariance between hidden state $X_{t+1}$ and observed state $Y_t$. The matrix $\Sigma$ denotes the covariance of the states in steady-state of the process $\{X_t\}$. It satisfies the Lyapunov equation (see eq. 172) and can be expressed in closed-form in the original state-space coordinates according to eq. 173.

For the autocovariance function the following symmetry condition holds:

$$C_{YY}(-\tau) = C_{YY}(\tau)^{\mathrm{T}}.$$

As can be seen in the autocovariance function, the correlations between observations over $\tau > 0$ time steps can be significantly larger for an LDS in steady state than for the previously formulated VAR and PVAR models.

Following a similar line of thought as in the forcing matrix concept from section 2.1 (see eq. 17), the LDS model according to eqs. 90 and 91 can be transformed into a more compact "forward innovation model" (e.g. van Overschee and de Moor 1996, de Cock 2002). In this context, "more compact" means that only a single noise source is used to model random performance fluctuations. The forward innovation model is an equivalent representation in the sense that the first-order and second-order statistics of the sequence of observations generated by the model in steady state are the same, i.e., the autocovariances $E\left[Y_t Y_{t+\tau}^{\mathrm{T}}\right]$ and cross-covariances $E\left[X_{t+1} Y_t^{\mathrm{T}}\right]$ are identical. The same property holds for the "backward form" and the corresponding "backward innovation form" that will be derived later in this section.

The forward innovation representation results from applying a Kalman filter (Kalman 1960; for a comprehensive consideration of theoretical and practical aspects see e.g. Kailath 1981, Honerkamp 2002 or Puri 2010) to the state-space model. In general, the Kalman filter operates recursively on time series of noisy input data from a CE project (or other dynamical systems) to calculate an estimate of the hidden system state that is statistically optimal. The filter is named after Rudolf E. Kálmán, who was one of the principal developers of its theoretical foundations. As will be shown in the analysis below, the algorithm follows a two-step procedure. In the first step – the so-called prediction step – the Kalman filter calculates an unbiased and linear estimate of the current state vector in conjunction with the covariances. Once the result of the next observation is obtained, the estimates are updated. The update of the state is done by weighting the previous state estimate and the measurement prediction error. These weights are determined in a way which assigns larger weights to state estimates with higher certainty.

We start the derivation of the forward innovation representation by defining the hidden-state variable as the conditional mean of the state $X_{t+1}$ given all measurements up to time $t$, that is

$$\hat{x}_{t+1|t} \coloneqq E[X_{t+1} | \{Y_t\}_0^t]$$

$$\coloneqq \int x_{t+1}\, f_\theta(x_{t+1} | \{y_t\}_0^t) dx_{t+1}.$$

Following the preferred notation, the term $\{Y_t\}_0^t$ represents the sequence of observations of task processing in the CE project that have been made across an interval of $t$ and are used to compute the conditional mean.

Using the state-space eqs. 90 and 91 and the fact that $\varepsilon_t$ has zero-mean, the state prediction is

$$\hat{x}_{t+1|t} = E[(A_0 \cdot X_t + \varepsilon_t) | \{Y_t\}_0^t]$$

$$= A_0 \cdot E[X_t | \{Y_t\}_0^t]$$
$$= A_0 \cdot \hat{x}_{t|t}.$$

and the state prediction error is given by

$$\begin{aligned}\tilde{X}_{t+1|t} &:= X_{t+1} - \hat{x}_{t+1|t} \\ &= A_0 \cdot X_t + \varepsilon_t - A_0 \cdot \hat{x}_{t|t} \\ &= A_0 \cdot \tilde{X}_{t|t} + \varepsilon_t.\end{aligned}$$

The standard Kalman filter calculates $\hat{x}_{t|t}$, which is an unbiased and linear Minimum Mean Square Error (MMSE, see e.g. Honerkamp 2002) estimate of the state vector $X_t$, given the current sequence of observations $\{Y_t\}_0^t$.

The state prediction covariance $\Phi_{t+1|t}$ is computed as follows:

$$\begin{aligned}\Phi_{t+1|t} &:= E\left[\tilde{X}_{t+1|t}\tilde{X}_{t+1|t}^{\mathrm{T}} \middle| \{Y_t\}_0^t\right] \\ &= A_0 \cdot \Phi_{t|t} \cdot A_0^{\mathrm{T}} + C.\end{aligned}$$

The predicted measurements, defined as

$$\hat{y}_{t+1|t} := E[Y_{t+1} | \{Y_t\}_0^t]$$

follow from the observation (measurement) equation and the fact that $\nu_t$ has zero mean:

$$\hat{y}_{t+1|t} = H \cdot \hat{x}_{t+1|t}.$$

The measurement prediction error,

$$\tilde{Y}_{t+1|t} := Y_{t+1} - \hat{y}_{t+1|t}$$

also simply called “innovation”, is given by

$$\begin{aligned}\tilde{Y}_{t+1|t} &= H \cdot X_{t+1} + \nu_{t+1} - H \cdot \hat{x}_{t+1|t} \\ &= H \cdot \tilde{X}_{t+1|t} + \nu_{t+1}.\end{aligned}$$

Moreover, the innovation covariance is given by

$$\begin{aligned}S_{t+1|t} &:= E\left[\tilde{Y}_{t+1|t}\tilde{Y}_{t+1|t}^{\mathrm{T}} \middle| \{Y_t\}_0^t\right] \\ &= H \cdot E\left[\tilde{X}_{t+1|t}\tilde{X}_{t+1|t}^{\mathrm{T}} \middle| \{Y_t\}_0^t\right] \cdot H^{\mathrm{T}} + V \\ &= H \cdot \Phi_{t+1|t} \cdot H^T + V \qquad (104)\end{aligned}$$

and the covariance between state and measurement is

$$\begin{aligned}Q_{t+1|t} &:= E\left[\tilde{X}_{t+1|t}\tilde{Y}_{t+1|t}^{\mathrm{T}} \middle| \{Y_t\}_0^t\right] \\ &= \Phi_{t+1|t} \cdot H^{\mathrm{T}}.\end{aligned}$$

Defining the filter gain,

$$\begin{aligned}W &:= Q_{t+1|t} \cdot S_{t+1|t}^{-1} \\ &= \Phi_{t+1|t} \cdot H^{\mathrm{T}} \cdot S_{t+1|t}^{-1}\end{aligned}$$

the update of the state is the cited MMSE estimate, which is given for Gaussian random variables in closed form as

$$\hat{x}_{t+1|t+1} = \hat{x}_{t+1|t} + W \cdot \tilde{Y}_{t+1|t}.$$

Now the forward innovation representation can be written down: From

$$\begin{aligned}\hat{x}_{t+1|t} &= A_0 \cdot \hat{x}_{t|t} \\ &= A_0\left(\hat{x}_{t|t-1} + W \cdot \tilde{Y}_{t|t-1}\right) \\ &= A_0 \cdot \hat{x}_{t|t-1} + A_0 \cdot W \cdot \tilde{Y}_{t|t-1}\end{aligned}$$

with the definition of the Kalman gain

$$\begin{aligned}K &:= A_0 \cdot W \\ &= A_0 \cdot \Phi_{t+1|t} \cdot H^{\mathrm{T}} \cdot S_{t+1|t}^{-1}\end{aligned} \tag{105}$$

and the more convenient notation

$$\begin{aligned}X_{t+1}^{f} &:= \hat{x}_{t+1|t} \\ \eta_t &:= \tilde{Y}_{t|t-1},\end{aligned}$$

we obtain the simultaneous system of equations:

$$X_{t+1}^{f} = A_0 \cdot X_t^{f} + K \cdot \eta_t \tag{106}$$

$$Y_t = H \cdot X_t^{f} + \eta_t. \tag{107}$$

In this alternative representation form, the time-independent Kalman gain matrix $K$ can be interpreted as forcing matrix of the state process noise $\eta_t$ (cf. eq. 17), which is driven by the single-source measurement prediction error. The time-dependent Kalman gain matrix will be calculated in the next section.

An explicit calculation of the state prediction covariance in steady-state, using the fact that $\hat{x}_{t+1|t} = E[X_{t+1}|\{Y_t\}_0^t]$ is already an expected value, leads to

$$\begin{aligned}\Phi_{t+1|t} &= E[\tilde{X}_{t+1|t}\tilde{X}_{t+1|t}^{\mathrm{T}}|\{Y_i, i \leq t\}] \\ &= E\left[\left(X_{t+1} - \hat{x}_{t+1|t}\right)\left(X_{t+1} - \hat{x}_{t+1|t}\right)^{\mathrm{T}}\middle|\{Y_i, i \leq t\}\right] \\ &= E[X_{t+1}X_{t+1}^{\mathrm{T}}|\{Y_i, i \leq t\})] - E[\left(\hat{x}_{t+1|t}\hat{x}_{t+1|t}\right)^{\mathrm{T}}] \\ &= \Sigma - \Sigma^{f}.\end{aligned} \tag{108}$$

$\Sigma$ is the covariance of the original state variable. In steady state it satisfies the famous Lyapunov equation (cf. eq. 172):

$$\Sigma = A_0 \cdot \Sigma \cdot A_0^{\mathrm{T}} + C.$$

The Lyapunov equation is explained in great detail in Lancaster and Tismenetsky (1985) and will also be discussed in section 4.1.1. $\Sigma^f$ is the covariance of the state variable in the forward innovation representation, that is the conditional mean of the state $x_{t+1}$ given all measurements from the infinite past up to time $t$. $\Sigma^f$ satisfies another Lyapunov equation (cf. eq. 172):

$$\Sigma^f = A_0 \cdot \Sigma^f \cdot A_0^{\mathrm{T}} + K \cdot S \cdot K^{\mathrm{T}}.$$

The entries of $\Sigma^f$ can be determined by solving an algebraic Ricatti equation (see eq. 214), which will be introduced in the implicit formulation of a complexity solution in section 4.2.2.

In steady-state $\Phi_{t+1|t}$ converges to $\Phi$ and so $S_{t+1|t}$ approaches the constant covariance matrix $S$, which was used in the above Lyapunov equation. With the autocovariance of the observable variables in the innovation representation

$$\begin{aligned} C_{YY}(0) &= E\left[Y_t Y_t^{\mathrm{T}}\right] \\ &= H \cdot \Sigma^f \cdot H^{\mathrm{T}} + S \end{aligned}$$

we arrive at an expression for the Kalman gain that is equivalent to the solution in the work of de Cock (2002):

$$K = \left(A_0 \cdot \Sigma \cdot H^{\mathrm{T}} - A_0 \cdot \Sigma^f \cdot H^{\mathrm{T}}\right)\left(C_{YY}(0) - H \cdot \Sigma^f \cdot H^{\mathrm{T}}\right)^{-1}. \tag{109}$$

It is obvious that the single-source random performance fluctuations that drive the state process in the forward innovation form can be expressed as

$$\eta_t \sim \mathcal{N}(0_q, S)$$

with covariance

$$S = C_{YY}(0) - H \cdot \Sigma^f \cdot H^{\mathrm{T}}. \tag{110}$$

Finally, let us show that both representations have the same autocovariances:

$$\begin{aligned} C_{YY}(0) &= E\left[Y_t Y_t^{\mathrm{T}}\right] \\ &= H \cdot \Sigma \cdot H^{\mathrm{T}} + V \\ C_{YY}^f(0) &= E\left[Y_t Y_t^{\mathrm{T}}\right] \ (Y_t \text{ generated based on } X_t^f) \\ &= H \cdot \Sigma^f \cdot H^{\mathrm{T}} + S \end{aligned}$$

and same cross-covariance between hidden and observable states

$$\begin{aligned} G &= E\left[X_{t+1} Y_t^{\mathrm{T}}\right] \\ &= A_0 \cdot \Sigma \cdot H^{\mathrm{T}} \\ G^f &= E\left[X_{t+1}^f \left(Y_t^f\right)^{\mathrm{T}}\right] \\ &= A_0 \cdot \Sigma^f \cdot H^{\mathrm{T}} + K \cdot S. \end{aligned}$$

To show that $C_{YY}(0) = C_{YY}^f(0)$, we simply insert eq. 104 into eq. 108:

$$S = H\left(\Sigma - \Sigma^f\right)H^{\mathrm{T}} + V.$$

Rearranging the above equality proves the statement. Secondly, rearranging the definition of the Kalman gain (eq. 105)

$$K = \left(A_0 \cdot \Sigma \cdot H^{\mathrm{T}} - A_0 \cdot \Sigma^f \cdot H^{\mathrm{T}}\right)S^{-1}$$

yields

$$K \cdot S = A_0 \cdot \Sigma \cdot H^{\mathrm{T}} - A_0 \cdot \Sigma^f \cdot H^{\mathrm{T}},$$

from which $G = G^f$ follows immediately.

In conclusion, the covariance matrices of the hidden states differ among both representations and it holds that $\Sigma \neq \Sigma^f$, in general. Furthermore, the noise covariances in the observed processes differ in the two representations and for arbitrary dynamics we have $V \neq S$. However, the autocovariances and covariances of the observed processes remain unchanged and the cross-covariance between hidden and observable states are the same. The parameters of the state-space model are $(A_0, H, C, V)$ whereas in the innovations representation the parameters are $(A_0, H, K, S)$. As we will show in section 4.2.2, the differences do not necessarily lead to a conflict in the different result for the evaluation of emergent complexity according to de Cock (2002) and this work: The explicit computation of the complexity measure *EMC* in section 4.2.1 leads to a result which depends only on the combined quantity $C_{YY}(0) = R_{YY}(0) = H \cdot \Sigma \cdot H^T + V$ and $G = A_0 \cdot \Sigma \cdot H$.

Concerning the analytical complexity evaluation that will be presented in section 4.2, it is also helpful to formulate a complementary “backward model” in which the autocovariances of the observed process and the cross-covariance between hidden and observable states are also left unchanged in steady state (van Overschee and de Moor 1996, de Cock 2002). In this model the recursive state eq. 90 runs not forward but backward in time. Due to the backward recursion the backward model is formulated by considering the MMSE estimate of $X_t$ given $X_{t+1}$:

$$\begin{aligned} \hat{x}_{t|t+1} &:= E[X_t | X_{t+1}] \\ &= E\left[X_t X_{t+1}^{\mathrm{T}}\right]\left(E\left[X_{t+1} X_{t+1}^{\mathrm{T}}\right]\right)^{-1} X_{t+1}, \end{aligned}$$

where the last equation holds true because all random variables are Gaussian. From the state-space representation according to eq. 90 we compute

$$\begin{aligned} E\left[X_t X_{t+1}^{\mathrm{T}}\right] &= E[X_t (A_0 \cdot X_t + \varepsilon_t)^{\mathrm{T}}] \\ &= \Sigma \cdot A_0^{\mathrm{T}}, \end{aligned}$$

and due to the satisfied stationary condition for the state covariance $E\left[X_{t+1} X_{t+1}^{\mathrm{T}}\right] = \Sigma$ we can express the MMSE estimate of $X_t$ given $X_{t+1}$ as

$$\hat{x}_{t|t+1} = \Sigma \cdot A_0^{\mathrm{T}} \cdot \Sigma^{-1} \cdot X_{t+1}.$$

Now, we define the error

$$\varepsilon_{t|t+1} := X_t - \hat{x}_{t|t+1}.$$

and the backward state as

$$\bar{X}_{t-1} := \Sigma^{-1} \cdot X_t. \tag{111}$$

(please note that the hat symbol denotes the hidden state variable of the backward model and not the means). Transposing the second to last equation and inserting the above one, we obtain the recursion for the hidden state in the backward model

$$\begin{aligned} \bar{X}_{t-1} &= \Sigma^{-1}\left(\hat{x}_{t|t+1} + \varepsilon_{t|t+1}\right) \\ &= \Sigma^{-1}\left(\Sigma \cdot A_0^{\mathrm{T}} \cdot \Sigma^{-1} \cdot X_{t+1} + \varepsilon_{t|t+1}\right) \end{aligned}$$

$$= A_0^{\mathrm{T}} \cdot \bar{X}_t + \bar{\varepsilon}_t$$
$$= \bar{A}_0 \cdot \bar{X}_t + \bar{\varepsilon}_t \quad (112)$$

with the definition of the backward dynamical operator

$$\bar{A}_0 \coloneqq A_0^{\mathrm{T}} \quad (113)$$

and the error term

$$\bar{\varepsilon}_t \coloneqq \Sigma^{-1} \cdot \varepsilon_{t|t+1}. \quad (114)$$

The output equation in backward form is obtained by considering the MMSE estimate of $Y_t$ given $X_{t+1}$:

$$\begin{aligned}
\hat{y}_{t|t+1} &\coloneqq E[Y_t|X_{t+1}] \\
&= E\left[Y_t X_{t+1}^{\mathrm{T}}\right]\left(E\left[X_{t+1} X_{t+1}^{\mathrm{T}}\right]\right)^{-1} X_{t+1} \\
&= E\left[(H \cdot X_t + v_t)\left(A_0^{\mathrm{T}} \cdot X_t + \varepsilon_t\right)^{\mathrm{T}}\right] \Sigma^{-1} \cdot X_{t+1} \\
&= H \cdot \Sigma \cdot A_0^{\mathrm{T}} \cdot \Sigma^{-1} \cdot X_{t+1} \\
&= H \cdot \Sigma \cdot A_0^{\mathrm{T}} \cdot \bar{X}_t \\
&= H \cdot \Sigma \cdot \bar{A}_0 \cdot \bar{X}_t.
\end{aligned}$$

Re-arranging the error equation

$$\bar{v}_t = Y_t - \hat{y}_{t|t+1}$$

we obtain

$$\begin{aligned}
Y_t &= \hat{y}_{t|t+1} + \bar{v}_t \\
&= H \cdot \Sigma \cdot \bar{A}_0 \cdot \bar{X}_t + \bar{v}_t \\
&= \bar{H} \cdot \bar{X}_t + \bar{v}_t, \quad (115)
\end{aligned}$$

where the backward output operator was defined as

$$\bar{H} = H \cdot \Sigma \cdot \bar{A}_0.$$

The joint covariance matrix of the zero-mean Gaussian processes $\{\bar{\varepsilon}_\mu\}$ and $\{\bar{v}_\nu\}$ is defined as

$$E\left[\begin{pmatrix} \bar{\varepsilon}_\mu \\ \bar{v}_\mu \end{pmatrix} \begin{pmatrix} \bar{\varepsilon}_\nu^{\mathrm{T}} & \bar{v}_\nu^{\mathrm{T}} \end{pmatrix}\right] = \begin{pmatrix} \bar{C} & \bar{S}_{\bar{\varepsilon}\bar{v}} \\ \bar{S}_{\bar{\varepsilon}\bar{v}}^{\mathrm{T}} & \bar{V} \end{pmatrix} \cdot \{\delta_{\mu\nu}\}.$$

Let $\bar{\Sigma}$ denote the covariance of the states in steady-state of the backward process $\{\bar{X}_t\}$. According to the definition from eq. 111, $\bar{\Sigma}$ can be expressed as the inverse of the forward state covariance:

$$\bar{\Sigma} = \Sigma^{-1}.$$

Since in the backward form the variables $\bar{\varepsilon}_t$ and $\bar{v}_t$ and their past histories are also independent of state $\bar{X}_t$, the backward state covariance matrix $\bar{\Sigma}$ also satisfies the Lyapunov equation (cf. eq. 172)

$$\bar{\Sigma} = A_0^{\mathrm{T}} \cdot \bar{\Sigma} \cdot A_0 + \bar{C} \quad \Leftrightarrow \quad \Sigma^{-1} = A_0^{\mathrm{T}} \cdot \Sigma^{-1} \cdot A_0 + \bar{C}$$

and can be expressed in closed-form similar to eq. 173. Hence, we can express the individual covariances of the zero-mean Gaussian processes as:

$$\bar{C} = \Sigma^{-1} - A_0^{\mathrm{T}} \cdot \Sigma^{-1} \cdot A_0$$
$$\bar{V} = C_{YY}(0) - G^{\mathrm{T}} \cdot \Sigma^{-1} \cdot G.$$

The autocovariance function is

$$C_{YY}(\tau) = \begin{cases} \bar{H} \cdot \bar{\Sigma} \cdot \bar{H}^{\mathrm{T}} + \bar{V} = G^{\mathrm{T}} \cdot \Sigma^{-1} \cdot G + \bar{V} & \tau = 0 \\ \bar{H} \cdot \bar{A}_o^{\tau-1} \cdot \bar{G} = G^{\mathrm{T}} \cdot \bar{A}_o^{\tau-1} \cdot G & \tau > 0 \end{cases},$$

where the cross-covariance between hidden and observed state is given by

$$\bar{G} = E\left[\bar{X}_{t-1} Y_t^{\mathrm{T}}\right]$$
$$= \bar{A}_0 \cdot \bar{\Sigma} \cdot \bar{H}^{\mathrm{T}} + \bar{S}_{\bar{\varepsilon}\bar{v}}.$$

Due to the definition of the backward state from eq. 111 the cross-covariance can be simply written as

$$\bar{G} = H^{\mathrm{T}}.$$

We can also develop a corresponding backward innovation form. The derivation of the backward innovation form follows exactly the same procedure that gave the forward innovation representation and we therefore only present the essential steps. The backward oriented hidden-state variable $\hat{\bar{x}}_{t-1|t}$ is defined as the conditional mean of the state $\bar{x}_{t-1}$ given all measurements from the last time step $T$ down to time step $t$, that is

$$\hat{\bar{x}}_{t-1|t} \coloneqq E[\bar{X}_{t-1} | \{Y_t\}_t^T]$$
$$\coloneqq \int \bar{x}_{t-1} f_{\bar{\theta}}(\bar{x}_{t-1} | \{y_t\}_t^T) d\bar{x}_{t-1}.$$

The state retrodiction (backward oriented state prediction) is

$$\hat{\bar{x}}_{t-1|t} = E[(\bar{A}_0 \cdot \bar{X}_t + \bar{\varepsilon}_t) | \{Y_t\}_t^T]$$
$$= \bar{A}_0 \cdot E[\bar{X}_t | \{Y_t\}_t^T]$$
$$= \bar{A}_0 \cdot \hat{\bar{x}}_{t|t}.$$

and the state retrodiction error is

$$\tilde{\bar{X}}_{t-1|t} \coloneqq \bar{X}_{t-1} - \hat{\bar{x}}_{t-1|t}$$
$$= \bar{A}_0 \cdot \bar{X}_t + \bar{\varepsilon}_t - \bar{A}_0 \cdot \hat{\bar{x}}_{t|t}$$
$$= \bar{A}_0 \cdot \tilde{\bar{X}}_{t|t} + \bar{\varepsilon}_t.$$

The state retrodiction covariance $\bar{\Phi}_{t+1|t}$ is

$$\bar{\Phi}_{t-1|t} \coloneqq E\left[\tilde{X}_{t-1|t} \tilde{X}_{t-1|t}^{\mathrm{T}} \middle| \{Y_t\}_t^T\right]$$
$$= \bar{A}_0 \cdot \bar{\Phi}_{t|t} \cdot \bar{A}_0^{\mathrm{T}} + \bar{C}.$$

The measurements retrodiction is

$$\hat{y}_{t-1|t} \coloneqq E[Y_{t-1} | \{Y_t\}_t^T]$$
$$= \bar{H} \cdot \hat{\bar{x}}_{t-1|t}$$

The measurement retrodiction error $\tilde{\bar{Y}}_{t-1|t} \coloneqq Y_{t-1} - \hat{\bar{y}}_{t-1|t}$ is given by

$$\bar{\tilde{\tilde{Y}}}_{t-1|t} = \bar{H} \cdot \bar{X}_{t-1} + \bar{v}_{t-1} - \bar{H} \cdot \hat{\hat{x}}_{t-1|t}$$
$$= \bar{H} \cdot \bar{\tilde{\tilde{X}}}_{t-1|t} + \bar{v}_{t-1}.$$

The innovation covariance is given by

$$\bar{S}_{t-1|t} := E\left[\bar{\tilde{\tilde{Y}}}_{t-1|t} \bar{\tilde{\tilde{Y}}}^{\mathrm{T}}_{t-1|t} \middle| \{Y_t\}_t^T\right]$$
$$= \bar{H} \cdot E\left[\bar{\tilde{\tilde{X}}}_{t-1|t} \bar{\tilde{\tilde{X}}}^{\mathrm{T}}_{t-1|t} \middle| \{Y_t\}_t^T\right] \cdot \bar{H}^{\mathrm{T}} + \bar{V}$$
$$= \bar{H} \cdot \bar{\Phi}_{t-1|t} \cdot \bar{H}^T + \bar{V}. \tag{116}$$

The covariance between state and measurement is

$$\bar{Q}_{t-1|t} := E\left[\bar{\tilde{\tilde{X}}}_{t-1|t} \bar{\tilde{\tilde{Y}}}^{\mathrm{T}}_{t-1|t} \middle| \{Y_t\}_t^T\right]$$
$$= \bar{\Phi}_{t-1|t} \cdot \bar{H}^{\mathrm{T}}.$$

Based on the filter gain

$$\bar{W} := \bar{Q}_{t-1|t} \cdot \bar{S}^{-1}_{t-1|t}$$
$$= \bar{\Phi}_{t-1|t} \cdot \bar{H}^{\mathrm{T}} \cdot \bar{S}^{-1}_{t-1|t}$$

the update of the state can be written as

$$\hat{\hat{x}}_{t-1|t-1} = \hat{\hat{x}}_{t-1|t} + \bar{W} \cdot \bar{\tilde{\tilde{Y}}}_{t-1|t}.$$

From the state retrodiction

$$\hat{\hat{x}}_{t-1|t} = \bar{A}_0 \cdot \hat{\hat{x}}_{t|t}$$
$$= \bar{A}_0\left(\hat{\hat{x}}_{t|t+1} + \bar{W} \cdot \bar{\tilde{\tilde{Y}}}_{t|t+1}\right)$$
$$= \bar{A}_0 \cdot \hat{\hat{x}}_{t|t+1} + \bar{A}_0 \cdot \bar{W} \cdot \bar{\tilde{\tilde{Y}}}_{t|t+1}$$

using the Kalman gain

$$\bar{K} = \bar{A}_0 \cdot \bar{W}$$
$$= \bar{A}_0 \cdot \bar{\Phi}_{t-1|t} \cdot \bar{H}^{\mathrm{T}} \cdot \bar{S}^{-1}_{t-1|t}, \tag{117}$$

we finally arrive at the simultaneous system of equations:

$$\bar{X}^b_{t-1} = \bar{A}_0 \cdot \bar{X}^b_t + \bar{K} \cdot \bar{\eta}_t \tag{118}$$
$$Y_t = \bar{H} \cdot \bar{X}^b_t + \bar{\eta}_t, \tag{119}$$

where we have used the more convenient notation

$$X^b_{t-1} := \hat{\hat{x}}_{t-1|t}$$
$$\bar{\eta}_t := \bar{\tilde{\tilde{Y}}}_{t-1|t}$$

and the definitions $\bar{A}_0 = A_0^{\mathrm{T}}$ and $\bar{H} = G^{\mathrm{T}} = \left(G^f\right)^{\mathrm{T}}$ as before.

Because $\hat{\hat{x}}_{t-1|t} = E[\bar{X}_{t-1}|\{Y_t\}_t^T]$ is an expected value, the state retrodiction covariance in steady-state can be expressed as

$$\begin{aligned}
\overline{\Phi}_{t-1|t} &= E[\tilde{\bar{X}}_{t-1|t}\tilde{\bar{X}}_{t-1|t}^{\mathrm{T}}|\{Y_i, i \geq t\}] \\
&= E\left[\left(\bar{X}_{t-1} - \hat{\bar{x}}_{t-1|t}\right)\left(\bar{X}_{t-1} - \hat{\bar{x}}_{t-1|t}\right)^{\mathrm{T}}\middle|\{Y_i, i \geq t\}\right] \\
&= E[\bar{X}_{t-1}\bar{X}_{t-1}^{\mathrm{T}}|\{Y_i, i \geq t\})] - E[\left(\hat{\bar{x}}_{t-1|t}\hat{\bar{x}}_{t-1|t}\right)^{\mathrm{T}}] \\
&= \bar{\Sigma} - \bar{\Sigma}^b. \qquad (120)
\end{aligned}$$

In steady state $\bar{\Sigma}$ and $\bar{\Sigma}^b$ satisfy the Lyapunov equations:

$$\bar{\Sigma} = \bar{A}_0 \cdot \bar{\Sigma} \cdot \bar{A}_0^{\mathrm{T}} + \bar{C}$$

$$\bar{\Sigma}^b = \bar{A}_0 \cdot \bar{\Sigma}^b \cdot \bar{A}_0^{\mathrm{T}} + \bar{K} \cdot \bar{S} \cdot \bar{K}^{\mathrm{T}}.$$

The entries of $\bar{\Sigma}^b$ can also be calculated by solving an algebraic Ricatti equation (see eq. 216 in section 4.2.2.).

As before, in steady-state $\overline{\Phi}_{t-1|t}$ converges to $\overline{\Phi}$ and $\bar{S}_{t-1|t}$ to $\bar{S}$. Based on the autocovariance

$$\begin{aligned}
C_{YY}(0) &= E\left[Y_t Y_t^{\mathrm{T}}\right] \\
&= \bar{H} \cdot \bar{\Sigma}^b \cdot \bar{H}^{\mathrm{T}} + \bar{S}
\end{aligned}$$

we can express the backward Kalman gain (cf. de Cock 2002) as:

$$\begin{aligned}
\bar{K} &= \left(\bar{A}_0 \cdot \bar{\Sigma} \cdot \bar{H}^{\mathrm{T}} - \bar{A}_0 \cdot \bar{\Sigma}^b \cdot \bar{H}^{\mathrm{T}}\right)\left(C_{YY}(0) - \bar{H} \cdot \bar{\Sigma}^b \cdot \bar{H}^{\mathrm{T}}\right)^{-1} \\
&= \left(H^{\mathrm{T}} - A_0^{\mathrm{T}} \cdot \bar{\Sigma}^b \cdot G\right)\left(C_{YY}(0) - G^{\mathrm{T}} \cdot \bar{\Sigma}^b \cdot G\right)^{-1}, \qquad (121)
\end{aligned}$$

where we have used the previous definitions $\bar{A}_0 = A_0^{\mathrm{T}}$ and $\bar{H} = G^{\mathrm{T}}$. We define $G^b \coloneqq \bar{H}$. It holds that $G^b = G^{\mathrm{T}} = \left(G^f\right)^{\mathrm{T}}$. The autocovariance function therefore can be expressed as

$$C_{YY}(\tau) = \begin{cases} \bar{H} \cdot \bar{\Sigma}^b \cdot \bar{H}^{\mathrm{T}} + \bar{S} = G^{\mathrm{T}} \cdot \bar{\Sigma}^b \cdot G + \bar{S} & \tau = 0 \\ \bar{H} \cdot \bar{A}_o^{\tau-1} \cdot \bar{G} = G^{\mathrm{T}} \cdot (A_o^{\tau-1})^{\mathrm{T}} \cdot H^{\mathrm{T}} & \tau > 0 \end{cases}.$$

The random performance fluctuations in the backward innovation model can therefore be expressed as

$$\bar{\eta}_t = \mathcal{N}(\xi; 0_q, \bar{S})$$

with covariance

$$\bar{S} = C_{YY}(0) - G^{\mathrm{T}} \cdot \bar{\Sigma}^b \cdot G. \qquad (122)$$

After comprehensively analyzing different forward and backward representations of an LDS, we will direct our attention toward a robust technique for estimating the independent parameters from data. An iterative maximum likelihood estimation technique for minimizing the deviation of the data from the predictions of the forward model will be presented in the next section. It is not difficult to transform the parameter estimates into the alternative forms by using the above definitions.

## 2.9 Maximum Likelihood Parameter Estimation with Hidden State Variables

Minimizing the deviation of observations from the model's predictions is equivalent to maximizing the likelihood of the sequence of observations conditioned on the model structure and independent parameters. Unfortunately, to date the only known methods for carrying out a maximum likelihood estimation of the parameters of an LDS have been iterative, and therefore computationally very demanding. The objective function to be maximized is the logarithmic probability of the fixed sequence of observations $\{y_t\}_0^T = (y_0, y_1, \dots, y_T)$ given an underlying parameterized LDS model with parameter vector $\theta = [A_0 \quad H \quad C \quad V \quad \pi_0 \quad \Pi_0]$:

$$\mathcal{L}(\theta) = \ln f_\theta[y_0, \dots, y_T] \to \max \tag{123}$$

The objective function is also known as the log-likelihood function. The log-likelihood function can be obtained from the joint *pdf* $f_\theta[x_0, \dots, x_T, y_0, \dots, y_T]$ of the LDS (eq. 93) through marginalization. The marginal *pdf* $f_\theta[y_0, \dots, y_T]$ can be calculated by integrating out the hidden states of the process (eq. 94) and can be simplified by factorization:

$$\begin{aligned}\mathcal{L}(\theta) &= \ln f_\theta[y_0, \dots, y_T] \\ &= \ln \int_{\mathbb{X}^p} \dots \int_{\mathbb{X}^p} f_\theta[x_0, \dots, x_T, y_0, \dots, y_T]\, dx_0 \dots dx_T \\ &= \ln \int_{\mathbb{X}^p} \dots \int_{\mathbb{X}^p} f_\theta[x_0] \prod_{t=0}^{T-1} f_\theta[x_{t+1}|x_t] \prod_{t=0}^{T} f_\theta[y_t|x_t]\, dx_0 \dots dx_T.\end{aligned} \tag{124}$$

While the gradient and Hessian of the log-likelihood function are usually very difficult to compute, it is relatively easy to calculate the logarithmic joint probability and its expected value for a particular setting of the independent parameters. In the following we will use a specific algorithm developed by Ghahramani and Hinton (1996) to indirectly optimize the log-likelihood of the observations by iteratively maximizing expectations. The general principle behind the Gharamani-Hinton algorithm is the expectation-maximization (EM) principle. The EM principle dates back to Dempster, Laird, and Rubin (1977) and can be regarded as a special kind of quasi-Newton algorithm. The searching direction of the algorithm has a positive projection on the gradient of the log-likelihood. Each iteration alternates between two steps, the estimation (E) and the maximization (M). The maximization step maximizes an expected log-likelihood function for a given estimate of the parameters that is recalculated in each iteration by the expectation step. As Roweis and Gharahmani (1999) show, we can use any parameterized probability distribution $g_{\breve{\theta}}[x_0, \dots, x_T]$ – not necessarily multivariate normal – over the hidden state variables to obtain a lower bound on the log-likelihood function $\mathcal{L}(\theta)$:

$$\begin{aligned}&\ln \int_{\mathbb{X}^p} \dots \int_{\mathbb{X}^p} f_\theta[x_0, \dots, x_T, y_0, \dots, y_T]\, dx_0 \dots dx_T \\ &\qquad = \ln \int_{\mathbb{X}^p} \dots \int_{\mathbb{X}^p} g_{\breve{\theta}}[x_0, \dots, x_T] \frac{f_\theta[x_0, \dots, x_T, y_0, \dots, y_T]}{g_{\breve{\theta}}[x_0, \dots, x_T]}\, dx_0 \dots dx_T \\ &\geq \int_{\mathbb{X}^p} \dots \int_{\mathbb{X}^p} g_{\breve{\theta}}[x_0, \dots, x_T] \ln \frac{f_\theta[x_0, \dots, x_T, y_0, \dots, y_T]}{g_{\breve{\theta}}[x_0, \dots, x_T]}\, dx_0 \dots dx_T \\ &= \int_{\mathbb{X}^p} \dots \int_{\mathbb{X}^p} g_{\breve{\theta}}[x_0, \dots, x_T] \ln f_\theta[x_0, \dots, x_T, y_0, \dots, y_T]\, dx_0 \dots dx_T \\ &\qquad - \int_{\mathbb{X}^p} \dots \int_{\mathbb{X}^p} g_{\breve{\theta}}[x_0, \dots, x_T] \ \ln g_{\breve{\theta}}[x_0, \dots, x_T]\, dx_0 \dots dx_T \\ &= \mathcal{F}(g_{\breve{\theta}}, \theta).\end{aligned}$$

The inequality in the middle of the above expression is known as Jensen´s inequality. For greater clarity, the parameter vector $\breve{\theta}$ of the auxiliary distribution $g_{\breve{\theta}}[x_0, \dots, x_T]$ is not explicitly declared in the following. In the EM literature, the auxiliary distribution $g_{\tilde{\theta}}$ is referred to as the auxiliary function or simply the $Q$-function (Dellaert 2002).

If the "energy" of the complete configuration $(X_0, X_1, \ldots, X_T, Y_0, Y_1, \ldots, Y_T)$ is defined as

$$-\ln f_\theta[x_0, \ldots, x_T, y_0, \ldots, y_T],$$

then the lower bound $\mathcal{F}(g, \theta) \leq \mathcal{L}(\theta)$ is the negative of a quantity that is known in statistical physics as the "free energy" (Honerkamp 2002). The free energy is the expected energy under the distribution $g[x_0, \ldots, x_T]$ minus the differential entropy of that distribution (Neal and Hinton 1998, Roweis and Ghahramani 1999). The definition of the differential entropy is given in eq. 173 and will be used in later sections to evaluate emergent complexity of the modeled phase of the NPD project.

The EM algorithm alternates between maximizing the function $\mathcal{F}(g, \theta)$ by the auxiliary distribution $g$ and by the parameter vector $\theta$, while holding the other fixed. The iteration number is denoted by $k$. Starting from an initial parameter setting $\theta_0$ it holds that

$$\text{E} - \text{step:} \qquad g_{k+1} = \arg\max_g \mathcal{F}(g, \theta_k) \tag{125}$$

$$\text{M} - \text{step:} \quad \theta_{k+1} = \arg\max_\theta \mathcal{F}(g_{k+1}, \theta). \tag{126}$$

It can be shown that the maximum of the E-step is obtained when $g$ is exactly the conditional *pdf* $f_\theta$ of $(X_0, X_1, \ldots, X_T)$ given $(Y_0, Y_1, \ldots, Y_T)$:

$$g[x_0, \ldots, x_T] = f_\theta[x_0, \ldots, x_T | y_0, \ldots, y_T].$$

Hence, the maximum in the M-step results when the term

$$\int_{\mathbb{X}^p} \ldots \int_{\mathbb{X}^p} g[x_0, \ldots, x_T] \ln f_\theta[x_0, \ldots, x_T, y_0, \ldots, y_T] \, dx_0 \ldots dx_T$$

is maximized, since the differential entropy does not depend on the parameters $\theta$. Therefore, we can also express the EM algorithm in a single maximization step:

$$\text{M} - \text{step:} \;\; \theta_{k+1} = \arg\max_\theta \int_{\mathbb{X}^p} \ldots \int_{\mathbb{X}^p} f_{\theta_k}[x_0, \ldots, x_T | y_0, \ldots, y_T] \ln f_\theta[x_0, \ldots, x_T, y_0, \ldots, y_T] \, dx_0 \ldots dx_T. \tag{127}$$

In that sense the EM principle can be interpreted as coordinate ascent in $\mathcal{F}(g, \theta)$. At the beginning of each M-step it holds that $\mathcal{F}(g, \theta) = \mathcal{L}(\theta)$. Since the E-step does not change $\theta$, the likelihood is guaranteed not to decrease after each combined EM-step (Neal and Hinton 1998, Roweis and Ghahramani 1999). Therefore, in the EM algorithm the solutions to the filtering and smoothing problem that are incorporated in the conditional distribution $f_{\theta_k}[x_0, \ldots, x_T | y_0, \ldots, y_T]$ are applied to estimate the hidden states given the observations and the re-estimated model parameters. These virtually complete data points are used to solve for new model parameters. In Figure 6 a graphical illustration of the first three iteration steps of the EM procedure are shown. Each lower bound $\mathcal{F}(g, \theta)$ touches the objective function $\mathcal{L}(\theta)$ at the current estimate $\theta_k$ and is in that sense optimal. However, it is clear from the figure that the approached maximum of the objective function is only locally optimal and therefore a proper initial setting $\theta_0$ of the independent parameters is crucial for the effectiveness of the EM algorithm. We will return later in this section to this issue.

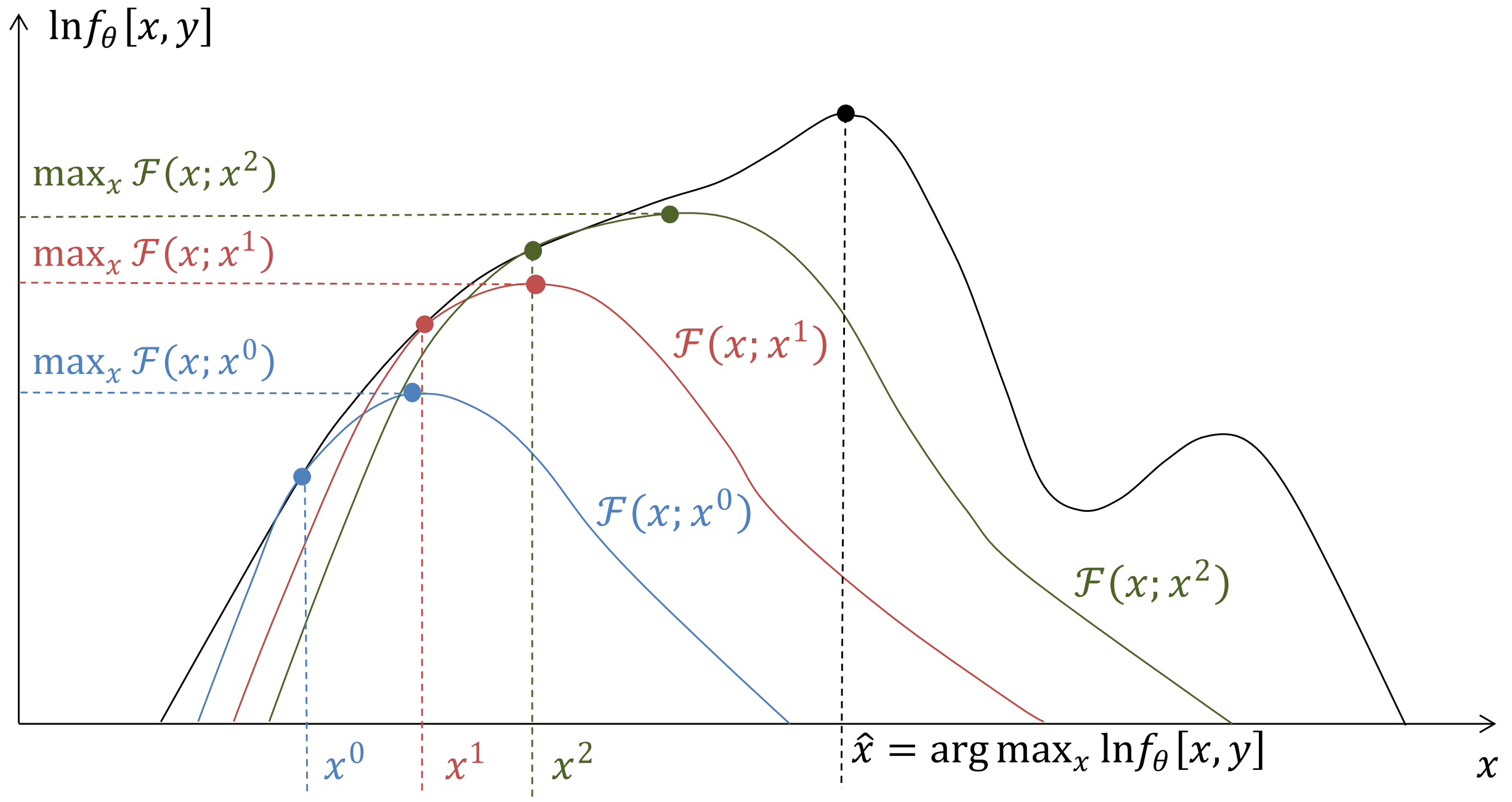

**Figure 6**

*Figure 6*. Graphical illustration of the Expectation-Maximization (EM) algorithm (Dempster, Laird, and Rubin 1977). The illustration was adapted from Streit (2006) and Wienecke (2013). The EM algorithm iterates between computing a lower bound and maximization. The figure shows the first three iterations in blue, red and green. In the case shown the algorithm converges to a local maximum of the objective function. For greater clarity, the sequences of hidden state variables $\{X_t\}_0^T$ and observations $\{Y_t\}_0^T$ are denoted by single vectors $x$ and $y$, respectively.

According to eq. 124 the joint probability $f_\theta[x_0,\dots,x_T,y_0,\dots,y_T]$ of hidden states and observations is multivariate normal and therefore the conditional distribution $f_{\theta_k}[x_0,\dots,x_T|y_0,\dots,y_T]$ also follows a multivariate normal distribution. There are reasonably efficient algorithms – the introduced Kalman filter and the corresponding smoother algorithm (see e.g. Shumway and Stoffer 1982) – for inferring this distribution for a given setting of the parameters. Given the estimated states obtained from the filtering and smoothing algorithms, it is usually easy to solve for new parameter values. For LDS, solving for new parameter values typically involves minimizing quadratic expressions such as eq. 128. This process is repeated using the new model parameters to infer the hidden states again, until the log-likelihood grows only very slowly.

The Ghahramani-Hinton algorithm makes the EM principle for LDS fully operational and can accurately estimate the complete set $\theta$ of independent parameters. The algorithm is guaranteed to converge to a local maximum of the log-likelihood function (Dempster et al., 1977). However, the convergence rate is typically linear in $T$ and polynomial in $q$ and it can take many iterations to reach a satisfactory predictive accuracy (for simplicity, we assume that $q \geq p$, Martens 2009). This can be a problem for long series of observations and a large state space, because each iteration involves recursive estimations for all hidden states over $T$ time steps. An alternative approach is the combined deterministic-stochastic subspace identification algorithm of van Overschee and de Moor (1996). The algorithm is called a subspace algorithm because it retrieves the system related matrices as subspaces of projected data matrices. The combined deterministic-stochastic subspace identification algorithm reduces the estimation problem to that of solving a large singular value decomposition problem (cf. canonical correlation analysis in section 4.1.3), subject to some heuristics in the form of user-defined weighting matrices. The user-defined weighting matrices define which part of the predictive information in the data it is important to retain. The algorithm allows for computing the non-steady-state Kalman filter sequence (cf. section 2.8) directly from the observation sequences, without knowledge of the system matrices. Because subspace identification is not iterative, it tends to be more efficient than EM. However, because it implicitly optimizes an objective function representing a sum of squared prediction errors instead of the log-likelihood function, and because implicitly uses a rather simple inference procedure for the hidden states, subspace identification is not statistically optimal. In particular, point estimates of hidden states lacking covariance information are obtained by conditioning on only the past $l$ data points, where $l$ is an independent parameter known as the "prediction horizon" (Martens 2009). In contrast, the EM algorithm estimates hidden states with mean and covariance information, conditioned on the entire observation sequence $\{y\}_0^T$ and the current estimate $\theta_k$ of the independent parameters in the $k$th iteration. Subspace-identification also tends to scale poorly to high-dimensional time series. When initialized properly the EM algorithm will find parameter settings with larger values of the log-likelihood function and lower prediction errors than those found by the deterministic-stochastic subspace identification algorithm. However, because of the speed of subspace identification, its ability to estimate the dimensionality of the state space, and the fact that it optimizes an objective that is similar to that of EM, subspace identifiation is often considered an excellent method for initializing iterative learning algorithms such as EM (Martens 2009). Smith and Robinson (2000) and Smith et al. (1999) have carried out comprehensive theoretical and empirical analyses of subspace identification and EM. Another standard approach for estimating LDS parameters are prediction error methods (Ljung 1999). In this approach, a one-step prediction-error objective is minimized. This is usually done via gradient-based optimization methods. In that sense these methods have a close kinship with the maximum likelihood method. However, prediction error methods are often based on a stationary predictor and for finite-length sequences of observation vectors therefore are not equivalent to the maximum likelihood principle, even if Gaussian noise is incorporated in the system equations. Prediction error methods are also embodied in a software package and therefore have become the dominant algorithms for system identification (Ljung 2000). But these methods do have some shortcomings. Typical implementations use either gradient descent and therefore require many iterations to converge, or use 2nd order optimization methods but then become impractical for large models (Martens 2010). Under certain circumstances the search for the parameters can be very laborious, involving search surfaces that may have many local minima. This parameter search is typically carried out using the damped Gauss-Newton method and therefore having good initial parameter values is of crucial importance for estimation accuracy. Because of the cited drawbacks of alternative methods and the theoretically very satisfactory formulation and solution of the objective function, we will direct our attention to the EM procedure in the following presentation.

According to eq. 127 the EM procedure requires the evaluation of the expected log-likelihood function of observations and hidden states for the particular setting $\theta_k$ of the independent parameters in the $k$th iteration:

$$\wp_k(\theta) = E_{\theta_k}\left[\ln f_\theta[x_0, \ldots, x_T, y_0, \ldots, y_T] \middle| \{y_t\}_0^T\right].$$

$\{y_t\}_0^T$ denotes the sequence of observations of task processing that have been made in the CE project across an interval of $T$ and are used to compute the expectations. In a similar manner $\{y_t\}_1^t$ denotes the subsequence of vectors that starts at time instant 1 and ends at instant $t \leq T$.

Considering the definition of joint *pdf* of the dynamic system from eqs. 95 and 96, the log-likelihood function over all hidden states and observations can be expressed as a sum of three simple quadratic terms:

$$\ln f_\theta[x_0,\dots,x_T,y_0,\dots,y_T]$$

$$= -(T+1)\frac{(p+q)}{2}\ln 2\pi - \frac{1}{2}\ln \mathrm{Det}[\Pi_0] - \frac{1}{2}T\ln \mathrm{Det}[C] - \frac{1}{2}(T+1)\ln \mathrm{Det}[V]$$

$$-\frac{1}{2}(x_0-\pi_0)^{\mathrm{T}}\cdot\Pi_0^{-1}\cdot(x_0-\pi_0) - \frac{1}{2}\sum_{t=1}^{T}(x_t - A_0\cdot x_{t-1})^{\mathrm{T}}\cdot C^{-1}\cdot(x_t - A_0\cdot x_{t-1})$$

$$-\frac{1}{2}\sum_{t=0}^{T}(y_t - H\cdot x_t)^{\mathrm{T}}\cdot V^{-1}\cdot(y_t - H\cdot x_t). \tag{128}$$

It can be shown that the expected log-likelihood function depends on three expected values related to the hidden state variables given the observations, namely $E_{\theta_k}\left[X_t\middle|\{Y_t\}_0^T\right]$, $E_{\theta_k}\left[X_tX_t^{\mathrm{T}}\middle|\{Y_t\}_0^T\right]$ and $E_{\theta_k}\left[X_tX_{t-1}^{\mathrm{T}}\middle|\{Y_t\}_0^T\right]$. Following the notation of the creators of the algorithm, we will use the variables

$$\hat{x}_t := E_{\theta_k}\left[X_t\middle|\{Y_t\}_0^T\right]$$
$$\hat{P}_t := E_{\theta_k}\left[X_tX_t^{\mathrm{T}}\middle|\{Y_t\}_0^T\right]$$
$$\hat{P}_{t,t-1} := E_{\theta_k}\left[X_tX_{t-1}^{\mathrm{T}}\middle|\{Y_t\}_0^T\right]$$

to encode these expectations. The variable $\hat{x}_t$ denotes the (re-)estimated state of the process $\{X_t\}$ at time instant $t$ given the complete series of observations $\{Y_t\}_0^T$. Interestingly, the state estimate $\hat{x}_t$ is not only based on past observations ($Y_0$,...,$Y_t$) but also on the future history (Y$y_t$,..., $Y_T$). This is in contrast to the classic Kalman filter in which only the estimates $E_{\theta_k}[X_t|\{Y_t\}_0^t]$ are considered (Puri 2010). The above first and second order statistics over the hidden states allow one to easily evaluate and optimize $\mathcal{L}(\theta)$ with respect to $\theta$ in the M-step. These quantities can be considered as the "full smoother" estimates of the Kalman smoother (Roweis and Ghahramani 1999, see below).

As explained in the M-step each of the independent parameters is re-estimated (see intuitive graphical representation in Figure 5). The new parameter estimate is obtained by maximizing the expected log-likelihood. To find a closed form of the maximum of the expected log-likelihood function the partial derivatives are calculated for each parameter from eq. 128. The following equations summarize the results (Ghahramani and Hinton 1996). The re-estimated quantities are indicated by the prime symbol.

Initial state location $\pi_0$:

$$\frac{\partial\wp_k}{\partial\pi_0} = (\hat{x}_0 - \pi_0)\Pi_0^{-1} = 0$$

$$\Rightarrow \quad \pi_0' = \hat{x}_0$$

Initial state covariance $\Pi_0$

$$\frac{\partial\wp_k}{\partial\Pi_0^{-1}} = \frac{1}{2}\Pi_0 - \frac{1}{2}\left(P_0 - \hat{x}_0\pi_0^{\mathrm{T}} - \pi_0\hat{x}_0^{\mathrm{T}} + \pi_0\pi_0^{\mathrm{T}}\right)$$

$$\Pi_0' = P_0 - \hat{x}_0\hat{x}_0^{\mathrm{T}}$$

Dynamical operator $A_0$:

$$\frac{\partial\wp_k}{\partial A_0} = -\sum_{t=1}^{T}C^{-1}\hat{P}_{t,t-1} + \sum_{t=1}^{T}C^{-1}A_0\,\hat{P}_{t-1} = 0$$

$$\Rightarrow \quad A_0' = \left(\sum_{t=1}^{T} \hat{P}_{t,t-1}\right)\left(\sum_{t=1}^{T} \hat{P}_{t-1}\right)^{-1}$$

State noise covariance $C$:

$$\frac{\partial \wp_k}{\partial C^{-1}} = \frac{T}{2} C - \frac{1}{2} \sum_{t=1}^{T} \left(\hat{P}_t - A_0 \hat{P}_{t-1,t} - \hat{P}_{t,t-1} A_0^{\mathrm{T}} + A_0 \hat{P}_{t-1} A_0^{\mathrm{T}}\right)$$

$$= \frac{T}{2} C - \frac{1}{2} \left(\sum_{t=1}^{T} \hat{P}_t - A_0' \sum_{t=1}^{T} P_{t-1,t}\right) = 0$$

$$\Rightarrow \quad C' = \frac{1}{T} \left(\sum_{t=1}^{T} \hat{P}_t - A_0' \sum_{t=1}^{T} \hat{P}_{t-1,t}\right)$$

Output operator $H$:

$$\frac{\partial \wp_k}{\partial H} = -\sum_{t=0}^{T} V^{-1} y_t \hat{x}_t^{\mathrm{T}} + \sum_{t=0}^{T} V^{-1} H \hat{P}_t = 0$$

$$\Rightarrow \quad H' = \left(\sum_{t=0}^{T} y_t \hat{x}_t^{\mathrm{T}}\right)\left(\sum_{t=0}^{T} P_t\right)^{-1}$$

Output noise covariance $V$:

$$\frac{\partial \wp_k}{\partial V^{-1}} = \frac{T+1}{2} V - \sum_{t=0}^{T} \left(\frac{1}{2} y_t y_t^{\mathrm{T}} - H \hat{x}_t y_t^{\mathrm{T}} + \frac{1}{2} H \hat{P}_t H^{\mathrm{T}}\right) = 0$$

$$\Rightarrow \quad V' = \frac{1}{T+1} \sum_{t=0}^{T} \left(y_t y_t^{\mathrm{T}} - H' \hat{x}_t y_t^{\mathrm{T}}\right)$$

To simplify the detailed description of the E-step, we will use three intermediate variables:

$$x_t^{\tau} \coloneqq E_{\theta_k}[X_t | \{Y_t\}_0^{\tau}]$$

$$\Sigma_t^{\tau} \coloneqq \mathrm{Var}_{\theta_k}[X_t | \{Y_t\}_0^{\tau}]$$

$$\Sigma_{t,t-1}^{\tau} \coloneqq E_{\theta_k}\left[X_t X_{t-1}^{\mathrm{T}} | \{Y_t\}_1^{\tau}\right] - x_t^{\tau} (x_t^{\tau})^{\mathrm{T}}.$$

The E-step consists of two sub-steps. The first sub-step is a forward recursion that uses the sequence of observations from $y_0$ to $y_t$ for parameter estimation. This forward recursion is the well-known Kalman filter which was introduced in the previous section. The second sub-step carries out a backward recursion that uses the observations from $y_T$ to $y_{t+1}$ (Rauch 1963). The combined forward and backward recursions are known as the Kalman smoother (Shumway and Stoffer 1982). The following Kalman-filter forward recursions hold for $t = 0$ to $T$:

$$x_t^{t-1} = A_0 x_{t-1}^{t-1}$$

$$\Sigma_t^{t-1} = A_0 \Sigma_{t-1}^{t-1} A_0^{\mathrm{T}} + C$$

$$K_t = \Sigma_t^{t-1} H^{\mathrm{T}} \left(H \Sigma_t^{t-1} H^{\mathrm{T}} + V\right)^{-1}$$

$$x_t^t = x_t^{t-1} + K_t(y_t - Hx_t^{t-1})$$
$$\Sigma_t = \Sigma_t^{t-1} - K_t H \Sigma_t^{t-1}.$$

It holds that $x_0^{-1} = \pi_0$ and $\Sigma_0^{-1} = \Pi_0$. $K_t$ is the time-dependent Kalman gain matrix (cf. section 2.8).

To compute $\hat{x}_t = x_t^T$ and $\hat{P}_t = \Sigma_t^T + x_t^T (x_t^T)^{\mathrm{T}}$ the following backward recursions have to be carried out from $t = 0$ to $t = T$(Shumway and Stoffer 1982):

$$J_{t-1} = \Sigma_{t-1}^{t-1} A_0^{\mathrm{T}} (\Sigma_t^{t-1})^{-1}$$
$$\hat{x}_{t-1} = x_{t-1}^{t-1} + J_{t-1}(\hat{x}_t - A_0 x_{t-1}^{t-1})$$
$$\hat{P}_{t-1} = \Sigma_{t-1}^{t-1} + J_{t-1}(\hat{P}_t - \Sigma_t^{t-1}) J_{t-1}^{\mathrm{T}}.$$

Moreover, if $t < T$ the conditional covariance $\hat{P}_{t,t-1} = \Sigma_{t,t-1}^T + x_t^T (x_{t-1}^T)^{\mathrm{T}}$ of the hidden states across two time steps can be obtained through the backward recursion:

$$\hat{P}_{t,t-1} = \Sigma_t^t J_{t-1}^{\mathrm{T}} + J_t(\hat{P}_{t+1,t} - A_0 \Sigma_t^t) J_{t-1}^{\mathrm{T}}.$$

The recursion is initialized with $\hat{P}_{T,T-1} = (I_q - K_T H) A_0 \Sigma_{T-1}^{T-1}$.

If not only a single sequence of empirically acquired observation vectors $y_t$ of work remaining is given but multiple realizations of the work processes had also been acquired in $N$ independent measurement trials, then the above equations can be easily generalized. The basic procedure involves calculating the expected values in the E-step for each sequence separately and summing up the individual quantities to accumulated expectations. The only difficulty here is estimating the initial state covariance. According to Ghahramani and Hinton (1996) we can define $\hat{x}_t^{[i]}$ as the state estimate of sequence $[i]$ at time instant $t$ and $\hat{x}_{N,t}$ as the mean estimate at the same time instant:

$$\hat{x}_{N,t} = \frac{1}{N} \sum_{i=1}^{N} \hat{x}_t^{[i]}.$$

Based on these estimates, we can then calculate the initial covariance as follows:

$$\Pi_0' = \hat{P}_0 - \hat{x}_{N,t} \hat{x}_{N,t}^{\mathrm{T}} + \frac{1}{N} \sum_{i=1}^{N} \left(\hat{x}_t^{[i]} - \hat{x}_{N,t}\right) \left(\hat{x}_t^{[i]} - \hat{x}_{N,t}\right)^{\mathrm{T}}.$$

In the M-step, the accumulated expectations are used to re-estimate the independent parameters.

An interesting way of significantly simplifying the EM algorithm is to replace the time-dependent matrices in the E- and M-steps with their steady-state values. There are efficient methods for finding these matrices (e.g. the doubling algorithm, Anderson and Moore 1979). Martens (2009) improved the efficiency of the EM algorithm of Ghahramani and Hinton (1996) by using a steady-state approximation, which simplifies inference of the hidden state and by using the fact that the M-step requires only a small set of expected second order statistics that can be approximated without doing complete inference for each $x_t$ as shown above. Martens´ experiments show that the resulting approximate EM algorithm performs nearly as well as the EM algorithm given the same number of iterations (Martens 2009). Since the calculations required per iteration of the approximate EM do not depend on $T$, it can be much more computationally efficient when $T$ is large.

In application areas with no or only shallow prior knowledge about the dynamic dependency structures, the (true or approximate) EM iterations are usually initialized with a setting $\theta_0$ of the independent parameters in which Gaussian random numbers are assigned to the initial state, the dynamical operator and the output operator. In a similar manner the diagonal and off-diagonal entries of the corresponding covariance matrices are set to Gaussian random numbers at the start of the iterations. Clearly, randomizing the initial covariance matrices has to be done under the constraint of matrix symmetry. Let $\kappa$ denote the number of iterations that were calculated using the EM algorithm in a specific modeling and simulation environment. The re-estimated setting of the independent parameters in the (last) $\kappa$th iteration is denoted by $\hat{\theta}_\kappa$. We assume that the independent parameters were re-estimated sufficiently often. Sufficiently often means that the log-likelihood grew only very slowly in the final iterations and a stable local optimum was found. In many practical cases 20 to 30 iterations are sufficient to reach a stable optimum.

It is important to note that there can be a substantial problem for system identification when using the EM procedure of Ghahramani and Hinton (1996) and later developments. If we simply estimate the parameters of a project without any constraints for the parameter space, the procedure lacks identifiability, i.e., an infinite number of parameterizations $\hat{\theta}_\kappa$ exists that yield the same maximum log-likelihood. Therefore, the estimated initial state, initial covariance, dynamical operator, state covariance, output operator and output covariance depend on the particular initial setting $\theta_0$ of the independent parameters that is used in the first iteration and there are many different groups of system matrices ($A_0$ and $H$) which can produce a given data set. To ensure the method does not lack identifiability, imposing the following constraints in the expectation and maximizations steps will suffice (see Yamaguchi et al., 2007):

- $C = I_q$
- $H^{\mathrm{T}} \cdot V \cdot H = \Lambda_{\mathrm{V}} := \mathrm{Diag}[\lambda_1, \ldots, \lambda_m]$
- an arbitrarily signed condition for all elements in a particular setting $\eta_i = \left(\eta_i^{(1)}, \ldots, \eta_i^{(m)}\right)^{\mathrm{T}}$ is assumed to be given on the parameter space, where $H^{\mathrm{T}} := \left(\eta_1, \ldots, \eta_p\right)$.

A corresponding algorithm for LDS with constraints was introduced by Yamaguchi et al. (2007) in their work on finding module-based gene networks. However, if emergent complexity of CE projects with hidden state variables has to be evaluated by using a parameterized LDS without constraints (see section 4.2), the lack of identifiability is usually not a critical issue, because we can simply average the complexity metric over a large number of valid parameter vectors $\hat{\theta}_\kappa$. The settings must have been obtained in independent simulation runs of the EM algorithm for different random initial settings $\theta_0$ of the independent parameters. We will return to the important issue of identifiability at the end of the next section. In a recent paper Papadopoulos and Digalakis (2010) developed a new identification procedure based on the EM framework for a broad family of identifiable state-space models and gave a complete solution of maximum likelihood estimation for general linear state space models.

If not only the coefficients and covariance matrices of an LDS have to be estimated from data but also the dimensionality of the hidden state process $\{X_t\}$, a good trade-off between the predictive accuracy gained by increasing the dimension of independent parameters and the danger of overfitting the model to random fluctuations and not to rules that generalize to other datasets has to be found. In an analogous manner to the model selection procedure that was introduced in section 2.4, information-theoretic or Bayesian criteria can be used to evaluate candidate LDS models. An alternative method is to use the previously cited combined deterministic-stochastic subspace identification algorithm of van Overschee and de Moor (1996) to estimate the dimensionality of the state space and to initialize the EM algorithm accordingly. However, in the following we focus on the Schwarz-Bayes information criterion (SBC, cf. eq. 43) because of its close theoretical connection to the minimum description length principle. This principle aims to select the model with the briefest recording of all relevant attribute information and builds on the intuitive notion that model fitting is equivalent to finding an efficient encoding of the data. However, in searching for an efficient code, it is important to not only consider the number of bits required to describe the deviations of the data from the model's predictions, but also the number of bits required to specify the independent parameters of the model (Bialek et al. 2001).The minimum description length principle will be elaborated in section 3.2.2.

According to the work of Yamaguchi et al. (2007), the SBC can be defined for an LDS with dimension $q$ of the hidden states as:

$$\begin{aligned} SBC(q) &= -\frac{2}{T}\ln f_{\hat{\theta}_\kappa}[y_0,\dots,y_T] + \frac{\ln T}{T}(q^2+qp) \\ &= -\frac{2}{T}\ln\prod_{t=0}^{T} f_{\hat{\theta}_\kappa}[y_t|y_{t-1},\dots,y_0] + \frac{\ln T}{T}(q^2+qp) \\ &= -\frac{2}{T}\sum_{t=0}^{T}\ln f_{\hat{\theta}_\kappa}[y_t|y_{t-1},\dots,y_0] + \frac{\ln T}{T}(q^2+qp). \end{aligned} \tag{129}$$

It holds that $f_{\hat{\theta}_\kappa}[y_0|y_{-1}] = f_{\hat{\theta}_\kappa}[y_0]$. The term $\ln f_{\hat{\theta}_\kappa}[y_0,\dots,y_T]$ denotes the best estimate of the local maximum of the log-likelihood function. For a converging estimation process, the best estimate is obtained in the $\kappa$th iteration of the EM algorithm, and the particular setting of the parameters is $\hat{\theta}_\kappa = [\hat{A}_0^{(\kappa)} \quad \hat{H}^{(\kappa)} \quad \hat{C}^{(\kappa)} \quad \hat{V}^{(\kappa)} \quad \hat{\pi}_0^{(\kappa)} \quad \hat{\Pi}_0^{(\kappa)}]$. According to the analysis in section 2.4, the quantity $-2/T\sum_{t=0}^{T}\ln f_{\hat{\theta}_\kappa}[y_t|y_{t-1},\dots,y_0]$ reflects the prediction error of the parameterized LDS and therefore represents the goodness-of-fit of data and model. The second term $\ln T/T\,(q^2+qp)$ penalizes model complexity. Equivalent to the definition of the SBC for a vector autoregression model of order $n$ in eq. 43, the factor $(q^2+qp)$ represents the number of freely estimated parameters of the candidate model. We prefer this formulation, because it can be traced back to our information-theoretic approach to complexity evaluation (section 3.2.3).

The log-likelihood $f_{\hat{\theta}_\kappa}[y_0,\dots,y_T]$ can be written in a simple parametric form (Martens 2009). To derive this form, we define the conditional covariance of the observations at time step $t$ as

$$\begin{aligned} \hat{S}_t &= \mathrm{Var}_{\hat{\theta}_\kappa}[Y_t|Y_{t-1},\dots,Y_0] \\ &= \hat{H}^{(\kappa)}\cdot\mathrm{Var}_{\hat{\theta}_\kappa}\left[X_t\middle|\{Y_t\}_0^{t-1}\right]\cdot\left(\hat{H}^{(\kappa)}\right)^{\mathrm{T}} + \hat{V}^{(\kappa)} \\ &= \hat{H}^{(\kappa)}\cdot\Sigma_t^{t-1}\cdot\left(\hat{H}^{(\kappa)}\right)^{\mathrm{T}} + \hat{V}^{(\kappa)} \\ &= \hat{H}^{(\kappa)}\cdot\hat{A}_0^{(\kappa)}\cdot\Sigma_{t-1}^{t-1}\cdot\left(\hat{A}_0^{(\kappa)\mathrm{T}}\right)\cdot\left(\hat{H}^{(\kappa)}\right)^{\mathrm{T}} + \hat{V}^{(\kappa)}. \end{aligned}$$

It holds that $\Sigma_0^{-1} = \hat{\Pi}_0^{(\kappa)}$ for $t = 0$.

It is obvious that

$$f_{\hat{\theta}_\kappa}[y_t|y_{t-1},\dots,y_0] = \mathcal{N}(y_t;\bar{y}_t,\hat{S}_t),$$

where

$$\bar{y}_t = E_{\hat{\theta}_\kappa}\left[Y_t \middle| \{Y_t\}_0^{t-1}\right] = \hat{H}^{(\kappa)} x_t^{t-1}.$$

It holds that $x_0^{-1} = \hat{\pi}_0^{(\kappa)}$ for $t = 0$.

Hence, the log-likelihood of the sequence of observations can be expressed by the equation:

$$\begin{aligned}
\ln f_{\hat{\theta}_\kappa}[y_0, \dots, y_T] &= \sum_{t=0}^{T} \ln f_{\hat{\theta}_\kappa}[y_t | y_{t-1}, \dots, y_0] \\
&= -\frac{1}{2}\sum_{t=0}^{T}\left(p \ln 2\pi + \ln \mathrm{Det}[\hat{S}_t] + \left(y_t - \hat{H}^{(\kappa)} x_t^{t-1}\right)^{\mathrm{T}} \left(\hat{S}_t\right)^{-1} \left(y_t - \hat{H}^{(\kappa)} x_t^{t-1}\right)\right) \\
&= -\frac{1}{2}\left((T+1)p \ln 2\pi + \ln \mathrm{Det}\left[\hat{\Pi}_0^{(\kappa)}\right] + \left(y_0 - \hat{H}^{(\kappa)}\hat{\pi}_0^{(\kappa)}\right)^{\mathrm{T}} \left(\hat{\Pi}_0^{(\kappa)}\right)^{-1} \left(y_0 - \hat{H}^{(\kappa)}\hat{\pi}_0^{(\kappa)}\right)\right) \\
&\quad -\frac{1}{2}\sum_{t=1}^{T}\left(\ln \mathrm{Det}[\hat{S}_t] + \left(y_t - \hat{H}^{(\kappa)} x_t^{t-1}\right)^{\mathrm{T}} \left(\hat{S}_t\right)^{-1} \left(y_t - \hat{H}^{(\kappa)} x_t^{t-1}\right)\right).
\end{aligned}$$

In the second summand in the last row of the above equation we have $x_t^{t-1} = E_{\hat{\theta}_\kappa}\left[x_t \middle| \{y\}_0^{t-1}\right] = \hat{A}_0^{(\kappa)} x_{t-1}^{t-1}$ and $\hat{S}_t = \mathrm{Var}_{\hat{\theta}_\kappa}\left[Y_t \middle| \{y\}_0^{t-1}\right] = \hat{H}^{(\kappa)} \hat{A}_0^{(\kappa)} \Sigma_{t-1}^{t-1} \left(\hat{A}_0^{(\kappa)}\right)^{\mathrm{T}} \left(\hat{H}^{(\kappa)}\right)^{\mathrm{T}} + \hat{V}^{(\kappa)}$ for $t > 0$.

The number of dimensions $q_{opt}$ of the hidden states of the LDS is considered as the optimal one if it is assigned minimum scores, that is

$$q_{opt} = \arg\min_q SBC(q). \tag{130}$$

## 2.10 Project Management Example Revisited

To demonstrate the more complex concept of a state process with hidden variables and validate the developed project model with field data, the above recursions were implemented into the Mathematica® modeling and simulation environment. The estimation routines were also verified through the Bayesian Net Toolbox for Matlab (BNT), which was developed by Murphy (1998). The field data came from the previous case study of NPD in the small German industrial company (see section 2.5, Schlick et al. 2012). For simplicity, we also focus in the following on the first two overlapping development tasks of project A, “conceptual sensor design” (task 1) and “design of circuit diagram” (task 2), and model only their overlapping range. Due to the barcode-based labor time system in the company, only the series of the time-on-task could be used as an objective observation to estimate the evolution of the hidden state process. According to section 2.5, the conceptual sensor design had reached a completion level of 39.84% when the design of the circuit diagram began. The observed work remaining at the initial time step is therefore $y_0 = (0.6016 \quad 1.0000)^{\mathrm{T}}$

Moreover, we assumed that each development task is accompanied by two latent subtasks representing horizontal and lateral communication. Due to the small size of the company, diagonal communication was not relevant. Under these conditions a six dimensional state process seems to adequate to capture the essential dynamics of the development project and we have $\mathrm{Dim}[x_t] = 6$. We term the corresponding

LDS model an LDS(6,2), because the hidden states are six dimensional and the observations two dimensional. We used the implemented estimation routines to re-estimate the independent parameters in 20 iterations of the EM algorithm. We chose an initial setting $\theta_0$ of the independent parameters, in which Gaussian random numbers were assigned to the initial state, the dynamical operator and the output operator. The covariance matrices were set to a multiple of the identity matrix.

The estimated initial state is

$$\hat{\pi}_0^{(20)} = \begin{pmatrix} -0.6310 \\ -0.0288 \\ -0.0494 \\ 0.0996 \\ 0.0887 \\ -0.1252 \end{pmatrix} \tag{131}$$

and the estimated initial covariance is

$$\hat{\Pi}_0^{(20)} = \{10^{-4}\} \begin{pmatrix} 2.59 & 2.34 & -0.49 & -0.16 & -2.69 & 3.14 \\ 2.34 & 2.41 & -0.35 & -0.09 & -2{,}34 & 2.99 \\ -0.49 & -0.35 & 0.18 & 0.07 & 0.52 & -0.40 \\ -0.16 & -0.09 & 0.07 & 0.27 & -0.40 & 0.07 \\ -2.69 & -2.34 & 0.52 & -0.40 & 4.22 & -3.69 \\ 3.14 & 2.99 & -0.40 & 0.07 & -3.69 & 0.45 \end{pmatrix}. \tag{132}$$

The estimated dynamical operator of the state process is given by

$$\hat{A}_0^{(20)} = \begin{pmatrix} 0.7529 & -0.2087 & -0.0181 & 0.8733 & 0.7012 & -0.0790 \\ 0.1867 & 0.4082 & -0.4242 & -0.1072 & 0.1469 & -0.7013 \\ 0.0825 & 0.1707 & 0.9819 & -0.3177 & -0.2657 & -0.3348 \\ 0.0639 & 0.2200 & 0.1778 & 0.6581 & -0.1411 & 0.2725 \\ 0.3227 & -0.6127 & -0.6896 & -1.2472 & 0.1161 & 0.3758 \\ -0.0109 & -0.0904 & 0.0083 & 0.8295 & 0.4847 & 0.0324 \end{pmatrix} \tag{133}$$

and the estimated covariance matrix of the normally distributed random variable $\varepsilon_t$ related to the state process is given by

$$\hat{C}^{(20)} = \{10^{-4}\} \begin{pmatrix} 2.18 & 1.54 & -0.75 & -0.75 & -0.81 & 1.66 \\ 1.54 & 2.21 & -0.24 & -0.28 & -0.43 & 1.43 \\ -0.75 & -0.24 & 0.91 & 0.49 & 0.42 & 0.13 \\ -0.75 & -0.28 & 0.49 & 0.98 & -1.08 & 0.02 \\ -0.81 & -0.43 & 0.42 & -1.08 & 4.42 & -1.42 \\ 1.66 & 1.43 & 0.13 & 0.02 & -1.42 & 2.9 \end{pmatrix} \tag{134}$$

Furthermore, the output operator

$$\hat{H}^{(20)} = \begin{pmatrix} -1.4273 & 0.4736 & 0.1767 & -1.9618 & -0.7579 & 0.1132 \\ -1.4239 & 0.5222 & -2.5623 & 0.4574 & 0.4166 & 0.7412 \end{pmatrix} \tag{135}$$

was computed. The estimated covariance matrix $\hat{V}$ of the normally distributed random variable $\eta_t$ related to the observation process is given by

$$\hat{V}^{(20)} = \{10^{-5}\} \begin{pmatrix} 1.53 & 0.25 \\ 0.25 & 7.13 \end{pmatrix}. \tag{136}$$

Based on the parameterized LDS model, an additional Monte Carlo simulation was carried out within the Mathematica® software environment. One thousand separate and independent simulation runs were calculated.

In an analogous way to Figures 3 and 4, Figure 7 shows the list plots of the empirically acquired work remaining for both tasks as well as the means and 95% confidence intervals of simulated time series of task processing, which were calculated for the overlapping range over 50 weeks. The stopping criterion of $\delta = 0.02$ was left unchanged and is plotted as a dashed line at the bottom of the chart. Interestingly, according to Figure 7 all 95% confidence intervals of the simulated work remaining of both tasks include the empirical data points from the real project before the stopping criterion is met. Furthermore, the confidence intervals are small. Therefore, the LDS model can be fitted much better to the data than the rather simple VAR(1) and VAR(2) models from section 2.5 (compare Figures 3 and 4 to Figure 7). For instance, in the parameterized VAR(1) model only 49 out of 50 95% confidence intervals for task 1 and 47 out of 50 intervals for task 2 covered the real data points of work remaining. It is important to point out that this high goodness-of-fit only holds under the assumption of a six dimensional state space.

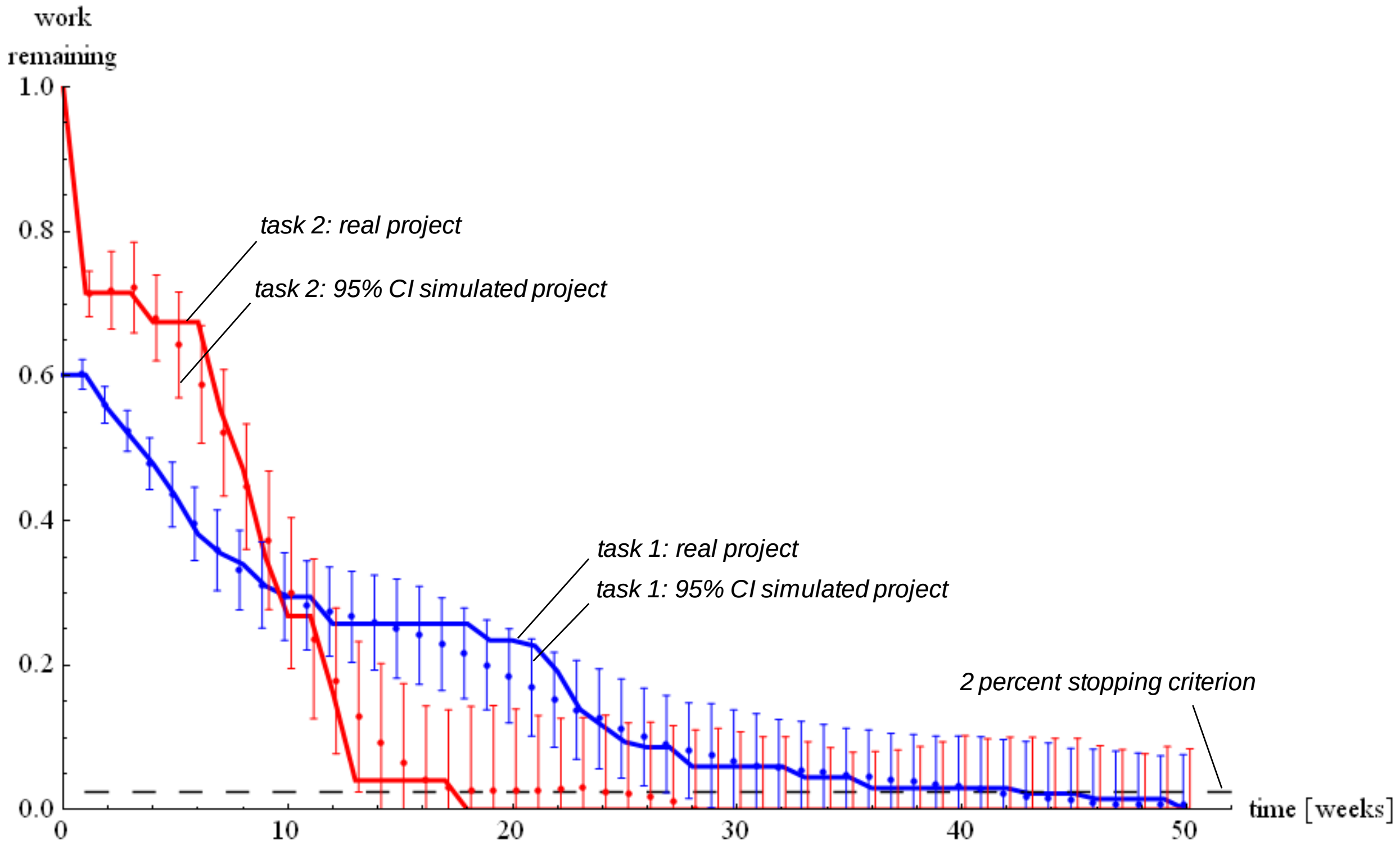

**Figure 7**

*Figure 7*. List plot of work remaining in the real and simulated NPD projects. In an analogous way to Figures 3 and 4, only the overlapping range of the first two tasks is shown in conjunction with the means of simulated traces of task processing as note points and 95% confidence intervals as error bars. The Monte Carlo simulation was based on the simultaneous system of equations 90 and 91. A six dimensional state vector was used to represent the state of the project ($LDS(6,2)$ model). The recursion formula for maximum likelihood parameter estimation based on Expectation-Maximization are given in the text. A total of 1000 separate and independent runs were calculated. Note points have been offset to distinguish the error bars. The stopping criterion of 2% is marked by a dashed line at the bottom of the plot.

If the parameter space is collapsed into three dimensions, for instance, the goodness-of-fit is usually not better than with the vector autogression model of first order. The six dimensional state process is also able to accurately predict the quite abrupt completion of task 2 in week 18, because the mean work remaining at this point in time is equal to the stopping criterion. The root-mean-square deviation between the predicted work remaining in task 1 and the field data is $\mathrm{RMSD}_{\mathrm{task1}} = 0.037$ and therefore 20% lower than the deviation for the VAR(1) model. For task 2, the deviation is $\mathrm{RMSD}_{\mathrm{task2}} = 0.055$. This value is 48% lower than the deviation obtained for the autoregression model of first order (see section 2.5, Schlick et al. 2012).

If the dimension of the hidden states cannot be specified by the model developer based on knowledge about the task and communication structure in the project as shown before, the Schwarz-Bayesian criterion (SBC) is a theoretically convincing alternative, because it favors the candidate model which is a posteriori most probable. The drawback is that the criterion is only valid in a setting with large sample size and therefore must be carefully applied in our case. We systematically varied the dimensions of the hidden states between two and six ($\mathrm{Dim}[x_t] = \{2,3,4,5,6\}$) and calculated the corresponding $SBC(q)$ values based on eq. 129. For each candidate model, 20 iterations of the EM algorithm were computed to estimate the independent parameters. As before, we used an initial setting $\theta_0$ of the independent parameters for each model in which Gaussian random numbers were assigned to the initial state, the dynamical operator and the output operator. The covariance matrices were set to a multiple of the identity matrix.

The model selection procedure showed that SBC is minimal for an LDS with a four dimensional state process and we have $SBC\left(q_{opt} = 4\right) = -20.444$. For this model the estimates of the independent parameters are as follows:

Initial state:

$$\hat{\pi}_0^{(20)} = \begin{pmatrix} 0.1337 \\ 0.2083 \\ -0.3291 \\ 1.0971 \end{pmatrix} \tag{137}$$

Initial covariance:

$$\hat{\Pi}_0^{(20)} = \{10^{-5}\} \begin{pmatrix} 2.68 & -6.97 & 1.82 & -9.40 \\ -6.97 & 22.79 & -5.89 & 31.31 \\ 1.82 & -5.89 & 2.95 & -8.18 \\ -9.40 & 31.31 & -8.18 & 43.20 \end{pmatrix} \tag{138}$$

Dynamical operator:

$$\hat{A}_0^{(20)} = \begin{pmatrix} 0.7596 & 0.1521 & -0.8117 & -0.2609 \\ 0.0709 & 0.7950 & -0.0233 & -0.5082 \\ 0.0703 & -0.0441 & 1.1015 & 0.1212 \\ 0.0387 & -0.14958 & -0.2686 & 0.1135 \end{pmatrix} \tag{139}$$

State noise covariance:

$$\hat{C}^{(20)} = \{10^{-4}\} \begin{pmatrix} 3.15 & -1.90 & -0.52 & -1.67 \\ -1.90 & 3.55 & 0.18 & 3.79 \\ -0.52 & 0.18 & 0.92 & -0.43 \\ -1.67 & 3.79 & -0.43 & 5.12 \end{pmatrix} \tag{140}$$

Output operator:

$$\hat{H}^{(20)} = \begin{pmatrix} -0.4465 & -1.3177 & -0.0023 & 0.8521 \\ 0.6695 & -1.2236 & 0.1059 & 1.0940 \end{pmatrix} \tag{141}$$

Observation noise covariance:

$$\hat{V}^{(20)} = \{10^{-5}\} \begin{pmatrix} 1.17 & 0.12 \\ 0.12 & 8.76 \end{pmatrix} \quad (142)$$

Moreover, one thousand separate and independent simulation runs were calculated to determine the root mean square deviation between the predicted work remaining in task 1 and the field data. The result is $\mathrm{RMSD}_{\mathrm{task1}} = 0.034$. This empirical deviation is almost 10% lower value that was computed for the $\mathrm{LDS}(6{,}2)$ model and more than 25% lower than the corresponding values for the $\mathrm{VAR}(1)$ and $\mathrm{VAR}(2)$ models (section 2.5). Interestingly, for task 1 the predictive accuracy of the $\mathrm{LDS}(4{,}2)$ model is higher than the $\mathrm{LDS}(6{,}2)$ model, even though the state space was reduced from six to four dimensions. For task 2, the empirical deviation is $\mathrm{RMSD}_{\mathrm{task2}} = 0.70$ and therefore approximately 35% and 42% lower than the values obtained for the first- and second-order regression models respectively (section 2.5). Compared to the previously analyzed LDS, the value is approximately 21% larger. We conclude that the predictive accuracy of an LDS with a four dimensional state process whose independent parameters were estimated by the introduced EM algorithm can not be Pareto-inferior to an $\mathrm{LDS}(6{,}2)$ model, and the predictions of the work remaining can be almost as accurate as for the more complex model. For both tasks the total root mean square deviation is only 13% higher. The predictive accuracy of the $\mathrm{LDS}(4{,}2)$ model is significantly higher than the accuracy of the $\mathrm{VAR}(1)$ and $\mathrm{VAR}(2)$ models for both tasks and shows the Pareto-superiority of the approach with hidden state variables. It is important to point out that these conclusions only hold if the independent parameters are estimated using a favorable initial setting. We will return to this issue in the sensitivity analyses.

Figure 8 shows the list plots of the time series from the real project as well as the means and 95% confidence intervals of simulated task processing for the $\mathrm{LDS}(4{,}2)$ model. As before, the stopping criterion of $\delta = 0.02$ is plotted as a dashed line at the bottom of the chart. For both tasks, the means and 95% confidence intervals follow a pattern that is very similar to the $\mathrm{LDS}(6{,}2)$ model (Figure 7). Therefore, all confidence intervals of the simulated work remaining include the data points from the real project before the stopping criterion is met. The only important difference between both models with hidden state variables is that for task 2 the lower dimensional model shows an oscillatory behavior of the means of work remaining after the real completion of the task in week 18. The means first undershoot the abscissa from week 18 to 21, then overshoot it from week 22 to week 32 and finally follow a smoothly decaying geometric series until week 50 (see bottom of Figure 8). Interestingly, these oscillations lead to an average system behavior that makes it possible to predict almost perfectly without using a stopping criterion the abrupt completion of task 2 in week 18.

Finally, we carried out different sensitivity analyses for the LDS models. The importance of these analyses should not be underestimated, given that the EM procedure of Ghahramani and Hinton (1996) and the later developments cited lack identifiability and therefore an infinite number of parameterizations $\hat{\theta}_\kappa$ theoretically exist that yield the same maximum log-likelihood. In the first sensitivity analysis, we investigated the sensitivity of the predictive accuracy of the LDS models in terms of the root-mean-square deviation between the predicted work remaining and the field data for both development tasks. For each model, 200 independent trials of repeated project execution were simulated and analyzed. In each trial, the independent parameters were estimated through 20 iterations of the EM algorithm with an initial setting $\theta_0$ of the independent parameters, in which Gaussian random numbers were assigned to the initial state, the dynamical operator and the output operator. The covariance matrices were set to a multiple of the identity matrix. Based on the estimated independent parameters, 1000 separate and independent simulation runs of project dynamics were computed and the root-mean-square deviations for each case were calculated. The complete dataset was used to calculate the mean $\mathrm{RMSD}^{\mathrm{all}}_{\mathrm{task1}}$ and $\mathrm{RMSD}^{\mathrm{all}}_{\mathrm{task2}}$ over all trials. These overall means were compared to the corresponding $\mathrm{RMSD}_{\mathrm{task1}}$ and $\mathrm{RMSD}_{\mathrm{task2}}$ values that were obtained in the previous Monte Carlo simulation. By doing so, we could assess whether the parameterized LDS(6,2) model according to eqs. 131 to 136 and the LDS(4,2) model according to eqs. 137 to 142 have an above-average predictive accuracy compared to alternative models from the same class. If badly conditioned covariance matrices occurred during the simulation and therefore the EM results possibly contained significant numerical errors, the complete trial was recalculated.

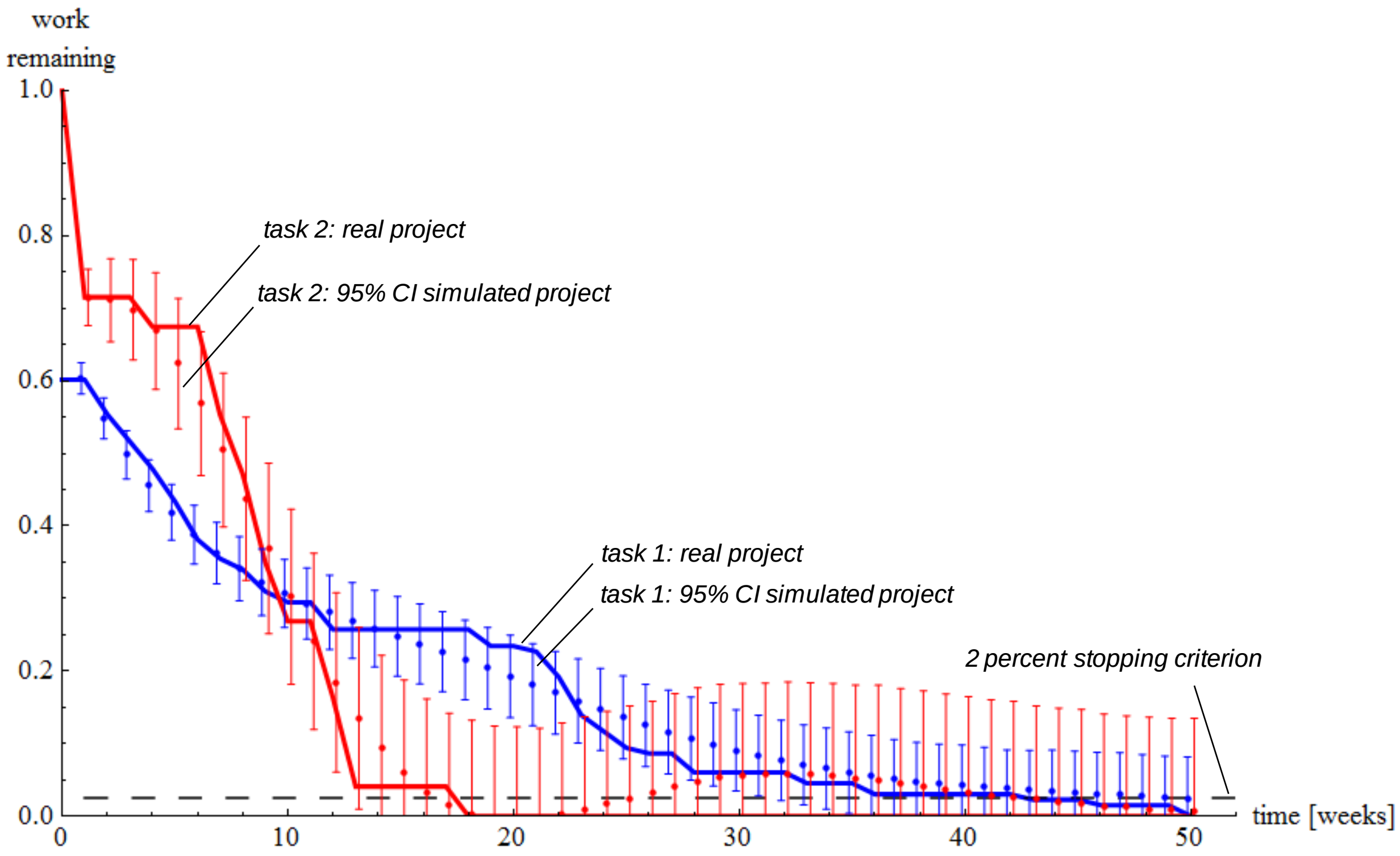

**Figure 8**

*Figure 8*. List plot of work remaining in the real and simulated NPD projects. In an analogous way to Figures 3, 4 and 7, only the overlapping range of the first two tasks is shown in conjunction with the means of simulated traces of task processing as note points and 95% confidence intervals as error bars. The Monte Carlo simulation was based on the simultaneous system of equations 90 and 91. In contrast to Figure 7, only a four dimensional state vector was used to represent the work remaining of the project (known as the $LDS(4,2)$ model). The recursion formula for maximum likelihood parameter estimation based on Expectation-Maximization are given in the text. A total of 1000 separate and independent runs were calculated. Note points have been offset to distinguish the error bars. The stopping criterion of 2% is marked by a dashed line at the bottom of the plot.

For the LDS(6,2) model, an overall $\mathrm{RMSD}_{\mathrm{task1}}^{\mathrm{all}} = 0.0398$ was computed for task 1 and $\mathrm{RMSD}_{\mathrm{task2}}^{\mathrm{all}} = 0.0746$ for task 2. Both overall deviations are larger than the $\mathrm{RMSD}_{\mathrm{task1}}$ and $\mathrm{RMSD}_{\mathrm{task2}}$ values from the previous simulation. Therefore, the predictive accuracy of the LDS(6,2) model shown is higher than the accuracy of an average model. An additional analysis of the empirical cumulative distribution function showed for task 1 that the $\mathrm{RMSD}_{\mathrm{task1}}$ value is equivalent to the second percentile of the distribution. In other words, the $\mathrm{RMSD}_{\mathrm{task1}}$ value is exceeded by 98% of all 200 simulated trials. For task 2 the $\mathrm{RMSD}_{\mathrm{task2}}$ value is equivalent to the 32nd percentile of the distribution. The relative accuracy deviation is approximately 7% for task 1 and 26% for task 2 and therefore negligible for almost every application in project management. The LDS(4,2) model lead to an overall $\mathrm{RMSD}_{\mathrm{task1}}^{\mathrm{all}} = 0.0455$ for task 1 and an overall $\mathrm{RMSD}_{\mathrm{task2}}^{\mathrm{all}} = 0.1065$ for task 2. Similar to the LDS(6,2) model, both values are larger than the previously calculated $\mathrm{RMSD_{task1}}$ and $\mathrm{RMSD_{task2}}$ values. However, the predictive accuracy of the $\mathrm{LDS}(4,2)$ model shown (eqs. 137 to 142) is much higher than the accuracy of average models and demonstrates unexpected predictive power for such a low-dimensional model. The empirical cumulative distribution function showed for task 1 that the $\mathrm{RMSD_{task1}}$ value represents approximately a fourth percentile and for task 2 the $\mathrm{RMSD_{task2}}$ value a 13th percentile. For task 1 the relative accuracy deviation is approximately 28% and for task 2 approximately 36% for the benefit of the given model. These values can be relevant for project controlling. For application in project management, it therefore makes sense to estimate the independent parameters in independent runs of the EM iterations with a randomized initial setting $\theta_0$ of the independent parameters and to select the model with the most favorable properties for the given application. Desired properties can include the highest total predictive accuracy, the highest predictive accuracy in certain tasks, etc.

The second sensitivity analysis evaluated the sensitivity of emergent complexity associated with the parameterized LDS models as constructive representations of a complex NPD project. The dependent variable to evaluate emergent complexity was the effective measure complexity (EMC) according to Grassberger (1986). The EMC is an information-theoretic quantity that measures the mutual information between the infinite past and future histories of a stationary stochastic process. Section 3.2.3 will formally introduce the EMC and will give a detailed explanation of its interrelationships to other key invariants of stochastic processes. In the sense of an information-theory learning curve, EMC measures the amount of apparent randomness at small observation windows during the stochastic process that can be resolved by considering correlations among blocks with increasing length. For a completely randomized work process with independent and identically distributed state variables, the apparent randomness cannot reduced by any means, and it therefore holds that EMC=0. For all other processes that have a persistent internal organization, EMC is strictly positive. NPD projects with more states and larger correlation length are assigned higher complexity values. If optimal predictions are influenced by events in the arbitrarily distant past, EMC can also diverge. However, this is not relevant for the sensitivity analysis of the linear systems. As in the first sensitivity analysis, 200 independent trials of repeated project execution were considered for each LDS model and the corresponding EMC values were calculated. In each trial, the independent parameters were re-estimated in 20 iterations of the EM algorithm. We used an initial setting $\theta_0$ of the independent parameters, in which Gaussian random numbers were assigned to the initial state, the dynamical operator and the output operator. The covariance matrices were set to a multiple of the identity matrix. The closed-form solution from eq. 203 was used to calculate the EMC values based on estimated independent parameters $\hat{\theta}_{20}$. For the $\mathrm{LDS}(6,2)$ model according to eqs. 131 to 136 the complexity value is $\mathrm{EMC_{LDS(6,2)}} = 2.8001$. The 200 independent trials lead to a mean complexity of $\mathrm{EMC_{LDS(6,2)}^{all}} = 2.8154$. The relative difference is less than 1% and therefore negliable for almost every application. An analysis of the empirical cumulative distribution function showed that the $\mathrm{EMC_{LDS(6,2)}}$ value is equivalent to the 52nd percentile of the empirical distribution and therefore very close to the median. Regarding the parameterized $\mathrm{LDS}(4,2)$ model (eqs. 137 to 142) a smaller complexity value of $\mathrm{EMC_{LDS(4,2)}} = 2.6694$ was obtained. The mean complexity over 200 trials was $\mathrm{EMC_{LDS(4,2)}^{all}} = 3.1254$. The relative difference is approximately 17% and therefore much larger than for $\mathrm{LDS}(6,2)$ model. This difference could be relevant for practical complexity evaluations. The $\mathrm{EMC_{LDS(4,2)}}$ value is equivalent to the eleventh percentile of the empirical distribution. This means that 89% of the models in the sample exceed the emergent complexity that is associated with the introduced dynamical operators and covariance matrices. Interestingly, the mean complexity related to the $\mathrm{LDS}(4,2)$ models is larger than the mean related to the higher-dimensional $\mathrm{LDS}(6,2)$ models. This shows that a stochastic process generated by an LDS with a comparatively low dimensionality of the hidden state space will not necessarily be less complex than higher dimensional representations.

In the third sensitivity analysis, we carried out an additional evaluation of the sensitivity of emergent complexity associated with the LDS models. However, we did not use the estimated independent parameters $\hat{\theta}_{20}$ to directly calculate the EMC values based on the closed-form solution from eq. 203. Instead we used them to generate data with the parameterized model. The data consisted of ten independent cases of the work remaining for both development tasks over $T = 100$ time steps. In line with the procedure used for the previous sensitivity analyses, 200 independent trials of repeated data generation were considered for each LDS model. In each trial, the ten independent cases of work remaining were input into the EM algorithm, and 20 iterations were calculated based on an initial setting $\theta_0$, in which Gaussian random numbers were assigned to the initial state, the dynamical operator and the output operator. The covariance matrices were set to a multiple of the identity matrix. After the 20th iteration, the re-estimated independent parameters $\hat{\theta}'_{20}$ were used to calculate the EMC values based on eq. 203. We hypothesized that although the introduced EM procedure lacks identifiability, the emergent complexity can be accurately estimated fusing data from the "true" $\mathrm{LDS}(6,2)$ and $\mathrm{LDS}(4,2)$ models by averaging over EMC values which were obtained in repeated Monte Carlo trials. This is a pivotal hypothesis, because if it proved unverifiable, the introduced maximum likelihood estimation procedure could produce unreliable complexity evaluations in the application domain. Concerning the $\mathrm{LDS}(6,2)$ model from eqs. 131 to 136, the mean complexity value that was estimated from the data is $\mathrm{EMC}_{\mathrm{LDS}(6,2)}^{\mathrm{all,data}} = 2.9465$. The relative difference between this value and the reference $\mathrm{EMC}_{\mathrm{LDS}(6,2)}$ value is approximately 5% and therefore slightly larger than the difference obtained in the purely analytical evaluation. An analysis of the empirical cumulative distribution function showed that the $\mathrm{EMC}_{\mathrm{LDS}(6,2)}$ value is equivalent to the 20th percentile of the empirical distribution of the $\mathrm{EMC}_{\mathrm{LDS}(6,2)}^{\mathrm{all,data}}$ values and therefore not so close to the median. For the parameterized $\mathrm{LDS}(4,2)$ model (eqs. 137 to 142) the corresponding mean complexity is $\mathrm{EMC}_{\mathrm{LDS}(4,2)}^{\mathrm{all,data}} = 2.9168$. The relative difference is approximately 9%. This value is slightly larger than the one obtained in the purely analytical evaluation. However, the relative differences are small for both models and seem to be negliable for most applications. The $\mathrm{EMC}_{\mathrm{LDS}(4,2)}$ value is equivalent to the eight percentile of the empirical distribution. Interestingly, the data-driven sensitivity analysis leads for the $\mathrm{LDS}(6,2)$ model to a mean complexity that is higher than for the alternative model. This result is inconsistent with the second sensitivity analysis. For application in project management it therefore makes sense to determine the EMC values directly by utilizing the closed-form solution from eq. 203 and not to rely on indirect estimation methods. Finally –and most importantly– the results show that although the EM procedure lacks identifiability, emergent complexity can be accurately estimated from data simply by averaging over EMC values which were obtained in repeated Monte Carlo trials. Additional methods of dealing with divergent and inconsistent findings by introducing auxiliary concepts seem to be unnecessary.

In conclusion, in terms of emergent complexity the introduced $\mathrm{LDS}(6,2)$ model seems to be a very good dynamical representation of all models within this class that are parameterized based on the EM algorithm. The emergent complexity associated with the alternative $\mathrm{LDS}(4,2)$ model shown is within the first quartile and therefore too low to be representative. However, we preferred to present that model with reduced state space dimensionality because the root-mean-square deviation between the predicted work remaining and the field data is within the first quantiles for both tasks, and for task 1 it is also lower than the deviation obtained of the $\mathrm{LDS}(6,2)$ model. This finding holds even though lower complexity values are assigned to the $\mathrm{LDS}(4,2)$ model. This is a good example of how emergent complexity in the sense of Grassberger´s theory cannot simply be evaluated by classic predictability measures. A good model not only has to be able to make accurate predictions; it must also use an internal state representation that is not unnecessarily complex. We will return to this important issue in the next section.

# 3 EVALUATION OF COMPLEXITY IN NEW PRODUCT DEVELOPMENT

The term “complexity” stems from the Latin word “complexitas”, meaning comprehensive or inclusive. In current language usage, it is the opposite of simplicity, but this interpretation does not appear to be underpinned by any explicit concept. Various disciplines have studied the concepts and principles of complexity in basic and applied scientific research. Several frameworks, theories and measures have been developed, depending on differing views of complexity among disciplines. An objective evaluation of structural and dynamic complexity in NPD would benefit system designers and managers, because it would be enable them to compare and optimize different systems in analytical and experimental studies. To obtain a comprehensive view of organizational, process and product elements and their interactions in the product development environment, a thorough review of the notion of complexity has to start by organizational theory (section 3.1). The literature on organizational theory demonstrates that the complexity of NPD projects results from different sources and the consideration of the underlying organizational factors and their interrelationships is important for successful project management (Kim and Wilemon 2009). However, we hypothesize that factor-based approaches are not sufficient to evaluate emergent complexity in open organizational systems with cooperative work processes and therefore the complexity theories and measures of basic scientific research must also be taken into account (cf. Amaral and Uzzi 2007). They can provide deep and consistent insights into emergent phenomena of systems and dynamic complexity of cooperation (section 3.2). Selected measures can also be used to optimize project organizational design (Schlick et al. 2009). The formalized measures build upon our intuition that a system is complex if it is difficult to describe and predict efficiently. A comprehensive overview of this concept including detailed descriptions and illustrations can be found in Shalizi (2006), Prokopenko et al. (2007) and Nicolis and Nicolis (2007). For effective complexity management in NPD, the product-oriented measures from theories of systematic engineering design are also relevant (section 3.3). Seminal work in this field has been done by Suh (2005) on the basis of information-theoretic quantities. These quantities are also the foundation of statistical complexity measures from basic scientific research, which means that Suh´s complexity theory and recent extensions of it (see Summers and Shah 2010) must be discussed in the light of the latest theoretical developments. Moreover, the literature that has been published around the design structure matrix (Steward 1981) as a dependency modeling technique has to be considered. This literature also provides a firm foundation for mathematical modeling of cooperative work in NPD projects. In general, we try to restrict our analyses to mature scientific theories because of their objectivity and construct validity.

## 3.1 Approaches from Organizational Theory

According to Murmann (1994) and Griffin (1997) complexity in the product development environment is determined by the number of (different) parts in the product and the number of embodied product functions. Kim and Wilemon (2003) developed a complexity assessment template covering these and other important “sources”. The first source in their assessment template is “technological complexity”, which can be divided into “component integration” and “technological newness”. The second source is the “market (environmental) complexity” that results from the sensitivity of the project’s attributes to market changes. “Development complexity” is the third source and is generated when different design decisions and components have to be integrated, qualified suppliers have to be found and supply chain relationships have to be managed. The fourth source is “marketing complexity” resulting from the problems of bringing the product to market. “Organizational complexity” is the fifth source because projects usually require intensive cooperation and involve many areas of the firm. In large-scale engineering projects other companies are also involved. Their coordination leads to “intraorganizational complexity”, the sixth source. In order to validate and prioritize complexity sources, Kim and Wilemon (2009) carried out an extensive empirical investigation. An analysis of exploratory field interviews with 32 project leaders and team members showed that technological challenges, product concept/customer requirement ambiguities and organizational complexity are major issues promoting complexity in NPD. The perceived dominant source was technological challenges, since about half of the respondents noted technological difficulties encountered in attempting to develop a product using an unproven technique or process.

Hölttä-Otto and Magee (2006) developed a project complexity framework based on the seminal work of Summers and Shah (2003). They identified three dimensions: the product itself (artifact), the project mission (design problem), and the tasks required to develop the product (process). The key indicators for each of these complexities are size, interactions and stretch (solvability). They conducted interviews in five divisions of large corporations competing in different industries in the North American market. The results show that the effort estimation is primarily based on the scale and the stretch of the project and, surprisingly, not on subsystem interactions. Tatikonda and Rosenthal (2000) focus on the task-dimension and relate project complexity to the nature, quantity and magnitude of the organizational subtasks and subtasks interactions required by a project.

A recent literature review and own empirical work about elements contributing to complexity in large engineering projects was published by Bosch-Rekveldt et al. (2011). The analysis of the literature sources and 18 semi-structured interviews in which six completed projects were studied in depth led to the development of the TOE framework. The framework covers 50 different elements, which are grouped into three main categories: "technical complexity" (T), "organizational complexity" and "environmental complexity" (E). Additional subcategories of TOE are defined on a lower level: "goals", "scope", tasks", "experience", "size", "resources", "project team", "trust", "stakeholders", "location", "market conditions", and "risks", showing that organizational and environmental complexity are more often linked with softer, qualitative aspects. Interestingly, Bosch-Rekveldt et al. (2011) distinguish between project complexity and project management (or managerial) complexity. Project management complexity is seen as a subset of project complexity. Various normative organizing principles to cope with managerial complexity can be found in the standard literature on project management (e.g. Shtub et al. 2006; Kerzner 2009). If, for instance, managerial complexity is low, project management within the classic functional organizational units of the company is usually most efficient and cross-functional project organization types can create unnecessary overhead. However, if coordination needs between functional, spatial and temporal boundaries are high, a matrix organization is often a better practice allowing development projects to be staffed with specialists from throughout the organization (Shtub et al. 2006). For large engineering projects with a long duration a pure project organization is often the preferred type in industry as the project is cared for full-temporally by a team that is fully responsible for the entire extent. Specific sources of managerial complexity and their impact on performance were also examined in the literature, e.g. communication across functional boundaries (Carlile 2002), cross-boundary coordination (Kellogg et al. 2006), spatial and temporal boundaries in globally distributed projects (Cummings et al. 2009) and the effects of a misalignment in the geographic configuration of globally distributed teams (O'Leary and Mortensen 2010). Maylor et al. (2008) developed an integrative model of perceived managerial complexity in project-based operations. Based on a multistage empirical study elements of complexity were identified and classified under the dimensions of "mission", "organization", "delivery", "stakeholder", and "team".

The literature review shows that there is a large variety of nomenclatures and definitions for the sources of complexity in NPD projects. However, the underlying factors have not yet been integrated into a single objective and valid framework. According to Lebcir (2011) there is an urgent need for a new, non-confusing, and comprehensive framework that is derived from the extensive body of knowledge. He suggests a framework in which "project complexity" is decomposed into "product complexity" and "innovation". Product complexity refers to structural complexity (see section 3.3) and is determined by "product size" in terms of the number of elements (components, parts, sub-systems, functions) in the product and "product interconnectivity", representing the level of linkages between elements. On the other hand, innovation refers to "product newness" and "project uncertainty". Product newness represents the degree of redesign of the product compared to previous generations of the same or similar product. Project uncertainty represents the fact that methods and capabilities are often not clearly defined at the starting point of a project. The results of a dynamic simulation indicate that an increase in uncertainty has a significant impact on the development time. The other factors tend to increase development time as they increase, but their impact is not significantly different in projects involving medium or high levels of these factors.

The complexity templates and frameworks from organization theory are especially beneficial for project management because they help to focus managerial intervention on empirically validated performance-shaping factors. It must be criticized, though, that without a quantitative theory of emergent complexity it is almost impossible to identify the essential variables and their interrelationships. Furthermore, it is very difficult to consolidate them into one consistent complexity metric. In the literature very few authors, such as Mihm et al. (2003, 2010), Rivkin and Siggelow (2003, 2007), Braha and Bar-Yam (2007) build upon quantitative scientific concepts for the analysis of complex sociotechnical systems. Mihm et al. (2003) presents analytical results from random matrix theory predicting that the larger the project, as measured by components or interdependencies, the more likely are problem-solving oscillations and the more severe they become—failure rates grow exponentially. In the work of Rivkin and Siggelkow (2003, 2007) the famous biological evolution theory of Kauffman and the NK model are used for studying organizations as systems of interacting decisions. Different interaction patterns such as block diagonal, hierarchical, scale-free, and so on are integrated into a simulation model to identify local optima. The results show that by holding the total number of interactions among decisions fixed, a shift in the pattern can alter the number of local optima by more than an order of magnitude. In a similar fashion Mihm et al. (2010) uses a statistical model and Monte-Carlo experiments to explore the effect of an organizational hierarchy on search solution stability, quality and speed. Their results show that assigning a lead function "anchoring" a solution speeds up problem solving, local solution choice should be delegated to the lowest hierarchical level and organizational structure matters little at the middle management level, but it matters at the "front line" — front-line groups should be kept small. Braha and Bar-Yam (2007) examine the statistical properties of networks of people engaged in distributed development and discuss their significance. The autoregression models of cooperative work that were introduced in section 2 (eq. 4 and 24) are quite closely related to their dynamical model. However, there are important differences: the VAR(1) models are defined over a continuous range of state values and can therefore represent different kinds of cooperation relationships as well as precedence relations (e.g. overlapping); each task is nonequally influenced by other tasks; and finally, correlations $\rho_{ij}$ between performance fluctuations among tasks $i$ and $j$ can be captured.

## 3.2 Approaches from Basic Scientific Research

### 3.2.1 Algorithmic Complexity

Historically, the most important measure from basic scientific research is algorithmic complexity, dating to the great mathematicians Kolmogorov, Solomonoff and Chaitin. They developed independently a measure known today as the "Kolmogorov–Chaitin complexity" (Chaitin 1987, Li and Vitányi 1997).

Considering an information processing system, the complexity of the intricate mechanisms of a nontrivial system can be evaluated using output signals and symbols that are communicated to an intelligent observer. In that sense, complexity is manifested to an observer through the complicated way in which events unfold in time and organize in state space. According to Nicolis and Nicolis (2007), characteristic signatures of such spatiotemporal complexity are nonrepetitiveness, a pronounced variability extending over many place and time scales, and sensitivity to initial conditions and to the other parameters. Furthermore, a given system can generate a variety of dependencies of this kind associated with the different states simultaneously available. If the transmitted output of a complex system is symbolic, it can be concatenated in the form of a data string $x$ and may be sequentially stored in a computer file for a post-hoc analysis. The symbols are typically chosen from a predefined alphabet $\mathcal{X}$. If the output is a time- or space-continuous signal, it can be effectively encoded with methods of symbolic dynamics (Lind and Marcus 1995, Nicolis and Nicolis 2007). The central idea of Kolmogorov, Solomonoff and Chaitin is that a generated string is "complex" if it is difficult for the observer to describe. The observer can describe the string by writing a computer program that reproduces it. The difficulty of description is measured by the length of the computer program on a Universal Turing Machine $U$. If $x$ is transformed into binary form, the algorithmic complexity of $x$, termed $K_U(x)$, is the length of the shortest program with respect to $U$ that will print $x$ and then halt. According to Chaitin (1987) an additional requirement is that the string $x$ has to be encoded by a prefix code $d(x)$. A prefix code is a type of code system, that has no valid code word that is a prefix (start substring) of any other valid code word in the set. The complete definition of the Kolmogorov–Chaitin complexity is:

$$K_U(x) = \min\{|d(p)|: U(p) = x\}.$$

In that sense, $K_U(x)$ is a measure of the computational resources needed to specify the data string $x$ in the language of $U$. We can directly apply this algorithmic complexity concept to project management by breaking down the whole work in the project into fine-grained activities $a_i$ and labeling the activities unambiguously by discrete events $e_i$ from a predefined set $\mathcal{X}$ ($i = 1, \ldots, |\mathcal{X}|$). During project execution it is recorded when activity $a_i$ is finished successfully and this is indicated by scheduling the corresponding event $e_i$. The sequence of scheduled events $x = \left(e_{j(o)}, e_{j(1)} \ldots\right)$ ($e_{j(i)} \in \mathcal{X}, j(\tau) \in \{1, \ldots, |\mathcal{X}|\}, \tau = 0,1, \ldots$) encodes how the events unfold in time and organize in a goal-directed workflow. The index $j(\tau)$ can be interpreted as a pointer to the event $e$ that occurred at position $\tau$ in the data sequence $x$. It is obvious that a simple periodic work process whose activities are processed in strict cycles, like in an assembly line, is not complex because we can store a sample of the period and write a program that repeatedly outputs it. At the opposite end of the complexity range in the algorithmic sense, a completely unpredictable work process without a purposeful internal organization cannot be described in any meaningful way except by storing every feature of task processing, because we cannot identify any persisting structure to utilize for a shorter description. This example shows very clearly that the algorithmic complexity is not a good measure for emergent complexity in NPD projects, because it is maximal in the case of purely random task processing. Intuitively, such a state of "amnesia", in which no piece of information from the project history is valuable for improving the forecasts of the project manager and the CE team members, is not really complex. The algorithmic complexity cannot uncover the important long-range interactions between tasks or evaluate multilayer interactions in the hierarchy of an organization either. An additional conceptual weakness of the algorithmic complexity measure and its later refinements is that it aims for an exact description of patterns. Many of the details of any configuration are just random fluctuations (noise) from different sources such as human performance variability. Clearly, it is impossible to identify regularities from noise that generalize to other datasets from the same complex system; to assess complexity, the underlying regularities and rules shaping system dynamics must be in focus. These regularities and rules must be separated from noise through specific selection principles. Therefore, a statistical representation is necessary that refers not to individual patterns but to a joint ensemble generated by a complex system in terms of an information source. In complex systems, the deterministic and probabilistic dimensions become two facets of the same reality: the limited predictability of complex systems (in the sense of the traditional description of phenomena), forces the need for an alternative view, and the probabilistic description enables sorting out regularities of a new kind. On the other hand, far from being applied in a heuristic manner, in which observations have to fit certain preexisting laws imported from classical statistics, the probabilistic description we deal with here is "intrinsic" (Nicolis and Nicolis 2007), meaning that it is generated by the underlying system dynamics. Depending on the scale of the phenomenon, a complex system may have to develop mechanisms for controlling randomness to sustain a global behavioral pattern or, in contrast, to thrive on randomness and to acquire in a transient manner the variability and flexibility needed for its evolution between two such configurations. Besides these significant conceptual weaknesses, a fundamental computational problem is that $K_U(x)$ cannot be calculated exactly. We can only approximate it "from above", which is the subject of the famous Chaitin theorem (Chaitin 1987).

### 3.2.2 Stochastic Complexity

The most prominent statistical complexity measure is Rissanen's (1989, 2007) stochastic complexity. It is rooted in the construction of complexity penalties for model selection (see procedure for VAR($n$) model in section 2.3), where a good trade-off between the prediction accuracy gained by increasing the number of independent parameters and the danger of overfitting the model to random fluctuations and not regularities that generalize to other datasets has to be found. In an early paper, Wallace and Boulton (1968) hypothesized that this trade-off could best be resolved by selecting the model with "the briefest recording of all attribute information." To the best of our knowledge, Akaike (1974) developed the first quantitative step along this line of thought based on information theory. His complexity term – named after its inventor Akaike Information Criterion (AIC, see section 2.3) − was created more or less ad hoc without bearing an explicit complexity theory in mind. Generally speaking, the AIC is defined as

$$AIC = -2\ln\left(l(\hat{\theta}_{1,\mathcal{T}})\right) + 2k, \tag{143}$$

where $l(\hat{\theta}_{1,\mathcal{T}})$ is the maximized value of the likelihood function $l(.)$ for the estimated statistical model, and $k$ denotes the number of free parameters in the model. In the above definition the dependency of the criterion on the number of data points is only implicit through the likelihood function. For a $VAR(n)$ model it can be made explicit as shown in eq. 42 from section 2.3. Akaike´s fundamental ideas were developed systematically by Rissanen in a series of papers and books starting from 1978. Rissanen (1989, 2007) emphasizes that fitting a statistical model to data is equivalent to finding an efficient encoding of those data and that in searching for an efficient code we need to measure not only the number of bits required to describe the deviations of the data from the model's predictions, but also the number of bits required to specify the independent parameters of the model (Bialek et al. 2001). This specification has to be done with a level of precision that is supported by the data.

To clarify this theoretically convincing concept, it is assumed that we had carried out a work sampling study in a complex CE project with many and intensive cooperation relationships among the development teams. Based on a large number of observations the proportion of time spent by the developers in predefined categories of activity $\mathcal{X} = \{x_1, \ldots, x_m\}$ (e.g. sketching, drawing, calculating, communicating etc.) was estimated with high statistical accuracy. In addition to the observations at random times a comprehensive longitudinal observation of workflows of different development teams was carried out in a specific project phase at regular intervals. The observations were made in $R$ independent trials and encoded by the same categories of activity $\mathcal{X}$. We define the $r$-th workflow in the specific project phase in a formal manner as a data string $x_r^{\mathcal{T}} = \left(x_{j_r(0)}, \dots, x_{j_r(\mathcal{T})}\right)$ of length $\mathcal{T}$ ($x_{j_r(\tau)} \in \mathcal{X}, j_r(\tau) \in \{1, \dots, |\mathcal{X}|\}, \tau = 0,1, \dots, \mathcal{T}, r = 1, \dots, R$). In a similar manner as in the previous section the index $j_r(\tau)$ can be interpreted as a pointer to activity $x_{j_r(\tau)} \in \mathcal{X}$ observed at time instant $\tau$ in the $r$-th data string $x_r^{\mathcal{T}}$. All empirically acquired data strings are stored in a database of ordered sequences $DB = \left\{x_1^{\mathcal{T}}, \dots, x_R^{\mathcal{T}}\right\}$. We aim at developing an integrative workflow model that can be used for prediction and evaluation of development activities in the project phase based on the theory of discrete random processes. Therefore, we start by defining a finite one-dimensional random process $(X_0, \dots, X_{\mathcal{T}})$ of discrete state variables. In terms of information theory the process communicates to an observer how the development activities unfold and organize in time. The statistical properties of the random process can be described by a joint probability mass function (*pmf*):

$$f_{(X_0,\dots,X_{\mathcal{T}})}\left(x_{j(0)}, \dots, x_{j(\mathcal{T})}\right) = P\left(X_0 = x_{j(0)}, \dots, X_{\mathcal{T}} = x_{j(\mathcal{T})}\right) \qquad j(\tau) \in \{1, \dots, |\mathcal{X}|\}.$$

The joint *pmf* is a function indicating the joint probability that the discrete random variables in the sequence $(X_0, \dots, X_{\mathcal{T}})$ are exactly equal to some observed data sequence $\left(x_{j(o)}, \dots, x_{j(\mathcal{T})}\right)$. Without limiting the generality of the approach, the joint *pmf* can be factorized over all $\mathcal{T}$ time steps using Bayes' theorem of conditional probability as:

$$P\left(X_0 = x_{j(0)}, \dots, X_{\mathcal{T}} = x_{j(\mathcal{T})}\right) = P\left(X_0 = x_{j(0)}\right) \prod_{\tau=2}^{\mathcal{T}} P\left(X_\tau = x_{j(t)} \middle| X_{\tau-1} = x_{j(\tau-1)}, \dots, X_1 = x_{j(1)}\right).$$

The above decomposition of the joint *pmf* into predictive conditional distributions $P\left(X_\tau = x_{j(t)} \middle| X_{\tau-1} = x_{j(\tau-1)}, \dots, X_1 = x_{j(1)}\right)$ with correlations of increasing length $\tau$ can capture interactions between activities of long range and evaluate multilayer interactions in the hierarchy and therefore holds true under any circumstances of cooperation relationships in the specific phase of the CE project. It is assumed that there are persistent workflow patterns in the project phase and we can express them by a reduced dependency structure capturing only short correlations, e.g. by using a Markov chain of order $n \ll \mathcal{T}$ or an equivalent dynamic Bayesian network (see Gharahmani 2001). In that sense the reduced dependency structure reflects only the essential signature of spatiotemporal coordination in the project phase on specific time scales. In the simplest case, only transitions between two consecutive development activities must be taken into account and a Markov chain of first order is an adequate candidate model to capture the essential dependencies. In this model the conditional probability distribution of development activities at the next time step – and in fact all future steps – depends only on the current activity and not on past instances of the process. Hence, the joint probability can be expressed as:

$$P\left(X_0 = x_{j(0)}, \dots, X_{\mathcal{T}} = x_{j(\mathcal{T})}\right) = P\left(X_0 = x_{j(0)}\right) \prod_{\tau=1}^{\mathcal{T}} P\left(X_\tau = x_{j(\tau)} \middle| X_{t-1} = x_{j(\tau-1)}\right).$$

After the model structure has been defined by the above factorization of the joint *pmf*, we have to specify the free parameters. Following the notation of the previous sections we denote the parameter vector by $\theta \in \mathbb{R}^k$. Due to the apparent "memorylessness" of the Markov chain of first order, only the initial distribution

$$\pi_0 = \left(P(X_0 = x_1), \dots, P\left(X_0 = x_{|\mathcal{X}|}\right)\right) \in [0; 1]^{|\mathcal{X}|}$$

of the probability mass over the state space $\mathcal{X}$ and the transition probabilities

$$P = (p_{ij}) = \begin{pmatrix} P(X_\tau = x_1 | X_{\tau-1} = x_1) & P(X_\tau = x_2 | X_{\tau-1} = x_1) & \dots \\ P(X_\tau = x_1 | X_{\tau-1} = x_2) & P(X_\tau = x_2 | X_{\tau-1} = x_2) & \\ \vdots & & \ddots \end{pmatrix} \in [0;1]^{|\mathcal{X}|^2}.$$

between consecutive activities are relevant for making good predictions. Hence, we have the ordered pair of independent parameters:

$$\theta_1 = [\pi_0 \quad P\,].$$

Please note that only $(|\mathcal{X}| - 1)$ parameters of the initial distribution $\pi_0$ and $|\mathcal{X}|(|\mathcal{X}| - 1)$ of the transition matrix $P$ are freely estimated parameters, because a legitimate *pmf* has to be formed and the constraints

$$\sum_{i=1}^{|\mathcal{X}|} \pi_0^{(i)} = 1 \quad \text{and} \quad \forall i\colon \sum_{j=1}^{|\mathcal{X}|} p_{ij} = 1$$

have to be satisfied.

We can use Maximum Likelihood Estimation (MLE, see chapter 2.4) to minimize the deviations of the empirically acquired data sequences from the model's predictions (see e.g. Papoulis and Pillai 2002, Shalizi 2006). In other words, the goodness of fit is maximized. The maximum likelihood estimate of the parameter pair $\theta_1$ is denoted by $\hat{\theta}_{1,\mathcal{T}}$. MLE was pioneered by R.A. Fisher (cf. Edwards 1972) under a repeated-sampling paradigm and is the most prominent estimation technique. As an estimation principle, maximum likelihood is supported by $\hat{\theta}_{1,\mathcal{T}}$´s asymptotic efficiency in a repeated sampling setting under mild regularity conditions and its attainment of the Cramer-Rao information lower bound in many exponential family examples in the finite-sample case (Hansen and Yu 2001). For a first-order Markov chain, the estimate $\hat{\theta}_{1,\mathcal{T}}$ can be determined by solving the objective function:

$$\begin{aligned} \hat{\theta}_{1,\mathcal{T}} &= \arg\max_{\theta_1} \prod_{r=1}^{R} P\big(X_0 = x_{j_r(0)} | \theta_1\big) \prod_{\tau=1}^{\mathcal{T}} P\big(X_\tau = x_{j_r(\tau)} \big| X_{t-1} = x_{j_r(\tau-1)}, \theta_1\big) \\ &= \arg\max_{(\pi_0, P)} \prod_{r=1}^{R} P\big(X_0 = x_{j_r(0)} | \pi_0\big) \prod_{\tau=1}^{\mathcal{T}} P\big(X_\tau = x_{j_r(\tau)} \big| X_{t-1} = x_{j_r(\tau-1)}, P\big) \end{aligned}$$

Please note that the objective function is only valid if all $R$ data sequences had been acquired in independent trials.

Due to the inherent memorylessness of the first-order Markov chain, this model usually is not expressive enough to capture the essential correlations between activities in a project phase. Consequently, a second-order Markov chain is considered as a second candidate model with expanded memory. For this model, the joint probability can be expressed as:

$$\begin{aligned} P\big(X_0 = x_{j(0)}, \dots, X_\mathcal{T} = x_{j(\mathcal{T})}\big) &= P\big(X_0 = x_{j(0)}\big) P\big(X_1 = x_{j(1)} \big| X_0 = x_{j(0)}\big) \\ &\quad \cdot \prod_{\tau=2}^{\mathcal{T}} P\big(X_\tau = x_{j(\tau)} \big| X_{\tau-1} = x_{j(\tau-1)}, X_{t-2} = x_{j(\tau-2)}\big). \end{aligned}$$

Clearly, the predictive distribution $P\left(X_\tau = x_{j(t)} \middle| X_{\tau-1} = x_{j(\tau-1)}, X_{t-2} = x_{j(\tau-2)}\right)$ cannot only capture first-order transitions between activities but also conditional transitions of the process covering two time steps. To parameterize this extended chain, three quantities are required: The initial distribution

$$\pi_0 = \left(P(X_0 = x_1), \dots, P\left(X_0 = x_{|\mathcal{X}|}\right)\right) \in [0;1]^{|\mathcal{X}|},$$

the transition probabilities between consecutive activities at the first two time steps

$$P_0 = \left(p_{0,ij}\right) = \begin{pmatrix} P(X_1 = x_1|X_0 = x_1) & P(X_1 = x_2|X_0 = x_1) & \dots \\ P(X_1 = x_1|X_0 = x_2) & P(X_1 = x_2|X_0 = x_2) & \\ \vdots & & \ddots \end{pmatrix} \in [0;1]^{|\mathcal{X}|^2}$$

and the transition probabilities for the next activity given both preceeding activities at arbitrary time steps

$$P = \left(p_{ij}\right) = \begin{pmatrix} p(x_1|x_1, x_1) & p(x_1|x_1, x_2) & \dots & p\left(x_1\middle|x_1, x_{|\mathcal{X}|}\right) \\ p(x_1|x_2, x_1) & p(x_1|x_2, x_2) & \dots & p\left(x_1\middle|x_2, x_{|\mathcal{X}|}\right) \\ \vdots & \vdots & & \vdots \\ p\left(x_1\middle|x_{|\mathcal{X}|}, x_1\right) & p\left(x_1\middle|x_{|\mathcal{X}|}, x_2\right) & & p\left(x_1\middle|x_{|\mathcal{X}|}, x_{|\mathcal{X}|}\right) \\ p(x_2|x_1, x_1) & p(x_2|x_1, x_2) & & p\left(x_2\middle|x_1, x_{|\mathcal{X}|}\right) \\ p(x_2|x_2, x_1) & p(x_2|x_2, x_2) & & p\left(x_2\middle|x_2, x_{|\mathcal{X}|}\right) \\ \vdots & \vdots & & \vdots \\ \vdots & \vdots & & \vdots \\ p\left(x_{|\mathcal{X}|}\middle|x_{|\mathcal{X}|}, x_1\right) & p\left(x_{|\mathcal{X}|}\middle|x_{|\mathcal{X}|}, x_2\right) & & p\left(x_{|\mathcal{X}|}\middle|x_{|\mathcal{X}|}, x_{|\mathcal{X}|}\right) \end{pmatrix} \in [0;1]^{|\mathcal{X}|^3}.$$

In the above matrix the shorthand notation $p\left(x_i\middle|x_j, x_k\right) = P\left(X_\tau = x_i\middle|X_{\tau-1} = x_j, X_{\tau-2} = x_k\right)$ was used. Hence, we have the parameter triple

$$\theta_2 = [\pi_0 \quad P_0 \quad P].$$

In this triple $(|\mathcal{X}| - 1)$ parameters of the initial distribution $\pi_0$, $|\mathcal{X}|(|\mathcal{X}| - 1)$, parameters of the initial transition matrix $P_0$ and $|\mathcal{X}|^2(|\mathcal{X}| - 1)$ of the general transition matrix $P$ are freely estimated parameters, because a legitimate *pmf* has to be formed. The ordered pair $[\pi_0 \quad P_0]$ can be regarded as the starting state of the chain. We denote the maximum likelihood estimate for the parameterized model by $\hat{\theta}_{2,\mathcal{T}}$. The corresponding objective function is:

$$\begin{aligned} \hat{\theta}_{2,\mathcal{T}} &= \arg\max_{\theta_2} \prod_{r=1}^{R} P\left(X_0 = x_{j_r(0)}|\theta_2\right) P\left(X_1 = x_{j_r(1)}\middle|X_0 = x_{j_r(0)}, \theta_2\right) \\ &\quad \cdot \prod_{\tau=2}^{\mathcal{T}} P\left(X_\tau = x_{j_r(\tau)}\middle|X_{t-1} = x_{j_r(\tau-1)}, X_{t-2} = x_{j_r(\tau-2)}, \theta_2\right) \\ &= \arg\max_{(\pi_0, P_0, P)} \prod_{r=1}^{R} P\left(X_0 = x_{j_r(0)}|\pi_0\right) P\left(X_1 = x_{j_r(1)}\middle|X_0 = x_{j_r(0)}, P_0\right) \\ &\quad \cdot \prod_{\tau=2}^{\mathcal{T}} P\left(X_\tau = x_{j_r(\tau)}\middle|X_{t-1} = x_{j_r(\tau-1)}, X_{t-2} = x_{j_r(\tau-2)}, P\right). \end{aligned}$$

It is not difficult to prove that the solution of the objective functions for Markov chains of first and second order (and also all higher orders) is equivalent to the relative frequencies of observed subsequences of activity in the database (Papoulis and Pillai 2002). In other words, the MLE results can be obtained by simple frequency counting of data substrings of interest. Let the #-operator be an unary counting operator that counts the number of times the data string $(x_{j(o)}x_{j(1)} \ldots)$ in the argument occurred in the database $DB = \{x_1^{\mathcal{T}}, \ldots, x_R^{\mathcal{T}}\}$. Then the MLE yields

$$\hat{\pi}_0 = \left\{\frac{1}{R}\right\}\left(\#(x_1)_{\tau=0}, \ldots, \#(x_{|\mathcal{X}|})_{\tau=0}\right)$$

$$\hat{P} = \left\{\frac{1}{R\mathcal{T}}\right\}\begin{pmatrix} \#(x_1x_1) & \#(x_1x_2) & \cdots \\ \#(x_2x_1) & \#(x_2x_2) & \\ \vdots & & \ddots \end{pmatrix}.$$

for the first-order Markov chain and

$$\hat{\pi}_0 = \left\{\frac{1}{R}\right\}\left(\#(x_1)_{\tau=0}, \ldots, \#(x_{|\mathcal{X}|})_{\tau=0}\right)$$

$$\hat{P}_0 = \left\{\frac{1}{R\mathcal{T}}\right\}\begin{pmatrix} \#(x_1x_1)_{\tau=0} & \#(x_1x_2)_{\tau=0} & \cdots \\ \#(x_2x_1)_{\tau=0} & \#(x_2x_2)_{\tau=0} & \\ \vdots & & \ddots \end{pmatrix}$$

$$\hat{P} = \left\{\frac{1}{R(\mathcal{T}-1)}\right\}\begin{pmatrix} \#(x_1x_1x_1) & \#(x_2x_1x_1) & \cdots & \#(x_{|\mathcal{X}|}x_1x_1) \\ \#(x_1x_2x_1) & \#(x_2x_2x_1) & \cdots & \#(x_{|\mathcal{X}|}x_2x_1) \\ \vdots & \vdots & & \vdots \\ \#(x_1x_{|\mathcal{X}|}x_1) & \#(x_2x_{|\mathcal{X}|}x_1) & & \#(x_{|\mathcal{X}|}x_{|\mathcal{X}|}x_1) \\ \#(x_1x_1x_2) & \#(x_2x_1x_2) & & \#(x_{|\mathcal{X}|}x_1x_2) \\ \#(x_1x_2x_2) & \#(x_2x_2x_2) & & \#(x_{|\mathcal{X}|}x_2x_2) \\ \vdots & \vdots & & \vdots \\ \vdots & \vdots & & \vdots \\ \#(x_1x_{|\mathcal{X}|}x_{|\mathcal{X}|}) & \#(x_2x_{|\mathcal{X}|}x_{|\mathcal{X}|}) & & \#(x_{|\mathcal{X}|}x_{|\mathcal{X}|}x_{|\mathcal{X}|}) \end{pmatrix}$$

for the second-order chain. To estimate the initial state probabilities $\hat{\pi}_0$ only the observations $\left(\#(x_1)_{\tau=0}, \ldots, \#(x_{|\mathcal{X}|})_{\tau=0}\right)$ in first time step $\tau = 0$ must be counted. To calculate the initial transition matrix $P_0$ of the Markov chain of second order only the data points in the first two time steps have to be considered, and we therefore use $\#(x_. x_.)_{\tau=0}$ to indicate the number of all leading substrings of length two. The estimate of the initial state distribution can be refined by using the data from the cited work sampling study.

The above solutions show that in a complex CE project manifesting its intrinsic complexity already in a single project phases by a rich body of data sequences with higher-order correlations, the data can usually be predicted much better with a second-order Markov chain than with a first-order model. This is due to the simple fact that the second-order chain has additional $|\mathcal{X}|^2(|\mathcal{X}|-1)$ free parameters to encode specific activity patterns and therefore a much larger memory capacity. By inductive reasoning we can proceed nesting Markov models of increasing order *n*

$$P\left(X_0 = x_{j(0)}, \ldots, X_{\mathcal{T}} = x_{j(\mathcal{T})}\right) =$$
$$P\left(X_0 = x_{j(0)}\right)P\left(X_1 = x_{j(1)} \middle| X_0 = x_{j(0)}\right) \ldots P\left(X_{n-1} = x_{j(n-1)} \middle| X_0 = x_{j(0)}, \ldots, X_{n-2} = x_{j(n-2)}\right)$$
$$\cdot \prod_{\tau=n}^{\mathcal{T}} P\left(X_\tau = x_{j(\tau)} \middle| X_{\tau-n} = x_{j(\tau-n)}, \ldots, X_{\tau-1} = x_{j(\tau-1)}\right). \quad (144)$$

and capture more and more details of the workflows. Formally speaking, the $n$-th order Markov model is the set of all $n$-th order Markov chains, i.e. all statistical representations that are equipped with a starting state and satisfy the above factorization of the joint *pmf*. Given the order $n$ of the chain, the probability distribution of $X_\tau$ depends only on the $n$ observations preceding $\tau$. However, beyond an order that is supported by the data, we start to "not see the forest for the trees" and incrementally fit the model to random fluctuations that do not generalize to other datasets from the same project phase.

In order to avoid this kind of overfitting, the maximum likelihood paradigm has to be extended, because for a parametric candidate model of interest, the likelihood function only reflects the conformity of the model to the data. As the complexity of the model is increased and more freely estimated parameters are included, the model usually becomes more capable of adapting to specific characteristics of the data. Therefore, selecting the parameterized model that maximizes the likelihood often leads to choosing the most complex model in the candidate set. The minimum description length (MDL) principle of Rissanen (1989, 2007) provides a natural safeguard against overfitting by using the briefest encoding of not only the attribute information related to the data sequences but also to the independent parameters of the candidate models. In general, let $\theta$ be a parameter vector of model class

$$\mathcal{M} = \{P(X_0 = x_{j(0)}, \ldots, X_{\mathcal{T}} = x_{j(\mathcal{T})} | \theta) \colon \theta \in \Theta \subset \mathbb{R}^k\}.$$

The sequence of discrete state variables $(X_0, \ldots, X_{\mathcal{T}} | \theta)$ forms a one-dimensional random process encoding an ensemble of histories that can be sufficiently explained by the dependency structure and the independent parameters of a candidate model within this class. By using a candidate model with specific structure and parameters, the joint *pmf* can usually be decomposed into predictive distributions whose conditional part does not scale with the length of the sequence and therefore does not need an exponentially growing number of freely estimated parameters.

As shown before, a model $\mathcal{M}^{(1)}$ from class $\mathcal{M}$ with parameter vector $\theta$ assigns a certain probability

$$p_\theta(x^{\mathcal{T}}) = P(X_0 = x_{j(0)}, \ldots, X_{\mathcal{T}} = x_{j(\mathcal{T})} | \theta) \tag{145}$$

to a data sequence $(x_{j(0)}, \ldots, x_{j(\mathcal{T})})$ of interest. If we take the definition of the Shannon information content

$$I[x] := \log_2 \frac{1}{P(X = x)}, \tag{146}$$

then the likelihood function $p_\theta(x^{\mathcal{T}})$ can be transformed into an information-theory loss function $L$

$$\begin{aligned} L[\theta, x^{\mathcal{T}}] &= I[p_\theta(x^{\mathcal{T}})] \\ &= \log_2 \frac{1}{P(X_0 = x_{j(0)}, \ldots, X_{\mathcal{T}} = x_{j(\mathcal{T})} | \theta)} \\ &= -\log_2 P(X_0 = x_{j(0)}, \ldots, X_{\mathcal{T}} = x_{j(\mathcal{T})} | \theta). \end{aligned} \tag{147}$$

According to eq. 144 we can interpret $L[\theta, x^{\mathcal{T}}]$ as the loss incurred when predicting $X_\tau$ based on the conditional distribution $P(X_\tau = x_{j(\tau)} | X_{\tau-n} = x_{j(\tau-n)}, \ldots, X_{\tau-1} = x_{j(\tau-1)})$, and when the actual outcome turned out to be $x_{j(\tau)}$ (Grünwald 2007). The loss is measured using a logarithmic scale. In the predictive view MLE aims at minimizing the accumulated logarithmic loss. We denote the maximum likelihood estimate by the member $\hat{\theta}_{\mathcal{T}}$. In the sense of information theory, minimizing the loss can also be thought of as minimizing the encoded length of the data based on an adequate prefix code $d(x)$. Shannon´s famous source coding theorem (see e.g. Cover and Thomas 1991) tells us that for an ensemble $X$ there exists a prefix code $d(x)$ with expected length $L[d(x), X]$ satisfying

$$-\sum_{x \in \mathcal{X}} P(X = x) \log_2 P(X = x) \leq L[d(x), X] < -\sum_{x \in \mathcal{X}} P(X = x) \log_2 P(X = x) + 1.$$

The term on the left of the inequality is the " information entropy" (see eq. 149). It measures in [bits] the amount of freedom of choice in the coding process. This fundamental quantity will be explained in detail in the next section. A beautifully simple algorithm for finding a prefix code with minimal expected length is the Huffman coding algorithm (see e.g. Cover and Thomas 1991). In this algorithm the two least probable data points in $\mathcal{X}$ are taken and assigned the longest codewords. The longest codewords are of equal length and differ only in the last digit. Next, these two symbols are combined into a new single symbol and the procedure is repeated. Since each recursion reduces the size of the alphabet by one, the algorithm will have assigned strings to all symbols after $|\mathcal{X}| - 1$ steps. Following the predictive view, we can obtain an intuitive interpretation of the logarithmic loss in terms of coding: the codelength needed to encode the data sequence $\left(x_{j(0)}, \dots, x_{j(\mathcal{T})}\right)$ with prefix code $d(x)$ based on the distribution $P(.)$ is just the accumulated logarithmic loss incurred when $P(.)$ is used to sequentially predict the $\tau$-th outcome on the basis of the previous $(\tau - 1)$ observations (Grünwald 2007).

It is obvious that this interpretation is incomplete; we have an encoded version of the data, but we have not said what the encoding scheme for the member $\hat{\theta}_{\mathcal{T}}$ is. Thus, the total description length $DL$ must be divided into two parts,

$$DL[x^{\mathcal{T}}, \theta, \Theta] = L[\theta, x^{\mathcal{T}}] + D[\theta, \Theta],$$

where $D[\theta, \Theta]$ is the number of bits needed to specify the member within the model class $\mathcal{M}$. The two parts of description length are usually obtained in a sequential two-stage encoding process (see Hansen and Yu 2001). In the first stage, the description length $D[\hat{\theta}_{\mathcal{T}}, \Theta]$ for the best-fitting member $\hat{\theta}_{\mathcal{T}}$ is calculated. The $\hat{\theta}_{\mathcal{T}}$´s maximizing the goodness-of-fit can be obtained both by MLE and Bayesian estimation. In the second stage, the description length of data $L\left[\hat{\theta}_{\mathcal{T}}, x^{\mathcal{T}}\right]$ is determined on the basis of the parameterized distribution $p_{\hat{\theta}_{\mathcal{T}}}\left(x^{\mathcal{T}}\right)$.

It is obvious that the model related to $D[\theta, \Theta]$ represents the part of the description, which can be generalized, whilst $L[\theta, x^{\mathcal{T}}]$ includes the noisy part that does not generalize to other datasets. If $D[\theta, \Theta]$ assigns short code words to simple models, we have the desired tradeoff: we can reduce the part of the data that looks like noise only by using a more elaborate model. Such an assignment provides a natural safeguard against overfitting. The minimum description length (MDL) principle of Rissanen (1989, 2007) allows the selection of the model that minimizes the total description length:

$$\theta_{MDL} \coloneqq \arg\min_{\theta} DL[x^{\mathcal{T}}, \theta, \Theta].$$

Minimizing the total description length is apparently a consistent principle in the light of maximum likelihood estimation, because if we want to maximize the joint probability $DL[x^{\mathcal{T}}, \theta, \Theta]$ we need to calculate the probability of the coincidence of the observed data and the different models and choose the maximizing model. It is important to point out that in MDL, one is never concerned with actual encodings but only with codelength functions, e.g. $L[d(x), X]$ for an ensemble $X$ encoded by a prefix code $d(x)$ (Grünwald 2007). The stochastic complexity $C_{SC}$ of the joint ensemble $X^{\mathcal{T}}$ with reference to the model class $\mathcal{M}$ is simply the MDL:

$$C_{SC}[x^{\mathcal{T}}, \Theta] \coloneqq \min_{\theta} DL[x^{\mathcal{T}}, \theta, \Theta]. \qquad (148)$$

Under mild conditions for the underlying data-generating process in the model class $\mathcal{M}$, as we provide more data, $\theta_{MDL}$ will converge to the model that minimizes the generalization error.

Regarding our previous example of workflow modeling with Markov chains, we can construct for didactic purposes a simple but reasonable code for the $n$-th order Markov model $\mathcal{M}^{(n)}$ within the class $\mathcal{M}_{MM}$ of all finite-order Markov models. The parameter vector of the $n$-th order Markov model is denoted by $\theta_n$. Firstly, the model order has to be described. We can start with a straightforward explicit description for $n$ that is based on a binary prefix code with $\lceil \log_2 n \rceil$ zeros followed by a one. The encoding of $n$ can be done by using a simple uniform code for $\{1, \dots, 2^{\lceil \log_2 n \rceil}\}$. Therefore, we need approximately $2\lceil \log_2 n \rceil$+1 bits to describe the model order. By applying Huffman's algorithm here, we can also obtain a more efficient uniform code with a length function that is not greater than $\lfloor \log_2 n \rfloor$ for all values of $\{1,2, \dots, n\}$ but is equal to $\lfloor \log_2 n \rfloor$ for at least two values in this set. The function $\lfloor . \rfloor$ gives the integer part of the argument. Whereas we know from Shannon's source coding theorem that an expected length of such a code is optimal only for a true uniform distribution of the order of the model, this code is a reasonable choice when little is known about how the data were generated. Secondly, the $\sum_{i=0}^{n} |\mathcal{X}|^i (|\mathcal{X}| - 1) = |\mathcal{X}|^{n+1}$ best-fitting free parameters $\hat{\theta}_{n,\mathcal{T}}$ have to be described. We start by discretizing the range $[0; 1]$ of a single ensemble into equal cells of size δ and then apply Huffman´s algorithm. If we discretize the Cartesian product $\Theta_{MM} = [0; 1]^{|\mathcal{X}|^{n+1}}$ associated with the joint ensemble $X^{\mathcal{T}}$in the same fashion, the quantity $-\log_2\left(p\left([0; 1]^{|\mathcal{X}|^{n+1}}\right) \cdot \delta^{|\mathcal{X}|^{n+1}}\right) = -\log_2 p\left([0; 1]^{|\mathcal{X}|^{n+1}}\right) - |\mathcal{X}|^{n+1} \log_2 \delta$ can be viewed as the code length of a prefix code for $\hat{\theta}_{n,\mathcal{T}}$ (Hansen and Yu 2001). Here, the probability density $p$ can be regarded as an auxiliary density. It is used instead of the unknown true parameter-generating density $f$. Assuming a continuous uniform distribution with density $p(x) = 1$ for $x \in [0; 1]^{|\mathcal{X}|^{n+1}}$ (and $q(x) = 0$ otherwise), additional $|\mathcal{X}|^{n+1} \log_2 \delta$ bits are needed to describe the free parameters. In a compact parameter space, we can refine the description and choose for the precision $\delta = \sqrt{1/\mathcal{T}}$ for each effective dimension. Rissanen (1989) showed that this choice of precision is optimal in regular parametric families. The intuitive explanation is that $\sqrt{1/n}$ represents the magnitude of the estimation error in $\hat{\theta}_{\mathcal{T}}$ and therefore there is no need to encode the estimator with greater precision (Hansen and Yu 2001). When the uniform encoder is used, one needs a total of $(|\mathcal{X}|^{n+1}/2)\log_2 \mathcal{T}$ bits to communicate an estimated parameter $\hat{\theta}_{n,\mathcal{T}}$ of dimension $|\mathcal{X}|^{n+1}$. Putting both partial descriptions together leads to

$$D[\theta_n, \Theta_{MM}] = \log_2 n + \frac{|\mathcal{X}|^{n+1}}{2} \log_2 \mathcal{T}.$$

Interestingly, the formalized description of the $n$-th order Markov model is similar to the Schwarz-Bayes Criterion (SBC) for the $VAR(n)$ (eq. 43) and LDS (eq. 129) project models in the sense that model complexity is penalized with a factor that increases linearly in the number of free parameters and logarithmically in the number of observations in the joint ensemble. This is a clear and unambiguous indication that there are deep theoretical connections between different approaches to model selection. The predictive view of Markovian models provides us with a refined interpretation of model selection based on the MDL principle: given two candidate models $\mathcal{M}^{(1)}$ and $\mathcal{M}^{(2)}$, prefer the model that minimizes the accumulated prediction error resulting from a sequential prediction of future outcomes given all past histories (Grünwald 2007).

Regarded as a principle of model selection, MDL has proved very successful in many application areas (see e.g. Grünwald 2007, Rissanen 2007). Nevertheless, a part of this success comes from carefully tuning the model-coding term $D[\theta, \Theta]$ so that models that do not generalize well turn out to have long encodings. This is not illegitimate, but it relies on the intuition and knowledge of the human model builder. In that sense, the MDL principle has nothing to say about how to select the suggested family of model classes (Rissanen 2007). Whatever its merits as a model selection method, stochastic complexity is not a good metric of emergent complexity in open organizational systems for three reasons (sensu Shalizi 2006). 1) The dependence on the model-encoding scheme, which is very difficult to formulate in a valid form for project-based organizations. 2) The log-likelihood term, $L[\theta, x^{\mathcal{T}}]$, can be decomposed into additional parts, one of which is related to the entropy rate of the information-generating work processes ($h_\mu$, eq. 158) and so it reflects their intrinsic unpredictability, not their complexity. Other parts indicate the degree to which even the most accurate model in $\mathcal{M}$ is misspecified; for instance, through an improper choice of the coordinate system. Thus, it largely reflects our unconscious incompetence as modelers, rather than a fundamental characteristic of the process. 3) The stochastic complexity reflects the need to specify some particular organizational model and to formally represent this specification. This is necessarily part of the process of model development but seems to have no significance from a theoretical point of view. For instance, a sociotechnical system under study does not need to represent its organization; it just has it (Shalizi 2006).

### 3.2.3 Effective Measure Complexity and Forecasting Complexity

Motivated by theoretical weaknesses such as these, the German physicist Peter Grassberger (1986) developed a simple but highly satisfactory complexity theory: complexity is the amount of information required for optimal prediction. Let's first analyze why this approach is plausible and then how it can be made fully operational. In general, there is a limit to the accuracy of any prediction of a given sociotechnical system set by the characteristics of the system itself, e.g. free will of decision makers, unpredictable performance variability, limited precision of measurement, sensitive dependence on initial conditions, etc. Suppose we have a model that is maximally predictive, i.e. its predictions are at the theoretical limit of accuracy. Prediction is always a matter of mapping inputs to outputs. In our application context, the inputs are the encoded historical traces of task processing (work left to finalize a specific design, open design issues that need to be addressed before design release, etc.) and the outputs are the expected work remaining, as well as the accumulated key performance indicators. However, usually not all aspects of the entire past are relevant for making good predictions. In fact, if the task processing is strictly periodic with a predefined cycle time, one only needs to know which of the $\varphi$ phases the work process is in. For a completely randomized work process with independent and identically distributed (iid) state variables, the past is completely irrelevant for predicting the future. Because of this "memorylessness", the clever, evidence-based estimates of an experienced project manager on average do not outperform naïve guesses of the outcome based on means. If we ask how much information about the past is relevant in these two extreme cases, the correct answers are $\log_2(\varphi)$ and 0, respectively. It is intuitive that these cases are of low complexity, and more informative dynamics "somewhere in between" must be assigned high complexity values. In terms of Shannon's famous information entropy $H[.]$ the "randomness" of the output either is simply a constant (low-period deterministic process with small algorithmic complexity) or grows precisely linearly with the length (completely randomized process with large algorithmic complexity). Hence, it can be concluded that both cases share the feature that corrections to the asymptotic behavior do not grow with the size of the dataset (Prokopenko et al. 2007). Grassberger considered the slow approach of the entropy to its extensive limit as an indicator of complexity. In other words, the subextensive components growing less rapidly with time than a linear function are of special interest for complexity evaluation.

When dealing with a Markovian model, such as the VAR model of cooperative task processing formulated in section 2.2, only the present state of work remaining is relevant for predicting the future (see eq. 4), so the amount of information needed for optimal prediction is just equal to the amount of information needed to specify the current state. More formally, any predictor $g$ will translate the one-dimensional infinite past $X_{-\infty}^{-1}$=$(X_{-\infty}, X_{-\infty+1}, \dots, X_{-1})$ into an effective state $S = g[X_{-\infty}^{-1}]$ and then make its prediction on the basis of $S$. This is true whether or not $g[\cdot]$ is formally a state-space model as we have formulated. The amount of information required to specify the effective state in case of discrete-type random variables (or discretized continuous-type random variables) can be expressed by Shannon's information entropy $H[S]$ (Cover and Thomas 1991). We will return later in this section to this point and take $H[S]$ to be the statistical complexity $C_{GCY}$ of $g[\cdot]$ under the assumption of a minimal maximally predictive model of the stationary stochastic process $\{X_t\}$ ($t \in \mathbb{Z}$, see eq. 161).

Shannon's information entropy represents the average information content of an outcome. Formally, it is defined for a discrete-type random variable $X$ with values in the alphabet $\mathcal{X}$ and probability mass $P(.)$ function as

$$H[X] := -\sum_{x \in \mathcal{X}} P(X = x) \log_2 P(X = x). \tag{149}$$

The information entropy $H[.]$ is non-negative and measures in [bits] the amount of freedom of choice in the associated decision process or, in other words, the degree of randomness. If we focus on the set $\mathcal{M}$ of maximally predictive models, we can define what Grassberger called "the true measure complexity $C_\mu$ of the process" as the minimal amount of information needed for optimal prediction:

$$C_\mu := \min_{g \in \mathcal{M}} H[g[X_{-\infty}^{-1}]. \tag{150}$$

The true measure complexity is also termed "forecasting complexity" (Zambella and Grassberger 1988), because it is defined on the basis of maximally predictive models requiring the least average information content of the memory variable. We will use the term "forecasting complexity" in the following as it is well-established and more intuitive. Unfortunately, Grassberger provided no procedure for finding the maximally predictive models or for minimizing the information content. However, he did draw the following conclusion. A basic result of information theory, called "the data-processing inequality", says that for any pair of random variables $X$ and $Y$ (or pair of sequences of random variables) the mutual information $I[.,.]$ follows the rule

$$I[X, Y] \geq I[g[X], Y].$$

It is therefore impossible to extract more information from observations by processing than was in the sample to begin with. Since the state $S$ of the predictor is a function of the past, it follows that

$$I[X_{-\infty}^{-1}; X_0^{\infty}] \geq I[g[X_{-\infty}^{-1}]; X_0^{\infty}],$$

where $X_0^\infty = (X_0, X_1, \dots, X_\infty)$ represents the infinite future of the stochastic process including the "present" that is encoded in the observation $X_0$.

The mutual information $I[.;.]$ is another key quantity of information theory. It can be equivalently expressed on the basis of the joint probability mass function $P(.,.)$ as

$$I[X, Y] := \sum_{x \in \mathcal{X}} \sum_{y \in \mathcal{Y}} P(X = x, Y = y) \log_2 \frac{P(X = x, Y = y)}{P(X = x) P(Y = y)} \tag{151}$$

or the information entropy $H[.]$ as

$$\begin{aligned} I[X,Y] &= H[X] - H[X|Y] \\ &= H[Y] - H[Y|X] \\ &= H[X] + H[Y] - H[X,Y] \\ &= H[X,Y] - H[X|Y] - H[Y|X]. \end{aligned}$$

The mutual information is also non-negative and measures the amount of information that can be obtained about one random variable by observing another. It is symmetric in terms of these variables. The amount of information $I[A;B]$ shared by transmitted and received signals is often maximized by system designers by choosing the best transmission technique. Channel coding guarantees that reliable communication is possible over noisy communication channels, if the rate of information transmission is below a certain threshold that is termed "the channel capacity", defined as the maximum mutual information for the channel over all possible probability distributions of the signal (see Cover and Thomas 1991). According to Polani et al. (2006) mutual information should not be regarded as something that is transported from a sender to a receiver as a "bulk" quantity. Instead, the thorough observation of the intrinsic dynamics of information can provide deep insight into the inner structure of information, and maximization of information transfer through selected channels appears to be one of the main evolutionary processes (Bialek et al. 2001, Polani et al. 2006).

Presumably, for optimal predictors, the amounts of information $I[X_{-\infty}^{-1}; X_0^{\infty}]$ and $I[g[X_{-\infty}^{-1}]; X_0^{\infty}]$ are equal and the predictor's state is just as informative as the original data. This is the case for so-called "$\varepsilon$-machines", which are analyzed below. Otherwise, the model would be missing potential predictive power. Another basic inequality is that $H[X] \geq I[X;Y]$, i.e. no variable contains more information about another than it does about itself. Even for the maximally predictive models it therefore holds that $H[X_{-\infty}^{-1}] \geq I[X_{-\infty}^{-1}; X_0^{\infty}]$. Grassberger called the latter quantity $I[X_{-\infty}^{-1}, X_0^{\infty}]$—the mutual information between the infinite past and future histories of a stochastic process—the effective measure complexity (EMC):

$$\text{EMC} \coloneqq I[X_{-\infty}^{-1}; X_0^{\infty}]. \tag{152}$$

Please recall that EMC is defined with reference to infinite sequences of random variables and is therefore only valid for stationary stochastic processes. The same is true for the forecasting complexity. For the sequence $(\ldots, X_{-1}, X_0, X_1, \ldots)$ stationarity implies that the joint probability distribution $P(.,\ldots,.)$ associated with any finite block of *n* variables $X^n \coloneqq X_{t+1}^{t+n} = (X_{t+1}, \ldots, X_{t+n})$ is independent of *t* and only depends on the block length *n*. The independency of the joint probability distribution of *t* can limit the evaluation of NPD projects in industry as the dynamical dependencies between process and product can significantly change over time. In this case an alternative complexity measure—known as the "binding information" —developed by Abdallah and Plumbley (2010) should be taken into consideration as it can evaluate non-stationary processes of different kinds.

If optimal predictions of the stationary stochastic process are influenced by events in the arbitrarily distant past, the mutual information diverges and the measure EMC tends to infinity (see discussion of predictive information $I_{pred}$ below).

Shalizi and Crutchfield (2001) proved that the forecasting complexity gives an upper bound of the EMC:

$$\text{EMC} \leq C_{\mu}.$$

In terms of a communication channel, EMC is the effective information transmission rate of the process. The units are bits. $C_\mu$ is the memory stored in that channel. Hence, the inequality above means that the memory needed to carry out an optimal prediction of the future cannot be less than the information that is transmitted from the past $X_{-\infty}^{-1}$ to the future $X_0^\infty$ (by storing it in the present). However, the specification of how the memory has to be designed and managed cannot be derived on the basis of information-theory considerations. Instead, a constructive and more structural approach based on a theory of computation must be developed. A very satisfactory theory based on "causal states" was developed by Crutchfield and Feldman (2003). These causal states lead to the cited $\varepsilon$-machines, as well as the Grassberger–Crutchfield–Young statistical complexity $C_{GCY}$, which will be presented later in this section.

EMC can be estimated purely from historical data, without use of a generative stochastic model of cooperative task processing. If the data are generated by a model in a specific class but with unknown parameters, we can derive closed-form solutions for EMC, as will be shown in sections 4.1, 4.2 and 4.3 for a VAR(1) model (cf. eq. 190). The mutual information between the infinite past and future histories of a stochastic process has been considered in many contexts. It is termed, for example, excess entropy **E** (Crutchfield and Feldman 2003, Ellison et al. 2009, Crutchfield et al. 2010), predictive information $I_{pred}(n \to \infty)$ (Bialek et al. 2001), stored information (Shaw 1984) or simply complexity (Arnold 1996, Li 1991). Rissanen (1996, 2007) also refers to the part of stochastic complexity required for coding model parameters as model complexity. Hence, there should be a close connection between Rissanen's ideas of encoding a data stream based on generative models and Grassberger´s ideas of extracting the amount of information required for optimal prediction. In fact, if the data allow a description by a model with a finite number of independent parameters, then mutual information between the data and the parameters is of interest, and this is also the predictive information about all of the future (Bialek et al. 2001). Rissanen´s approach was further strengthened by a result of Vitányi and Li (2000) showing that an estimation of parameters using the MDL principle is equivalent to Bayesian parameter estimations with a "universal" prior (Li and Vitányi 1997). Since the mutual information between the infinite past and future histories can quantify the statistical dependence of cooperative work processes, it will be used in the following to evaluate emergent complexity in NPD projects.

In addition to $C_\mu$ and EMC, another key invariant of stochastic processes that was discovered much earlier is Shannon's source entropy rate:

$$h_\mu \coloneqq \lim_{\eta \to \infty} \frac{H[X^{n=\eta}]}{\eta}. \tag{153}$$

This limit exists for all stationary processes. The source entropy rate is the intrinsic randomness that cannot be reduced, even after considering statistics over longer and longer blocks of generating variables. The unit of $h_\mu$ is bits/symbol. It is also known as per-symbol entropy, thermodynamic entropy density, Kolmogorov–Sinai entropy or metric entropy. The source entropy rate is zero for periodic processes. Surprisingly, it is also zero for deterministic processes with infinite memory. The source entropy rate is larger than zero for irreducibly unpredictable processes like the cited iid process or Markov processes. The capacity of a communication channel must be larger than $h_\mu$ for error-free data transmission (Cover and Thomas 1991). Interestingly, the source entropy rate is related to the algorithmic complexity (section 3.1): $h_\mu$ is equal to the average length (per variable) of the minimal program with respect to $U$ that, when run, will cause the Universal Turing Machine to produce a typical configuration and then halt (Cover and Thomas 1991). In the above definition the variable $H[X^n]$ is the joint information entropy of length-$n$ blocks $(X_{t+1}, \dots, X_{t+n})$. This entropy is not the entropy of a finite string $x^n$ with length $n$; rather, it is the entropy of sequences with length $n$ drawn from mainly much longer or infinite output generated by the process in the steady state. The variable $n$ is the nonnegative order parameter and can be interpreted as an expanding observation window of length $n$ over the output. In the following, we will use the shorthand notation $H(n)$ to represent this kind of entropy, which is also termed Shannon block entropy (Grassberger 1986, Bialek et al. 2001). For discrete-type random variables the block entropy is defined as

$$
\begin{aligned}
H(n) &:= H[X^n] \\
&= H[X_{t+1}, \ldots, X_{t+n}] \\
&= -\sum_{\mathcal{X}} \ldots \sum_{\mathcal{X}} P\big(X_{t+1} = x_{j(t+1)}, \ldots, X_{t+n} = x_{j(t+n)}\big) \log_2 P\big(X_{t+1} = x_{j(t+1)}, \ldots, X_{t+n} = x_{j(t+n)}\big)
\end{aligned}
\tag{154}
$$

with

$$H(0) := 0. \tag{155}$$

The sums in eq. 154 run over all possible blocks of length $n$. The corresponding definition for continuous-type variables will be given in eq. 164. Interestingly, the length-$n$ approximation $h_\mu(n)$ of the entropy rate $\mathrm{h}_\mu$ can be defined as the two-point slope of the block entropy $\mathrm{H(n)}$:

$$h_\mu(n) := H(n) - H(n-1), \tag{156}$$

with

$$h_\mu(0) := \log_2 |\mathcal{X}|. \tag{157}$$

Vice versa, $h_\mu(n)$ is the discrete derivative of the block entropy with respect to the block length $n$. In that sense, the length-$n$ approximation is a dynamic entropy representing the entropy gain. It can be seen that the entropy gain can also be expressed as conditional entropy (cf. eq. 168)

$$h_\mu(n) := H[X_n | X^{n-1}].$$

In the limit of infinitely long blocks, it is equal to the source entropy rate

$$h_\mu = \lim_{\eta \to \infty} h_\mu(n = \eta). \tag{158}$$

In general $h_\mu(n)$ differs from the estimate $H(n)/n$ for any given $n$ but converges to the same limit, namely the source entropy rate $h_\mu$. According to Crutchfield and Feldman (2003) $h_\mu(n)$ typically overestimates $h_\mu$ at finite $n$, and each difference $h_n - h_\mu$ represents the difference between the entropy rate conditioned on $n$ measurements and the entropy rate conditioned on an infinite number of measurements. In that sense, it estimates the information-carrying capacity in blocks in which the difference is not actually random but arises from correlations. The difference $h_n - h_\mu$ can therefore be interpreted as the local predictability. These local “overestimates” can be used to define a universal learning curve $\Lambda(n)$ (Bialek et al. 2001) as

$$\Lambda(n) := h_\mu(n) - h_\mu\,, \;\; n \geq 1. \tag{159}$$

EMC is simply the discrete integral of $\Lambda(n)$ with respect to the block length $n$, which controls the speed of convergence of the dynamic entropy to its limit (Crutchfield et al. 2010):

$$\mathrm{EMC} := \sum_{n=1}^{\infty} \Lambda(n). \tag{160}$$

In the sense of a learning curve, EMC measures the amount of apparent randomness at small block length $n$ that can be "explained away" by considering correlations among blocks with increasing length $n+1, n+2...$ . Grassberger (1986) analyzed the manner in which $h_\mu(n)$ approaches its limit $h_\mu$, noting that for certain classes of stochastic processes with long-range correlations, the convergence can be very slow and this is an indicator of complexity. Moreover, he found that the convergence can be so slow, that EMC is infinite. These phenomena were analyzed in great detail by Bialek et al. (2001). To do so, they defined the predictive information $I_{pred}(n)$ $(n \geq 1)$ as the mutual information between a block of length $n$ and the infinite future following the block:

$$\begin{aligned} I_{pred}(n) &:= \lim_{\eta \to \infty} I[X_{-n}^{-1}; X_0^{\eta}] \\ &= \lim_{\eta \to \infty} H(n) + H(\eta) - H(n+\eta). \end{aligned}$$

Bialek et al. (2001) showed that even if $I_{pred}(n)$ diverges as $n$ tends to infinity, the manner of its divergence allows conclusions to be drawn about the learnability of the underlying process. They also emphasized that the predictive information is the subextensive component of the entropy:

$$H(n) = nh_\mu + I_{pred}(n).$$

From the above equation, it can be seen that the sum of the first $n$ terms of the discrete integral of the universal learning curve $\Lambda(n)$, that is, $H(n) - nh_\mu$, is equal to $I_{pred}(n)$ (Abdallah and Plumbley 2010):

$$I_{pred}(n) = \sum_{i=1}^{n} \Lambda(i).$$

As expected, $I_{pred}(n)$ (as well as EMC) is zero for an iid process. According to Bialek et al. (2001), it is positive in all other cases and grows with time less rapidly than a linear function (subextensive). $I_{pred}(n)$ may either stay finite or grow infinitely with time. If it stays finite, no matter how long we observe the past of a process, we gain only a finite amount of information about the future. This holds true, for instance, for the cited periodic processes after the period $\varphi$ has been identified. A longer period results in larger complexity values and $I_{pred}(n \to \infty) = \text{EMC} = \log_2(\varphi)$. For some irregular processes, the best predictions may depend only on the immediate past, e.g. in our Markovian model of task processing or generally when evaluating a system far away from phase transitions or symmetry breaking. In these cases, $I_{pred}(n \to \infty) = \text{EMC}$ is also small and is bound by the logarithm of the number of accessible states. Systems with more accessible states and larger memories are assigned larger complexity values. On the other hand, if $I_{pred}(n)$ diverges and optimal predictions are influenced by events in the arbitrarily distant past, then the rate of growth may be slow (logarithmic) or fast (sublinear power). If the acquired data allow us to infer a model with a finite number of independent parameters, or identify a set of generative rules that can be described by a finite number of parameters, then $I_{pred}(n)$ grows logarithmically with the size of the sample. The coefficient of this divergence counts the dimensionality of the model space (i.e. the number of relevant independent parameters). Sublinear power-law growth can be associated with infinite parameter models or with nonparametric models, such as continuous functions with smoothness constraints. Typically these cases occur where predictability over long time scales is governed by a progressively more detailed description as more data points are observed.

To make the previously introduced key invariant $C_\mu$ (forecasting complexity, eq. 150) of a stochastic process operational in terms of a theory of computation and to clarify its relationship to the other key invariant EMC (effective measure complexity, eq. 160) by using a structurally rich model and not just a purely mathematical representation of a communication channel, we refer in what follows to the seminal work of Crutchfield and Young (1989, 1990) on computational mechanics. They provided a procedure for finding the minimal maximally predictive model and its causal states by means of an $\varepsilon$-machine (Ellison et al. 2009, Crutchfield et al. 2010). The general goal of building an $\varepsilon$-machine is to find a constructive representation of a nontrivial process that not only allows good predictions on the basis of the stored predictive information but also uncovers the essential mechanisms producing a system's behavior. To build a minimal maximally predictive model of a stationary stochastic process, we can formally define an equivalence relation $x_{-\infty}^{-1} \sim \hat{x}_{-\infty}^{-1}$ that groups all process histories that give rise to the same prediction:

$$x_{-\infty}^{-1} \sim \hat{x}_{-\infty}^{-1} :\Leftrightarrow \{P(X_0^{\infty} | X_{-\infty}^{-1} = x_{-\infty}^{-1}) = P(X_0^{\infty} | X_{-\infty}^{-1} = \hat{x}_{-\infty}^{-1})\}.$$

Hence, for the purpose of forecasting, two different past sequences of observations are considered equivalent if they result in the same predictive distribution. The above equivalence relation determines the process's causal state, which partitions the space $X_{-\infty}^{-1}$ of pasts into sets that are predictively equivalent. The causal state $\varepsilon(x_{-\infty}^{-1})$ of $x_{-\infty}^{-1}$ is its equivalence class

$$\varepsilon(x_{-\infty}^{-1}) \coloneqq \{\hat{x}_{-\infty}^{-1} : x_{-\infty}^{-1} \sim \hat{x}_{-\infty}^{-1}\},$$

and the causal state function $\varepsilon(.)$ defines a deterministic sufficient memory $\mathcal{M}_\varepsilon$ (see Shalizi and Crutchfield 2001, Löhr 2012). The set of memory states of the $\varepsilon$-machine is simply the set of causal states

$$\mathcal{M}_\varepsilon \coloneqq \{\varepsilon(x_{-\infty}^{-1}) : x_{-\infty}^{-1} \in \mathcal{X}^{\mathbb{N}}\}.$$

The set $\mathcal{X}$ represents the finite alphabet on which the stationary stochastic process is defined. The set of causal states $\mathcal{M}_\varepsilon$ does not need to be countable and can therefore represent either discrete or continuous state spaces. Shalizi and Crutchfield (2001) showed that the equivalence relation $x_{-\infty}^{-1} \sim \hat{x}_{-\infty}^{-1}$ is minimally sufficient and unique. Hence, it allows for highest compression of the data, while containing all the relevant information on local dynamics. For practical purposes, longer and longer histories are analyzed, from $x_{-L}^{-1}$ up to a predefined maximum length $L = L_{max}$, and the partition into classes for a fixed future horizon $X_0^t$ is obtained. In principle, we start at the most coarse-grained level, grouping together those histories that have the same predictive distribution for the next observable $X_0$, and then refine the partition. The refinement is recursively carried out by further subdividing the classes using the predictive distributions of the next two observables $X_0^1$, the next three observables $X_0^2$, etc.

After all causal states have been identified, an $\varepsilon$-machine of a stationary stochastic process can be constructed. To simplify the definition of the forecasting complexity $C_\mu$, we start by using an informal representation in the form of a stochastic output automaton that is expressed by the causal state function $\varepsilon$, a set of transition matrices $\mathcal{J}$ for the states defined by $\varepsilon$, and the start state $s_0$. The start state is unique. Given the current state $s \in \mathcal{M}_\varepsilon$ of the automaton, a transition to the next state $s' \in \mathcal{M}_\varepsilon$ is determined by the output symbol (or measurement) $x \in \mathcal{X}$. State-to-state transitions are probabilistic and therefore must be represented for each output symbol $x$ by a separate transition matrix $T^{(x)} \in \mathcal{J}$. Each row and column of the transition matrices in the set $\mathcal{J}$ stands for an individual causal state. A stochastic output automaton can also be transformed into an equivalent edge-emitting hidden Markov model (Löhr 2012). A hidden Markov model is a universal machine that is defined over a set of non-observable internal states $\mathcal{M}_\varepsilon$. It therefore does not directly reveal its internal mechanisms to external observers; it only expresses them indirectly through emitted symbols. The emitted symbols are edge-labels of the hidden states. The model can be formally represented by the tuple $(\mathcal{M}_\varepsilon, \mathcal{X}, \pi, \{T^{(x)}\})$. The start state of the hidden Markov model is not unique but determined by an initial probability distribution $\pi$. Depending on the current internal state $s_t$, at each time step $t$ a transition to the new internal state $s_{t+1}$ is made and an output symbol $x_{t+1}$ from the alphabet $\mathcal{X}$ is emitted. The corresponding entry $T_{ij}^{(x)}$ of the transition matrix $T^{(x)}$ gives the probability $P(S_{t+1} = s_{t+1}, X_{t+1} = x_{t+1} | S_t = s_t)$ of transitioning from current state $s_t$ indexed by $i$ to the next $s_{t+1}$ indexed by $j$ on "seeing" measurement $x$. This operation may be also thought of as a weighted random walk on the associated graphical model (Travers and Crutchfield 2011): from the current state $s_t$, the next state $s_{t+1}$ is determined by selecting an outgoing edge from current state $s_t$ according to their probabilities. After a transition has been selected, the model moves to the new state and outputs the symbol current state $x$ labeling the edge. The transition matrices are usually non-symmetric. From the theory of Markov processes (see e.g. Puri 2010) it is well known that in steady state the probability distribution over the hidden states is independent of the initial-state distribution. Edge-emitting hidden Markov models can also be expressed by an initial probability distribution $\pi$, by a state process $\{S_t\}$ and by an output process $\{X_t\}$, which means that they are theoretically similar to the continuous-type linear dynamical systems that were analyzed in section 2.8. However, continuous-type linear dynamical systems usually do not possess the property of "unifliarity" (see below) and therefore cannot be used to directly calculate the entropy rate of the process.

To obtain the transition matrices $T^{(x)}$, one can parse the data sequence of interest in a sequential manner, identify all causal state transitions defined by $\varepsilon$ over histories $x_0^t$ and $x_0^{t+1}$, and estimate the transition probabilities $P(S', X = x_{t+1} | S)$ using frequency counting (MLE, see chapter 2.4) or Bayesian methods. The transition probabilities allow calculation of an invariant probability distribution $P(S)$ over the causal states. This probability is obtained as the normalized principal eigenvector of the transition matrix $T = \sum_{x \in \mathcal{X}} T^{(x)}$ (Ellison et al. 2009). The matrix $T$ is stochastic and $\sum_{j=1}^{|\mathcal{M}_\varepsilon|} T_{ij} = 1$ holds for each $i$.

Interestingly, causal states have a Markovian property in that they render the past and future statistically independent. In other words, they shield the future from the past:

$$P(X_{-\infty}^{-1}, X_0^{\infty} | S) = P(X_{-\infty}^{-1} | S)\, P(X_0^{\infty} | S).$$

Moreover, they are optimally predictive in the sense that knowing what causal state a process is in is as good as having the entire past: $P(X_0^{\infty} | S) = P(X_0^{\infty} | X_{-\infty}^{-1})$. Causal shielding is therefore equivalent to the fact that the causal states capture all of the information shared between past and future. Hence, $I[S, X_0^{\infty}] = \mathrm{EMC}$. Out of all maximally predictive models $\mathcal{M}$ for which $I[\mathcal{M}, X_0^{\infty}] = \mathrm{EMC}$, the $\varepsilon$-machine captures the minimal amount of information that a stationary stochastic process must store in order to communicate all excess entropy from the past to the future. In that sense the $\varepsilon$-machine is as close to perfect determinism as any rival that has the same predictive power (Jänicke and Scheuermann 2009). The minimal amount of information that a stationary stochastic process $X_{-\infty}^{\infty} = (\ldots, X_{-1}, X_0, X_1, \ldots)$ must store is the Shannon information contained in the causal states—the forecasting complexity— and it holds that

$$C_\mu(X_{-\infty}^{\infty}) = H[S].$$

Because of its significance in complex systems science, the forecasting complexity is also termed Grassberger–Crutchfield–Young statistical complexity $C_{GCY}$ (Shalizi 2006). It should not be confused with Rissanen's stochastic complexity $C_{SC}$ from eq. 148, because the underlying concepts are based on a true theory of computation. We have (Ellison et al. 2009)

$$C_{GCY} = H[S] \leq H[\mathcal{M}]$$

$$C_{GCY} = -\sum_{\{s\epsilon\mathcal{M}_S\}} P(S)\log_2 P(S) \leq H[\mathcal{M}]. \quad (161)$$

As we have argued, the causal states are an objective property of the stochastic process under consideration and therefore the associated statistical complexity $C_{GCY}$ does not depend on our ineptness as modelers or our (possibly poor) means of description. It is equal to the length of the shortest description of the past that is relevant to the actual dynamics of the system. As was shown above, for iid sequences it is exactly 0, and for periodic sequences it is $\log_2(\varphi)$. A detailed description of an algorithm achieving an $\varepsilon$-machine reconstruction and calculation of $C_{GCY}$ for one-dimensional and two-dimensional time series can be found in Shalizi and Shalizi (2004, 2003).

Moreover, the entropy rate $h_\mu$ can be directly calculated on the basis of a process's $\varepsilon$-machine (Ellison et al. 2009) because of unifilarity:

$$\begin{aligned} h_\mu &= H[X|S] \\ &= -\sum_{s\epsilon\mathcal{M}_S} P(S) \sum_{xs'\epsilon\mathcal{X}\mathcal{M}_S} T^{(x)}_{ss'} \log_2 \sum_{s'\epsilon\mathcal{M}_S} T^{(x)}_{ss'}. \end{aligned}$$

Unifilarity means that from the start state $s_0$ of the process, each generated sequence of observations corresponds to exactly one sequence of causal states. In a hidden Markov model representation of an $\varepsilon$-machine this property can be verified easily. For each hidden state, each emitted symbol appears on at most one edge. In the above equation, we used the shorthand notation $T^{(x)}_{ss'}$ to denote the matrix entry $T^{(x)}_{ij}$ corresponding to causal state $s$ in row $i$ and causal state $s'$ in column $j$ of the transition matrix associated with output symbol $x$. The probability $P(S)$ denotes the asymptotic probability of the causal states.

An interesting extension of Grassberger's effective measure complexity has been developed recently by Ball et al. (2010). These authors also quantify strong emergence within an ensemble of histories of a complex system in terms of mutual information between past and future history, but focus on the part of the information that persists across an interval of time $\tau > 0$. In that sense, we can specify the "persistent mutual information" as a complexity measure in its own right evaluating the deficit in the information entropy in the joint history compared with that of past and future taken independently. Formally, the persistent mutual information can be defined on the basis of the EMC (eq. 152) extended by the lead time $\tau$ to evaluate the persistent part as

$$\mathrm{EMC}(\tau) := I[X_{-\infty}^{-1}; X_\tau^\infty], \quad (162)$$

where $X_{-\infty}^{-1}$ designates the history of the stochastic process from an infinite past to present time, and $X_\tau^\infty$ is the corresponding future of the system from the later time $\tau$ onwards. The key distinguishing feature of the definition above is the exclusion of information on $X_0^{\tau-1}$, that is the intervening time interval of length $\tau$. For continuous state variables, $\mathrm{EMC}(\tau)$ has the merit of being independent of continuous changes of the variable, so long as they preserve time labeling (Ball et al. 2010). $\mathrm{EMC}(\tau)$ is known to be a Lyapunov function for the process, so that it decays with increasing lead time (Ay 2010). According to the recent analysis of several multivariate information measures of James et al. (2011) the persistent mutual information has an especially interesting interpretation: it represents the Shannon information that is communicated from the past to the future, but does not flow through the current sequence of observations $X_0^{\tau-1}$. If this amount of information is larger than zero, a complete description of the process requires storing all the information from the past that is relevant for generating future behavior by an internal configuration. The internal configuration is necessary to keep track of the state information, because the current sequence of observations $X_0^{\tau-1}$ can only capture features of shorter term correlation and is therefore not expressive enough to capture all features that are relevant for forecasting. In the words of James et al. (2011): “This is why we build models and cannot rely on only collecting observation sequences.”

### 3.3 Complexity Measures from Theories of Systematic Engineering Design

The most prominent complexity theory in the field of systematic engineering design has been developed by Suh (2005) on the basis of his famous axiomatic design theory. His theory aims at providing a systematic way of designing products and large-scale systems, as well as to determine the best designs among those proposed. Suh’s complexity theory is based on his well-known axiomatic design theory. He defines complexity in the functional domain rather than in the physical domain of the design world. In the functional domain, uncertainty is measured through information-theory quantities. Alternative approaches to evaluate complexity in the field of NPD that are not based on information-theory variables (e.g. Lindemann et al. 2009) are not considered in the following, because they can only evaluate structural and not time-dependent complexity.

In Suh’s axiomatic design theory, the product to be developed and the problem to solve the design issues are coupled through functional requirements (FRs) and design parameters (DPs). He proposed two axioms for design: the independence and the information axioms. The independence axiom states that the FRs should be maintained by the designer or design team independent of each other. When there are two or more FRs, the design solution must be such that each of the FRs can be satisfied without affecting any of the other FRs. This means that a correct set of DPs is to be chosen so as to satisfy the FRs and maintain their independence. If the independence can be maintained for all FRs, the design is termed “uncoupled”. An uncoupled design is an optimal solution in the sense of the theory. Once the FRs are established, the next step in the design process is the conceptualization process, which occurs during the mapping process going from the functional to the physical domain.

The conceptualization process may produce several designs, all of which may be satisfactory in terms of the independence axiom. Even for the same task defined by a given set of FRs, it is likely that different product developers will come up with different designs, because there are many solutions satisfying a given set of $m$ FRs ($FR_1, \dots, FR_m$). The information axiom provides a quantitative measure of the merits of a given design, and is thus useful in selecting the best design among those that are acceptable. The information axiom is formulated within an information-theory framework and states that the best design is that with the highest probability of success. Following the definition of the Shannon information content in eq. 146 the information content $I_i$ for a given functional requirement $FR_i$ ($1 \leq i \leq m$) is expressed as the logarithmic probability $p_i$ of satisfying this specific FR:

$$\begin{aligned} I_i &= \log_2 \frac{1}{p_i} \\ &= -\log_2 p_i. \end{aligned}$$

In the general case of $m$ specified FRs, the information content $I_{sys}$ for the entire system under study is

$$I_{sys} = -\log_2 P(X^m),$$

where $P(X^m)$ denotes the joint probability that all $m$ FRs are satisfied. When all FRs are statistically independent, as in an uncoupled design, the information content $I_{sys}$ can be decomposed into independent summands and expressed as

$$
\begin{aligned}
I_{sys} &= \sum_{i=1}^{m} I_i \\
&= -\sum_{i=1}^{m} \log_2 p_i.
\end{aligned}
$$

When not all FRs are statistically independent (in the so called "decoupled design"), there holds

$$
I_{sys} = -\sum_{i=1}^{m} \log_2 p_{i|\{j\}} \qquad \text{for } \{j\} = \{1, \dots, i-1\}.
$$

In the above equation $p_{i|\{j\}}$ is the conditional probability of satisfying $FR_i$ given that all other correlated $\{FR_j\}_{j=1,\dots,i-1}$ are also satisfied. It is assumed that the FRs are ordered according to their number of correlations. The information axiom states that the best design is that with smallest $I_{sys}$, because the least amount of information in the sense of Shannon's theory is required to achieve the design goals. When all probabilities are one, the information content is zero and the design is optimal in the sense of the axiom. Conversely, when one or more probabilities are zero, the information required is infinite and the system has to be redesigned to satisfy the information axiom.

The probability of success $p_i$ can be determined by the intersection of the design range defined by the product developers to satisfy the FRs and the ability of the system to produce the part within the specified range. This probability can be computed by specifying the design range (*r*) for the FR and by determining the system range (*sr*) that the proposed design can provide to satisfy the FR. The lower bound of the specified design range for functional requirement $FR_i$ is denoted by $r^l[FR_i]$, and the upper bound by $r^u[FR_i]$. The system range can be modeled in statistical terms on the basis of a *pdf* (section 2.1). The *pdf* is specified over the theoretically feasible state space. The system *pdf* is denoted by $f_{sys}[FR_i]$. The overlap between the design and system ranges is called "the common range" (*cr*), and this is the only range where the FR is satisfied. Consequently, the area $A_{cr}$ under the system *pdf* within the common range is the design's probability of achieving the specified goal. Hence, the information content $I_i$ can be expressed as

$$
\begin{aligned}
I_i &= -\log_2 A_{cr} \\
&= -\log_2 \int_{r^l[FR_i]}^{r^u[FR_i]} f_{sys}[FR_i] \, dFR_i.
\end{aligned}
$$

Suh (2005) calls a design complex when its probability of success is low and hence the information content $I_{sys}$ required to satisfy the FRs is high. Complex designs often arise when there are many components, because as their number increases through functional decomposition, the probability that some of them do not meet the specified requirements also increases, such as when the interfaces between components introduce additional errors. In order to govern the design process toward more effective, efficient and robust large-scale systems, a dedicated complexity axiom is defined that simply states "reduce the complexity of a system" (Suh 2005). The quantitative measure for complexity in the sense of this axiom is the information content, which was defined in the above equations. The rationale behind the axiom is that complex systems may require more information to make the system function. Therefore, Suh (2005) ties the notion of complexity to the design range for the FRs—the tighter the design range, the more difficult it becomes to satisfy the FRs. An uncoupled design is likely to be least complex. However, the complexity of a decoupled design can be high because of so-called "imaginary complexity" if we do not understand the system. It is not really complex, but it appears to be so because of our lack of understanding of generalized or physical functions.

According to Suh (2005) complexity can also be a function of time if the system range changes as a function of time. In this case, two types of time-dependent complexity must be distinguished: time-dependent combinatorial complexity and time-dependent periodic complexity. Time-dependent combinatorial complexity is defined as the complexity that increases as a function of time because of a continued expansion in the number of possible combinations of FRs and DPs in time, which may lead to chaotic behavior or system failure. It occurs because future events occur randomly in time and cannot be predicted, even though they depend on the current state. On the other hand, periodic complexity is defined as the complexity that only exists in a finite time period, resulting in a finite and limited number of probable combinations. Concerning a system subjected to combinatorial complexity, Suh (2005) concludes that the uncertainty of future outcomes continues to grow over time, and as a result, the system cannot have long-term stability and reliability. In the case of systems with periodic complexity, it is supposed that the system is deterministic and can renew itself over each period. Therefore, he concludes that a stable and reliable system must be periodic. It is obvious that a system with time-dependent combinatorial complexity can be changed to one with time-dependent periodic complexity by defining a set of functions repeating periodically. This can be achieved temporally, geometrically, thermally, electrically and by other constructive means. In conclusion, engineered systems in NPD should have small time-independent real and imaginary complexities and no time-dependent combinatorial complexity. If the system range must change as a function of time, the developer should be able to introduce time-dependent periodic complexity. These criteria need to be satisfied regardless of the size of the system and the number of FRs and DPs specified for the system.

Although Suh's complexity theory is grounded in axiomatic design theory and has been successfully applied in different domains, our criticism is that product and design problems are evaluated irrespective of the work processes, which are needed to decompose the FRs and DPs. The decomposition is a highly cooperative process that must be taken into account to satisfy all specified FRs on time and to avoid cycles of continuing revisions. Furthermore, the fact that Suh uses the information content I_sys directly as a complexity measure can be subject to criticism. I_sys is a simple additive measure that only represents the encoded length of the design in terms of binary design decisions. It does not take into account the encoding scheme. However, both parts of the description of a design are important because the description can always be simplified by formulating more complicated design rules, more complex standard components or interfaces (cf. section 3.2.2). Finally, Suh (2005) does not define specific measures for time-dependent complexity.

El-Haik and Yang (1999) have extended Suh's theory by representing the imaginary part of complexity through the differential entropy (section 4) associated with the joint pdf of FRs with three components of variability, vulnerability and correlation. These components evaluate the product design according to the vector of DPs (see Summers and Shah 2010). Although this approach is able to assess the mapping from the FRs to the DPs through an analysis of the topological structure of the design structure matrix (Browning 2001, see discussion below) and the variability of the design parameters (measured by the differential entropy of the joint pdf of DPs), the dynamics of the development processes in terms of a work transformation matrix (WTM, section 2.2) are not taken into account. An alternative view introduced by Braha and Maimon (1998) suggests that complexity is a fundamental characteristic of the information content within either the product or the process. They introduce two measures that quantify either the structural representation of the information or the functional probability of achieving the specified requirements. The measures are able to compare products and processes at different levels of abstraction. The process is nominally defined as mapping between the product and problem, where the coupling determines process complexity. The size of the process is defined as the summation over the number of instances of operators (relationships) and operands (entities). A process instance is a sequence of the instances of operands and operators. The average information content of sequences can be evaluated on the basis of the block entropy (eq. 16). As the design takes on different types of representations through the development stages, the average information contained changes. Braha and Maimon (1998) suggest that the ratio of the amount of average information content between the initial and current states is a measure of the current abstraction level. The effort required to move between abstraction levels is inversely proportional to this ratio. The proportionality constant is the information content of the current state. Summers and Shah (2010) follow these lines of thought and propose a process size complexity measure that includes the vocabulary of the specific representation for the problem, the product, the development process and the four operators available for sequencing the states of the design evolution. The measure is defined as

$$Cx_{size_process} := \left(M^o + C^o + P_{op}\right) \ln \left|idv + ddv + dr + mg + a_{op} + e_{op} + s_{op} + r_{op}\right|.$$

In the above definition the size of the vocabulary is represented by the total number of possible primitive modules ($M^o$), possible relations between these modules ($C^o$) and possible operators and operands ($P_{op}$). The additional parameters denote the variables whose values are controlled by the designer ($idv$), are derived from the independent design parameters, other dependent variables and design relations ($ddv$), are constraints that dictate the association between the other design variables ($dr$), are used to determine how well the current design configuration meets the goals ($mg$) plus the four operators available for sequencing the states. Although the concepts based on information contents are appealing, the fact that the development process is only analyzed on different hierarchical description levels, not on the basis of an explicit state-space model of cooperative work opens it to criticism, because it does not take into account dynamic entropies in the sense of Grassberger's theory. Furthermore, in real design problems, it is difficult to identify all operators and operands in advance and to specify valid sequences leading from one level of abstraction to the next.

In addition to methods for measuring characteristics of the design based on information-theoretic quantities, a large body of literature has been published around the design structure matrix (Steward 1981) as a dependency modeling technique supporting complexity management by focusing attention on the elements of a system and the dependencies through which they are related. Recent surveys can be found in the textbooks of Lindemann et al. (2009) or Eppinger and Browning (2012). Browning (2001) distinguishes two basic DSMs types: static and time-based. Static DSMs represent either product components or teams in an organization existing simultaneously. Time-based DSMs either represent dynamic activities indicating precedence relationships or design parameters that change as a function of time. Generated static DSMs are usually analyzed for structural characteristics or by clustering algorithms (e.g. Rogers et al. 2006), whilst time-based DSMs are typically used to optimize workflows based on sequencing, tearing and banding algorithms (e.g. Gebala and Eppinger 1991; Maurer 2007). Kreimeyer et al. (2008) reviews and discusses a comprehensive set of metrics that can be applied to assess the structure of engineering design processes encoded by DSMs (and other forms). According to Browning´s taxonomy, the WTM as dynamical operator of state eq. 4 is a static task-based DSM, because the development tasks are processed concurrently and continuing feedback/feed forward loops are modeled through the off-diagonal elements. The majority of work on complexity management with static DSMs focuses on the concept of modularity in identifying cluster structures (see Baldwin and Clark 2000). This work has been very influential in academia and industry. An important limitation, however, is a purely static view of the product structure and, consequently, of the task structure and the interactions among them. A task processing on different time scales corresponding to different autonomous task processing rates cannot be represented. Recent publications indicate that technical dependencies in product families tend to be volatile and therefore coordination needs among development tasks can evolve over time (e.g. Cataldo et al. 2006; Cataldo et al. 2008; Sosa 2008). When those evolving coordination needs are not adequately managed, significant misalignments of organizational structure and product architecture can occur that have a negative effect on product quality (Gokpinar et al. 2010). An effective method for dealing with volatility of dependencies is to use different WTMs for different phases of the project in which no task is theoretically processed independently of the others. Furthermore, at the transition points between phases additional task-mapping matrices can be specified. By doing so, the number of tasks as well as the kind and intensity of coordination needs can be adapted. It is also possible to specify phase-dependent covariances of performance fluctuations. In many NPD projects the performance fluctuations tend to be larger for late development stages that are close to the desired start of production. Another limitation of the concept of product modularity is that the organizational patterns of a development project (e.g. communication links, team co-membership) not necessarily mirror the technical dependency structures (Sosa et al. 2004). The literature review of Colfer and Baldwin (2010) shows that the “mirroring hypothesis” was supported in only 69% of the cases. Support for the hypothesis was strongest in the within-firm sample, less strong in the across-firm sample, and relatively weak in the open collaborative sample. In that sense WTMs and covariance matrices represent dynamic dependency structures in their own right. They must be related to product components or development teams through additional multiple domain mapping matrices (Danilovic and Browning 2007) and cannot be substituted by the traditional modeling elements.

An approach to measure structural complexity based on static component-based DSMs that is formally similar to our analysis in the spectral basis (see sections 2.3 and 4.2) has been introduced by Sinha and de Weck (2009, 2012). The parameters of their metric $C_{SW}$ are related to the complexity of each of the $n$ components in the product (represented by the $\alpha_i$’s) and to the complexity of each of the $m$ interfaces (connections) between a pair of components (represented by the $\beta_{ij}$’s). Moreover, a normalization factor $\gamma$ is introduced. The definition is (Denman et al. 2011, Sinha and de Weck 2012):

$$C_{SW} := \sum_{i=1}^{n} \alpha_i + \left( \sum_{i=1}^{n} \sum_{j=1}^{n} \beta_{ij} A_{ij} \right) \gamma E(A)$$

The normalization factor $\gamma$ is taken as $1/n$ and used to map the $n$ different components onto a comparable scale. The matrix $A$ is an adjacency matrix that corresponds to the component-based DSM of the product as follows:

$$A = (a_{ij}) = \begin{cases} 1 & \forall (i,j): (i \neq j) \wedge (i,j) \in \Upsilon \\ 0 & otherwise \end{cases} .$$

The exogenous variable $\Upsilon$ represents the set of connected product components. In that sense, the adjacency matrix is simply a binary form of the component-based DSM, in which ones are placed in the cells with marks and zeros elsewhere. The diagonal elements of $A$ are zero. The underlying concept of the metric $C_{SW}$ is that in order to develop the individual components, a non-zero complexity is involved. This complexity can vary across components and is represented by the $\alpha_i$´s, the so-called component complexity estimate (Sinha and de Weck 2012). Similar arguments hold true for the complexity $\beta_{ij}$ of each interface, the so-called final interface complexity (Sinha and de Weck 2012). If there are multiple types of interface between two components (energy flow, material flow, control action flow etc.), large beta coefficients are assigned since it would require more effort to implement them compared to a simpler (univariate) connection. An important aspect is the correlation among the component complexity Estimate and final interface complexity that can vary depending on the kind of product. For large-scale mechanical systems, the $\beta_{ij}$´s are often much smaller than the $\alpha_i$´s and $\alpha_j$´s. However, in micro or nanoscale systems it can be the opposite, because it is often much more difficult to develop the interfaces. The different interface complexities can be captured using a multiplicative model

$$\beta_{ij} = f_{ij}\alpha_i\alpha_j,$$

where $f_{ij}$ stands for the interface complexity factor (Sinha and de Weck 2012). Finally, the term $E(A)$ represents the graph energy of the adjacency matrix $A$. The graph energy is defined as the sum of the singular values $\sigma_i$ of the orthogonal vectors:

$$E(A) := \sum_{i=1}^{n} \sigma_i \ ,$$

where

$$A = U\Sigma_A V^{\mathrm{T}}$$

$$\Sigma_A = \mathrm{Diag}[\sigma_i].$$

The graph energy is invariant under isomorphic transformations (Weyuker 1988) and therefore highly objective.

Ameri and Summers (Ameri et al. 2008, Summers and Ameri 2008) developed a complementary connectedness measure and an algorithm for assessing design connectivity complexity based on graphical models. In the graphical models, the development tasks are nodes of a graph and connected through variable dependency. The algorithm manipulates the graph in terms of connectivity. This manipulation starts by eliminating all unary relations as these do not contribute to the connectivity complexity of the graph. Once the unary relations have been removed, the score keeping variables are initialized. From this point forward, the graph connectivity algorithm is a recursive algorithm that is applied against all subgraphs that are generated in the process. A cumulative score is maintained to quantify the connectedness of the whole structure (see Summers and Shah 2010). This approach seems to have significant limitations for assessing emergent complexity in NPD. The graph of development tasks is recursively decomposed into subgraphs, which tears apart potentially important indirect connections. Furthermore, “vicious” cycles of activities emerging from either short or long feedbacks cannot be evaluated adequately. Consequently, we will not consider the design connectivity complexity in the following.

The information-theory and dependency-structure-based complexity metrics from theories of systematic engineering design are undoubtedly beneficial in facilitating studies that require the use of equivalent but different design problems and in comparing computer-aided design automation tools. Nevertheless, in the following analytical section will bring the EMC into focus according to Grassberger’s seminal theoretical work (1986), because it can measure self-generated complexity and it provides a foundation to derived closed-form solutions of different strengths from first principles. Furthermore, EMC stresses the dynamic nature of cooperative work in NPD projects and it can be calculated efficiently from generative models or from historical data.

Also very interesting for applications in project management is the later-formulated persistent mutual information $\mathrm{EMC}(\tau)$. This is partly because of its intimate relationship with the famous Lyapunov function (Nicolis and Nicolis 2007) of a process, and partly because the generated complexity "landscape" becomes more and more informative as the lead time increases. However, this phenomenon goes beyond the scope of the paper and will be analyzed in detail in future work. To lay the analytical foundations for future studies of emergent complexity we will present closed-form solutions of the persistent mutual information for the developed vector autoregression models and linear dynamical systems in the corresponding sections. These solutions are generalized from the closed-form expressions for $\mathrm{EMC} = \mathrm{EMC}(\tau = 0)$ which will be presented in the beginning of sections 4.1, 4.2 and 4.3 (see eqs. 175, 181, 190 and 193).

More details on complexity measures from statistical physics, information theory and computer science are presented in Shalizi (2006), Prokopenko et al. (2007), Nicolis and Nicolis (2007), Ellison et al. (2009) and Crutchfield et al. (2010). A focused review of complexity measures for the evaluation of human–computer interaction including two empirical validation studies can be found in Schlick et al. (2006 and 2010).

# 4 MODEL-BASED EVALUATION OF EMERGENT COMPLEXITY

The main problem that has to be addressed in this paper is that Grassberger (1986) defined the EMC on the basis of information-theory variables for the evaluation of stochastic processes with discrete states and did not generalize the concepts to continuous-state processes like the developed VAR model of cooperative task processing according to state eq. 4. However, Bialek et al. (2001), de Cock (2002), Bialek (2003) and Ellison et al. (2009) pioneered the generalization of Grassberger's theory and measures toward continuous systems in their works, and we can build upon their results. Their analyses show that we must primarily consider the so-called "differential block entropy" (eq. 164) and the corresponding continuous-type mutual information as a basic information-theory quantities.

In general, the differential entropy extends the basic idea of Shannon's classic information entropy as a universal measure of uncertainty about a discrete-type random variable with known probability mass function over the finite alphabet $\mathcal{X}$ to a $p$-dimensional continuous-type variable $X$ with a probability density function $f[x]$ (*pdf*, see previous sections) whose support is a set $\mathbb{X}^p$. The differential entropy is defined as:

$$H[X] := -\int_{\mathbb{X}^p} f[x] \log_2 f[x]\, dx. \tag{163}$$

The differential block entropy (cf. eq. 154) is defined in an analogous manner as:

$$H(n) := H[X^n] = \int_{\mathbb{X}^p} \dots \int_{\mathbb{X}^p} f[x_1, \dots, x_n] \log_2 f[x_1, \dots, x_n]\, dx_1 \dots\, dx_n. \tag{164}$$

In the above equation $f[x_1, \dots, x_n]$ denotes the joint *pdf* of the vectors $(X_1, \dots, X_n)$ with support $\mathbb{X}^{np}$.

The information entropy of a discrete-type random variable is non-negative and can be used as a measure of average surprisal. This is slightly different for a continuous-type variable, whose differential entropy can take any value from $-\infty$ to $\infty$ and is only used to measure only changes in uncertainty (Cover and Thomas 1991, Papoulis and Pillai 2002). For instance, the differential entropy of a continuous random variable $X$ that is uniformly distributed from 0 to $a$ (and whose *pdf* is therefore $f[x] = 1/a$ from 0 to $a$, and 0 elsewhere) is $\log_2 a$. For $a < 1$ the differential entropy is negative and can get arbitrarily small as $a$ approaches 0. The differential entropy measures the entropy of a continuous distribution relative to the uniformly distributed one. For a Gaussian distribution with a variance of $\sigma^2$ the differential entropy is $H[X] = 1/2 \log_2 \sigma^2 + const$. Thus the differential entropy can be regarded as a generalization of the familiar notion of variance. With a normal distribution, the differential entropy is maximized for a given variance. An additional subtlety is that the continuous entropy can be negative or positive depending on the coordinate system used for encoding the state vectors. This also holds true for the differential block entropy. However, it can be proven that the complexity measure EMC calculated on the basis of dynamic entropies (cf. eqs. 159 and 160) is always positive and may exist even in cases where the block entropies diverge. Under the assumption of an underlying VAR model, for instance, a closed-form solution for the EMC can be derived that is simply a ratio of determinants of covariance matrices (cf. eqs. 175 and 186), which in most industrial case studies is a real number that is much larger than zero. In this case, the generalized complexity measure can be interpreted similarly to discrete-state processes. Furthermore, it can be proven that for finite complexity values EMC is independent of the basis in which the state vectors of work remaining are represented, and is invariant under linear transformations of the state-space coordinates for any regular transformation matrix (Schneider and Griffies 1999). This independency is due to the mere fact that the measure can be expressed as the continuous-type mutual information $I[X_{-\infty}^{-1}; X_0^{\infty}]$ between the infinite past and future histories of a stochastic process where the mutual information $I[.;.]$ between the sequences $X_1^n = (X_1, \dots, X_n)$ and $Y_1^m = (Y_1, \dots, Y_m)$ of random vectors with support $\mathbb{X}^{nq}$ and $\mathbb{Y}^{mp}$ is defined as

$$I[X_1^n; Y_1^m] = \int_{\mathbb{X}^q} \dots \int_{\mathbb{Y}^p} f[x_1, \dots, x_n, y_1, \dots, y_m] \log_2 \frac{f[x_1, \dots, x_n, y_1, \dots, y_m]}{f[x_1, \dots, x_n] f[y_1, \dots, y_m]} \, dx_1 \dots dx_n dy_1 \dots dy_m \,. \quad (165)$$

For a bivariate Gaussian with a correlation coefficient of $\rho$ there is $I[X;Y] = 1/2 \log_2(1-\rho^2)$. In that sense the mutual information can be viewed as a generalized covariance. Kraskov et al. (2004) published a simple proof that the mutual information as defined in eq. 165 is not only invariant under linear transformations but also with respect to arbitrary reparameterizations based on smooth and uniquely invertible maps $x_1' = x_1'(x_1), \dots, x_n' = x_n'(x_n), y_1' = y_1'(y_1), \dots, y_m' = y_m'(y_m)$ . Therefore, $I[.;.]$ provides a measure of statistical dependence between variables that is independent of the subjective choice of the measurement instrument. The analyses of Bialek et al. (2001) and other researchers show that the generalized measure is a valid, expressive and consistent quantity for evaluating emergent complexity in open systems.

The generalization of the EMC to project organizations that are modeled by continuous state variables will be carried out step-by-step in the following sections. Some of the calculations are mathematically quite involved, but the interested reader will find that they lay important foundations for complexity analyses of open organizational systems, not only NPD. In section 4.1, we start by calculating closed-form solutions with different strength for the vector autoregression models that were introduced in sections 2.1, 2.2 and 2.5. These models do not have “hidden” state variables and therefore are quite easy to analyze in information-theoretic terms. To simplify the analysis a generalized solution for a VAR(1) process that does not refer to a specific family of *pdf*s of the random performance fluctuations is calculated in section 4.1. We will use this generalized solution to derive closed-form solutions for the original state space (section 4.1.1) and the spectral basis (section 4.1,2) under the assumption of Gaussian behavior. Furthermore, a very compact closed-form solution will be obtained through a canonical correlation analysis (section 4.1.3). To clarify the concept of emergent complexity polynomial-based solutions for simple projects with two and three tasks are presented in section 4.1.4. This section also includes a short analytical study of minimizing complexity under productivity constraints. Moreover, lower bounds are put on the EMC in section 4.1.5. In section 4.2, an additional closed-form solution for a Markov process with hidden variables (section 2.8) is calculated. This solution is quite difficult to interpret, because the underlying dynamic model of cooperative work can generate long-range correlations among state variables. Nevertheless, the solution shows that Grassberger´s theory can be applied in a straightforward manner to a larger model class that is especially rich for applications in project management.

## 4.1 Closed-form Solutions for Vector Autoregression Models of Cooperative Work

To obtain analytical results, it is assumed that the parameterized VAR(1) process $\{X_t\}$ is strict-sense stationary (Puri, 2010) and therefore all its statistical properties (especially the first and second moments) are invariant to a shift in the chosen time origin. Let $f_\theta[x_{t+1}, \dots, x_{t+n}]$ $(t \epsilon \mathbb{Z}, n \in \mathbb{N})$ be the joint *pdf* of the block of random vectors $(X_{t+1}, \dots, X_{t+n})$ generating the stochastic process, and let $f_\theta[x_{t+n} | x_{t+1}, \dots, x_{t+n-1}]$ be the conditional density of vector $X_{t+n}$ given vectors $X_{t+1}, \dots, X_{t+n-1}$. We use the shorthand notation $f[.]$ and $f[.|.]$ in the following to denote these density functions. Due to strict sense stationarity the joint distribution of any sequence of samples does not depend on the sample's placement:

$$f[x_{t+1},\dots,x_{t+n}] = f[x_{t+1+\tau},\dots,x_{t+n+\tau}] \quad (t\epsilon\,\mathbb{Z}, n\,\epsilon\,\mathbb{N}, \tau \geq 0)$$

We can use the index $\upsilon$ instead of $t$ to express the shift-invariance. Therefore, $f[x_{\upsilon+1},\dots,x_{\upsilon+n}]$ denotes the joint *pdf* and $f[x_{\upsilon+n}|x_{\upsilon+1},\dots,x_{\upsilon+n-1}]$ denotes the conditional density of the process in the steady state. The conditional density is given by (cf. Billingsley 1995):

$$f[x_{\upsilon+n}|x_{\upsilon+1},\dots,x_{\upsilon+n-1}] = \frac{f[x_{\upsilon+1},\dots,x_{\upsilon+n}]}{f[x_{\upsilon+1},\dots,x_{\upsilon+n-1}]}.$$

Since the considered VAR(1) process is a Markov process (eq. 13), the conditional density simplifies to

$$f[x_{\upsilon+n}|x_{\upsilon+1},\dots,x_{\upsilon+n-1}] = f[x_{\upsilon+n}|x_{\upsilon+n-1}] = \frac{f[x_{\upsilon+n-1},x_{\upsilon+n}]}{f[x_{\upsilon+n-1}]}, \tag{166}$$

and the strict stationarity condition implies (Brockwell and Davis 1991)

$$f[x_{\upsilon+n}|x_{\upsilon+n-1}] = f[x_{\upsilon}|x_{\upsilon-1}] = f[x_2|x_1] \text{ and } f[x_{\upsilon+n-1}] = f[x_{\upsilon}] = f[x_1] \;\; \forall \upsilon \geq 2\,. \tag{167}$$

Furthermore, we assume that ergodicity holds, and the complexity measure can be conveniently derived using stochastic calculus based on an ensemble average or an infinite number of realizations of the unpredictable performance fluctuations (see Puri, 2010). To compute the EMC for the introduced VAR(1) process in the steady state, please recall from eq. 152 that

$$\text{EMC} = I[X_{-\infty}^{-1}; X_0^{\infty}]\,.$$

According to the definition of the mutual information $I[.\,;.]$ from eq. 167, we can write the information that is communicated from the past to the future as

$$I[X_{-\infty}^{-1}; X_0^{\infty}] = \int_{\mathbb{X}^p} \dots \int_{\mathbb{X}^p} f[x_{-\infty}^{-1}, x_0^{\infty}] \log_2 \frac{f[x_{-\infty}^{-1}, x_0^{\infty}]}{f[x_{-\infty}^{-1}] f[x_0^{\infty}]}\, dx_{-\infty}^{-1}\, dx_0^{\infty}. \tag{168}$$

In the above equation the shorthand notation $f[x_{-\infty}^{-1}, x_0^{\infty}] = f[x_{-\infty}, x_{-\infty+1},\dots,x_{-1},x_0,x_1,,\dots,x_{\infty-1},x_{\infty}]$, $f[x_{-\infty}^{-1}] = f[x_{-\infty},x_{-\infty+1},\dots,x_{-1}]$, $f[x_0^{\infty}] = f[x_0,x_1,\dots,x_{\infty-1},x_{\infty}]$, $dx_{-\infty}^{-1} = dx_{-\infty}dx_{-\infty+1}\dots dx_{-1}$ and $dx_0^{\infty} = dx_0 dx_1 \dots dx_{\infty}$ was used. Due to the Markov property (eq. 167) the joint *pdfs* can be factorized:

$$\begin{aligned}
f[x_{-\infty}^{-1}, x_0^{\infty}] &= f[x_{-\infty}] f[x_{-\infty+1}|x_{-\infty}] \dots f[x_{-1}|x_{-2}] f[x_0|x_{-1}] f[x_1|x_0] \dots f[x_{\infty}|x_{\infty-1}] \\
f[x_{-\infty}^{-1}] &= f[x_{-\infty}] f[x_{-\infty+1}|x_{-\infty}] \dots f[x_{-1}|x_{-2}] \\
f[x_0^{\infty}] &= f[x_0] f[x_1|x_0] \dots f[x_{\infty}|x_{\infty-1}].
\end{aligned}$$

Hence, we can simplify the mutual information:

$$\begin{aligned}
I[X_{-\infty}^{-1}; X_0^{\infty}] &= \int_{\mathbb{X}^p} \dots \int_{\mathbb{X}^p} f[x_{-\infty}^{-1}, x_0^{\infty}] \log_2 \frac{f[x_0|x_{-1}]}{f[x_0]}\, dx_{-\infty}^{-1}\, dx_0^{\infty} \\
&= \int_{\mathbb{X}^p} \dots \int_{\mathbb{X}^p} f[x_{-\infty}^{-1}, x_0^{\infty}] \log_2 f[x_0|x_{-1}]\, dx_{-\infty}^{-1}\, dx_0^{\infty} \\
&\quad - \int_{\mathbb{X}^p} \dots \int_{\mathbb{X}^p} f[x_{-\infty}^{-1}, x_0^{\infty}] \log_2 f[x_0]\, dx_{-\infty}^{-1}\, dx_0^{\infty} \\
&= \int_{\mathbb{X}^p} \int_{\mathbb{X}^p} \log_2 f[x_0|x_{-1}]\, dx_0 dx_{-1} \int_{\mathbb{X}^p} \dots \int_{\mathbb{X}^p} f[x_{-\infty}^{-1}, x_0^{\infty}]\, dx_{-\infty} \dots dx_{-2} dx_1 \dots dx_{\infty}
\end{aligned}$$

$$- \int_{\mathbb{X}^p} \log_2 f[x_0]\, dx_0 \int_{\mathbb{X}^p} \dots \int_{\mathbb{X}^p} f[x_{-\infty}^{-1}, x_0^{\infty}]\ dx_{-\infty} \dots dx_{-1} dx_1 \dots dx_{\infty}. \qquad (169)$$

On the basis of the definitions of the marginal density functions

$$f[x_0] = \int_{\mathbb{X}^p} \dots \int_{\mathbb{X}^p} f[x_{-\infty}^{-1}, x_0^{\infty}]\ dx_{-\infty} \dots dx_{-1} dx_1 \dots dx_{\infty}$$

$$f[x_{-1}, x_0] = \int_{\mathbb{X}^p} \dots \int_{\mathbb{X}^p} f[x_{-\infty}^{-1}, x_0^{\infty}]\, dx_{-\infty} \dots dx_{-2} dx_1 \dots dx_{\infty}$$

we can conclude that

$$\begin{aligned} I[X_{-\infty}^{-1}; X_0^{\infty}] &= \int_{\mathbb{X}^p} \int_{\mathbb{X}^p} f[x_{-1}, x_0] \log_2 f[x_0|x_{-1}]\, dx_{-1} dx_0 - \int_{\mathbb{X}^p} f[x_0] \log_2 f[x_0] dx_0 \\ &= \int_{\mathbb{X}^p} \int_{\mathbb{X}^p} f[x_0|x_{-1}]\, f[x_{-1}] \log_2 f[x_0|x_{-1}]\, dx_{-1} dx_0 - \int_{\mathbb{X}^p} f[x_0] \log_2 f[x_0] dx_0, \qquad (170) \end{aligned}$$

or equivalently

$$I[X_{-\infty}^{-1}; X_0^{\infty}] = \int_{\mathbb{X}^p} \int_{\mathbb{X}^p} f[x_1|x_0]\, f[x_0] \log_2 f[x_1|x_0]\, dx_0 dx_1 - \int_{\mathbb{X}^p} f[x_0] \log_2 f[x_0] dx_0.$$

It is obvious that the second summand is the differential entropy of the random variable $X_0$ with probability density function $f[x_0]$. The first summand represents the entropy of the random variable $X_1$ conditioned on the variable $X_0$ taking a value in the support $\mathbb{X}^p$. The first summand therefore represents a conditional entropy that is obtained by averaging over all possible values for $X_0$.

### 4.1.1 Closed-form Solutions in Original State Space

To calculate the EMC on the basis of the generalized solution from eq. 170 in the coordinates of the original state space $\mathbb{R}^p$, we must find the *pdf* of the generated stochastic process in the steady state. Let the $p$-dimensional random vector $X_1$ be normally distributed with location $\mu_1 = A_0 \cdot x_0$ and covariance $\Sigma_1 = C$ (eq. 4 and 9), that is $X_1 \sim \mathcal{N}(x; A_0 \cdot x_0, C)$. Starting with this random vector the project evolves according to state eq. 4. Due to the strictly stationary behavior for $t \to \infty$ a joint probability density is formed that is invariant under shifting the origin. Hence, for the locus we must have

$$\mu = A_0 \cdot \mu + E[\varepsilon_t] = A_0 \cdot \mu \qquad (171)$$

and for the variance

$$\Sigma = A_0 \cdot \Sigma \cdot A_0^{\mathrm{T}} + \mathrm{Var}[\varepsilon_t] = A_0 \cdot \Sigma \cdot A_0^{\mathrm{T}} + C. \qquad (172)$$

It follows from eq. 171 that $\mu$ must be an eigenvector corresponding to the eigenvalue 1 of the WTM $A_0$. Clearly, if the modeled project is asymptotically stable and the modulus of the largest eigenvalue of $A_0$ is less than 1, no such eigenvector can exist. Hence, the only vector that satisfies this equation is the zero vector, indicating that there is no remaining work:

$$\mu(t \to \infty) = 0_p.$$

Eq. 172 is known as the Lyapunov equation (see e.g. Lancaster and Tismenetsky 1985, cf. eq. 3). Let $\lambda_1(A_0), \dots, \lambda_p(A_0)$ be the eigenvalues of WTM $A_0$ ordered by magnitude. If $|\lambda_1(A_0)| < 1$, the solution of the Lyapunov equation can be written as an infinite power series (Lancaster and Tismenetsky 1985):

$$\Sigma = \sum_{k=0}^{\infty} A_0^k \cdot C \cdot \left(A_0^{\mathrm{T}}\right)^k. \qquad (173)$$

It can also be expressed using the Kronecker product ⊗:

$$\mathrm{vec}[\Sigma] = [I_{p^2} - A_0 \otimes A_0]^{-1}\,\mathrm{vec}[C\,].$$

$\Sigma$ is also positive-semidefinite. In the above equation it is assumed that $I_{p^2} - A_0 \otimes A_0$ is invertible, vec[.] is the vector function which was already used for the derivation of the least square estimators in section 2.7, and $I_{p^2}$ is the identity matrix of size $p^2 \times p^2$.

Under the assumption of Gaussian behavior, it is not difficult to find different closed-form solutions. Recalling that the random vector $X_0$ in steady state is normally distributed with location $\mu$ and covariance $\Sigma$, it follows from textbooks (e.g. Cover and Thomas 1991) that the differential entropy as the second summand in eq. 170 can be expressed as

$$\begin{aligned} -\int_{\mathbb{X}^p} f[x_0] \log_2 f[x_0] dx_0 &= -\int_{\mathbb{R}^p} \mathcal{N}(x_0; \mu, \Sigma) \log_2 \mathcal{N}(x_0; \mu, \Sigma) dx_0 \\ &= \log_2\left((2\pi)^{p/2}\sqrt{\mathrm{Det}[\Sigma]}\right) + \frac{p}{2}. \end{aligned}$$

For the calculation of the conditional entropy (first summand in eq. 170), the following insight is helpful. Given a value $x_0$, the distribution of $X_1$ is a normal distribution with location $A_0 \cdot x_0$ and covariance $C$. Hence, the conditional entropy is simply equal to minus the differential entropy of that distribution. For Gaussian distributions, the differential entropy is independent of the locus. Therefore, for the conditional entropy it holds that

$$\begin{aligned} &\int_{\mathbb{X}^p}\int_{\mathbb{X}^p} f[x_1|x_0] f[x_0] \log_2 f[x_1|x_0]\, dx_0 dx_1 \\ &= \int_{\mathbb{R}^p}\int_{\mathbb{R}^p} \mathcal{N}(x_1; A_0x_0, C)\mathcal{N}(x_0; \mu, \Sigma)\log_2\mathcal{N}(x_1; A_0x_0, C)\, dx_0 dx_1 \\ &= \int_{\mathbb{R}^p} \mathcal{N}(x_1; A_0x_0, C) \log_2\mathcal{N}(x_1; A_0x_0, C) dx_1 \\ &= \log_2\left(\frac{1}{(2\pi)^{p/2}\sqrt{\mathrm{Det}[C]}}\right) - \frac{p}{2}. \end{aligned}$$

It follows for the VAR(1) project model that

$$\begin{aligned} \mathrm{EMC} &= \log_2\left(\frac{1}{(2\pi)^{p/2}\sqrt{\mathrm{Det}[C]}}\right) - \frac{p}{2} + \log_2\left((2\pi)^{p/2}\sqrt{\mathrm{Det}[\Sigma]}\right) + \frac{p}{2} \\ &= \frac{1}{2}\log_2\left(\frac{\mathrm{Det}[\Sigma]}{\mathrm{Det}[C]}\right) = \frac{1}{2}\log_2(\mathrm{Det}[\Sigma]) - \frac{1}{2}\log_2(\mathrm{Det}[C]) = \frac{1}{2}\log_2 \mathrm{Det}[\Sigma \cdot C^{-1}]. \qquad (174) \end{aligned}$$

According to the above equation, the EMC can be decomposed additively into dynamic and pure-noise parts. The dynamic part represents the differential entropy of the state variables. If the noise is isotropic, the dynamic part completely decouples from the noise, as will be shown in eqs. 178 and 179 (Ay et al. 2010). If the solution of the Lyapunov equation (eq. 173) is substituted into the above equation, we can write the desired first closed-form solution as

$$\mathrm{EMC} = \frac{1}{2}\log_2\left(\frac{\mathrm{Det}\left[\sum_{k=0}^{\infty} A_0^k \cdot C \cdot \left(A_0^{\mathrm{T}}\right)^k\right]}{\mathrm{Det}[C]}\right). \tag{175}$$

The covariance matrices above are positive-semidefinite. Under the assumption that they are of full rank, the determinants are positive, and the range of the EMC is [0,+∞]. Interestingly, we can reshape the above solution so that it can be interpreted in terms of Shannon's famous "Gaussian channel" (cf. eq. 190 and the associated discussion) as

$$\mathrm{EMC} = \frac{1}{2}\log_2 \mathrm{Det}\left[I_p + \left(\sum_{k=1}^{\infty} A_0^k \cdot C \cdot \left(A_0^{\mathrm{T}}\right)^k\right) \cdot C^{-1}\right]. \tag{176}$$

If the covariance $C$ is decomposed into an orthogonal forcing matrix $K$ and a diagonal matrix $\Lambda_K$ as shown in eq. 16, the determinant in the denominator of eq. 175 can be replaced by $\mathrm{Det}[C] = \mathrm{Det}[\Lambda_K]$.

We can also separate the noise component $K \cdot \Lambda_K \cdot K^{\mathrm{T}}$ in the sum and reshape the determinant in the numerator as follows:

$$\begin{aligned}
\mathrm{EMC} &= \frac{1}{2}\log_2\left(\frac{\mathrm{Det}\left[\sum_{k=0}^{\infty} A_0^k \cdot K \cdot \Lambda_K \cdot K^{\mathrm{T}} \cdot \left(A_0^{\mathrm{T}}\right)^k\right]}{\mathrm{Det}[\Lambda_K]}\right)\\
&= \frac{1}{2}\log_2\left(\frac{\mathrm{Det}\left[\sum_{k=1}^{\infty} A_0^k \cdot K \cdot \Lambda_K \cdot K^{\mathrm{T}} \cdot \left(A_0^{\mathrm{T}}\right)^k + K \cdot \Lambda_K \cdot K^{\mathrm{T}}\right]}{\mathrm{Det}[\Lambda_K]}\right)\\
&= \frac{1}{2}\log_2\left(\frac{\mathrm{Det}[K] \cdot \mathrm{Det}\left[K^{\mathrm{T}} \cdot \left(\sum_{k=1}^{\infty} A_0^k \cdot K \cdot \Lambda_K \cdot K^{\mathrm{T}} \cdot \left(A_0^{\mathrm{T}}\right)^k\right) \cdot K + \Lambda_K\right] \cdot \mathrm{Det}[K^{\mathrm{T}}]}{\mathrm{Det}[\Lambda_K]}\right)\\
&= \frac{1}{2}\log_2\left(\frac{\mathrm{Det}\left[K^{\mathrm{T}} \cdot \left(\sum_{k=1}^{\infty} A_0^k \cdot K \cdot \Lambda_K \cdot K^{\mathrm{T}} \cdot \left(A_0^{\mathrm{T}}\right)^k\right) \cdot K + \Lambda_K\right]}{\mathrm{Det}[\Lambda_K]}\right)
\end{aligned}$$

Moreover, because $\Lambda_K$ is diagonal, taking $\mathrm{Tr}[\log_2(\Lambda_K)]$ is equivalent to $\log_2(\mathrm{Det}[\Lambda_K])$ and we have

$$\mathrm{EMC} = \frac{1}{2}\log_2 \mathrm{Det}[A_0^{\prime} + \Lambda_K] - \frac{1}{2}\sum_{i=1}^{p} \log_2 \lambda_i(C), \tag{177}$$

where $A_0^{\prime} := K^{\mathrm{T}} \cdot \left(\sum_{k=1}^{\infty} A_0^k \cdot K \cdot \Lambda_K \cdot K^{\mathrm{T}} \cdot \left(A_0^{\mathrm{T}}\right)^k\right) \cdot K$.

If the noise is isotropic, that is, the variances along the independent directions are equal ($C = \{\sigma^2\} \cdot I_p$), and therefore correlations $\rho_{ij}$ (eq. 28) between performance fluctuations do not exist, we obtain a surprisingly simple solution:

$$\begin{aligned}
\mathrm{EMC} &= \frac{1}{2}\log_2 \mathrm{Det}\left[\sum_{k=0}^{\infty} A_0^k \cdot \left(A_0^{\mathrm{T}}\right)^k\right]\\
&= \frac{1}{2}\log_2 \mathrm{Det}\left[\left(I_p - A_0 \cdot A_0^{\mathrm{T}}\right)^{-1}\right]
\end{aligned}$$

$$= -\frac{1}{2}\log_2 \mathrm{Det}\left[I_p - A_0 \cdot A_0^{\mathrm{T}}\right]. \tag{178}$$

The above solution is based on the von Neumann series that generalizes the geometric series to matrices (cf. section 2.2).

If the matrix $A_0$ is diagonalizable, it can be decomposed into eigenvectors $\vartheta_i(A_0)$ in the columns $S_{:i}$ of $S$ (eq. 20) and written as $A_0 = S \cdot \Lambda_S \cdot S^{-1}$. $\Lambda_S$ is a diagonal matrix with eigenvalues $\lambda_i(A_0)$ along the principal diagonal.

Hence, if $C = \{\sigma^2\} \cdot I_p$ and $A_0$ is diagonalizable, the EMC from eq. 178 can be fully simplified:

$$\begin{aligned}\mathrm{EMC} &= \frac{1}{2}\log_2 \prod_{i=1}^{p} \frac{1}{1-\lambda_i(A_0)^2} \\ &= \frac{1}{2}\sum_{i=1}^{p} \log_2 \frac{1}{1-\lambda_i(A_0)^2} \\ &= -\frac{1}{2}\sum_{i=1}^{p} \log_2\left(1-\lambda_i(A_0)^2\right). \end{aligned} \tag{179}$$

Both closed-form solutions that were obtained under the assumption of isotropic noise only depend on the dynamical operator $A_0$, and therefore the dynamic part of the project can be seen to decouple completely from the random performance fluctuations. Under these circumstances the argument $(1-\lambda_i(A_0)^2)$ of the binary logarithmic function can be interpreted as the damping coefficient of design mode $\phi_i = \left(\lambda_i(A_0), \vartheta_i(A_0)\right)$ (see section 2.1).

Similarly, for a project in which all $p$ development tasks are processed independently at the same autonomous processing rate $a$, the dynamic part completely decouples from the performance fluctuations under arbitrary correlations of noise vector components. In this non-cooperative environment with minimum richness of temporal and structure-organizational dependencies, we simply have $A_0 = \mathrm{Diag}[a, \dots, a]$. For EMC, it therefore holds that

$$\begin{aligned}\mathrm{EMC} &= \frac{1}{2}\log_2\left(\frac{\mathrm{Det}\left[\sum_{k=0}^{\infty}(\mathrm{Diag}[a,\dots,a])^k \cdot C \cdot \left(\mathrm{Diag}[a,\dots,a]^{\mathrm{T}}\right)^k\right]}{\mathrm{Det}[C]}\right) \\ &= \frac{1}{2}\log_2\left(\frac{\mathrm{Det}\left[C \cdot \sum_{k=0}^{\infty}(\mathrm{Diag}[a,\dots,a])^k \cdot (\mathrm{Diag}[a,\dots,a])^k\right]}{\mathrm{Det}[C]}\right) \\ &= \frac{1}{2}\log_2\left(\mathrm{Det}\left[\sum_{k=0}^{\infty}(\mathrm{Diag}[a^2,\dots,a^2])^k\right]\right) \\ &= \frac{1}{2}\log_2\left(\mathrm{Det}\left[\mathrm{Diag}\left[\frac{1}{1-a^2},\dots,\frac{1}{1-a^2}\right]\right]\right) \\ &= -\frac{p}{2}\log_2(1-a^2). \end{aligned} \tag{180}$$

An additional closed-form solution in which the EMC can be expressed in terms of the dynamical operator $A_0$ and a so-called prewhitened operator $W$ was formulated by DelSole and Tippett (2007) and Ay (2010). Using $\mathrm{Det}[A]/\mathrm{Det}[B] = \mathrm{Det}[A \cdot B^{-1}]$ and the Lyapunov eq. 172 we can write

$$\frac{\mathrm{Det}[C]}{\mathrm{Det}[\Sigma]} = \mathrm{Det}\left[\left(\Sigma - A_0 \cdot \Sigma \cdot A_0^{\mathrm{T}}\right) \cdot \Sigma^{-1}\right] = \mathrm{Det}\left[I_p - A_0 \cdot \Sigma \cdot A_0^{\mathrm{T}} \cdot \Sigma^{-1}\right].$$

Defining

$$W := \Sigma^{-\frac{1}{2}} \cdot A_0 \cdot \Sigma^{\frac{1}{2}}$$

we obtain

$$\frac{\mathrm{Det}[C]}{\mathrm{Det}[\Sigma]} = \mathrm{Det}\left[I_p - W \cdot W^{\mathrm{T}}\right],$$

where $\mathrm{Det}\left[I_p - A \cdot N \cdot A^{-1}\right] = \mathrm{Det}\left[I_p - N\right]$ and $\Sigma = \Sigma^{\mathrm{T}}$ were used. Hence, we obtain the EMC also as

$$\mathrm{EMC} = -\frac{1}{2} \log_2 \mathrm{Det}\left[I_p - W \cdot W^{\mathrm{T}}\right]. \tag{181}$$

According to DelSole and Tippett (2007) the application of the dynamical operator $W$ can be regarded as a whitening transformation of the state-space coordinates of the dynamical operator $A_0$ by means of the covariance matrix $\Sigma$.

Concerning the evaluation of the persistent mutual information—represented by the variable $\mathrm{EMC}(\tau)$ — of a vector autoregressive process, section 3.2.3 showed that this can be expressed by the continuous-type mutual information $I[.;.]$ as

$$\begin{aligned}\mathrm{EMC}(\tau) &= I[X_{-\infty}^{-1}; X_\tau^\infty] \\ &= \int_{\mathbb{X}^p} \dots \int_{\mathbb{X}^p} f[x_{-\infty}^{-1}, x_\tau^\infty] \log_2 \frac{f[x_{-\infty}^{-1}, x_\tau^\infty]}{f[x_{-\infty}^{-1}] f[x_\tau^\infty]}\, dx_{-\infty}^{-1}\, dx_\tau^\infty.\end{aligned}$$

The independent parameter $\tau \geq 0$ denotes the lead time. The term $f[x_{-\infty}^{-1}]$ designates the joint *pdf* of the infinite one-dimensional history of the stochastic process. Likewise, $f[x_\tau^\infty]$ designates the corresponding *pdf* of the infinite future from time $\tau$ onward. We used the shorthand notation $f[x_{-\infty}^{-1}, x_\tau^\infty] = f[x_{-\infty}, x_{-\infty+1}, \dots, x_{-1}, x_\tau, x_{\tau+1},, \dots, x_{\infty-1}, x_\infty]$, $f[x_{-\infty}^{-1}] = f[x_{-\infty}, \dots, x_{-1}]$, $f[x_\tau^\infty] = f[x_\tau, \dots, x_\infty]$, $dx_{-\infty}^{-1} = dx_{-\infty} \dots dx_{-1}$ and $dx_\tau^\infty = dx_\tau \dots dx_\infty$. Informally, for positive lead times the quantity $I[X_{-\infty}^{-1}; X_\tau^\infty]$ can be interpreted as the information that is communicated from the past to the future, but does not flow through the current sequence of observations $X_0^{\tau-1}$. Assuming strict stationarity, the joint *pdfs* are invariant under shifting the origin. Due to the Markov property of the VAR(1) model they can be factorized as follows:

$$\begin{aligned}f[x_{-\infty}^{-1}, x_\tau^\infty] &= \int_{\mathbb{X}^p} \dots \int_{\mathbb{X}^p} f[x_{-\infty}^{-1}, x_0^\infty]\, dx_0 \dots dx_{\tau-1} \\ &= f[x_{-\infty}] f[x_{-\infty+1}|x_{-\infty}] \dots f[x_{-1}|x_{-2}] f[x_{\tau+1}|x_\tau] \dots f[x_\infty|x_{\infty-1}] \\ &\quad \times \int_{\mathbb{X}^p} \dots \int_{\mathbb{X}^p} f[x_0|x_{-1}] \dots f[x_\tau|x_{\tau-1}] dx_0 \dots dx_{\tau-1}\end{aligned}$$

$$f[x_{-\infty}^{-1}] = f[x_{-\infty}] f[x_{-\infty+1}|x_{-\infty}] \dots f[x_{-1}|x_{-2}]$$

$$f[x_\tau^\infty] = f[x_\tau]f[x_{\tau+1}|x_\tau] \dots f[x_\infty|x_{\infty-1}].$$

Hence, we can simplify the mutual information as follows:

$$I[X_{-\infty}^{-1}; X_\tau^\infty] = \int_{\mathbb{X}^p} \dots \int_{\mathbb{X}^p} f[x_{-\infty}^{-1}, x_\tau^\infty] \log_2 \frac{\int_{\mathbb{X}^p} \dots \int_{\mathbb{X}^p} f[x_0|x_{-1}] \dots f[x_\tau|x_{\tau-1}] dx_0 \dots dx_{\tau-1}}{f[x_\tau]} \, dx_{-\infty}^{-1} \, dx_\tau^\infty.$$

According to the famous Chapman-Kolmogorov equation (Papoulis and Pillai, 2002) it holds that:

$$\int_{\mathbb{X}^p} \dots \int_{\mathbb{X}^p} f[x_0|x_{-1}] \dots f[x_\tau|x_{\tau-1}] dx_0 \dots dx_{\tau-1} = f[x_\tau|x_{-1}]$$

Hence, we have

$$\begin{aligned}
I[X_{-\infty}^{-1}; X_\tau^\infty] = & \int_{\mathbb{X}^p} \int_{\mathbb{X}^p} \log_2 f[x_\tau|x_{-1}] \, dx_{-1} \, dx_\tau \int_{\mathbb{X}^p} \dots \int_{\mathbb{X}^p} f[x_{-\infty}^{-1}, x_\tau^\infty] \, dx_{-\infty}^{-2} \, dx_{\tau+1}^\infty \\
& - \int_{\mathbb{X}^p} \log_2 f[x_\tau] dx_\tau \int_{\mathbb{X}^p} \dots \int_{\mathbb{X}^p} f[x_{-\infty}^{-1}, x_\tau^\infty] \, dx_{-\infty}^{-1} \, dx_{\tau+1}^\infty \\
& = \int_{\mathbb{X}^p} \int_{\mathbb{X}^p} f[x_{-1}, x_\tau] \log_2 f[x_\tau|x_{-1}] \, dx_{-1} dx_\tau - \int_{\mathbb{X}^p} f[x_\tau] \log_2 f[x_\tau] \, dx_\tau \\
& = \int_{\mathbb{X}^p} f[x_{-1}] dx_{-1} \int_{\mathbb{X}^p} f[x_\tau|x_{-1}] \log_2 f[x_\tau|x_{-1}] \, dx_\tau - \int_{\mathbb{X}^p} f[x_\tau] \log_2 f[x_\tau] \, dx_\tau.
\end{aligned}$$

For a VAR(1) process the transition function is defined as

$$f[x_\tau|x_{-1}] = \mathcal{N}\left(x_\tau; A_0^\tau \cdot x_{-1}, C(\tau)\right),$$

with the lead-time dependent covariance

$$\begin{aligned}
C(\tau) &= A_0 \cdot C(\tau-1) \cdot A_0^{\mathrm{T}} + C \\
&= \sum_{k=0}^{\tau} A_0^k \cdot C \cdot \left(A_0^{\mathrm{T}}\right)^k.
\end{aligned}$$

We find the solution

$$\begin{aligned}
\mathrm{EMC}(\tau) &= \frac{1}{2} \log_2 \mathrm{Det}[\Sigma] - \frac{1}{2} \log_2 \mathrm{Det}[C(\tau)] \\
&= \frac{1}{2} \log_2 \frac{\mathrm{Det}[\Sigma]}{\mathrm{Det}[C(\tau)]} \\
&= \frac{1}{2} \log_2 \mathrm{Det}\left[\Sigma \cdot \left(C(\tau)\right)^{-1}\right].
\end{aligned}$$

The solution can also be expressed as the logarithm of the variance ratio (Ay 2010):

$$\mathrm{EMC}(\tau) = \frac{1}{2} \log_2 \frac{\mathrm{Det}[\Sigma]}{\mathrm{Det}\left[\Sigma - A_0^{\tau+1} \cdot \Sigma \cdot \left(A_0^{\mathrm{T}}\right)^{\tau+1}\right]} \tag{182}$$

noting that $C = \Sigma - A_0 \cdot \Sigma \cdot A_0^{\mathrm{T}}$. As in section 4.1 we can rewrite the above solution on the basis of the dynamical operator $A_0$ and lead-time dependent prewhitened operator $W(\tau)$ (eq. 181; DelSole and Tippett 2007, Ay 2010) as

$$\mathrm{EMC}(\tau) = -\frac{1}{2}\log_2 \mathrm{Det}\left[I_p - W(\tau)\cdot W(\tau)^{\mathrm{T}}\right],$$

with

$$W(\tau) = \Sigma^{-\frac{1}{2}} \cdot A_0^{\tau+1} \cdot \Sigma^{\frac{1}{2}}. \tag{183}$$

### 4.1.2 Closed-form Solutions in the Spectral Basis

In this section, we calculate additional solutions in which the dependence of the EMC on the anisotropy of the noise is made explicit. These most expressive solutions are much easier to interpret, and to derive them we work in the spectral basis (cf. eq. 20). According to Neumaier and Schneider (2001), the steady-state covariance matrix $\Sigma'$ in the spectral basis can be calculated on the basis of the transformed covariance matrix of the performance fluctuations $C' = S^{-1} \cdot C \cdot ([S^{\mathrm{T}}]^*)^{-1}$ (eq. 26) as

$$\Sigma' = \begin{pmatrix} \frac{{c'_{11}}^2}{1-\lambda_1\overline{\lambda_1}} & \frac{\rho'_{12}c'_{11}c'_{22}}{1-\lambda_1\overline{\lambda_2}} & \cdots \\ \frac{\rho'_{12}c'_{11}c'_{22}}{1-\lambda_2\overline{\lambda_1}} & \frac{{c'_{22}}^2}{1-\lambda_2\overline{\lambda_2}} & \cdots \\ \vdots & \vdots & \ddots \end{pmatrix}. \tag{184}$$

In the above equation, the $\rho'_{ij}$ 's are the transformed correlations, which were defined in eq. 28 for a WTM $A_0$ with arbitrary structure and in eq. 32 for $A_0$'s that are symmetric. The ${c'_{ii}}^2$ 's (cf. eq. 6) and $\rho'_{ij}c'_{ii}c'_{jj}$'s (cf. eq. 7) are the scalar-valued variance and covariance covariance components of $C'$ in the spectral basis:

$$C' = \begin{pmatrix} {c'_{11}}^2 & \rho'_{12}c'_{11}c'_{22} & \cdots \\ \rho'_{12}c'_{11}c'_{22} & {c'_{22}}^2 & \cdots \\ \vdots & \vdots & \ddots \end{pmatrix}. \tag{185}$$

The transformation into the spectral basis is a linear transformation of the state-space coordinates (see eq. 26) and therefore does not change the mutual information being communicated from the past into the future by the stochastic process. Hence, the functional form of the closed-form solution from eq. 174 holds, and the EMC can be calculated as the variance ratio (Schneider and Griffies 1999; de Cock 2002):

$$\mathrm{EMC} = \frac{1}{2}\log_2\left(\frac{\mathrm{Det}[\Sigma']}{\mathrm{Det}[C']}\right) = \frac{1}{2}\log_2 \mathrm{Det}\left[\Sigma' \cdot C'^{-1}\right]. \tag{186}$$

The basis transformation does not change the positive-definiteness of the covariance matrices. Under the assumption that the matrices are of full rank, the determinants are positive. The determinant $\mathrm{Det}[\Sigma']$ of the covariance matrix $\Sigma'$ can be regarded as a generalized variance of the stationary process in the spectral basis, while $\mathrm{Det}[C']$ represents the generalized variance of the prediction error after the basis transformation. In terms of information theory, the variance ratio can be interpreted as the (entropy lost and) information gained when the modeled project is in the steady state, and the state is observed by the project manager with predefined "error bars", which cannot be under-run because of the inherent prediction error. The inverse $C'^{-1}$ is the so-called "concentration matrix" or "precision matrix". The variance ratio can also be interpreted in a geometrical framework (de Cock 2002). It is well known that the volume $\mathrm{Vol}[.]$ of the parallelepiped spanned by the rows or columns of a covariance matrix, e.g. $\Sigma'$, is equal to the value of its determinant:

$$\mathrm{Vol}\big[\mathrm{parallelepiped}[\Sigma']\big] = \mathrm{Det}[\Sigma'].$$

In that sense the inverse variance ratio $\mathrm{Det}[C']/\mathrm{Det}[\Sigma']$ represents the factor by which the volume of the parallelepiped referring to the dynamical part due to state transitions can be collapsed due to the state observation by the project manager leading to a certain information gain.

An important finding is that the scalar-valued variance and covariance components of the noise part are not relevant for the calculation of the EMC. This follows from the definition of a determinant (see eq. 195). The calculated determinants of $\Sigma'$ and $C'$ just give rise to the occurrence of the factor $\prod_{n=1}^{p} c'_{nn}$, which cancels out:

$$\mathrm{Det}\big[\Sigma' \cdot C'^{-1}\big] = \mathrm{Det}[\Sigma'] \cdot \mathrm{Det}\big[C'^{-1}\big] = \frac{\mathrm{Det}[\Sigma']}{\mathrm{Det}[C']}.$$

Hence, we can also calculate with the "normalized" covariance matrices $\Sigma_N'$ and $C_N'$:

$$\Sigma_N' = \begin{pmatrix} \frac{1}{1-|\lambda_1|^2} & \frac{\rho_{12}'}{1-\lambda_1\overline{\lambda_2}} & \cdots \\ \frac{\rho_{12}'}{1-\lambda_2\overline{\lambda_1}} & \frac{1}{1-|\lambda_2|^2} & \cdots \\ \vdots & \vdots & \ddots \end{pmatrix} \tag{187}$$

$$C_N' = \begin{pmatrix} 1 & \rho_{12}' & \cdots \\ \rho_{12}' & 1 & \cdots \\ \vdots & \vdots & \ddots \end{pmatrix}. \tag{188}$$

It can be proved that the normalized covariance matrices are also positive-semidefinite. If they are furthermore not rank deficient, inconsistencies of the complexity measure do not occur. According to Shannon's classic information-theory findings about the capacity of a Gaussian channel (Cover and Thomas 1991), the normalized covariance matrix $\Sigma_N'$ can be decomposed into summands as follows:

$$\Sigma_N' = C_N' + \begin{pmatrix} \frac{1}{1-|\lambda_1|^2} - 1 & \frac{\rho_{12}'}{1-\lambda_1\overline{\lambda_2}} - \rho_{12}' & \cdots \\ \frac{\rho_{12}'}{1-\lambda_2\overline{\lambda_1}} - \rho_{12}' & \frac{1}{1-|\lambda_2|^2} - 1 & \cdots \\ \vdots & \vdots & \ddots \end{pmatrix}.$$

The second summand in the above equation is defined as $\Sigma_N''$. This matrix can be simplified:

$$\Sigma_N'' = \begin{pmatrix} \frac{|\lambda_1|^2}{1-|\lambda_1|^2} & \rho_{12}' \frac{\lambda_1\overline{\lambda_2}}{1-\lambda_1\overline{\lambda_2}} & \cdots \\ \rho_{12}' \frac{\lambda_2\overline{\lambda_1}}{1-\lambda_2\overline{\lambda_1}} & \frac{|\lambda_2|^2}{1-|\lambda_2|^2} & \cdots \\ \vdots & \vdots & \ddots \end{pmatrix}. \tag{189}$$

We obtain the most expressive closed-form solution based on the signal-to-noise ratio $\mathrm{SNR} := \Sigma_N'' \cdot {C_N'}^{-1}$:

$$\mathrm{EMC} = \frac{1}{2} \log_2 \mathrm{Det}\big[I_p + \Sigma_N'' \cdot {C_N'}^{-1}\big]. \tag{190}$$

The SNR can be interpreted as the ratio of the variance $\Sigma_N''$ of the signal in the spectral basis that is generated by cooperative task processing and the effective variance $C_N'$ of the performance fluctuations or noise. The variance of the signal drives the stochastic process to a certain extent and can be reinforced through the structural organization of the project. The effective noise is in the same units as the input $x_t$.

This is called “referring the noise to the input” and is a standard method in physics for characterizing detectors, amplifiers and other devices (Bialek 2003). Clearly, if one builds a photodetector it is not so useful to quote the noise level at the output in volts; one wants to know how this noise limits the ability to detect dim lights. Similarly, when we characterize an NPD project that uses a stream of reports to encode a quasicontinuous workflow, we don’t want to know the variance in the report rate; we want to know how fluctuations in the response of the project organization limit precision in estimating the real work progress (signal), and this amounts to defining an effective noise level in the units of the signal itself. In the present case, this is just a matter of “dividing” generalized variances, but in reality it is a fairly complex task. According to Sylvester’s determinant theorem, we can swap the factors in the second summand:

$$\mathrm{Det}\left[I_p + \Sigma_N'' \cdot {C_N'}^{-1}\right] = \mathrm{Det}\left[I_p + {C_N'}^{-1} \cdot \Sigma_N''\right].$$

The obtained closed-form solution in the spectral basis has at most only $(p^2 - p)/2 + p = p(p+1)/2$ independent parameters, namely the eigenvalues $\lambda_i(A_0)$ of the WTM and the correlations $\rho'_{ij}$ in the spectral basis, and not a maximum of the approximately $p^2 + (p^2 - p)/2 + p = p(3p+1)/2$ parameters encoded in both the WTM $A_0$ and the covariance matrix $C$ (eq. 176). In other words, through a transformation into the spectral basis we can identify the essential variables influencing emergent complexity in the sense of Grassberger’s theory and reduce the dimensionality of the problem in many cases by the factor $(3p+1)/(p+1)$.

Furthermore, these independent parameters are easy to interpret. The eigenvalues $\lambda_i(A_0)$ represent the essential temporal dependencies of the modeled project in terms of effective productivity rates on linearly independent scales determined by the eigenvectors $\vartheta_i(A_0)$ $(i = 1 \ldots p)$. The effective productivity rates depend only on the design modes $\phi_i$ of the WTM $A_0$ and therefore reflect the project’s organizational design. The lower the effective productivity rates because of slow task processing or strong task couplings, the less the design modes are “damped”, and hence the larger the project complexity. On the other hand, the correlations $\rho'_{ij}$ model the essential dependencies between the unpredictable performance fluctuations in open organizational systems that can give rise to an excitation of the design modes and their interactions. This excitation can compensate for the damping factors of the design mode. The $\rho'_{ij}$’s scale linearly with the variances $\lambda_i(C)$ along each independent direction of the fluctuation variable $\eta_t$: the larger the variances, the larger the correlations and the stronger the excitation (eq. 28). However, the scale factors are determined not only by a linear interference between design modes $\phi_i$ and $\phi_j$ caused by cooperative task processing but also by the weighted interference with fluctuation modes $\Psi_i$ and $\Psi_j$ caused by correlations among performance fluctuations (cf. eqs. 28 and 32). In other words, the strong emergent complexity of the modeled NPD project does not simply come from the least-damped design mode $\phi_1 = \left(\lambda_1(A_0), \vartheta_1(A_0)\right)$ because this mode may not be sufficiently excited, but rather is caused (at least theoretically) by a complete interference between all design and fluctuation modes. Emergent complexity in the sense of Grassberger’s theory is a holistic property of the structure and process organization and in most real cases cannot be reduced to singular properties of the project organizational design. This is a truly nonreductionist approach to complexity assessment insisting on the specific character of the organizational design as a whole.

Similarly to the previous section, we can obtain a closed-form solution for the persistent mutual information $\mathrm{EMC}(\tau)$ in the spectral basis. The transformation into the spectral basis is a linear transformation of the state-space coordinates and therefore does not change the persistent mutual information communicated from the past into the future by the stochastic process. Hence, in analogy to eq. 184 the variance ratio can also be calculated

$$\text{EMC}(\tau) = \frac{1}{2}\log_2 \frac{\text{Det}[\Sigma']}{\text{Det}\left[\Sigma' - \Lambda_S^{\tau+1}\cdot\Sigma'\cdot\left(\left[\Lambda_S^{\mathrm{T}}\right]^*\right)^{\tau+1}\right]}$$

in the spectral basis, where the diagonal matrix $\Lambda_S$ is the dynamical operator (eq. 24) as

$$\Lambda_S = \text{Diag}[\lambda_i(A_0)] \qquad 1 \leq i \leq p.$$

Because $\Lambda_S$ is diagonal, the solution in the spectral basis can be simplified to

$$\begin{aligned}
\text{EMC}(\tau) &= -\frac{1}{2}\log_2 \frac{\text{Det}\left[\Sigma' - \Lambda_S^{\tau+1}\cdot\Sigma'\cdot\left(\left[\Lambda_S^{\mathrm{T}}\right]^*\right)^{\tau+1}\right]}{\text{Det}[\Sigma']} \\
&= -\frac{1}{2}\log_2 \frac{\text{Det}\left[\Sigma' - \Lambda_S^{\tau+1}\cdot\Sigma'\cdot{\Lambda_S^*}^{\tau+1}\right]}{\text{Det}[\Sigma']} \\
&= -\frac{1}{2}\log_2 \text{Det}\left[I_p - \Lambda_S^{\tau+1}\cdot\Sigma'\cdot{\Lambda_S^*}^{\tau+1}\cdot{\Sigma'}^{-1}\right] \\
&= -\frac{1}{2}\log_2 \text{Det}\left[I_p - \Sigma'(\tau)\cdot{\Sigma'}^{-1}\right], \qquad (191)
\end{aligned}$$

with $\Sigma'(\tau) = \Lambda_S^{\tau+1}\cdot\Sigma'\cdot{\Lambda_S^*}^{\tau+1}$ $(\tau \geq 0)$.

As with the derivation of the expressive closed-form solution in section 4.2, the generalized variance term $\Sigma' - \Lambda_S^{\tau+1}\cdot\Sigma'\cdot\left(\left[\Lambda_S^{\mathrm{T}}\right]^*\right)^{\tau+1} = \Sigma' - \Lambda_S^{\tau+1}\cdot\Sigma'\cdot{\Lambda_S^*}^{\tau+1}$ in the denominator of the variance ratio can be written in an explicit matrix form:

$$\begin{aligned}
&\Sigma' - \Lambda_S^{\tau+1}\cdot\Sigma'\cdot{\Lambda_S^*}^{\tau+1} \\
&= \begin{pmatrix}
\frac{{c'_{11}}^2}{1-\lambda_1\overline{\lambda_1}} - \frac{{\lambda_1}^{\tau+1}c'_{11}\left(\overline{\lambda_1}\right)^{\tau+1}}{1-\lambda_1\overline{\lambda_1}} & \frac{\rho'_{12}c'_{11}c'_{22}}{1-\lambda_1\overline{\lambda_2}} - \frac{{\lambda_1}^{\tau+1}\rho'_{12}\sqrt{c'_{11}c'_{22}}\left(\overline{\lambda_2}\right)^{\tau+1}}{1-\lambda_1\overline{\lambda_2}} & \cdots \\
\frac{\rho'_{12}c'_{11}c'_{22}}{1-\lambda_2\overline{\lambda_1}} - \frac{{\lambda_2}^{\tau+1}\rho'_{12}\sqrt{c'_{11}c'_{22}}\left(\overline{\lambda_1}\right)^{\tau+1}}{1-\lambda_2\overline{\lambda_1}} & \frac{{c'_{22}}^2}{1-\lambda_2\overline{\lambda_2}} - \frac{{\lambda_2}^{\tau+1}c'_{22}\left(\overline{\lambda_2}\right)^{\tau+1}}{1-\lambda_2\overline{\lambda_2}} & \cdots \\
\vdots & \vdots & \ddots
\end{pmatrix} \\
&= \begin{pmatrix}
\left({c'_{11}}^2\right)\frac{1-|\lambda_1|^{2(\tau+1)}}{1-|\lambda_1|^2} & c'_{11}c'_{22}\frac{\rho'_{12}\left(1-{\lambda_1}^{\tau+1}\left(\overline{\lambda_2}\right)^{\tau+1}\right)}{1-\lambda_1\overline{\lambda_2}} & \cdots \\
c'_{11}c'_{22}\frac{\rho'_{12}\left(1-{\lambda_2}^{\tau+1}\left(\overline{\lambda_1}\right)^{\tau+1}\right)}{1-\lambda_2\overline{\lambda_1}} & \left({c'_{22}}^2\right)\frac{1-|\lambda_2|^{2(\tau+1)}}{1-|\lambda_2|^2} & \cdots \\
\vdots & \vdots & \ddots
\end{pmatrix},
\end{aligned}$$

noting that $C' = \Sigma' - \Lambda_S\cdot\Sigma'\cdot\Lambda_S^*$. It can proved that the covariances $c'_{ij}$ in the above matrix form are not relevant for the calculation of $\text{EMC}(\tau)$. This follows from the definition of a determinant (see eq. 195). When calculating the determinants of $\Sigma'$ and $\Sigma' - \Lambda_S^{\tau+1}\cdot\Sigma'\cdot{\Lambda_S^*}^{\tau+1}$, they just give rise to the occurrence of a factor $\prod_{n=1}^{p} c'_{nn}$, which cancels out in the variance ratio. Therefore, the persistent mutual information can also be calculated fusing normalized covariance matrices. The normalized covariance matrix of $\Sigma'$, termed $\Sigma'_N$, was defined in eq. 187. The normalized covariance matrix of $\Sigma' - \Lambda_S^{\tau+1}\cdot\Sigma'\cdot{\Lambda_S^*}^{\tau+1}$ is simply

$$\Sigma'_N - \Lambda_S^{\tau+1} \cdot \Sigma'_N \cdot \Lambda_S^{*\,\tau+1} = \begin{pmatrix} \frac{1-|\lambda_1|^{2(\tau+1)}}{1-|\lambda_1|^2} & \rho'_{12}\frac{\left(1-\lambda_1{}^{\tau+1}\left(\overline{\lambda_2}\right)^{\tau+1}\right)}{1-\lambda_1\overline{\lambda_2}} & \cdots \\ \rho'_{12}\frac{\left(1-\lambda_2{}^{\tau+1}\left(\overline{\lambda_1}\right)^{\tau+1}\right)}{1-\lambda_2\overline{\lambda_1}} & \frac{1-|\lambda_2|^{2(\tau+1)}}{1-|\lambda_2|^2} & \cdots \\ \vdots & \vdots & \ddots \end{pmatrix}.$$

Hence,

$$\begin{aligned} \mathrm{EMC}(\tau) &= -\frac{1}{2}\log_2 \frac{\mathrm{Det}\left[\Sigma'_N - \Lambda_S^{\tau+1} \cdot \Sigma'_N \cdot \Lambda_S^{*\,\tau+1}\right]}{\mathrm{Det}[\Sigma'_N]} \\ &= -\frac{1}{2}\log_2 \mathrm{Det}\left[I_p - \Lambda_S^{\tau+1} \cdot \Sigma'_N \cdot \Lambda_S^{*\,\tau+1} \cdot \Sigma_N'^{-1}\right] \\ &= -\frac{1}{2}\log_2 \mathrm{Det}\left[I_p - \Sigma'_N(\tau) \cdot \Sigma_N'^{-1}\right], \end{aligned} \tag{192}$$

with $\Sigma'_N(\tau) = \Lambda_S^{\tau+1} \cdot \Sigma'_N \cdot \Lambda_S^{*\,\tau+1}$ $(\tau \geq 0)$.

### 4.1.3 Closed-form Solution through Canonical Correlation Analysis

If the matrix $C'_N$ representing the inherent prediction error in the spectral basis is diagonal in the same coordinate system as the normalized covariance matrix $\Sigma'_N$ contributed by cooperative task processing, then the matrix product $\Sigma'_N \cdot C_N'^{-1} = \left(I_p + \Sigma''_N \cdot C_N'^{-1}\right)$ is diagonal, and simple reduction of emergent complexity to singular properties of the design modes $\phi_i = \left(\lambda_i(A_0), \vartheta_i(A_0)\right)$ and fluctuation modes $\Psi_i = \left(\lambda_i(C), k_i(C)\right)$ will work. In this case, the elements along the principal diagonal are the signal-to-noise ratios along each independent direction. Hence, the EMC is proportional to the sum of the log-transformed ratios, and these summands are the only independent parameters. However, in the general case we have to diagonalize the above matrix product in a first step to obtain an additional closed-form solution. This closed-form solution has the least number of independent parameters. In spite of its algebraic simplicity, the solution is not very expressive, because the spatiotemporal covariance structures of the open organizational system are not revealed. We will return to this point after presenting the solution.

Unfortunately, the diagonalization of the matrix product $\Sigma'_N \cdot C_N'^{-1}$ cannot be carried out through an eigendecomposition, because the product of two symmetric matrices is not necessarily symmetric itself. Therefore, the left and right eigenvectors can differ and do not form a set of mutually orthogonal vectors, as they would if the product was diagonal. Nevertheless, we can always rotate our coordinate system on the space of the output to make the matrix product diagonal (Schneider and Griffies 1999). To do this, we decompose $\Sigma'_N \cdot C_N'^{-1}$ into singular values (singular value decomposition, de Cock 2002) as

$$\Sigma'_N \cdot C_N'^{-1} = U \cdot \Lambda_{UV} \cdot V^{\mathrm{T}},$$

where

$$U \cdot U^{\mathrm{T}} = I_p \text{ and } V \cdot V^{\mathrm{T}} = I_p$$

and

$$\Lambda_{UV} = \mathrm{Diag}[\sigma'_i] \quad 1 \leq i \leq p.$$

The columns of $U$ are the left singular vectors; those of $V$ are the right singular vectors. The columns of $V$ can be regarded as a set of orthonormal "input" basis vectors for $\Sigma'_N \cdot {C'_N}^{-1}$; the columns of $U$ form a set of orthonormal "output" basis vectors. The diagonal values $\sigma'_i$ in matrix $\Lambda_{svd}$ are the singular values, which can be thought of as scalar "gain controls" by which each corresponding input is multiplied to give a corresponding output.

The $\sigma'_i$'s are the only independent parameters of the following closed-form solution (see eq. 193). These parameters can be directly calculated through a canonical correlation analysis (CCA) of the stochastic process (see de Cock 2002) and not necessarily through a transformation into the spectral basis. The CCA was introduced by Hotelling (1935) and is used for state-space identification. The goal is to find a suitable basis for cross-correlation between two random variables —in our case the infinite, one-dimensional sequences of random variables representing the past ($X_{-\infty}^{-1}$) and future ($X_0^{\infty}$) of the process in the steady state. More formally, given the ordered concatenation of the random variables representing the past history

$$X_{-\infty}^{-1} = (\cdots \quad X_{-2} \quad X_{-1}).$$

and the future history

$$X_0^{\infty} = (X_0 \quad X_1 \quad \cdots),$$

we seek an orthonormal base $(u_1, \ldots, u_m)$ for $X_{-\infty}^{-1}$ and another orthonormal base $(v_1, \ldots, v_m)$ for $X_0^{\infty}$ that have maximal correlations but are internally uncorrelated. Therefore, it must hold that $E(u_i, v_j) = \rho_i \delta_{ij}$, for $k, j \leq m$ (where $\delta_{ij}$ is the Kronecker delta according to eq. 10). The resulting basis variables $(u_1, \ldots, u_m)$ and $(v_1, \ldots, v_m)$ are termed the canonical variates and the correlation coefficients $\rho_i$ between the canonical variates are called the canonical correlations. The cardinality of the base depends on the informational structure of the process. The $\rho_i$'s are not to be confused with the introduced ordinary correlations $\rho_{ij}$ and $\rho'_{ij}$. The requirement that the $\rho_i$'s be nonnegative and ordered in decreasing magnitude makes the choice of bases unique if all canonical correlations are distinct. It is important to note that for a strict-sense stationary VAR(1) process $\{X_t\}$ only the $p$ leading canonical correlations $\rho_i$ of each pair $(X_{-\infty}^{-1}, X_0^{\infty})$ of subprocesses are non-zero and therefore the cardinality of the base is equal $p$ (de Cock 2002, Boets et al. 2007). This is due to the simple fact that the process is Markovian and therefore the amount of information that the past provides about the future can always be encoded in the probability distribution over the $p$-dimensional present state (assuming an efficient coding mechanism is used). Furthermore, because of strict-sense stationarity, all $\rho_i$'s are less than one.

The relationship between the singular values $\sigma'_i$ of $\Sigma'_N \cdot {C'_N}^{-1}$ and the canonical correlations $\rho_i$ in our case is as follows:

$$\sigma'_i = \frac{1}{1 - \rho_i^2} \qquad 1 \leq i \leq p.$$

Under the assumption that $\mathrm{Det}\left[\Sigma'_N \cdot {C'_N}^{-1}\right] > 0$, it is possible to prove that $\mathrm{Det}[U] \cdot \mathrm{Det}[V] = 1$. We can obtain the desired closed-form solution as follows:

$$\begin{aligned}\mathrm{EMC} &= \frac{1}{2} \log_2 \mathrm{Det}\left[\Sigma'_N \cdot {C'_N}^{-1}\right] \\ &= \frac{1}{2} \log_2 \mathrm{Det}[U \cdot \Lambda_{UV} \cdot V^{\mathrm{T}}]\end{aligned}$$

$$
\begin{aligned}
&= \frac{1}{2}\log_2(\mathrm{Det}[U]\cdot\mathrm{Det}[\Lambda_{UV}]\cdot\mathrm{Det}[V]) \\
&= \frac{1}{2}\log_2\mathrm{Det}[\Lambda_{UV}] \\
&= \frac{1}{2}\mathrm{Tr}[\log_2(\Lambda_{UV})] \\
&= \frac{1}{2}\sum_{i=1}^{p}\log_2\sigma_i' \\
&= \frac{1}{2}\sum_{i=1}^{p}\log_2\left(\frac{1}{1-\rho_i^2}\right) \\
&= -\frac{1}{2}\sum_{i=1}^{p}\log_2\left(1-\rho_i^2\right).
\end{aligned}
\tag{193}
$$

In spite of its algebraic simplicity, a main disadvantage of this closed-form solution with only $p$ independent parameters $\sigma_i'$ or $\rho_i^2$ is that both the temporal dependencies of the modeled project in terms of essential productivity rates (represented by the $\lambda_i$'s), and the essential organizational dependencies in terms of intensity of cooperative relationships exciting fluctuations (represented by the $\rho_{ij}'$ 's) are not explicit, but are compounded into correlation coefficients between the canonical variates. Therefore, it is impossible for the project manager to analyze and interpret the spatiotemporal covariance structures of the open organizational system and to identify countermeasures for coping with complexity.

A canonical correlation analysis over $\tau$ time steps leads to the following solution of the persistent mutual information:

$$
\begin{aligned}
\mathrm{EMC}(\tau) &= \frac{1}{2}\log_2\mathrm{Det}\left[\Sigma_N'\cdot\left(\Sigma_N' - \Lambda_S^{\tau+1}\cdot\Sigma_N'\cdot{\Lambda_S^*}^{\tau+1}\right)^{-1}\right] \\
&= \frac{1}{2}\log_2\mathrm{Det}[U(\tau)\cdot\Lambda_{UV}(\tau)\cdot V(\tau)^{\mathrm{T}}] \\
&= \frac{1}{2}\log_2(\mathrm{Det}[U(\tau)]\cdot\mathrm{Det}[\Lambda_{UV}(\tau)]\cdot\mathrm{Det}[V(\tau)]) \\
&= \frac{1}{2}\log_2\mathrm{Det}[\Lambda_{UV}(\tau)] \\
&= \frac{1}{2}\mathrm{Tr}\left[\log_2\left(\Lambda_{UV}(\tau)\right)\right] \\
&= \frac{1}{2}\sum_{i=1}^{p}\log_2\sigma_i'(\tau) \\
&= \frac{1}{2}\sum_{i=1}^{p}\log_2\left(\frac{1}{1-\left(\rho_i(\tau)\right)^2}\right) \\
&= -\frac{1}{2}\sum_{i=1}^{p}\log_2\left(1-\left(\rho_i(\tau)\right)^2\right),
\end{aligned}
\tag{194}
$$

where the matrix product $U(\tau)\cdot\Lambda_{svd}(\tau)\cdot V(\tau)^{\mathrm{T}}$ denotes the singular value decomposition of $\Sigma'_N\cdot(\Sigma'_N-\Lambda_S^{\tau+1}\cdot\Sigma'_N\cdot({\Lambda_S}^*)^{\tau+1})^{-1}$ as a function of the lead time $\tau$. The $\sigma'_i(\tau)$'s and $\rho_i(\tau)$'s represent, respectively, the singular values and canonical correlations given the lead time.

#### 4.1.4 Polynomial-based Solutions for Projects with Two and Three Tasks

We can also analyze the spatiotemporal covariance structure of $\Sigma'_N$ (eq. 187) in the spectral basis explicitly by recalling the definition of a determinant. If $B=(b_{ij})$ is a matrix of size $p$, then

$$\mathrm{Det}(B)=\sum_{\beta\in R_p}\mathrm{sgn}(\beta)\prod_{i=1}^{p}b_{i,\beta(i)} \tag{195}$$

holds. $R_p$ is the set of all permutations of $\{1,\dots,\mathrm{p}\}$. Thus, because of the regular structure of the matrix $\Sigma'_N$, $\mathrm{Det}[\Sigma'_N]$ is a sum of $p!$ summands. Each of these summands is a fraction, because it is a product of elements from $\Sigma'_N$, where exactly one entry is chosen from each row and column. The denominator of those fractions is a product consisting of p factors of $1-\lambda_i(\mathrm{A}_0)\lambda_i\overline{\lambda_j(\mathrm{A}_0)}$. The numerator is a product of 2, 3,…, $p$ factors $\rho'_{ij}$, or just 1 if the permutation is the identity. (The case of one factor cannot occur, because the amount of factors equals the amount of numbers changed by the permutation β, and there is no permutation that changes just one number). The coefficients $(i,j)$ of the factor $1-\lambda_i(\mathrm{A}_0)\overline{\lambda_j(\mathrm{A}_0)}$ in the denominator correspond to the coefficients $(k,l)$ of the factor $\rho'_{kl}$ in the numerator, i.e. $i=l$ and $j=k$, if $i\neq k$ holds. Otherwise, in the case that $i=k$, no corresponding factor is multiplied in the numerator, because the appropriate entry of $\Sigma'_N$ lies on the principal diagonal. Moreover, $1-\lambda_i(A_0)\overline{\lambda_j(A_0)}=1-|\lambda_i(A_0)|^2$ holds in that case.

These circumstances are elucidated for small NPD projects with only $p=2$ and $p=3$ tasks. For $p=2$ we have

$$\Sigma'_N=\begin{pmatrix}\dfrac{1}{1-|\lambda_1|^2} & \dfrac{\rho'_{12}}{1-\lambda_1\overline{\lambda_2}}\\ \dfrac{\rho'_{12}}{1-\lambda_2\overline{\lambda_1}} & \dfrac{1}{1-|\lambda_2|^2}\end{pmatrix},$$

hence,

$$\mathrm{Det}[\Sigma'_N]=\frac{1}{(1-|\lambda_1|^2)(1-|\lambda_2|^2)}-\frac{{\rho'_{12}}^2}{(1-\lambda_2\overline{\lambda_1})(1-\lambda_1\overline{\lambda_2})}.$$

For $p=3$ we have

$$\Sigma'_N=\begin{pmatrix}\dfrac{1}{1-|\lambda_1|^2} & \dfrac{\rho'_{12}}{1-\lambda_1\overline{\lambda_2}} & \dfrac{\rho'_{13}}{1-\lambda_1\overline{\lambda_3}}\\ \dfrac{\rho'_{12}}{1-\lambda_2\overline{\lambda_1}} & \dfrac{1}{1-|\lambda_2|^2} & \dfrac{\rho'_{23}}{1-\lambda_2\overline{\lambda_3}}\\ \dfrac{\rho'_{13}}{1-\lambda_3\overline{\lambda_1}} & \dfrac{\rho'_{23}}{1-\lambda_3\overline{\lambda_2}} & \dfrac{1}{1-|\lambda_3|^2}\end{pmatrix},$$

hence,

$$
\begin{aligned}
\mathrm{Det}[\Sigma'_N] = \quad & \frac{1}{(1-|\lambda_1|^2)(1-|\lambda_2|^2)(1-|\lambda_3|^2)} \\
& - \frac{{\rho'_{23}}^2}{(1-|\lambda_1|^2)(1-\lambda_3\overline{\lambda_2})(1-\lambda_2\overline{\lambda_3})} - \frac{{\rho'_{13}}^2}{(1-|\lambda_2|^2)(1-\lambda_3\overline{\lambda_1})(1-\lambda_1\overline{\lambda_3})} \\
& - \frac{{\rho'_{12}}^2}{(1-|\lambda_3|^2)(1-\lambda_1\overline{\lambda_2})(1-\lambda_2\overline{\lambda_1})} + \frac{\rho'_{12}\rho'_{13}\rho'_{23}}{(1-\lambda_1\overline{\lambda_2})(1-\lambda_2\overline{\lambda_3})(1-\lambda_3\overline{\lambda_1})} \\
& + \frac{\rho'_{12}\rho'_{13}\rho'_{23}}{(1-\lambda_2\overline{\lambda_1})(1-\lambda_3\overline{\lambda_2})(1-\lambda_1\overline{\lambda_3})} .
\end{aligned}
$$

The results for $C'_N$ are much simpler. From eqs. 187 and 188 it follows that the numerator is the same, whereas the denominator is just 1.

For $p = 2$ we have

$$
C'_N = \begin{pmatrix} 1 & \rho'_{12} \\ \rho'_{12} & 1 \end{pmatrix},
$$

hence,

$$
\mathrm{Det}[C'_N] = 1 - {\rho'_{12}}^2.
$$

For $p = 3$ we have

$$
C'_N = \begin{pmatrix} 1 & \rho'_{12} & \rho'_{13} \\ \rho'_{12} & 1 & \rho'_{23} \\ \rho'_{13} & \rho'_{23} & 1 \end{pmatrix},
$$

hence,

$$
\mathrm{Det}[C'_N] = 1 + 2\rho'_{12}\rho'_{13}\rho'_{23} - {\rho'_{12}}^2 - {\rho'_{13}}^2 - {\rho'_{23}}^2.
$$

Now, we suppose that all eigenvalues $\lambda_i(A_0)$ are real. Under this assumption $\mathrm{Det}[\Sigma'_N \cdot {C'}_N^{-1}]$ can be written as

$$
\frac{1}{(1-{\lambda_1}^2)(1-{\lambda_2}^2)} + \frac{{\rho'_{12}}^2}{1-{\rho'_{12}}^2} \cdot \frac{(\lambda_1 - \lambda_2)^2}{(1-{\lambda_1}^2)(1-{\lambda_2}^2)(1-\lambda_1\lambda_2)^2}, \tag{196}
$$

for $p = 2$, and as

$$
\begin{aligned}
& \prod_{i=1}^{3} \frac{1}{1-{\lambda_i}^2} + \frac{1}{\mathrm{Det}[C'_N]} \left( \frac{{\rho'_{12}}^2}{1-{\lambda_3}^2} \frac{(\lambda_1 - \lambda_2)^2}{(1-{\lambda_1}^2)(1-{\lambda_2}^2)(1-\lambda_1\lambda_2)^2} \right. \\
& + \frac{{\rho'_{13}}^2}{(1-{\lambda_2}^2)} \frac{(\lambda_1 - \lambda_3)^2}{(1-{\lambda_1}^2)(1-{\lambda_3}^2)(1-\lambda_1\lambda_3)^2} \\
& + \frac{{\rho'_{23}}^2}{(1-{\lambda_1}^2)} \frac{(\lambda_2 - \lambda_3)^2}{(1-{\lambda_2}^2)(1-{\lambda_3}^2)(1-\lambda_2\lambda_3)^2} + \frac{2\rho'_{12}\rho'_{13}\rho'_{23}}{(1-{\lambda_1}^2)(1-{\lambda_2}^2)(1-{\lambda_3}^2)} \\
& \left. \cdot \frac{(1-{\lambda_1}^2)(1-{\lambda_2}^2)(1-{\lambda_3}^2) - (1-\lambda_1\lambda_2)(1-\lambda_1\lambda_3)(1-\lambda_2\lambda_3)}{(1-\lambda_1\lambda_2)(1-\lambda_1\lambda_3)(1-\lambda_2\lambda_3)} \right)
\end{aligned} \tag{197}
$$

for $p = 3$.

For $p = 2$ tasks it is obvious that the first and second summand in eq. 196 can take only values in the range [0,+∞], and for different correlations $\rho_{12}' \in [-1; 1]$ the first summand $1/\left((1-\lambda_1{}^2)(1-\lambda_2{}^2)\right)$ is a lower bound. For $p = 3$ tasks it can also be proved that the second summand in eq. 197 can only take values in the range [0,+∞] in view of the definition of the covariance matrix. The first summand is also a lower bound. Interestingly, the coefficient $\rho_{12}'{}^2/(1-\rho_{12}'{}^2)$ in eq. 196 is equivalent to Cohen´s $f^2$, which is an appropriate effect size measure to use in the context of an F-test for ANOVA or multiple regression. By convention, in the behavioral sciences effect sizes of 0.02, 0.15, and 0.35 are termed small, medium, and large, respectively (Cohen 1988). The squared product-moment correlation $\rho'^2_{12}$ can also be easily interpreted within the class of linear regression models. If an intercept is included in a linear regression model, then $\rho_{12}'{}^2$ is equivalent to the well known coefficient of determination $R^2$. The coefficient of determination provides a measure of how well future outcomes are likely to be predicted by the statistical model of interest.

Moreover, interesting questions arise from the identification of these lower bounds. The answers will improve the understanding of the unexpectedly rich dynamics that even small open organizational systems can generate. The identified lower bounds can be reached, if and only if either the noise is isotropic, that is, for the corresponding covariance matrix in the original state-space coordinates the expression $C = \{\sigma^2\} \cdot I_p$ holds (see eq. 178), or the dynamical operator $A_0$ is symmetric and the column vectors of the forcing matrix $K$ are "aligned", in the sense that $A_0 = \{c\} \cdot K$ holds ($c \in \mathbb{R}$ or $c = \mathrm{Diag}[c_i]$ in general). More details about the interrelationship between $A_0$ and $K$ were presented earlier in section 2.3.

The first question we deal with is identifying the optimal distribution of eigenvalues $\lambda_i$ that minimizes emergent complexity according to the metric $\mathrm{EMC} = \frac{1}{2}\log_2 \mathrm{Det}\left[\Sigma_N' \cdot C'^{-1}_N\right]$ subject to the constraint that the mean total work $x_{tot} \in \mathbb{R}^+$ in the project is constant. In other words, we seek project organization designs that could on average process the same amount of work with minimum complexity. A closed-form solution of the mean vector $\bar{x}$ of the accumulated work for distinct tasks in an asymptotically stable project given the initial state $x_0$ can be calculated across an infinite time interval as $\bar{x} = \left(I_p - A_0\right)^{-1} \cdot x_0$ (see section 2.2). The mean total work $x_{tot} = \mathrm{Total}[\bar{x}]$ in the project is simply the sum of the vector components (eq. 12). For two tasks, the above question can be formulated as a constrained optimization problem:

$$\min_{(a_{11},a_{12},a_{21},a_{22})} \frac{1}{2}\log_2 \mathrm{Det}\left[\frac{1}{\left(1-\left(\lambda_1\left[\begin{pmatrix} a_{11} & a_{21} \\ a_{12} & a_{22}\end{pmatrix}\right]\right)^2\right)\left(1-\left(\lambda_2\left[\begin{pmatrix} a_{11} & a_{21} \\ a_{12} & a_{22}\end{pmatrix}\right]\right)^2\right)}\right]$$

$$\text{subject to Total}\left[\begin{pmatrix} 1-a_{11} & a_{12} \\ a_{21} & 1-a_{22}\end{pmatrix}^{-1} \cdot \begin{pmatrix} x_{01} \\ x_{02}\end{pmatrix}\right] = x_{tot}$$

For three tasks, the corresponding formulation would be:

$$\min_{\left((a_{ij})_{(i,j)\in\{1,2,3\}^3}\right)} \frac{1}{2}\log_2\mathrm{Det}\left[\frac{1}{\left(1-\left(\lambda_1\left[\begin{pmatrix} a_{11} & a_{12} & a_{13} \\ a_{21} & a_{22} & a_{23} \\ a_{31} & a_{32} & a_{33}\end{pmatrix}\right]\right)^2\right)(1-(\lambda_2[\ldots])^2)(1-(\lambda_3[\ldots])^2)}\right]$$

$$\text{subject to Total}\left[\begin{pmatrix} 1-a_{11} & a_{12} & a_{13} \\ a_{21} & 1-a_{22} & a_{23} \\ a_{31} & a_{32} & 1-a_{33}\end{pmatrix}^{-1}\cdot\begin{pmatrix} x_{01} \\ x_{02} \\ x_{03}\end{pmatrix}\right] = x_{tot} \quad .$$

In these equations, $\lambda_i[.]$ represents the *i*-th eigenvalue of the argument matrix. According to section 2.2 the function Total[...] computes the sum of the components of the argument vector. To solve the constrained optimization problems, the method of Lagrange multipliers is used. Unfortunately, this strategy allows finding closed-form solutions only under additional constraints. The first additional constraint is that only two tasks can be processed. Furthermore, both tasks have to be "uncoupled" and the corresponding off-diagonal elements $a_{12} = 0$ and $a_{21} = 0$ indicate the absence of cooperative relationships. Finally, the initial state is constrained to a setting in which both tasks are 100% to be completed, that is $x_0 = [1\ 1]^{\mathrm{T}}$, and in this case the total work must be larger than 2 ($x_{tot} > 2$). Under these constraints, it follows that the eigenvalues $\lambda_1(A_0)$ and $\lambda_2(A_0)$ are equal to the autonomous task-processing rates:

$$\lambda_1\left[\begin{pmatrix} a_{11} & 0 \\ 0 & a_{22}\end{pmatrix}\right] = a_{11} \text{ and } \lambda_2\left[\begin{pmatrix} a_{11} & 0 \\ 0 & a_{22}\end{pmatrix}\right] = a_{22}.$$

The closed-form solution of the constrained optimization problem is the piecewise-defined function:

$$\mathrm{EMC}_{min} = \begin{cases} \dfrac{2\cdot\log\left(\dfrac{{x_{tot}}^2}{(x_{tot}-1)}\right)}{\log(4)} - 2 & \text{if } 2 < x_{tot} \le 2+\sqrt{2} \\ \dfrac{\log(2x_{tot}-1)}{\log(4)} - 1 & \text{if } 2+\sqrt{2} < x_{tot} \end{cases}$$

The corresponding equations for the autonomous task processing rates (alias eigenvalues) are

$$a_{11}^{min} = \lambda_1^{min} = \begin{cases} \dfrac{x_{tot}-2}{x_{tot}} & \text{if } 2 < x_{tot} \le 2+\sqrt{2} \\ \dfrac{1}{x_{tot}-1-\sqrt{2+(x_{tot}-4)x_{tot}}} & \text{if } 2+\sqrt{2} < x_{tot} \end{cases}$$

$$a_{22}^{min} = \lambda_2^{min} = \begin{cases} \dfrac{x_{tot}-2}{x_{tot}} & \text{if } 2 < x_{tot} \le 2+\sqrt{2} \\ \dfrac{1}{x_{tot}-1+\sqrt{2+(x_{tot}-4)x_{tot}}} & \text{if } 2+\sqrt{2} < x_{tot} \end{cases}$$

When we analyze the above solutions, an interesting finding is that the value $x_{tot}^{\vDash} = 2+\sqrt{2} \sim 3.414$ of the total amount of work indicates a kind of "bifurcation point" in the complexity landscape. Below that point, minimum complexity values are assigned for an even distribution of both autonomous task processing rates (or eigenvalues); above it, minimum complexity values are attained, if and only if the difference between rates is

$$a_{11}^{min} - a_{22}^{min} = \frac{2\sqrt{2 + (x_{tot} - 4)x_{tot}}}{2x_{tot} - 1}.$$

This bifurcation behavior of an open organizational system in which only two uncoupled tasks are processed was unexpected. Figure 9 shows the bifurcation point in detail.

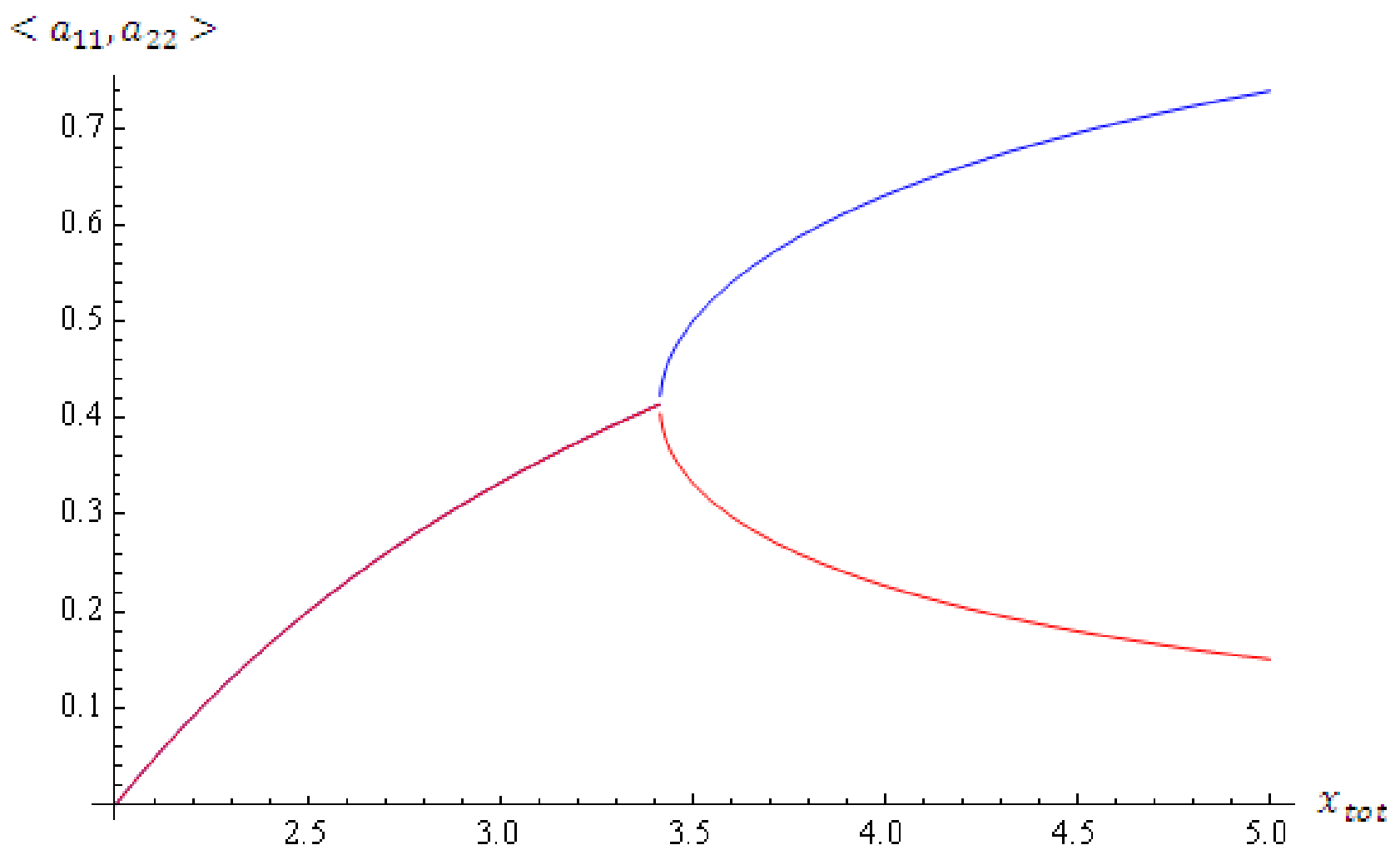

**Figure 9**

*Figure 9*. Plot of autonomous task-processing rates $a_{11}$ and $a_{22}$ leading to a minimum EMC subject to the constraint that the mean total work $x_{tot}$ in the project is constant. The underlying closed-form solution was calculated based on Lagrange multipliers. Note that the solution only holds under the assumption that the tasks are uncoupled and the initial state is $X_0 = [1,1]^T$, in which case $x_{tot}$ must be larger than 2.

### 4.1.5 Bounds on EMC

To calculate the lower bounds on EMC for an arbitrary number of tasks we can make use of Oppenheim's inequality (see Horn and Johnson 1985). Let $M$ and $N$ be positive-semidefinite matrices and let $M \circ N$ be the entry-wise product of these matrices (so-called "Hadamard product"). The Hadamard product of two positive-semidefinite matrices is again positive-semidefinite. Furthermore, if $M$ and $N$ are positive-semidefinite, then the following equality based on Oppenheim holds:

$$\text{Det}[M \circ N] \geq \left(\prod_{i=1}^{p} M_{[[i,i]]}\right) \text{Det}[N].$$

Let $M = \left(M_{[[i,j]]}\right) = \left(1/\left(1 - \lambda_i(A_0)\overline{\lambda_j(A_0)}\right)\right)$ be a Cauchy matrix $(1 \leq i, j \leq p)$. The elements along the principal diagonal of this matrix represent the "damping factor" $1 - |\lambda_i|^2$ of design mode $\phi_i$ , and the off-diagonal elements $1 - \lambda_i \overline{\lambda_j}$ are the damping factors between the interacting modes $\phi_i$ and $\phi_j$. We follow the convention that the eigenvalues are ordered in decreasing magnitude in rows. Let $N = C_N'$ be the normalized covariance matrix of the noise, as defined in eq. 188. Then the normalized covariance matrix of the signal $\Sigma_N'$ from eq. 187 can be written as the Hadamard product $\Sigma_N' = M \circ C_N'$. According to Oppenheim's inequality, the following inequality holds:

$$\text{EMC} = \frac{1}{2}\log_2\left(\frac{\text{Det}[\Sigma_N']}{\text{Det}[C_N']}\right) = \frac{1}{2}\log_2\left(\frac{\text{Det}[M \circ C_N']}{\text{Det}[C_N']}\right) \geq \frac{1}{2}\log_2\left(\frac{\left(\prod_{i=1}^{p} M_{[[i,i]]}\right)\text{Det}[C_N']}{\text{Det}[C_N']}\right)$$

$$= \frac{1}{2}\log_2\left(\prod_{i=1}^{p}\frac{1}{1-|\lambda_i|^2}\right)$$

$$= -\frac{1}{2}\sum_{i=1}^{p}\log_2(1-|\lambda_i|^2). \qquad (198)$$

The lower bound according to the above equation is equal to the closed-form solution for EMC that was obtained under the assumptions of isotropic noise ($C = \{\sigma^2\}I_p$) and $A_0$ being diagonalizable (see eq. 179). In other words, emergent complexity in NPD projects can be kept to a minimum, if the variances of the unpredictable performance fluctuations are equalized by purposeful interventions of the project manager and correlations between vector components are suppressed.

Next, because of the commutativity of the Hadamard product, it holds that

$$\mathrm{EMC} = \frac{1}{2}\log_2\left(\frac{\mathrm{Det}[\Sigma'_N]}{\mathrm{Det}[C'_N]}\right) = \frac{1}{2}\log_2\left(\frac{\mathrm{Det}[C'_N \circ M]}{\mathrm{Det}[C'_N]}\right) \geq \frac{1}{2}\log_2\left(\frac{\left(\prod_{i=1}^{p} C'_{N[[i,i]]}\right)\mathrm{Det}[M]}{\mathrm{Det}[C'_N]}\right)$$

$$= \frac{1}{2}\log_2\left(\frac{\mathrm{Det}[M]}{\mathrm{Det}[C'_N]}\right).$$

The determinant of the Cauchy matrix $M$ in the numerator can be written as (Krattenthaler 2005)

$$\mathrm{Det}[M] = \mathrm{Det}\begin{bmatrix} \frac{1}{1-|\lambda_1|^2} & \frac{1}{1-\lambda_1\overline{\lambda_2}} & \cdots \\ \frac{1}{1-\lambda_2\overline{\lambda_1}} & \frac{1}{1-|\lambda_2|^2} & \cdots \\ \vdots & \vdots & \ddots \end{bmatrix} = \frac{\prod_{i<j}^{p}(\lambda_i-\lambda_j)(\overline{\lambda_i}-\overline{\lambda_j})}{\prod_{i,j}^{p}(1-\lambda_i\overline{\lambda_j})}.$$

Hence,

$$\mathrm{EMC} = \frac{1}{2}\log_2\left(\frac{\mathrm{Det}[C'_N \circ M]}{\mathrm{Det}[C'_N]}\right)$$

$$\geq \frac{1}{2}\log_2\left(\frac{\prod_{i<j}^{p}(\lambda_i-\lambda_j)(\overline{\lambda_i}-\overline{\lambda_j})}{\prod_{i,j}^{p}(1-\lambda_i\overline{\lambda_j})\,\mathrm{Det}[C'_N]}\right)$$

$$= \frac{1}{2}\left(\sum_{i<j}^{p}\left(\log_2(\lambda_i-\lambda_j)+\log_2(\overline{\lambda_i}-\overline{\lambda_j})\right) - \sum_{i,j}^{p}\log_2(1-\lambda_i\overline{\lambda_j}) - \log_2\mathrm{Det}[C'_N]\right). \qquad (199)$$

The lower bound on the EMC in the above equation is only defined for a dynamical operator $A_0$ with distinct eigenvalues. Under this assumption, a particularly interesting property of the bound is that it includes not only the damping factors $(1-\lambda_i\overline{\lambda_j})$ inherent to the dynamical operator $A_0$ (as does the bound in eq. 198) but also the differences between eigenvalues $(\lambda_i-\lambda_j)$ and their complex conjugates $(\overline{\lambda_i}-\overline{\lambda_j})$. We can draw the conclusion that under certain circumstances, differences among effective productivity rates (represented by the $\lambda_i$'s) stimulate emergent complexity in NPD (cf. eqs. 196 and 197). Conversely, small complexity scores are assigned if the effective productivity rates are similar.

Additional analyses have shown that the lower bound defined in eq. 198 is tighter when the eigenvalues of the dynamical operator $A_0$ are of similar magnitudes. Conversely, the lower bound defined in eq. 199 comes closer to the true complexity values if the magnitudes of the eigenvalues are unevenly distributed.

Finally, it is also possible to put both upper and lower bounds on the EMC that are explicit functions of the dynamical operator $A_0$ and its dimension $p$. To find these bounds, we considered results for the determinant of the solution of the Lyapunov equation (eq. 172, cf. Mori et al. 1982). Let $\Sigma$ be the covariance matrix of the process in the steady state, and let the dominant eigenvalue of $A_0$ be less than 1 in magnitude. Then we have

$$\mathrm{Det}[\Sigma] \geq \frac{\mathrm{Det}[C]}{\left(1-(\mathrm{Det}[A_0])^{\frac{p}{2}}\right)^p}.$$

Moreover, if $A_0$ is diagonalizable and $\lambda_{max}\left[A_0^{\mathrm{T}} \cdot A_0\right] \cdot C - A_0 \cdot \Sigma \cdot A_0^{\mathrm{T}}$ is positive-semidefinite, then

$$\mathrm{Det}[\Sigma] \leq \frac{\mathrm{Det}[C]}{\left(1-\left(\lambda_{max}\left[A_0^{\mathrm{T}} \cdot A_0\right]\right)\right)^p},$$

where $\lambda_{max}\left[A_0^{\mathrm{T}} \cdot A_0\right]$ denotes the dominant eigenvalue of $A_0^{\mathrm{T}} \cdot A_0$. Based on eq. 174 we can calculate the following bounds:

$$-\frac{p}{2}\log_2\left(1-(\mathrm{Det}[A_0])^{\frac{p}{2}}\right) \leq \mathrm{EMC} \leq -\frac{p}{2}\log_2\left(1-\lambda_{max}\left[A_0^{\mathrm{T}} \cdot A_0\right]\right). \qquad (200)$$

The upper bound only holds if $A_0$ is diagonalizable and $\lambda_{max}\left[A_0^{\mathrm{T}} \cdot A_0\right] \cdot C - A_0 \cdot \Sigma \cdot A_0^{\mathrm{T}}$ is positive-semidefinite. If $C$ is diagonal, then $\lambda_{max}\left[A_0^{\mathrm{T}} \cdot A_0\right] \cdot C - A_0 \cdot \Sigma \cdot A_0^{\mathrm{T}}$ is always positive-semidefinite. Both bounds grow strictly monotonically with the dimension of the dynamical operator $A_0$. If the project is asymptotically stable, the lower bound grows linearly, and the upper bound grows slightly faster than linearly. In other words, the EMC assigns larger complexity values to projects with more tasks, if the task couplings are similar. One can also divide the measure by the dimension $p$ of the state space and compare the complexity of projects with different cardinalities.

## 4.2 Closed-form Solutions for Hidden Markov Models of Cooperative Work

### 4.2.1 Explicit Formulation

In an analogous manner to section 4.1 the EMC of a Linear Dynamical System (LDS, section 2.8) can be expressed by the continuous-type mutual information $I[.;.]$ as

$$\begin{aligned}\mathrm{EMC} &= I[Y_{-\infty}^{-1}; Y_0^{\infty}] \\ &= \int_{\mathbb{Y}^p} \cdots \int_{\mathbb{Y}^p} f[y_{-\infty}^{-1}, y_0^{\infty}] \log_2 \frac{f[y_{-\infty}^{-1}, y_0^{\infty}]}{f[y_{-\infty}^{-1}] f[y_0^{\infty}]}\, dy_{-\infty}^{-1}\, dy_0^{\infty}.\end{aligned}$$

The term $f[y_{-\infty}^{-1}]$ designates the joint *pdf* of the observable infinite one-dimensional history. $f[y_0^{\infty}]$ designates the corresponding *pdf* of the observable infinite future. Informally, the quantity $I[Y_{-\infty}^{-1}; Y_0^{\infty}]$ can be interpreted as the information that is communicated from the infinite past to the infinite future. Please note: in what follows, we will not use multiplication signs or dots. This is to conserve space and improve legibility of the formulas.

According to section 2.8 we have the simultaneous system of equations:

$$X_{t+1} = A_0 X_t + \varepsilon_t$$
$$Y_t = H X_t + \nu_t$$

with

$$\varepsilon_t = \mathcal{N}(\xi; 0_q, C)$$
$$\nu_t = \mathcal{N}(\zeta; 0_p, V).$$

It is assumed that the hidden Markov process $\{X_t\}$ is strict-sense stationary and in steady state a stable distribution $f[x_v]$ is formed. From the state-space model the following normal distributions can be deduced in steady state $t \to \infty$:

$$f[x_v] = \mathcal{N}(x_v; \mu, \Sigma)$$
$$f[x_0|x_{-1}] = \mathcal{N}(x_0; A_0 x_{-1}, C)$$
$$f[y_v|x_v] = \mathcal{N}(y_v; H x_v, V),$$

where $\mu$ and $\Sigma$ are the solutions to the steady-state conditions (eqs. 171 and 172)

$$\mu = A_0 \mu$$
$$\Sigma = A_0 \Sigma A_0^{\mathrm{T}} + C.$$

Interestingly, the Markov property holds in steady state for the observable process and we have

$$f[y_{-\infty}^{-1}, y_0^{\infty}] = f[y_{-\infty}] f[y_{-\infty+1}|y_{-\infty}] \dots f[y_{-1}|y_{-2}] f[y_0|y_{-1}] f[y_1|y_0] \dots f[y_\infty|y_{\infty-1}]$$
$$f[y_{-\infty}^{-1}] = f[y_{-\infty}] f[y_{-\infty+1}|y_{-\infty}] \dots f[y_{-1}|y_{-2}]$$
$$f[y_0^{\infty}] = f[y_0] f[y_1|y_0] \dots f[y_\infty|y_{\infty-1}].$$

Hence, the expression for the EMC reduces to

$$\begin{aligned}
I[Y_{-\infty}^{-1}; Y_0^{\infty}] &= \int_{\mathbb{Y}^p} \dots \int_{\mathbb{Y}^p} f[y_{-\infty}^{-1}, y_0^{\infty}] \log_2 \frac{f[y_0|y_{-1}]}{f[y_0]} dy_{-\infty}^{-1} dy_0^{\infty} \\
&= \int_{\mathbb{Y}^p} \dots \int_{\mathbb{Y}^p} f[y_{-\infty}^{-1}, y_0^{\infty}] \log_2 f[y_0|y_{-1}] dy_{-\infty}^{-1} dy_0^{\infty} \\
&\quad - \int_{\mathbb{Y}^p} \dots \int_{\mathbb{Y}^p} f[y_{-\infty}^{-1}, y_0^{\infty}] \log_2 f[y_0] dy_{-\infty}^{-1} dy_0^{\infty} \\
&= \int_{\mathbb{Y}^p} \int_{\mathbb{Y}^p} \log_2 f[y_0|y_{-1}]\, dy_{-1} dy_0 \left( \int_{\mathbb{Y}^p} \dots \int_{\mathbb{Y}^p} f[y_{-\infty}^{-1}, y_0^{\infty}]\, dy_{-\infty} \dots dy_{-2} dy_1 \dots dy_\infty \right) \\
&\quad - \int_{\mathbb{Y}^p} \log_2 f[y_0] dy_0 \left( \int_{\mathbb{Y}^p} \dots \int_{\mathbb{Y}^p} f[y_{-\infty}^{-1}, y_0^{\infty}]\, dy_{-\infty} \dots dy_{-1} dy_1 \dots dy_\infty \right).
\end{aligned}$$

Exploiting the relations for the marginal probability densities, we obtain:

$$I[Y_{-\infty}^{-1}; Y_0^{\infty}] = \int_{\mathbb{Y}^p} \int_{\mathbb{Y}^p} f[y_{-1}, y_0] \log_2 f[y_0|y_{-1}] dy_{-1} dy_0 - \int_{\mathbb{Y}^p} f[y_0] \log_2 f[y_0] dy_0$$

$$= \int_{\mathbb{Y}^p} f[y_{-1}] dy_{-1} \int_{\mathbb{Y}^p} f[y_0|y_{-1}] \log_2 f[y_0|y_{-1}] dy_0 - \int_{\mathbb{Y}^p} f[y_0] \log_2 f[y_0] dy_0.$$

The probability density for the observable variable $Y_t$ can then be expressed as:

$$f[y_t] = \int_{\mathbb{X}^p} f[y_t, x_t] dx_t$$

$$= \int_{\mathbb{X}^p} f[y_t|x_t] f[x_t] dx_t$$

$$= \int_{\mathbb{X}^p} \mathcal{N}(y_t; Hx_t, V) \mathcal{N}(x_t; \mu, \Sigma) dx_t.$$

In order to solve the above integral, it is useful to apply the following transformation formula for normal distributions:

$$\mathcal{N}(y; Hx, V)\mathcal{N}(x; \mu, \Sigma) = \mathcal{N}(y; H\mu, S)\mathcal{N}(x; \mu + W(y - H\mu), \Sigma - WSW^{\mathrm{T}})$$

with

$$S = H\Sigma H^{\mathrm{T}} + V \text{ and } W = \Sigma H^{\mathrm{T}} S^{-1}.$$

Hence, we obtain:

$$f[y_t] = \mathcal{N}(y_t; H\mu, H\Sigma H^{\mathrm{T}} + V).$$

For the calculation of $f[y_0|y_{-1}] = f[y_{-1}, y_0]/f[y_{-1}]$ we insert the hidden states $x_{-1}$ and $x_0$ and exploit the Markov property

$$f[y_0|y_{-1}] = \int_{\mathbb{X}^p} \int_{\mathbb{X}^p} \frac{f[y_{-1}, y_0, x_{-1}, x_0]}{f[y_{-1}]} dx_{-1} dx_0$$

$$= \int_{\mathbb{X}^p} \int_{\mathbb{X}^p} f[x_{-1}|y_{-1}] f[x_0|x_{-1}] f[y_0|x_0] dx_{-1} dx_0.$$

Because of Bayes theorem

$$f[x_{-1}|y_{-1}] = \frac{f[y_{-1}|x_{-1}] f[x_{-1}]}{f[y_{-1}]},$$

we find

$$f[y_0|y_{-1}] = \frac{1}{f[y_{-1}]} \int_{\mathbb{X}^p} \int_{\mathbb{X}^p} f[x_{-1}] f[y_{-1}|x_{-1}] f[x_0|x_{-1}] f[y_0|x_0] dx_{-1} dx_0$$

$$= \frac{1}{f[y_{-1}]} \int_{\mathbb{R}^p} \int_{\mathbb{R}^p} \mathcal{N}(x_{-1}; \mu, \Sigma) \mathcal{N}(y_{-1}; Hx_{-1}, V) \mathcal{N}(x_0; A_0 x_{-1}, C) \mathcal{N}(y_0; Hx_0, V) dx_{-1} dx_0 \quad .$$

First, we transform the first two Gaussians as:

$$\mathcal{N}(x_{-1}; \mu, \Sigma)\mathcal{N}(y_{-1}; Hx_{-1}, V) = \mathcal{N}(y_{-1}; H\mu, H\Sigma H^{\mathrm{T}} + V)\mathcal{N}(x_{-1}; \mu + W(y_{-1} - H\mu), \Sigma - WSW^{\mathrm{T}}),$$

with

$$S = H\Sigma H^{\mathrm{T}} + V \text{ and } W = \Sigma H^{\mathrm{T}} S^{-1}.$$

The first Gaussian on the right hand side cancels $f[y_{-1}]$. The second Gaussian on the right hand side together with the third Gaussian $\mathcal{N}(x_0; A_0 x_{-1}, C)$ from the previous expression for $f[y_0|y_{-1}]$ yields:

$$\begin{aligned}
&\mathcal{N}(x_0; A_0 x_{-1}, C)\mathcal{N}(x_{-1}; \mu + W(y_{-1} - H\mu), \Sigma - WSW^{\mathrm{T}}) \\
&\quad = \mathcal{N}(x_0; A_0(\mu + W(y_{-1} - H\mu)), A_0(\Sigma - WSW^{\mathrm{T}})A_0^{\mathrm{T}} + \mathrm{C})\mathcal{N}(x_{-1}; \bar{x}_{-1}, C')
\end{aligned}$$

with some inconsequential mean $\bar{x}_{-1}$ and covariance $C'$. After integration with respect to $x_{-1}$ we have:

$$f[y_0|y_{-1}] = \int_{\mathbb{R}^p} \mathcal{N}(x_0; A_0(\mu + W(y_{-1} - H\mu)), A_0(\Sigma - WSW^{\mathrm{T}})A_0^{\mathrm{T}} + C)\mathcal{N}(y_0; Hx_0, V) dx_0.$$

Again, by transforming the two Gaussians we can carry out easily the integration with respect to $x_0$ and obtain:

$$f[y_0|y_{-1}] = \mathcal{N}(y_0; HA_0(\mu + W(y_{-1} - H\mu)), H(A_0(\Sigma - WSW^T)A_0^{\mathrm{T}} + C)H^{\mathrm{T}} + V).$$

Using the fact that the differential entropy of a multivariate Gaussian distribution $\mathcal{N}(x; \mu, C)$ is given by

$$\int_{\mathbb{X}^p} \mathcal{N}(x; \mu, C) \log_2 \mathcal{N}(x; \mu, C)\, dx = \log_2 \left( (2\pi)^{p/2} \sqrt{\mathrm{Det}[C]} \right) + \frac{p}{2},$$

we can arrive at the result for the EMC:

$$I(Y_{-\infty}^{-1}; Y_0^{\infty}) = \frac{1}{2} \log_2 \mathrm{Det}[H\Sigma H^{\mathrm{T}} + V] - \frac{1}{2} \log_2 \mathrm{Det}[D], \tag{201}$$

with

$$\begin{aligned}
D &= H\left(A_0\left(\Sigma - \Sigma H^{\mathrm{T}}\left(H\Sigma H^{\mathrm{T}} + V\right)^{-\mathrm{T}} H\Sigma^{\mathrm{T}}\right)A_0^{\mathrm{T}} + C\right)H^{\mathrm{T}} + V \\
&= HA_0\left(\Sigma - \Sigma H^{\mathrm{T}}\left(H\Sigma H^{\mathrm{T}} + V\right)^{-\mathrm{T}} H\Sigma^{\mathrm{T}}\right)A_0^{\mathrm{T}} H^{\mathrm{T}} + HCH^{\mathrm{T}} + V \\
&= HA_0\Sigma A_0^{\mathrm{T}} H^{\mathrm{T}} - HA_0\Sigma H^{\mathrm{T}}\left(H\Sigma H^{\mathrm{T}} + V\right)^{-\mathrm{T}} H\Sigma^{\mathrm{T}} A_0^{\mathrm{T}} H^{\mathrm{T}} + HCH^{\mathrm{T}} + V \\
&= H\left(A_0\Sigma A_0^{\mathrm{T}} + C\right)H^{\mathrm{T}} - HA_0\Sigma H^{\mathrm{T}}\left(H\Sigma H^{\mathrm{T}} + V\right)^{-\mathrm{T}} H\Sigma^{\mathrm{T}} A_0^{\mathrm{T}} H^{\mathrm{T}} + V \\
&= H\Sigma H^{\mathrm{T}} + V - HA_0\Sigma H^{\mathrm{T}}\left(H\Sigma H^{\mathrm{T}} + V\right)^{-\mathrm{T}} H\Sigma^{\mathrm{T}} A_0^{\mathrm{T}} H^{\mathrm{T}}.
\end{aligned} \tag{202}$$

Note A: In the case of $H = I_q$ and $V = \{0\}I_q$ we obtain the same result as for the VAR model (see eq. 174).

Note B: For small covariance $V$, i.e. if the eigenvalues of $(H\Sigma H^T)^{-1}V$ lie inside the unit circle, we may expand

$$\begin{aligned}
\left(H\Sigma H^{\mathrm{T}} + V\right)^{-\mathrm{T}} &= \left(H\Sigma H^{\mathrm{T}}\right)^{-1}\left(I + \left(H\Sigma H^{\mathrm{T}}\right)^{-1} V\right)^{-1} \\
&\approx \left(H\Sigma H^{\mathrm{T}}\right)^{-1}\left(I - \left(H\Sigma H^{\mathrm{T}}\right)^{-1} V\right),
\end{aligned}$$

and arrive at an approximate expression for $D$:

$$D = H\left(A_0 H^{-1} H^{-\mathrm{T}} \Sigma^{-1} H^{-1} V H \Sigma A_0 + C\right) H^{\mathrm{T}} + V. \tag{203}$$

Assuming furthermore $V = \{\sigma_v^2\} I_q$ we obtain:

$$D = H\left(\{\sigma_v^2\} A_0 H^{-1} H^{-\mathrm{T}} A_0 + C\right) H^{\mathrm{T}} + \{\sigma_v^2\} I_q. \tag{204}$$

Following the procedure from section 4.1.2, we can also express EMC as the signal-to-noise ratio:

$$\begin{aligned}
I(Y_{-\infty}^{-1}; Y_0^{\infty}) &= -\frac{1}{2}\log_2 \mathrm{Det}\left[I_p - HA_0\Sigma H^{\mathrm{T}}\left(H\Sigma H^{\mathrm{T}} + V\right)^{-\mathrm{T}} H\Sigma^{\mathrm{T}} A_0^{\mathrm{T}} H^{\mathrm{T}}\left(H\Sigma H^{\mathrm{T}} + V\right)^{-1}\right] \\
&= \frac{1}{2}\log_2 \left(\mathrm{Det}\left[I_p - HA_0\Sigma H^{\mathrm{T}}\left(H\Sigma H^{\mathrm{T}} + V\right)^{-\mathrm{T}} H\Sigma^{\mathrm{T}} A_0^{\mathrm{T}} H^{\mathrm{T}}\left(H\Sigma H^{\mathrm{T}} + V\right)^{-1}\right]\right)^{-1} \\
&= \frac{1}{2}\log_2 \mathrm{Det}\left[\left(I_p - HA_0\Sigma H^{\mathrm{T}}\left(H\Sigma H^{\mathrm{T}} + V\right)^{-\mathrm{T}} H\Sigma^{\mathrm{T}} A_0^{\mathrm{T}} H^{\mathrm{T}}\left(H\Sigma H^{\mathrm{T}} + V\right)^{-1}\right)^{-1}\right] \\
&= \frac{1}{2}\log_2 \mathrm{Det}\left[\sum_{k=0}^{\infty}\left(HA_0\Sigma H^{\mathrm{T}}\left(H\Sigma H^{\mathrm{T}} + V\right)^{-\mathrm{T}} H\Sigma^{\mathrm{T}} A_0^{\mathrm{T}} H^{\mathrm{T}}\left(H\Sigma H^{\mathrm{T}} + V\right)^{-1}\right)^{k}\right] \\
&= \frac{1}{2}\log_2 \mathrm{Det}\left[I_p + \sum_{k=1}^{\infty}\left(HA_0\Sigma H^{\mathrm{T}}\left(H\Sigma H^{\mathrm{T}} + V\right)^{-\mathrm{T}} H\Sigma^{\mathrm{T}} A_0^{\mathrm{T}} H^{\mathrm{T}}\left(H\Sigma H^{\mathrm{T}} + V\right)^{-1}\right)^{k}\right].
\end{aligned} \tag{205}$$

The above derivation is based on the von Neumann series generated by the operator $HA_0\Sigma H^{\mathrm{T}}\left(H\Sigma H^{\mathrm{T}} + V\right)^{-\mathrm{T}} H\Sigma^{\mathrm{T}} A_0^{\mathrm{T}} H^{\mathrm{T}}\left(H\Sigma H^{\mathrm{T}} + V\right)^{-1}$. The von Neumann series generalizes the geometric series (cf. section 2.2). The infinite sum represents the signal-to-noise ratio.

In the above closed-form complexity solutions, the dependency of EMC on the covariance $C$ of the performance fluctuations is only implicit through the steady-state covariance matrix $\Sigma$. We can completely eliminate this dependency by shifting the informational structure contained in $C$ into the state transition matrix $A_0$ and the observation matrix $H$. To shift the informational structure we can utilize the linear transformations given in eqs. 99 and 100 and “whiten” the state vectors. Through the whitening the covariance of the performance fluctuations is represented by a normal distribution with location zero and identity covariance matrix $I_q$, i.e. $C' = I_q$. Hence, for the steady-state covariance $\Sigma'$ of the hidden state process (eq. 173), it holds that:

$$\begin{aligned}
\Sigma' &= \sum_{k=0}^{\infty} (A_0')^k \left({A_0'}^{\mathrm{T}}\right)^k \\
&= \left(I_q - A_0' {A_0'}^{\mathrm{T}}\right)^{-1}.
\end{aligned}$$

We have used the von Neumann solution again. The whitening operation is a linear transformation of the coordinates of the state process, and the closed-form solution from eqs. 201 and 202 is invariant under this transformation (and others). Hence, it holds that

$$I(Y_{-\infty}^{-1}; Y_0^{\infty}) = \frac{1}{2}\log_2 \mathrm{Det}\left[H'\left(I_q - A_0' {A_0'}^{\mathrm{T}}\right)^{-1} {H'}^{\mathrm{T}} + V\right] - \frac{1}{2}\log_2 \mathrm{Det}[D'],$$

with

$$D' = H'\left(I_q - A_0' {A_0'}^{\mathrm{T}}\right)^{-1} {H'}^{\mathrm{T}} + V - H' A_0'\left(I_q - A_0' {A_0'}^{\mathrm{T}}\right)^{-1} {H'}^{\mathrm{T}} \left(H' A_0'\left(I_q - A_0' {A_0'}^{\mathrm{T}}\right)^{-1} {H'}^{\mathrm{T}} + V\right)^{-\mathrm{T}}$$

$$\cdot H'\left(I_q - A_0' A_0'^{\mathrm{T}}\right)^{-\mathrm{T}} A_0'^{\mathrm{T}} H'^{\mathrm{T}}$$

$$A_0' = \Lambda_U^{-1/2} \cdot U^{\mathrm{T}} \cdot A_0 \cdot U \cdot \Lambda_{\mathrm{U}}^{1/2}$$

$$H' = U \cdot \Lambda_U^{1/2} \cdot H \,.$$

The rescaling factors $U$ and $\Lambda_U$ are given by the eigendecomposition of the covariance matrix $C$:

$$C = U \cdot \Lambda_U \cdot U^{-1}.$$

Recalling from section 2.8 that we can express the combined quantity $H\Sigma H^T + V$ more compactly as the autocovariance $C_{yy}(0)$ of the observation process (eq. 101) and the combined quantity $A_0 \Sigma H$ as the cross-covariance $G$ between hidden and observable states, the result for the EMC from eqs. 201 and 202 can now be written more compactly as

$$\begin{aligned} I(Y_{-\infty}^{-1}; Y_0^{\infty}) &= \frac{1}{2}\log_2 \mathrm{Det}[C_{YY}(0)] - \frac{1}{2}\log_2 \mathrm{Det}[C_{YY}(0) - HGC_{YY}^{-1}(0)G^{\mathrm{T}}H^{\mathrm{T}}] \\ &= \frac{1}{2}\log_2 \frac{\mathrm{Det}[C_{YY}(0)]}{\mathrm{Det}[C_{YY}(0) - HGC_{YY}^{-1}(0)G^{\mathrm{T}}H^{\mathrm{T}}]}, \end{aligned} \tag{206}$$

Because the autocovariance $C_{YY}(\tau)$ and the autocorrelation $R_{YY}(\tau)$ are equal in steady state (eq. 102), we can also express the complexity measure by the determinant ratio:

$$I(Y_{-\infty}^{-1}; Y_0^{\infty}) = \frac{1}{2}\log_2 \frac{\mathrm{Det}[R_{YY}(0)]}{\mathrm{Det}[R_{YY}(0) - HGR_{YY}^{-1}(0)G^{\mathrm{T}}H^{\mathrm{T}}]}.$$

The closed-form solution from eq. 201 in conjunction with eq. 202 allows us to develop homologous vector autoregression models of different expressiveness for the hidden Markov model. These models generate for $t \to \infty$ stochastic processes with equivalent effective measure complexity, but the state variables are completely observable and have dimensionality $p$ and not $q$. Therefore, the stochastic processes communicate the same amount of information from the infinite past to the infinite future without utilizing hidden variables and a state space of (possibly) higher dimensionality. In that sense, the homologous models reveal all correlations and structures during the observation time and do not possess any kind of crypticity (Ellison et al. 2005). We start by focusing on homologous VAR(1) models with dynamical operator $A_0^h$ and noise covariance $C^h$ that are defined over a $p$-dimensional space $\mathbb{R}^p$ of observable states $X_t^h$:

$$X_t^h = A_0^h \cdot X_{t-1}^h + \varepsilon_t^h \qquad t = 1, \ldots, T,$$

with

$$\varepsilon_t^h \sim \mathcal{N}(0_p, C^h).$$

Assuming that the noise of the homologous model is isotropic, i.e. $C^h \sim \mathcal{N}(0_p, \{\sigma_v^2\} I_p)$, we can construct a dynamical operator $A_0^h$ representing a large variety of cooperation relationships. The preferred structure of relationships must be determined in the specific application context of complexity evaluation. According to the analysis in section 4.1.1 only two constraints must be satisfied: 1) $A_0^h$ must be diagonalizable and 2) for the weighted sum of eigenvalues $\lambda_i(A_0^h)$, it must hold that (cf. eq. 179):

$$-\frac{1}{2}\sum_{i=1}^{p} \log_2\left(1 - \lambda_i(A_0^h)^2\right) = \frac{1}{2}\log_2 \mathrm{Det}[H\Sigma H^{\mathrm{T}} + V] - \frac{1}{2}\log_2 \mathrm{Det}[D]. \tag{207}$$

It is obvious that the most simple homologous model can be constructed by setting the autonomous task processing rates as diagonal elements of $A_0^h$ to the same rate $a$, i.e. $A_0^h = \mathrm{Diag}[a, \dots, a]$. For this structurally non-informative model, the corresponding stationary stochastic process communicates the same amount of information from the infinite past to the infinite future, if

$$a = \sqrt{1 - 2^{-\frac{1}{p}(\log_2 \mathrm{Det}[H\Sigma H^{\mathrm{T}} + V] - \log_2 \mathrm{Det}[D])}}.$$

The above equation also holds for homologous models with non-isotropic noise, because all vector components are processes at the same time scale. A different approach to develop a homologous VAR(1) model is to transfer the informational structure of the covariance matrix $D$ (eq. 202) of the hidden Markov model completely to the noise of the homologous model by setting $C^h \sim \mathcal{N}(0_p, D)$, and to adapt the dynamical operator $A_0^h$ accordingly. An effective way to adapt $A_0^h$ is to define it to be a symmetric matrix. The eigenvectors $\vartheta_i(A_0^h)$ of a symmetric dynamical operator are mutually orthogonal and have only real components. According to the analysis of the correlations between state vector components in the spectral basis from section 2.3, the off-diagonal elements $C_{[[i,j]]}^{h'}$ of the transformed covariance matrix are zero if the column vectors of the forcing matrix $K$ associated with $D$ (eq. 16) and the column vectors of the transformation matrix $S$ are pairwise collinear. Hence, we can adapt the dynamical operator as follows:

$$A_0^h = K \cdot \Lambda_{SK}$$
$$\Lambda_{SK} = \mathrm{Diag}[c_i],$$

where the forcing matrix is given by the eigendecomposition of $D$ as

$$D = K \cdot \Lambda_D \cdot K^{-1}.$$

The diagonal elements $c_i \in \mathbb{R}$ must be defined in such a way that the constraint of eq. 207 is satisfied. Finally, we can develop a homologous model that is defined over a one-dimensional state space. This model is termed an auto-regressive moving average (ARMA) model and is characterized by the following linear difference equation (see e.g. Puri 2010):

$$Y_t = \sum_{i=1}^{p} a_i \cdot Y_{t-i} + \sum_{j=1}^{q} b_i \cdot U_{t-j}. \qquad 208$$

The input of the model is Gaussian white noise with variance $\sigma^2 = 1$, i.e. $U \sim \mathcal{N}(0,1)$. This model is notated ARMA$(p, q)$ in the literature (please note that in this notation the variable $q$ does not denote the dimensionality of the observation vectors $Y_t$; it denotes the number of inputs $U_{t-j}$ driving the process). It is obvious that an ARMA$(p, q)$ model can be rewritten as either a VAR$(p)$ model of order $p$ (section 2.4) or an LDS$(p, 1)$ model (section 2.8) (see e.g. de Cock 2002). It is not difficult to show that for a stable and strictly minimum phase ARMA$(p, q)$ model the effective measure complexity is given by

$$\begin{aligned}\mathrm{EMC} &= \frac{1}{2}\log_2 \frac{\prod_{i,j=1}^{p,q} \left|1 - \alpha_i \bar{\beta}_j\right|}{\prod_{i,j=1}^{p} \left|1 - \alpha_i \bar{\alpha}_j\right| \prod_{i,j=1}^{q} \left|1 - \beta_i \bar{\beta}_j\right|} \\ &= \frac{1}{2}\left(\sum_{i,j}^{p,q} \log_2 \left|1 - \alpha_i \bar{\beta}_j\right| - \sum_{i,j}^{p} \log_2 \left|1 - \alpha_i \bar{\alpha}_j\right| + \sum_{i,j}^{q} \log_2 \left|1 - \beta_i \bar{\beta}_j\right|\right),\end{aligned}$$

where the variables $\alpha_1, \ldots, \alpha_p$ denote the roots of the polynomial $a(z) = z^p + a_1 z^{p-1} + \cdots + a_p$ and $\beta_1, \ldots, \beta_q$ the roots of the polynomial $b(z) = z^q + b_1 z^{q-1} + \cdots + b_q$ (see e.g. de Cock 2002). These polynomials are the results of the z-transform of the difference equation of the ARMA$(p, q)$ model. The well-known transfer function $H(z)$ from control theory is the quotient of these polynomials. Since the polynomials are real, the roots are all real or come in conjugate pairs. Hence, for the poles $\alpha_1, \ldots, \alpha_p$ and the zeros $\beta_1, \ldots, \beta_q$ of the transfer function $H(z)$ of the homologous ARMA$(p, q)$ model, it must hold that

$$\frac{1}{2}\left(\sum_{i,j}^{p,q} \log_2\left|1-\alpha_i\bar{\beta}_j\right| - \sum_{i,j}^{p} \log_2\left|1-\alpha_i\bar{\alpha}_j\right| + \sum_{i,j}^{q} \log_2\left|1-\beta_i\bar{\beta}_j\right|\right)$$
$$= \frac{1}{2}\log_2 \mathrm{Det}[H\Sigma H^{\mathrm{T}} + V] - \frac{1}{2}\log_2 \mathrm{Det}[D].$$

After considering different homologous models without hidden state variables, we return to the full LDS defined by the simultaneous system of equations from the beginning of this section. According to the theoretical analysis in section 3.2.3, the persistent mutual information that is associated with the stationary stochastic process of this model can be expressed by the continuous-type mutual information $I[.;.]$ that is communicated from the observable infinite one-dimensional past to the infinite future from time $\tau$ onward as

$$\begin{aligned}\mathrm{EMC}(\tau) &= I[Y_{-\infty}^{-1}; Y_\tau^\infty] \\ &= \int_{\mathbb{Y}^p} \cdots \int_{\mathbb{Y}^p} f[y_{-\infty}^{-1}, y_\tau^\infty] \log_2 \frac{f[y_{-\infty}^{-1}, y_\tau^\infty]}{f[y_{-\infty}^{-1}] f[y_\tau^\infty]}\, dy_{-\infty}^{-1}\, dy_\tau^\infty.\end{aligned}$$

The term $f[y_{-\infty}^{-1}]$ designates the joint *pdf* of the infinite one-dimensional past history. $f[y_\tau^\infty]$ designates the corresponding *pdf* of the infinite future history from time $\tau$ onward.

Informally, for positive lead times the quantity $I[Y_{-\infty}^{-1}; Y_\tau^\infty]$ can be interpreted as the information that is generated by the sequence of hidden states $X_{-\infty}^{-1}$, but that does not flow through the current sequence of observations $Y_0^{\tau-1}$.

By expanding the *pdf* of the state-space model from section 2.8

$$f[y_{-\infty}^{-1}, y_\tau^\infty] = \int_{\mathbb{Y}^p} \cdots \int_{\mathbb{Y}^p} f[y_{\infty}^{-1}, y_0^\infty] dy_0^{\tau-1},$$

and exploiting the Markov property, we obtain:

$$I[Y_{-\infty}^{-1}; Y_\tau^\infty] = \int_{\mathbb{Y}^p} \cdots \int_{\mathbb{Y}^p} f[y_{-\infty}^{-1}, y_\tau^\infty] \log_2 \frac{\int_{\mathbb{Y}^p} \cdots \int_{\mathbb{Y}^p} (f[y_0|y_{-1}] \ldots f[y_\tau|y_{\tau-1}]) dy_0^{\tau-1}}{f[y_\tau^\infty]} dy_{-\infty}^{-1} dy_\tau^\infty.$$

Using the Kolmogorov-Chapman equation for the Markov process, we have

$$\int_{\mathbb{Y}^p} \cdots \int_{\mathbb{Y}^p} (f[y_0|y_{-1}] \ldots f[y_\tau|y_{\tau-1}]) dy_0^{\tau-1} = f[y_\tau|y_{-1}],$$

and we obtain the result

$$
\begin{aligned}
I[Y_{-\infty}^{-1};Y_\tau^{\infty}] &= \int_{\mathbb{Y}^p}\int_{\mathbb{Y}^p}\log_2 f[y_\tau|y_{-1}]\,dy_{-1}dy_\tau\left(\int_{\mathbb{Y}^p}\cdots\int_{\mathbb{Y}^p} f[y_{-\infty}^{-1},y_\tau^{\infty}]dy_{-\infty}^{-2}dy_{\tau+1}^{\infty}\right)\\
&\quad-\int_{\mathbb{Y}^p}\log_2 f[y_\tau]\,dy_\tau\left(\int_{\mathbb{Y}^p}\cdots\int_{\mathbb{Y}^p} f[y_{-\infty}^{-1},y_\tau^{\infty}]dy_{-\infty}^{-1}dy_{\tau+1}^{\infty}\right)\\
&= \int_{\mathbb{Y}^p}\int_{\mathbb{Y}^p} f[y_{-1},y_\tau]\log_2 f[y_\tau|y_{-1}]\,dy_{-1}dy_\tau-\int_{\mathbb{Y}^p} f[y_\tau]\log_2 f[y_\tau]\,dy_\tau\\
&= \int_{\mathbb{Y}^p} f[y_{-1}]dy_{-1}\int_{\mathbb{Y}^p} f[y_\tau|y_{-1}]\log_2 f[y_\tau|y_{-1}]dy_\tau-\int_{\mathbb{Y}^p} f[y_\tau]\log_2 f[y_\tau]\,dy_\tau.
\end{aligned}
$$

For the calculation of $f[y_\tau|y_{-1}]=f[y_{-1},y_\tau]/f[y_{-1}]$ we can follow the procedure as above and use Bayes theorem to express $f[x_{-1}|y_{-1}]$. We have

$$
\begin{aligned}
f[y_\tau|y_{-1}] &= \int_{\mathbb{X}^p}\int_{\mathbb{X}^p}\frac{f[y_{-1},y_\tau,x_{-1},x_\tau]}{f[y_{-1}]}dx_{-1}dx_\tau\\
&= \int_{\mathbb{X}^p}\int_{\mathbb{X}^p} f[x_{-1}|y_{-1}]f[x_\tau|x_{-1}]f[y_\tau|x_\tau]\,dx_{-1}dx_\tau\\
&= \frac{1}{f[y_{-1}]}\int_{\mathbb{X}^p}\int_{\mathbb{X}^p} f[x_{-1}]\,f[x_{-1}|y_{-1}]f[x_\tau|x_{-1}]f[y_\tau|x_\tau]dx_{-1}dx_\tau.
\end{aligned}
$$

The only difference from the case of $\tau=0$ is in the density, $f[x_\tau|x_{-1}]$, which was previously calculated as

$$
\begin{aligned}
f[x_\tau|x_{-1}] &= \mathcal{N}(x_\tau;A_0^{\tau+1}x_{-1},C_{\tau+1})\\
C_\tau &= A_0C_{\tau-1}A_0^{\mathrm{T}}+C\\
&= \sum_{k=0}^{\tau-1}A_0^kC\left(A_0^{\mathrm{T}}\right)^k
\end{aligned}
$$

Carrying out the integrations as above, we obtain the result:

$$
I(Y_{-\infty}^{-1},Y_\tau^{\infty})=\frac{1}{2}\log_2\mathrm{Det}[H\Sigma H^{\mathrm{T}}+V]-\frac{1}{2}\log_2\mathrm{Det}[D_\tau], \tag{209}
$$

with

$$
D_\tau=H\left(A_0^{\tau+1}\left(\Sigma-\Sigma H^{\mathrm{T}}\left(H\Sigma H^{\mathrm{T}}+V\right)^{-\mathrm{T}}H\Sigma^{\mathrm{T}}\right)\left(A_0^{\mathrm{T}}\right)^{\tau+1}+C^{\tau+1}\right)H^{\mathrm{T}}+V. \tag{210}
$$

Using the combined quantity $C_{YY}(0)=H\Sigma H^T+V$ (eq. 101), the above closed-form solution can be written more compactly as

$$
I(Y_{-\infty}^{-1},Y_\tau^{\infty})=\frac{1}{2}\log_2\mathrm{Det}[C_{YY}(0)]-\frac{1}{2}\log_2\mathrm{Det}[D_\tau], \tag{211}
$$

with

$$
D_\tau=H\left(A_0^{\tau+1}\left(\Sigma-\Sigma H^{\mathrm{T}}C_{YY}^{-\mathrm{T}}H\Sigma^{\mathrm{T}}\right)\left(A_0^{\mathrm{T}}\right)^{\tau+1}+C^{\tau+1}\right)H^{\mathrm{T}}+V. \tag{212}
$$

### 4.2.2 Implicit Formulation

Interestingly, the closed-form solution of EMC that builds on Shannon´s fundamental notion of the signal-to-noise ratio (eq. 205) can also be written in a very expressive implicit form. The implicit form is easy to interpret because its independent parameters can be derived from solutions of fundamental equations. In order to derive the implicit form we work with the "forward innovation model" from section 2.8 (eqs. 106 and 107):

$$X_{t+1}^f = A_0 \cdot X_t^f + K \cdot \eta_t$$
$$Y_t = H \cdot X_t^f + \eta_t.$$

According to de Cock (2002), the mutual information of observable infinite one-dimensional past and future histories of the stationary process can be expressed as

$$I(Y_{-\infty}^{-1}; Y_0^{\infty}) = -\frac{1}{2}\log_2 \mathrm{Det}\left[I_q - \Sigma_f\left(G_z^{-1} + \Sigma^f\right)^{-1}\right]. \quad (213)$$

The covariance matrix $\Sigma^f$ is the solution of the Lyapunov equation (cf. eq. 172)

$$\Sigma^f = A_0 \Sigma^f A_0^{\mathrm{T}} + KSK^{\mathrm{T}}.$$

In the above Lyapunov equation

$$K = \left(G^f - A_0 \cdot \Sigma^f \cdot H^{\mathrm{T}}\right)\left(C_{YY}(0) - H \cdot \Sigma^f \cdot H^{\mathrm{T}}\right)^{-1}.$$

is the Kalman gain (eq. 105) and

$$S = C_{YY}(0) - H\Sigma^f H^{\mathrm{T}}.$$

is the covariance $S_{t+1|t}$ (eq. 104) of the single-source random performance fluctuations $\eta_t$ for $t \to \infty$. Hence, we have the following algebraic Ricatti equation for $\Sigma^f$ (van Overschee and de Moor 1996):

$$\Sigma^f = A_0 \Sigma^f A_0^{\mathrm{T}} + \left(G^f - A_0 \cdot \Sigma^f \cdot H^{\mathrm{T}}\right)\left(C_{YY}(0) - H \cdot \Sigma^f \cdot H^{\mathrm{T}}\right)^{-1}\left(\left(G^f\right)^{\mathrm{T}} - H \cdot \Sigma^f \cdot H^{\mathrm{T}}\right). \quad (214)$$

The additional covariance matrix $G_z$ from eq. 213 satisfies the Lyapunov equation

$$G_z = (A_0 - KH)^{\mathrm{T}} G_z (A_0 - KH) + H^{\mathrm{T}} S^{-1} H. \quad (215)$$

An important finding of de Cock (2002) is that the inverse aggregated covariance matrix $\left(G_z^{-1} + \Sigma^f\right)^{-1}$ is the solution of another Lyapunov equation

$$\bar{\Sigma}^b = \bar{A}_0 \bar{\Sigma}^b \bar{A}_0^{\mathrm{T}} + \bar{K}\bar{S}\bar{K}^{\mathrm{T}}$$
$$= A_0^{\mathrm{T}} \bar{\Sigma}^b A_0 + \bar{K}\bar{S}\bar{K}^{\mathrm{T}},$$

which is related to the backward innovation representation of the corresponding backward model (eqs. 118 and 119):

$$\bar{X}_{t-1}^b = \bar{A}_0 \cdot \bar{X}_t^b + \bar{K} \cdot \bar{\eta}_t$$
$$Y_t = \bar{H} \cdot \bar{X}_t^b + \bar{\eta}_t.$$

Substituting the Kalman gain $\bar{K}$ (eq. 121) and the noise covariance $\bar{S} = E\left[\bar{\eta}_t \bar{\eta}_t^{\mathrm{T}}\right]$ (eq. 122) in the Lyapunov equation for the backward innovation representation leads to the following algebraic Ricatti equation for the backward state covariance matrix:

$$\begin{aligned}
\bar{\Sigma}^b &= A_0^{\mathrm{T}}\bar{\Sigma}^b A_0 + \left(H^{\mathrm{T}} - A_0^{\mathrm{T}}\bar{\Sigma}^b G\right)\left(\left(H^{\mathrm{T}} - A_0^{\mathrm{T}}\bar{\Sigma}^b G\right)\left(C_{YY}(0) - G^{\mathrm{T}}\bar{\Sigma}^b G\right)^{-1}\right)^{\mathrm{T}} \\
&= A_0^{\mathrm{T}}\bar{\Sigma}^b A_0 + \left(H^{\mathrm{T}} - A_0^{\mathrm{T}}\bar{\Sigma}^b G\right)\left(C_{YY}(0) - G^{\mathrm{T}}\bar{\Sigma}^b G\right)^{-\mathrm{T}}\left(H^{\mathrm{T}} - A_0^{\mathrm{T}}\bar{\Sigma}^b G\right)^{\mathrm{T}} \\
&= A_0^{\mathrm{T}}\bar{\Sigma}^b A_0 + \left(H^{\mathrm{T}} - A_0^{\mathrm{T}}\bar{\Sigma}^b G\right)\left(C_{YY}(0) - G^{\mathrm{T}}\bar{\Sigma}^b G\right)^{-1}\left(H - G^{\mathrm{T}}\bar{\Sigma}^b A_0\right). \qquad (216)
\end{aligned}$$

Hence, the most intuitive solution is obtained (de Cock 2002):

$$\begin{aligned}
I(Y_{-\infty}^{-1}; Y_0^{\infty}) &= -\frac{1}{2}\log_2 \mathrm{Det}\left[I_q - \Sigma^f\left(G_z^{-1} + \Sigma^f\right)^{-1}\right] \\
&= -\frac{1}{2}\log_2 \mathrm{Det}\left[I_q - \Sigma^f\bar{\Sigma}^b\right]. \qquad (217)
\end{aligned}$$

According to Sylvester’s determinant theorem, this solution can equivalently be expressed based on the signal-to-noise ratio $\mathrm{SNR} = G_z \cdot \left(\left(\Sigma^f\right)^{-1}\right)^{-1} = \left(\left(\Sigma^f\right)^{-1}\right)^{-1} \cdot G_z$, and we have:

$$\begin{aligned}
I(Y_{-\infty}^{-1}; Y_0^{\infty}) &= -\frac{1}{2}\log_2 \mathrm{Det}\left[I_q - \bar{\Sigma}^b\Sigma^f\right] \\
&= \frac{1}{2}\log_2 \mathrm{Det}\left[I_q + G_z\Sigma^f\right] \\
&= \frac{1}{2}\log_2 \mathrm{Det}\left[I_q + \Sigma^f G_z\right].
\end{aligned}$$

The standard numerical approach to solve the forward Riccati eq. 214 is to solve the generalized eigenvalue problem

$$\begin{pmatrix} A_0^{\mathrm{T}} - H^{\mathrm{T}}\left(C_{YY}(0)\right)^{-1}G^f & 0 \\ -G^f\left(C_{YY}(0)\right)^{-1}\left(G^f\right)^{\mathrm{T}} & I_q \end{pmatrix}\begin{pmatrix} W_1 \\ W_2 \end{pmatrix} = \begin{pmatrix} I_q & -H^{\mathrm{T}}\left(C_{YY}(0)\right)^{-1}H \\ 0 & A_0 - G^f\left(C_{YY}(0)\right)^{-1}H \end{pmatrix}\begin{pmatrix} W_1 \\ W_2 \end{pmatrix}\Lambda$$

and compute the covariance matrix $\Sigma^f$ as

$$\Sigma^f = W_2 \cdot W_1^{-1},$$

see, e.g. van Overschee and de Moor (1996). The complementary backward Ricatti eq. 216 can be tackled by solving

$$\begin{pmatrix} A_0 - G\left(C_{YY}(0)\right)^{-1}H & 0 \\ -H^{\mathrm{T}}\left(C_{YY}(0)\right)^{-1}H & I_q \end{pmatrix}\begin{pmatrix} W_1 \\ W_2 \end{pmatrix} = \begin{pmatrix} I_q & -G\left(C_{YY}(0)\right)^{-1}H^{\mathrm{T}} \\ 0 & A_0^{\mathrm{T}} - H^{\mathrm{T}}\left(C_{YY}(0)\right)^{-1}\left(G^f\right)^{\mathrm{T}} \end{pmatrix}\begin{pmatrix} W_1 \\ W_2 \end{pmatrix}\Lambda$$

and computing the covariance matrix $\bar{\Sigma}_b$ as

$$\bar{\Sigma}^b = W_2 \cdot W_1^{-1}.$$

It is obvious that the same numerical function can be used in the preferred programming language to solve the above generalized eigenvalue problems. This function must be called for the forward Riccati equation with the argument $(A_0, H, G^f, C_{YY}(0))$, whilst for the backward Ricatti equation the argument must be $(A_0^{\mathrm{T}}, \left(G^f\right)^{\mathrm{T}}, H^{\mathrm{T}}, C_{YY}(0))$.

Similar to the canonical correlation analysis of the basic VAR(1) process in section 4.1.3, we can diagonalize the forward and backward state covariance matrices obtained by solving the algebraic Ricatti eqs. 214 and 216 simultaneously and bring them in a form called “stochastic balanced realization” (Desai and Pal 1984). A stochastic balanced representation is an innovations representation with state covariance matrix equal to the canonical correlation coefficient matrix for the sequence of observations. Let the eigendecomposition (cf. eq. 15) of the product of the state covariance matrices $\Sigma_f \bar{\Sigma}_b$ be given by the representation

$$\Sigma^f \bar{\Sigma}^b = M \Lambda_M^2 M^{-1}$$
$$\Lambda_M^2 = \mathrm{Diag}\left[\lambda_i(\Sigma^f \bar{\Sigma}^b)\right] \qquad 1 \leq i \leq q.,$$

where the eigenvector matrix $M$ is picked as

$$M = U_{\bar{\Sigma}^b} \Lambda_{\bar{\Sigma}^b}^{-1/2} U_{\Sigma^f} \Lambda_{\mathrm{M}}^{1/2},$$

the matrices $U_{\bar{\Sigma}_b}$ and $\Lambda_{\bar{\Sigma}_b}$ based on an additional eigendecomposition as

$$U_{\bar{\Sigma}^b} \Lambda_{\bar{\Sigma}^b} U_{\bar{\Sigma}^b}^{-1} = \bar{\Sigma}^b,$$

and

$$U_{\Sigma^f} \Lambda_{\mathrm{M}}^2 U_{\Sigma^f}^{-1} = \Lambda_{\bar{\Sigma}^b}^{1/2} U_{\bar{\Sigma}^b}^{-1} \Sigma^f \Lambda_{\bar{\Sigma}^b}^{1/2}.$$

Furthermore, let the forward state that is subject to the simultaneous diagonalization of the state covariance matrices be

$$X_t^d = T \cdot X_t^f$$

and the corresponding backward state be

$$\bar{X}_t^d = T^{-1} \cdot \bar{X}_{t-1}^b$$

with the coefficient of the similarity transformation

$$T = M^{\mathrm{T}},$$

then in steady state it holds for the expectations (Desai and Pal 1984) that:

$$E\left[X_t^d \left(X_t^d\right)^{\mathrm{T}}\right] = \Lambda_{\mathrm{M}} = E\left[\bar{X}_t^d \left(\bar{X}_t^d\right)^{\mathrm{T}}\right].$$

Hence, the stochastic balanced representation allows us to make the dependency of the effective measure complexity on the eigenvalues of the product of the state covariance matrices $\Sigma^f \bar{\Sigma}^b$ explicit:

$$\begin{aligned} I(Y_{-\infty}^{-1}; Y_0^{\infty}) &= -\frac{1}{2}\log_2 \mathrm{Det}\left[I_p - \Sigma^f \bar{\Sigma}^b\right] \\ &= -\frac{1}{2}\log_2 \mathrm{Det}\left[I_q - \Lambda_M^2\right] \\ &= -\frac{1}{2}\log_2 \prod_{i=1}^{q} \left(1 - \lambda_i\left(\Sigma^f \bar{\Sigma}^b\right)\right) \\ &= -\frac{1}{2}\log_2 \prod_{i=1}^{q} (1 - \rho_i{}^2) \end{aligned}$$

$$= -\frac{1}{2}\sum_{i=1}^{q} \log_2 (1 - {\rho_i}^2). \tag{218}$$

In the last line of the above equation the $\rho_i$´s represent the canonical correlations, which were already introduced in section 4.1.3 (cf. eq. 193) to analyze emergent complexity based on a reduced number of independent parameters. In other words, the eigenvalues of $\Sigma^f \bar{\Sigma}^b$ are just the squares of the canonical correlation coefficients between the canonical variates. However, it is important to note that in contrast to section 4.1.3 the infinite, one-dimensional random sequences representing the past ($X_{-\infty}^{-1}$) and future ($X_0^{\infty}$) of the hidden state process are not the subject of the canonical correlation analysis, but rather the canonical correlations between the pair ($Y_{-\infty}^{-1}, Y_0^{\infty}$) of past and future histories of the observation process are considered to evaluate complexity explicitly. Due to the potentially higher dimensionality of the state space of the hidden state process ($q > p$), all $q$ complexity-shaping summands $\log_2(1 - {\rho_i}^2)$ that can give rise to correlations between observations of the project state must therefore be considered. The reduced dimension of the observation process is usually not sufficient, because apart from in organizationally retarded cases not only the $p$ but also the $q$ leading canonical correlations are non-zero. The observation process is not necessarily Markovian and therefore the amount of information that the past provides about the future usually cannot be "stored" in the $p$-dimensional present state. However, because of strict-sense stationarity of the state process, all $\rho_i$'s are less than one. The canonical correlations $\rho_i$'s should not be confused with the ordinary correlations $\rho_{ij}$ and $\rho'_{ij}$, which were introduced in section 2.

As an alternative to the use of the stochastic balanced representation of Pal and Desai (1984), a minimum phase balancing based on the scheme of McGinnie (1994) could be carried out. The minimum phase balancing scheme allows us to find a forward innovation form of the LDS model in which the state covariance matrix $\Sigma^f$ (eq. 214) and the covariance matrix $G_z$ (eq. 215) are equal and diagonal. Let

$$\Lambda_P = \mathrm{Diag}[\sigma_i] \qquad 1 \le i \le q$$

be this diagonal matrix and $\sigma_i$ the minimum phase singular values of the dynamical system. Under these circumstances, we simply have

$$I(Y_{-\infty}^{-1}; Y_0^{\infty}) = -\frac{1}{2}\sum_{i=1}^{q} \log_2 (1 - {\sigma_i}^2). \tag{219}$$

# 5 VALIDITY ANALYSIS OF CLOSED-FORM SOLUTIONS IN CONJUNCTION WITH MATHEMATICAL MODELS OF COOPERATIVE WORK

Finally, the internal validity of the obtained closed-form solutions for the effective measure complexity (EMC) in conjunction with the mathematical models of cooperative work in NPD projects is analyzed. In order to analyze the internal validity we manipulate different independent variables (such as differences in productivity between developers, and in release periods between information about geometrical/topological entities) to see what effect it has on emergent complexity. The analysis is based on simple parameter studies and additional Monte Carlo experiments. Monte Carlo experiments are a special class of algorithms that rely on repeated random sampling of the work processes to compute their results. The repeated random sampling was carried out within a self-developed simulation environment. We will present and discuss the results of two validation studies with different objectives and different mathematical models of cooperative work.

## 5.1 Optimization of Project Organization Design

The objective of the first study is to design the project organization of an NPD project subjected to concurrent engineering for minimal emergent complexity. To evaluate emergent complexity the information-theoretic metric EMC is used in the spectral basis and different settings of task processing that can be represented by the basic VAR(1) model are considered. To simplify the calculations we developed efficient numerical functions based on the most expressive closed-form solution from eq. 190. Organizational optimization based on a formal complexity metric in conjunction with mathematical models of cooperative work is an application area that is particularly interesting, because complex sociotechnical systems can be purposefully designed, and established management principles and heuristics can be objectively evaluated. Especially in NPD projects requiring intensive cooperation, the classical principles and heuristics (e.g. constructing self-contained systems, Peters 1991; striving for decoupled design with minimum information content, Suh 2005; etc.) can fall short because they focus on the formalized design problem and product and tend to underestimate the effects of the cooperative problem solving process. As shown in the previous sections, the iterative and closely interacting work processes can induce unexpected performance variability and generate effects that cannot be trivially reduced to singular properties of the constituent tasks. Instead, these effects emerge as a result of multiple interactions and can lead to critical phenomena of emergent complexity such as the cited “design churns” (Yassine et al. 2003) or “problem-solving oscillations” (Mihm et al. 2003; Mihm and Loch 2006). Moreover, from a theoretical point of view, it is interesting to analyze whether EMC is not only valid for stochastic processes in the steady state but can also asses the “preasymptotic” behavior of project dynamics. It is obvious that different projects can have different preasymptotics, according to the speed and kind of convergence to that asymptote. Some properties that hold in the preasymptote of a complex project can be significantly different from those that take place in the long run, and we want to investigate in the following sections whether the relevant features, in terms of effort, are captured by the complexity metric. Moreover, a nonnegligible portion of NPD projects in industry show divergent work remaining that does not have an asymptote at all. Although divergent behavior of projects is critical from a practical point of view, it can be predicted easily within the framework of the developed theory of project dynamics and need not be analyzed further. This is because EMC simply assigns infinite complexity values to divergent projects as one would expect. For instance, it is not difficult to see that the infinite sum in eq. 175 diverges if the dominant eigenvalue $\lambda_1[A_0]$ of WTM $A_0$ has a magnitude larger than 1. Hence, the equation $|\lambda_1[A_0]| = 1$ denotes the bound between convergent and divergent projects under the given boundary conditions.

### 5.1.1 Unconstrained Optimization

We start the studies on optimizing project organization design by formulating an unconstrained optimization problem and solving it through a complete enumeration of organization designs satisfying certain boundary conditions. In a second step, a constrained optimization problem is formulated and solved by applying the same principle. The constraint is that the expected total work $x_{tot}$ according to eq. 12 is constant among the experimental conditions.

#### 5.1.1.1 Methods

The developed objective function in our first study quantifies the complexity of a given organization design under the dynamic regime of the introduced state equations (eq. 3 for original state space coordinates and eq. 24 for spectral basis). We seek to minimize emergent complexity by systematically choosing the optimal project organization design from within an allowed set. The elements of the set are distinct project organization designs that satisfy boundary conditions on productivity, cooperation relationships and performance variability. The set is complete in the sense that valid alternative organization designs with different asymptotic behavior do not exist. To simplify the problem formulation, the elements of the set are represented by WTMs and the corresponding covariance matrices. The covariance matrices are simply linear functions the WTMs.

In the first optimization study, we consider a CE project that involves different development teams. We focus on three small CE teams in the project whose work is coordinated by a system-integration engineer. Each CE team has three members, with each team member $i$ processing one development task $i$ with an autonomous task processing rate $a_{ii}$. The teams work on a component design level. A work transformation from one time step to the next represents one week of development. The vector components of the state variable $X_t$ represent the relative number of open issues that need to be resolved before final design release. The tasks in each team are "fully coupled" with respect to the components to be designed, and the corresponding off-diagonal elements $a_{ij}(i \neq j)$ of the WTM indicate a symmetric intensity of cooperative relationships that is encoded by the independent parameter $f_1 > 0$. To avoid additional reinforcement loops, it is assumed that the three CE teams are not directly cooperating. The average task processing rate of the developers is represented by the independent parameter $a \in [0; 1]$. The individual task processing rates must not be equal but can vary around the mean by an offset $\Delta a > 0$. There are three distinct productivity levels: 1) the most productive developers were able to process their tasks at rate $a_{ii} = a - \Delta a$; 2) the least productive developers processed their tasks at rate $a_{ii} = a + \Delta a$; and 3) averagely productive developers processed their tasks at rate $a_{ii} = a$. Because the three CE teams are not directly cooperating, boundary-spanning activities have to be coordinated by a 10th system-integration engineer ($i = 10$) who exchanged information directly with all nine developers. The productivity of this engineer is average, and $a_{10,10} = a$. The additional independent parameters $0 < f_2 \ll a$ and $0 < f_3 \ll a$ represent the strength of the forward and backward informational couplings between the nine developers and the system-integration engineer. In real development projects, for instance in the German automotive industry, the system-integration engineer is a member of a superordinate system-level team coordinating the development efforts on large scale (e.g. powertrain), but for the sake of simplicity, we ignore this additional management hierarchy in the following. We also do not consider other technical or organizational interfaces between teams. In addition to the cited boundary conditions holding on individual and team levels, we assumed that the mean task processing rate $\bar{a}$ of all individuals in the entire project is $a$. We also assume that the project is asymptotically stable and that the means converge to the fix point of no remaining work for all tasks. In order to guarantee asymptotic stability, the values of the independent parameters $a$, $\Delta a$, $f_1$, $f_2$ and $f_3$ must be carefully chosen, so that for all feasible project organization designs the dominant eigenvalue $\lambda_1[.]$ of the corresponding WTM has a magnitude smaller than 1. By doing so, only finite complexity values are assigned.

The optimization of project organization design aims to assign team members with different productivity levels ($a - \Delta a, a + \Delta a$ or $a$) to the three CE teams such that the emergent complexity in the sense of the EMC metric can be kept to a minimum. Under the given boundary conditions a total of 40320 assignments of team members can be distinguished. However, due the symmetry of cooperation relationships within teams, these assignments can be reduced to eight essential assignments and therefore the allowed set consists of only eight distinct work transformation matrices (WTMs). The additional assignments are just permutations of the eight basic WTMs and therefore are not relevant for evaluating complexity. The eight distinct WTMs can be ordered according to the total variance of the productivity rates over all three teams. In the following we also term the total variance the "diversity" of the project organization design, because it represents the accumulated deviation of the individual productivity rates from the mean rate $a$. Eq. 220 shows the first WTM $A_{01}$ from the allowed set, where the total variance of productivity rates is maximal. This is due to the fact that the mean task-processing rate $a$ holds not only for the entire project but also on the level of the three CE teams. For each team the variance of productivity rates is $\Delta a^2$. Hence, the total variance is $3\Delta a^2$.

team 1 team 2 team 3

$$A_{01} = \begin{pmatrix} a-\Delta a & f_1 & f_1 & 0 & 0 & 0 & 0 & 0 & 0 & f_3 \\ f_1 & a & f_1 & 0 & 0 & 0 & 0 & 0 & 0 & f_3 \\ f_1 & f_1 & a+\Delta a & 0 & 0 & 0 & 0 & 0 & 0 & f_3 \\ 0 & 0 & 0 & a-\Delta a & f_1 & f_1 & 0 & 0 & 0 & f_3 \\ 0 & 0 & 0 & f_1 & a & f_1 & 0 & 0 & 0 & f_3 \\ 0 & 0 & 0 & f_1 & f_1 & a+\Delta a & 0 & 0 & 0 & f_3 \\ 0 & 0 & 0 & 0 & 0 & 0 & a-\Delta a & f_1 & f_1 & f_3 \\ 0 & 0 & 0 & 0 & 0 & 0 & f_1 & a & f_1 & f_3 \\ 0 & 0 & 0 & 0 & 0 & 0 & f_1 & f_1 & a+\Delta a & f_3 \\ f_2 & f_2 & f_2 & f_2 & f_2 & f_2 & f_2 & f_2 & f_2 & a \end{pmatrix} \quad (220)$$

Eq. 221 shows the WTM $A_{08}$ as the last element of the allowed set.

team 1 team 2 team 3

$$A_{08} = \begin{pmatrix} a-\Delta a & f_1 & f_1 & 0 & 0 & 0 & 0 & 0 & 0 & f_3 \\ f_1 & a-\Delta a & f_1 & 0 & 0 & 0 & 0 & 0 & 0 & f_3 \\ f_1 & f_1 & a-\Delta a & 0 & 0 & 0 & 0 & 0 & 0 & f_3 \\ 0 & 0 & 0 & a & f_1 & f_1 & 0 & 0 & 0 & f_3 \\ 0 & 0 & 0 & f_1 & a & f_1 & 0 & 0 & 0 & f_3 \\ 0 & 0 & 0 & f_1 & f_1 & a & 0 & 0 & 0 & f_3 \\ 0 & 0 & 0 & 0 & 0 & 0 & a+\Delta a & f_1 & f_1 & f_3 \\ 0 & 0 & 0 & 0 & 0 & 0 & f_1 & a+\Delta a & f_1 & f_3 \\ 0 & 0 & 0 & 0 & 0 & 0 & f_1 & f_1 & a+\Delta a & f_3 \\ f_2 & f_2 & f_2 & f_2 & f_2 & f_2 & f_2 & f_2 & f_2 & a \end{pmatrix} \quad (221)$$

WTM $A_{08}$ represents an organization design with zero productivity variance on team level and therefore also zero total variance. It is obvious that this design has minimum diversity. In terms of human-centered organization design and management we have an extreme kind of "selective" team organization because CE team 1 only includes team members with maximum productivity, whilst team 3 consist only of people with low productivity. To conserve space we do not show all eight distinct WTMs explicitly but present the essential properties of the team organization in parametric form in Table 1.

Table 1. Overview of the eight distinct assignments of team members with different productivity levels ($a - \Delta a, a + \Delta a$ $or$ $a$) to the three CE teams. The WTMs $A_{01}$ and $A_{08}$ representing project organization designs with maximum total variance $3\Delta a^2$ and zero total variance of autonomous productivity rates are shown explicitly in eqs. 220 and 221. The mean productivity rate over all three CE teams is always $a$.

| WTM | mean productivity | | | variance productivity | | |
|---|---|---|---|---|---|---|
| | team 1 | team 2 | team 3 | team 1 | team 2 | team 3 |
| $A_{01}$ | $a$ | $a$ | $a$ | $\Delta a^2$ | $\Delta a^2$ | $\Delta a^2$ |
| $A_{02}$ | $a$ | $a - 1/3\,\Delta a$ | $a + 1/3\,\Delta a$ | $\Delta a^2$ | $4/3\,\Delta a^2$ | $1/3\,\Delta a^2$ |
| $A_{03}$ | $a - 1/3\,\Delta a$ | a | $a + 1/3\,\Delta a$ | $4/3\,\Delta a^2$ | 0 | $4/3\,\Delta a^2$ |
| $A_{04}$ | a | $a - 1/3\,\Delta a$ | $a - 1/3\,\Delta a$ | $\Delta a^2$ | $1/3\,\Delta a^2$ | $4/3\,\Delta a^2$ |
| $A_{05}$ | $a + 1/3\,\Delta a$ | $a + 1/3\,\Delta a$ | $a - 2/3\,\Delta a$ | $4/3\,\Delta a^2$ | $1/3\,\Delta a^2$ | $1/3\,\Delta a^2$ |
| $A_{06}$ | $a - 2/3\,\Delta a$ | $a$ | $a + 2/3\,\Delta a$ | $1/3\,\Delta a^2$ | $\Delta a^2$ | $1/3\,\Delta a^2$ |
| $A_{07}$ | $a - 2/3\,\Delta a$ | $a - 1/3\,\Delta a$ | $a + \Delta a$ | $1/3\,\Delta a^2$ | $1/3\,\Delta a^2$ | 0 |
| $A_{08}$ | $a - \Delta a$ | $a$ | $a + \Delta a$ | 0 | 0 | 0 |

For all eight project organization designs in Table 1, we assumed that the standard deviation $c_{ii}$ of performance fluctuations (eq. 6) influencing task $i$ in the project is proportional to the task processing rate $a_{ii}$ (i $= 1, \ldots, 10$): the faster a task is processed, the less it is perturbed, and the smaller the standard deviation ($\sqrt{c_{ii}} \sim a_{ii}$). The $c_{ii}$'s are the elements along the principal diagonal of covariance matrix $C_{i \in \{1,\ldots,8\}}$. The proportionality constant is $r = 0.02$. We also assumed that correlations between the performance variability of tasks do not exist and that the covariance matrices are diagonal. For the WTMs $A_{01}$ and $A_{08}$ we therefore have the following representations:

$$C_1 = \{r^2\} \begin{pmatrix} a-\Delta a & 0 & 0 & 0 & 0 & 0 & 0 & 0 & 0 & 0 \\ 0 & a & 0 & 0 & 0 & 0 & 0 & 0 & 0 & 0 \\ 0 & 0 & a-\Delta a & 0 & 0 & 0 & 0 & 0 & 0 & 0 \\ 0 & 0 & 0 & a-\Delta a & 0 & 0 & 0 & 0 & 0 & 0 \\ 0 & 0 & 0 & 0 & a & 0 & 0 & 0 & 0 & 0 \\ 0 & 0 & 0 & 0 & 0 & a+\Delta a & 0 & 0 & 0 & 0 \\ 0 & 0 & 0 & 0 & 0 & 0 & a-\Delta a & 0 & 0 & 0 \\ 0 & 0 & 0 & 0 & 0 & 0 & 0 & a & 0 & 0 \\ 0 & 0 & 0 & 0 & 0 & 0 & 0 & 0 & a+\Delta a & 0 \\ 0 & 0 & 0 & 0 & 0 & 0 & 0 & 0 & 0 & a \end{pmatrix} \tag{222}$$

$$C_8 = \{r^2\} \begin{pmatrix} a-\Delta a & 0 & 0 & 0 & 0 & 0 & 0 & 0 & 0 & 0 \\ 0 & a-\Delta a & 0 & 0 & 0 & 0 & 0 & 0 & 0 & 0 \\ 0 & 0 & a-\Delta a & 0 & 0 & 0 & 0 & 0 & 0 & 0 \\ 0 & 0 & 0 & a & 0 & 0 & 0 & 0 & 0 & 0 \\ 0 & 0 & 0 & 0 & a & 0 & 0 & 0 & 0 & 0 \\ 0 & 0 & 0 & 0 & 0 & a & 0 & 0 & 0 & 0 \\ 0 & 0 & 0 & 0 & 0 & 0 & a+\Delta a & 0 & 0 & 0 \\ 0 & 0 & 0 & 0 & 0 & 0 & 0 & a+\Delta a & 0 & 0 \\ 0 & 0 & 0 & 0 & 0 & 0 & 0 & 0 & a+\Delta a & 0 \\ 0 & 0 & 0 & 0 & 0 & 0 & 0 & 0 & 0 & a \end{pmatrix}. \tag{223}$$

We assumed that all ten parallel tasks were initially 100% incomplete and the initial state is

$$x_0 = \begin{pmatrix} 1 \\ 1 \\ 1 \\ 1 \\ 1 \\ 1 \\ 1 \\ 1 \\ 1 \\ 1 \end{pmatrix}. \tag{224}$$

According to the paradigm of lean management, it is hypothesized that for non-negligible productivity differences between the developers, “productivity leveling” at the team level minimizes complexity. Productivity leveling at the team level means that in each of the three teams, members with high productivity ($a - \Delta a$), low productivity ($a + \Delta a$) and average productivity ($a$) directly cooperate and that the average task processing rate $a$ does not only hold for the whole project but also for the three CE teams. Such an assignment was shown in WTM $A_{01}$ (eq. 220). Productivity leveling is a self-developed concept that generalizes the popular concept of production leveling (see e.g. Liker 2004) to NPD projects and knowledge-based service systems. Production leveling, also known as production smoothing, is a method for improving efficiency. It was vital to developing production efficiency in the Toyota Production System and lean production (Liker 2004). The goal is to produce parts, components and modules at a constant rate so that further processing and assembly can also be carried out at a constant rate with small variance. If a later process step varies its output in terms of timing and quality, the variance of these fluctuations will increase as it moves up the line towards the earlier processes. This phenomenon is known as demand amplification and it can also spill over into the complete supply chain, leading to the well-known bullwhip effect (see e.g. Sterman 2000). Productivity leveling also aims to improve efficiency in an organization. The general idea, however, is not to process the tasks at a constant rate, which is unfeasible in complex work systems, but rather to utilize techniques of diversity management and to find optimal (or near optimal) assignments of team members with different productivity and competency levels. By doing so, the project work can be done effectively and efficiently without design churns or other critical emergent phenomena characteristic of complex sociotechnical systems. It also makes it possible to keep developing human knowledge, skills and abilities. Productivity leveling is not a demand-driven technique in the way that production leveling is. Rather, it is a holistic approach to designing interactions between humans, tasks and products/services that considers performance fluctuations as an opportunity to innovate and learn. It aims to increase awareness of emergent phenomena characteristic in open organizational systems and to leverage from them the greatest advantage for the individuals an work teams. In the light of this concept, we can reformulate the above hypothesis and posit that maximum productivity diversity in teams leads to minimum emergent complexity. In the framework of the developed theory and models of cooperative work, both aspects are just two sides of the same coin.

To verify these complementary hypotheses, all distinct eight assignments of team members to the three development teams were analyzed. For each assignment the value of the complexity measure EMC was calculated on the basis of eq. 190. The base set of independent parameters was $\theta_1 = [a = 0.9 \quad f_1 = 0.01 \quad f_2 = 0.01 \quad f_3 = 0.005]$. The productivity offset $\Delta a$ was varied systematically on levels $\Delta a_1 = 0.001$ (small difference) and $\Delta a_2 = 0.01$ (large difference). The small productivity difference served as the baseline condition. In addition to EMC as an innovative information-theory key performance indicator (KPI) in project management, we computed the classic KPIs "project duration" and "total work" based on a sample of 10000 independent replications for all valid assignments. Clearly, the larger the mean project duration $\bar{T}$ or total work $\bar{x}_{tot}$ and the corresponding standard deviations, the lower the performance under the given organizational boundary conditions. According to section 2, the total work $x_{tot}$ represents the total effort involved in completing the project deliverables. Both classic KPIs can be determined easily in Monte Carlo experiments by counting the time and effort needed to process all ten tasks, until the cited stopping criterion is met. The time units are [weeks], the effort units are [work measurement units], abbreviated as [wmu]. The [wmu] refer to the units of the vector components of the state variable $X_t$ and therefore represent the relative number of open issues that need to be resolved before final design release. We also calculated the expected total work $x_{tot}$ analytically according to eq. 61. The expected total work is by definition accumulated over an infinite past history and therefore does not take the stopping criterion of the Monte Carlo experiments into account. The stopping criterion for the simulated projects was that a maximum of 5% of the work remained for all tasks.

The Mathematica software package from Wolfram Research was used to carry out the Monte Carlo experiments and to compute the dependent variables analytically.

#### 5.1.2.1 Results and Discussion

Figure 10 shows a typical run of the Monte Carlo simulation for WTM $A_{01}$ (eq. 220) and for parameter vector $\theta_1 = [a = 0.9 \quad f_1 = 0.04 \quad f_2 = 0.01 \quad f_3 = 0.005 \quad \Delta a = 0.01]$. Even though the chosen productivity offset $\Delta a_2 = 0.01$ is large in this run and the intensity of cooperative relationships ($f_1 = 0.04$) is high, the time series of work remaining are smooth and do not show heavy performance fluctuations around means. Figure 11 shows the corresponding result for WTM $A_{08}$ according to eq. 221 and the same parameter vector. It is obvious that such an "unbalanced" organization design leads to larger fluctuations and most importantly to additional correlations between work processes on team level.

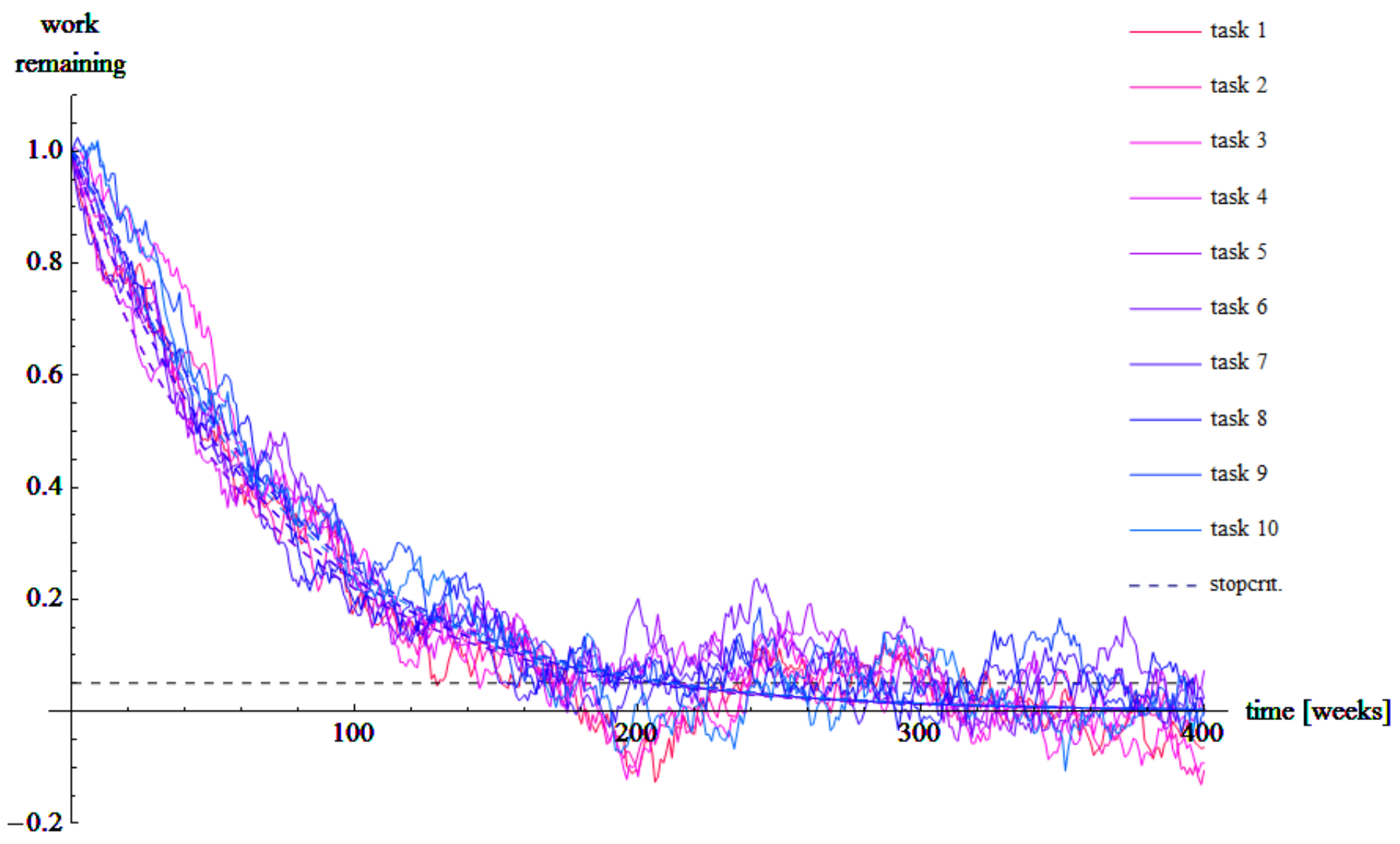

**Figure 10**

*Figure 10*. List plot of work remaining in a simulated NPD project, in which the mean task-processing rate $a$ holds not only for the entire project but also on the level of the three CE teams (see WTM $A_{01}$ in eq. 220). The simultaneous processing of all ten development tasks is shown. The data are based on a single run of the Monte Carlo experiment. The plot also shows means of simulated time series of task processing as dashed curves. The Monte Carlo experiment was based on state eq. 4. The parameters were $a = 0.9$, $f_1$ =0.04, $f_2 = 0.01$, $f_3 = 0.005$ and $\Delta a = 0.01$. The stopping criterion of 5% is marked by a dashed line at the bottom of the plot.

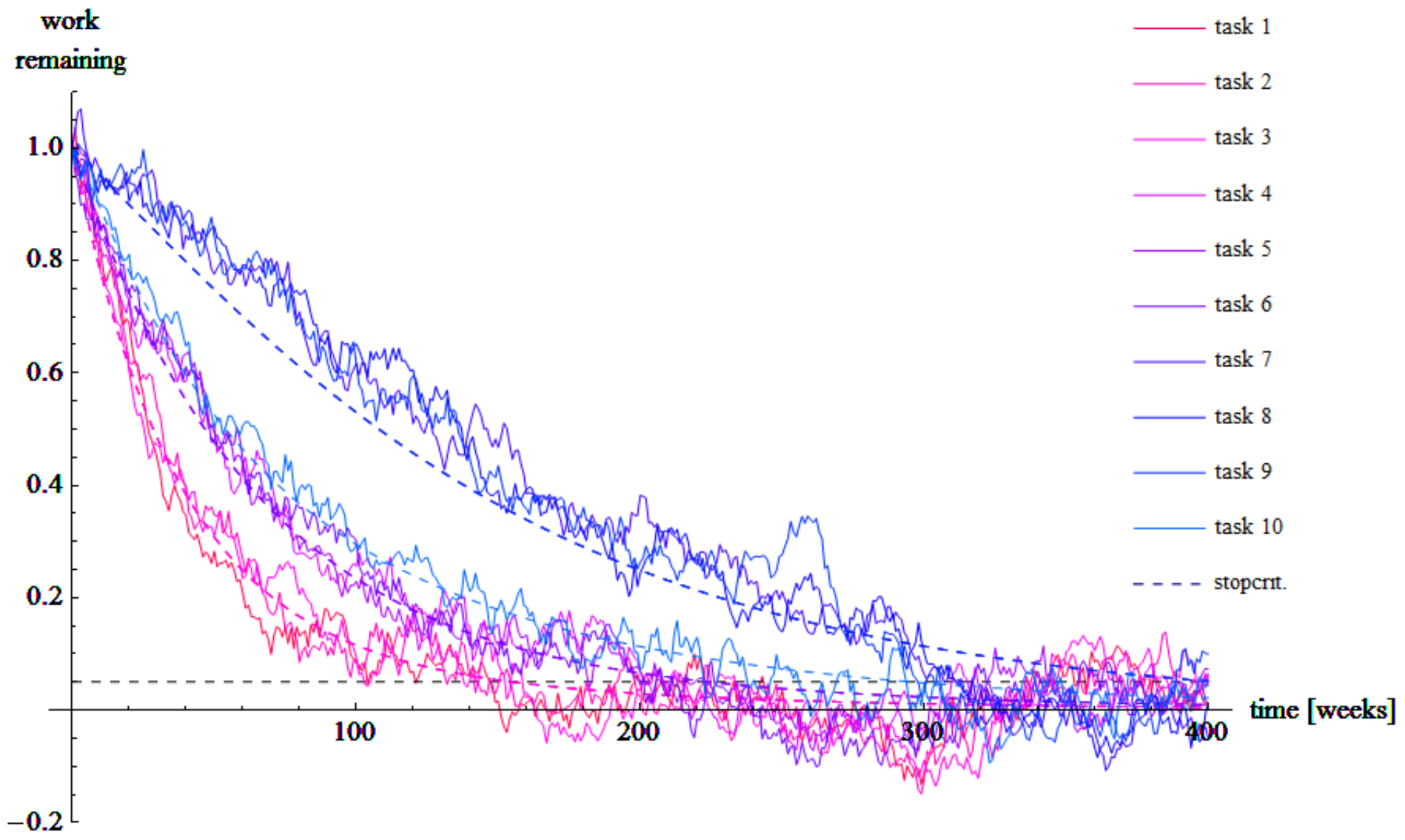

**Figure 11**

*Figure 11*. List plot of work remaining in a simulated NPD project, in which the mean task-processing rate $a$ holds only for the team 2 but but not teams 1 and 3 (see WTM $A_{08}$ in eq. 221). The simultaneous processing of all ten development tasks is shown. The data are based on a single run of the Monte Carlo experiment. The plot also shows means of simulated time series of task processing as dashed curves. The Monte Carlo experiment was based on state eq. 4. The parameters were $a = 0.9$, $f_1$ =0.04, $f_2 = 0.01$, $f_3 = 0.005$ and $\Delta a = 0.01$. The stopping criterion of 5% is marked by a dashed line at the bottom of the plot.

The analytical analyses show that for a small productivity offset ($\Delta a_1 = 0.001$) but high intensity of cooperative relationships ($f_1$ = 0.04), the organization design has little influence on complexity. The lowest complexity value under these conditions is $\mathrm{EMC}(\mathrm{A}_{01}) = 14.205$ and the largest value is $\mathrm{EMC}(\mathrm{A}_{01}) = 14.2068$. The corresponding expected total work is $x_{tot}(\mathrm{A}_{01}) = 680.851$ and $x_{tot}(\mathrm{A}_{08}) = 682.092$. The maximum difference in the complexity variable EMC among the valid assignments is only 0.00163. Surprisingly, this holds, although the intensity of cooperative relationships is close to the bound of project divergence. Interestingly, in case of such a small productivity offset the Monte Carlo experiments show that the mean project duration differs by only 0.65% among the eight valid assignments. The shortest mean project duration $\bar{T}(\mathrm{A}_{01})$ is 213.019 [weeks]. It is obtained for WTM $\mathrm{A}_{01}$, which was assigned the minimum complexity value. The standard deviation is 36.50 [weeks]. The largest mean project duration is $\bar{T}(\mathrm{A}_{08})$ =213.73 [weeks]. As expected, it is obtained for WTM $\mathrm{A}_{08}$ representing a project with maximum complexity. The standard deviation is 37.05 [weeks]. If the project manager is able to “level out” the productivity of all team members with such a small offset and if the diversity is virtually zero, the results show that the project organization design has little effect. This finding holds for all KPIs and arbitrary mean task processing rates $a$, because all tasks are processed on very similar time scales. However, when the productivity offset is increased to $\Delta a_2 = 0.01$ —ceteris paribus— the complexity differences among the eight distinct assignments of team members grow significantly. The most important finding from the analysis of these assignments is that an assignment with leveled productivity at the team level and therefore maximum diversity of productivity within teams leads to minimal complexity and it holds that $\mathrm{EMC}(\mathrm{A}_{01}) = 14.266$. This supports the cited “productivity leveling” hypothesis and the complementary “productivity diversity” hypothesis. The expected total work is $x_{tot}(\mathrm{A}_{01}) = 701.939$. The histogram of the project duration that was computed on the basis of a sample of 10000 simulated projects with minimal complexity is shown in Figure 12. The sample mean is $\bar{T}(A_{01}) =$ 220.435 [weeks]. The standard deviation is 38.45 [weeks]. The corresponding histogram of the total work $x_{tot}$ in the simulated projects is shown in Figure 13. The sample mean is $\bar{x}_{tot}(A_{01}) =$ 675.888 [weeks] and the standard deviation is 55.988 [weeks].

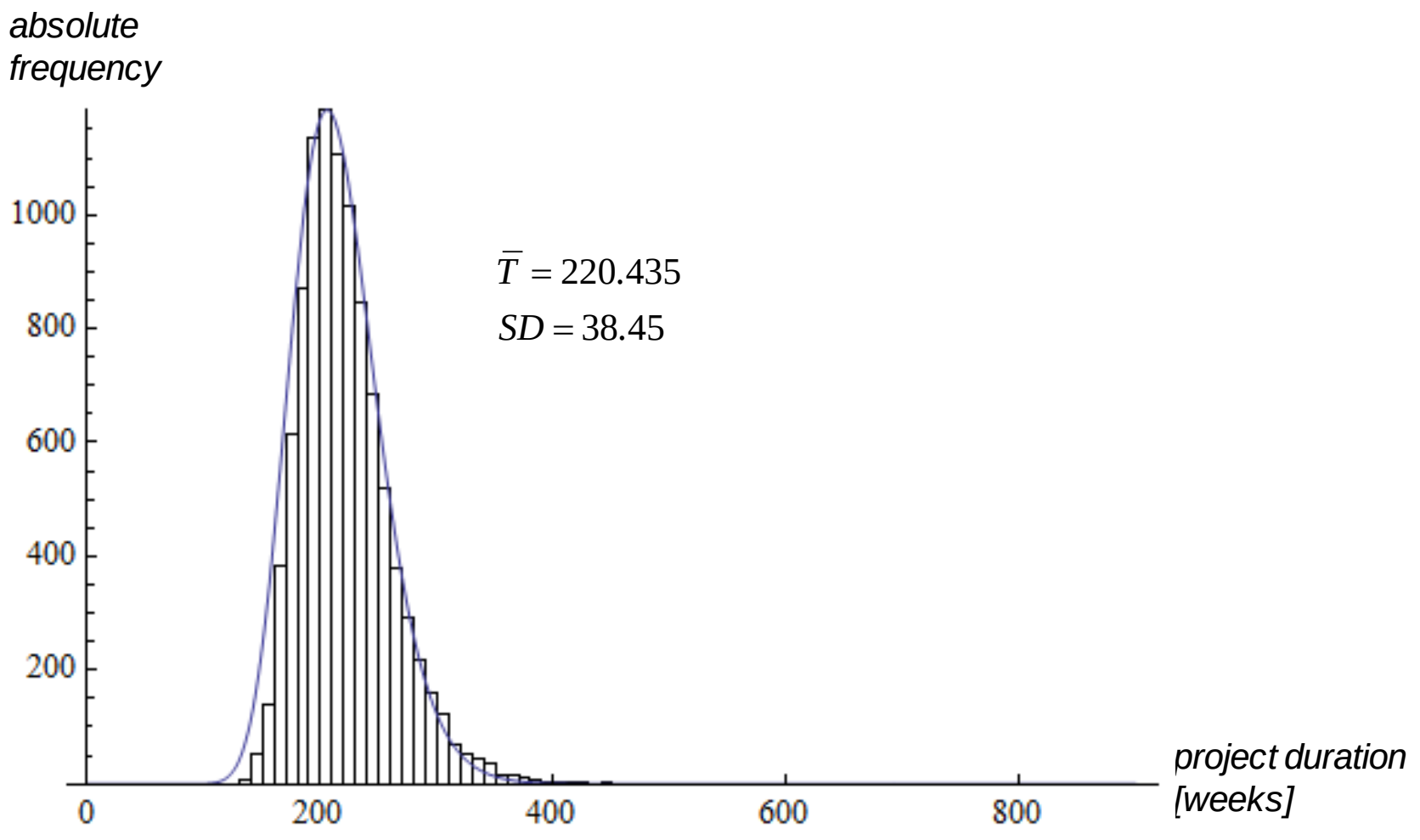

**Figure 12**

*Figure 12*. Histogram of the project duration calculated for the project organization design with minimal complexity. This design is encoded by WTM $A_{01}$ in eq. 220. The sample consisted of 10000 independent runs. In these runs all tasks were initially 100% incomplete. The Monte Carlo experiment was based on state eq. 4. The parameters were $a = 0.9$, $f_1$ =0.04, $f_2 = 0.01$, $f_3 = 0.005$ and $\Delta a = 0.01$. The effective measure complexity is $EMC_{A01} = 14.266$.

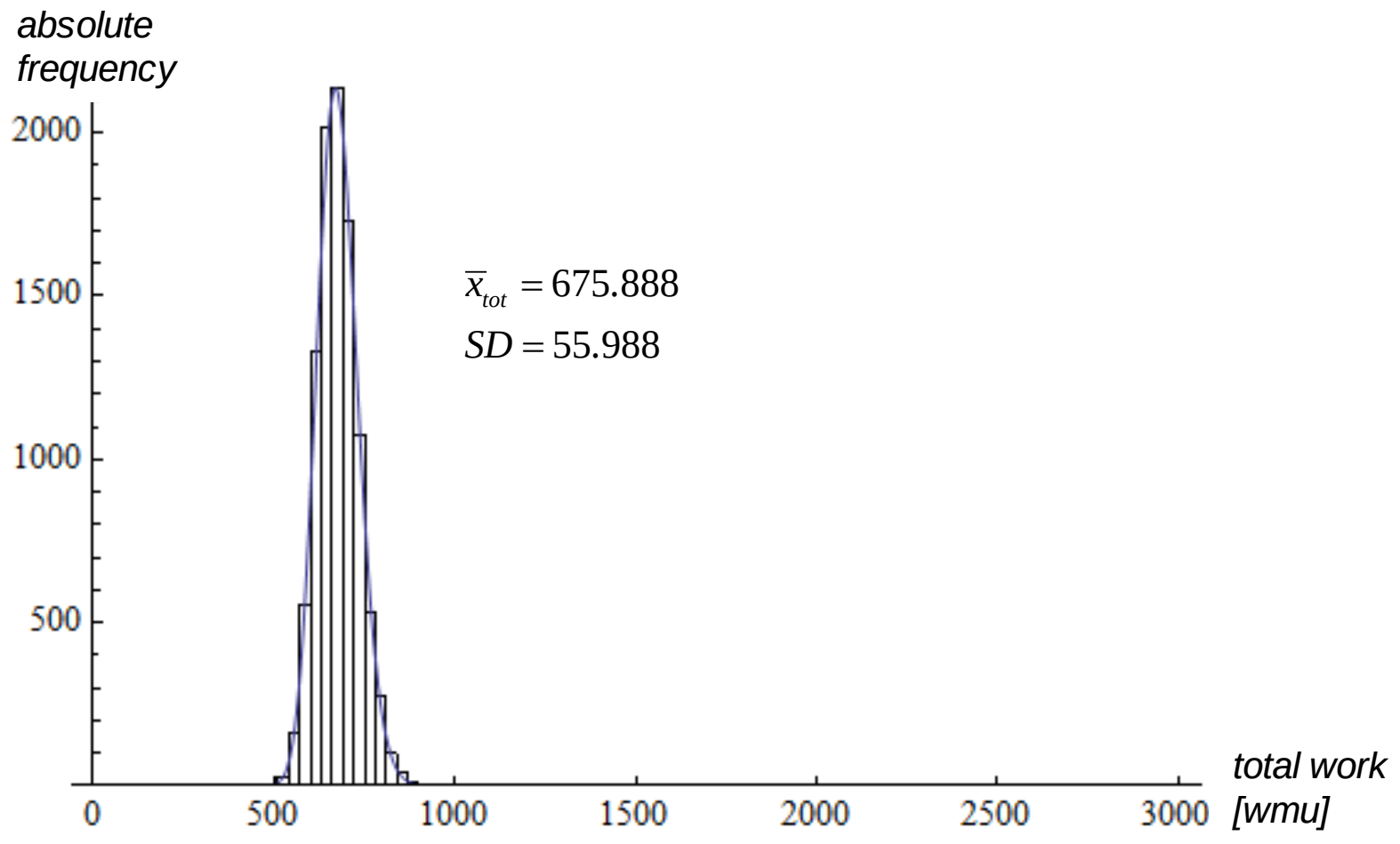

**Figure 13**

*Figure 13*. Histogram of the total work calculated for the project organization design with minimal complexity. This design is encoded by WTM $A_{01}$ in eq. 220. The sample consisted of 10000 independent runs. In these runs all tasks were initially 100% incomplete. The Monte Carlo experiment was based on state eq. 4. The parameters were $a = 0.9$, $f_1 = 0.04$, $f_2 = 0.01$, $f_3 = 0.005$ and $\Delta a = 0.01$. The effective measure complexity is $EMC_{A01} = 14.266$. The expected total work is $x_{tot}(A_{01}) = 680.851$.

Conversely, CE team building toward low diversity (above- or below-average productivity) at the team level significantly increases complexity. An extreme example of organization design with zero diversity was shown in WTM $A_{08}$ (eq. 221). In this case, the calculated complexity value is at a maximum, with $\mathrm{EMC}_{A08} = 14.468$. The corresponding expected total work is $x_{tot}(\mathrm{A}_{08}) = 886.207$. The histogram of the project duration that was calculated in the Monte Carlo experiments is shown in Figure 14. The mean grows from $\bar{T}(A_{01}) = 220.467$ [weeks] in the case of maximum diversity (Figure 13) to $\bar{T}(A_{08}) = 344.091$ [weeks] in the extremely nondiverse case shown. Furthermore, the standard deviation increases from 38.449 to 86.34 [weeks], and therefore the risk of schedule overruns grows significantly. The growth of means and standard deviation of the project duration in the Monte Carlo experiments is not unexpected because all three teams and the system-integration engineer have to wait for the members of team 3 to finish their work. Therefore, in spite of performance fluctuations the project duration is largely determined by the least productive team. The corresponding histogram of the total work $x_{tot}$ is shown in Figure 15. The sample mean is $\bar{x}_{tot}(A_{08}) = 855.171$ [weeks] and the standard deviation is 99.93 [weeks].

For the other six WTMs representing cases of team diversity in between the extremes, the means and standard deviations of the project duration, as well as the total work grow monotonically with EMC. Hence, the formal complexity metric is a good predictor for both KPIs.

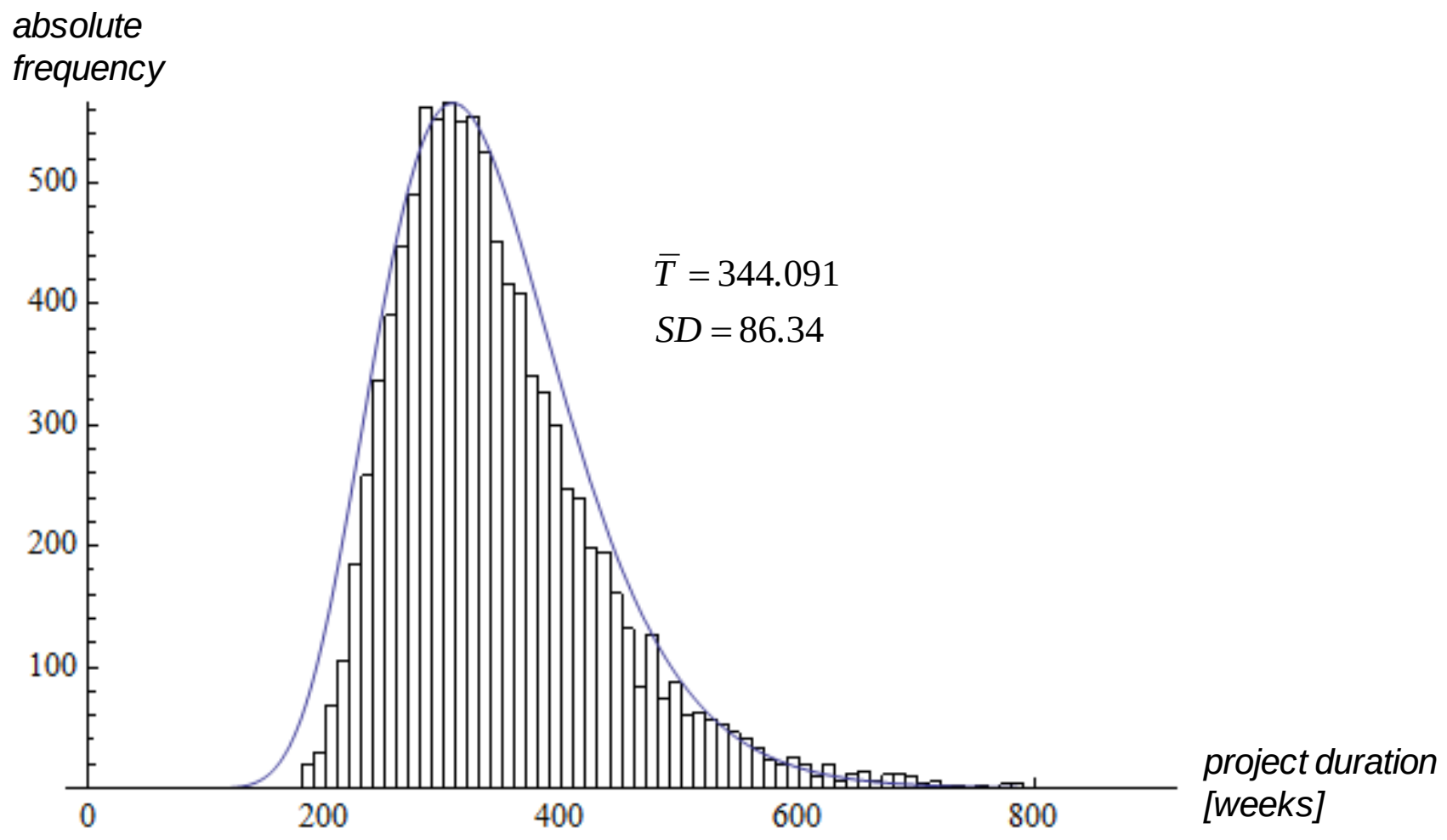

**Figure 14**

*Figure 14*. Histogram of the project duration calculated for the project organization design with third-lowest complexity. This design is encoded by WTM $A_{08}$ in eq. 221. Simulation conditions and parameters are the same as in fig. 8. The effective measure complexity is $EMC_{A02} = 14.468$.

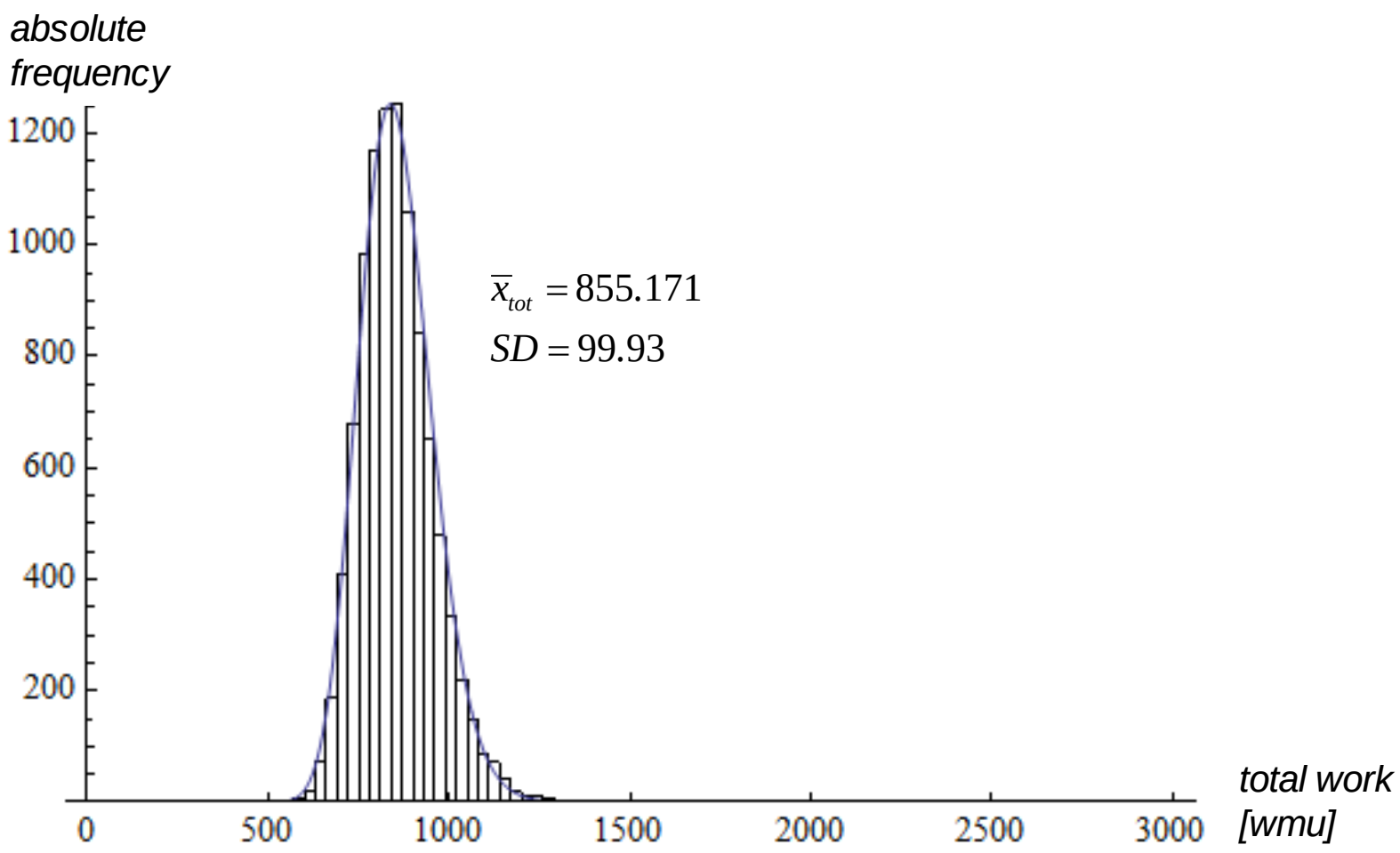

**Figure 15**

*Figure 15*. Histogram of the total work calculated for the project organization design with third-lowest complexity. This design is encoded by WTM $A_{08}$ in eq. 221. Simulation conditions and parameters are the same as in fig. 9. The effective measure complexity is $EMC_{A01} = 14.468$.

Additional analyses have shown that the larger the emergent complexity, the greater the evolution toward a stable solution of the project duration at any particular instance can differ from the average unperturbed behavior. As a result, projects that in the absence of random performance fluctuations would converge smoothly to the desired goal state can deviate significantly from this path. When the complexity is low, convergence to zero remaining work is smooth, and the project duration is approximately normally distributed with small variance (Figure 12). Above certain complexity thresholds, however, the distribution undergoes a transition to a long-tailed log-normal form and implies a possible project duration significantly greater than average (Figure 14). This finding is somehow counterintuitive, because the project state equation is linear and does not incorporate multiplicative noise as the one developed by Huberman and Wilkinson (2005). However, because of the necessary stopping criterion that must be assigned by the project manager, significant deviations from normality can occur, and lead times far from the average are quite likely. Under certain circumstances these deviations can also accelerate processing of the tasks. Accelerated processing means that the mean project duration in the Monte Carlo experiments is much shorter than the expected project duration. The expected project duration can be determined analytically by summing the expected state vectors for increasing time intervals $T$ (eq. 11), until all vector components are smaller than the stopping criterion. However, the standard deviation also grows monotonically with the acceleration factor, which increases the risk of not meeting the schedule.

### 5.1.2 Constrained Optimization

After presentating and discussion of the results of the basic unconstrained optimization problem in project organization design, we move on to formulating and solving an associated constrained optimization problem. The constraint is that the expected total work $x_{tot}$ remains on a constant level among the different assignments of individuals to the three development teams. The constraint is satisfied by systematic intervention in the strength $f_3$ of the backward informational couplings between the nine developers and the system-integration engineer. We only considered a setting in which the productivity offset $\Delta a$ was large ($\Delta a = \Delta a_2 = 0.01$). The base set of independent parameters therefore was $\theta_2 = [a = 0.9 \quad f_1 = 0.01 \quad f_2 = 0.01 \quad \Delta a = 0.01]$. The WTMs $A_{01}$ to $A_{08}$ were ordered by emergent complexity. Hence, WTM $A_{01}$ represents the organization design that leads to minimum emergent complexity in the sense of the EMC metric and WTM $A_{08}$ to maximum complexity. This order corresponds to an ordering by the total variance of autonomous task processing rates over all three development teams (see Table 1). We start by presenting analytical complexity results and go on to present the results of the Monte Carlo experiments.

#### 5.1.2.1 Methods

As in the previous study, the developed objective function in the constraint optimization represents the complexity of a given organization design under the dynamic regime of the state equations 3 and eq. 24. We seek to minimize complexity by systematically choosing the project organization design from the introduced eight distinct assignments under the constraint that the expected total work $x_{tot}$ according to eq. 12 is equal 701.939 [wmu]. This expected total work corresponds to the minimum value that was identified in the previous study for the project organization design with minimum emergent complexity. This design is encoded by WTM $A_{01}$ (eq. 220) and is characterized by a maximum diversity of autonomous productivity rates in the three development teams. Starting with the base level $f_3 = 0.005$ of the strength of the backward informational couplings between the nine developers and the system-integration engineer, the feedback strength was reduced incrementally for project organization designs with less diversity of autonomous productivity rates until the required total amount of work $x_{tot} = 701.939$ was reached. In other words, the independent parameter was adjusted by systematic algorithmic intervention of the experimenter so that the total expected effort in the project did not change under the eight distinct organization designs. To keep the total work constant the independent parameter $f_3$ was adjusted by a self-developed iterative method so that it did not deviate more than $10^{-6}$ [wmu] from the correct value $x_{tot} = x_{tot}(A_{01}) = 701.939$. The time a scale was not modified.

Following the previous procedure, we assumed that the standard deviation $c_{ii}$ of performance fluctuations (eq. 6), which influence task $i$ in the project is proportional to the task processing rate with proportionality constant $r = 0.02$. Hence, the covariance matrices must not be modified (see eq. 222 for organization design encoded by WTM $A_{01}$ and eq. 223 for organization design encoded by WTM $A_{08}$).

Other correlations between vector components were not considered. The initial state was not changed and is given by eq. 224.

The Mathematica software package from Wolfram Research was used to carry out the analytical calculations and the Monte Carlo experiments. The stopping criterion for the Monte Carlo experiments was that at a maximum of 5% of work remained for all tasks in the simulated projects. In addition to EMC as an innovative information-theory KPI, the classic quantities "project duration" and "total work" were used to evaluate performance. To calculate these KPIs, 10,000 independent runs were considered for each project organization design.

The results of the Monte Carlo experiments were also analyzed by inferential statistical methods. Therefore, additional samples based on 100 independent runs were drawn and the corresponding test statistics for the project duration and total work were calculated. To simplify the analysis, only the project organization designs with maximum diversity of autonomous productivity rates within teams (see WTM $A_{01}$ in eq. 217) and zero diversity (see WTM $A_{08}$ in eq. 220) were considered. We hypothesized that lower values of the complexity metric EMC lead to a significantly lower expected project duration. The level of significance was set to $\alpha = 0.05$. To evaluate this hypothesis the Kruskal-Wallis (see e.g. Field 2009) location equivalence test was used. The Kruskal-Wallis test performs a hypothesis test on the project duration data with null hypothesis $H_{0,T}$ that the true location parameters of the samples are equal, i.e. $\mu_T(A_{01}) = \mu_T(A_{08})$, and alternative hypothesis $H_{a,T}$ that at least one is different. The test is a non-parametric method and is based on ranks. It assumes an identically shaped and scaled distribution for each factor level, except for any difference in medians. The parametric equivalence of the Kruskal-Wallis test is the popular one-way analysis of variance (ANOVA, see e.g. Field 2009). The ANOVA could not be used to evaluate the null hypothesis, because the project duration is not normally distributed (see Figure 18). Furthermore, we hypothesized that different values of the complexity metric do not lead to significantly different means of the total work in the Monte Carlo experiments. The rationale behind this (possibly slightly counterintuitive) hypothesis is that we have formulated a constrained optimization problem, in which the analytically obtained expected total work $x_{tot}$ is deliberately kept constant under the different organizational conditions and this systematic intervention should not lead to significant differences of the total effort in the simulated projects. Hence, we formulate the null hypothesis $H_{0,x_{tot}}$ that the true location parameters of the samples are equal, i.e. $\mu_{x_{tot}}(A_{01}) = \mu_{x_{tot}}(A_{08})$.

We also carried out goodness-of-fit hypothesis tests to evaluate the differences between the distributions of performance data for both project organization designs. The null hypothesis $H_{0,gof}$ was that performance data drawn from a sample with maximum diversity in autonomous task processing rates do not come from a different distribution than the data that was obtained for zero diversity. The alternative hypothesis $H_{a,gof}$ is that the data comes from a different distribution. The well-known Kolmogorov-Smirnov test was used to evaluate the hypothesis (see e.g. Field 2009). The level of significance was also set to $\alpha = 0.05$.

#### 5.1.2.2 Results and Discussion

In order to satisfy the constraint $x_{tot} = 701.939$ that was imposed on the total effort involved in completing the deliverables, the strength $f_3$ of the backward informational couplings between the nine developers and the system-integration engineer had to be reduced by a minimum value of $0.00027$ for an organization design in which the mean productivity of team 1 is average, the mean autonomous task processing rates of team 2 are $\Delta a$ below average and the mean autonomous task processing rates of team 3 are $\Delta a$ above average. The associated WTM is not shown, but readers can easily construct it if they wish. Interestingly, the maximum reduction of the backward coupling strength was necessary for the organization design with zero diversity (see WTM $A_{08}$ in eq. 220). In this case the reduction was 0.00303. A list plot of the reductions of the backward coupling strength is shown in Figure 16. The results show that the lower the productivity diversity within teams (following the order of the WTMs $A_{01}$ to $A_{08}$ from left to right in the figure), the more the backward coupling strength must be reduced to satisfy the constraint.

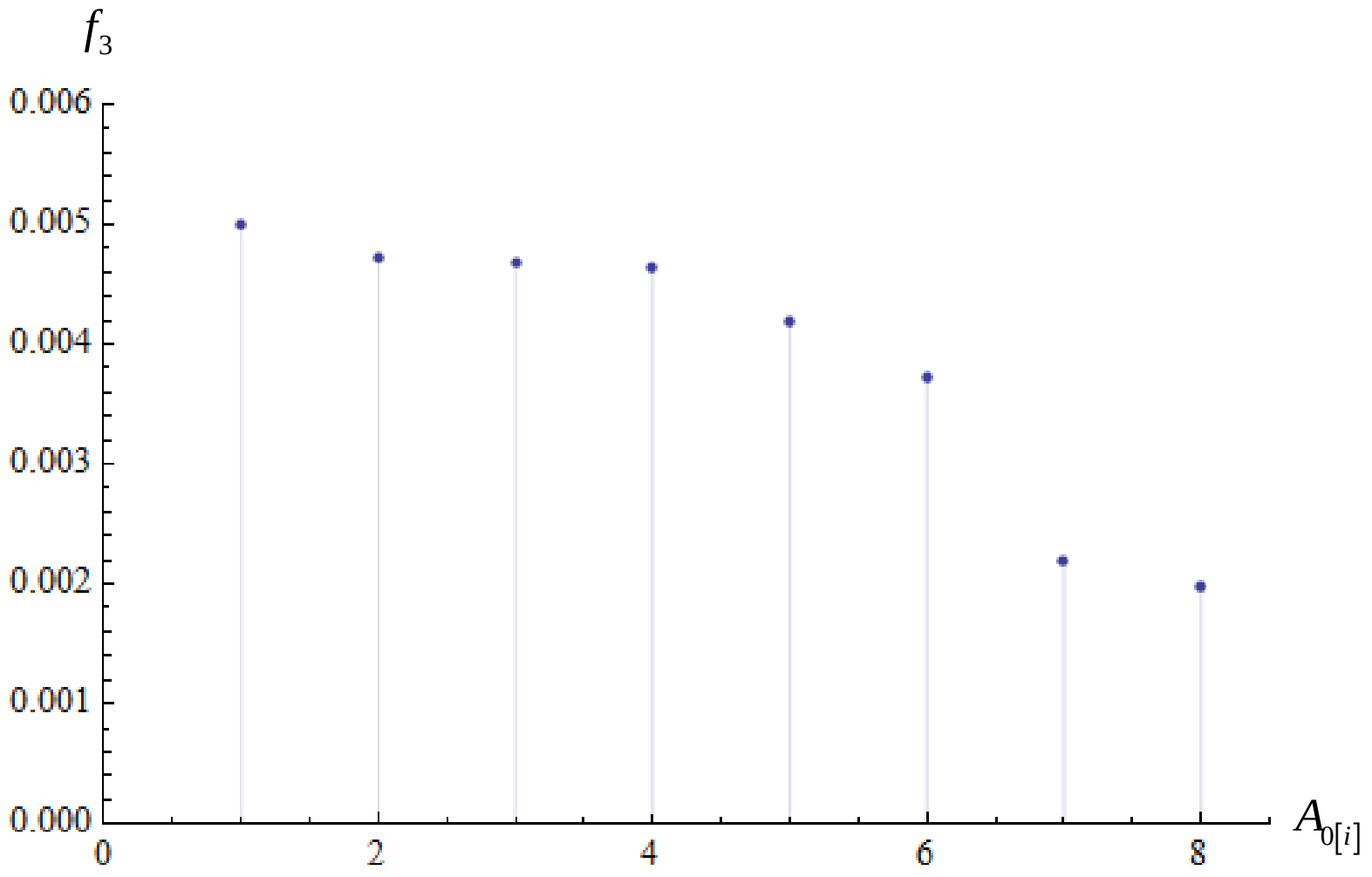

**Figure 16**

*Figure 16*. Adjustments of original coupling parameter $f_3 = 0.005$ (see eq. 220) that were necessary to satisfy the constraint on the expected total work $x_{tot}(A_{01}) = \ldots = x_{tot}(A_{08}) = 701.939$ for the eight distinct project organization designs. The additional independent parameters were not changed.

The values of the complexity metric EMC that correspond to the reduction of the backward coupling strength $f_3$ are shown in Figure 17. Interestingly, the constrained optimization leads to complexity values that increase with decreasing strength of backward informational couplings between the nine developers and the system-integration engineer. Hence, the ordering of the project organization designs by emergent complexity that was determined in the previous section does not change after imposing the constraint on the total effort.

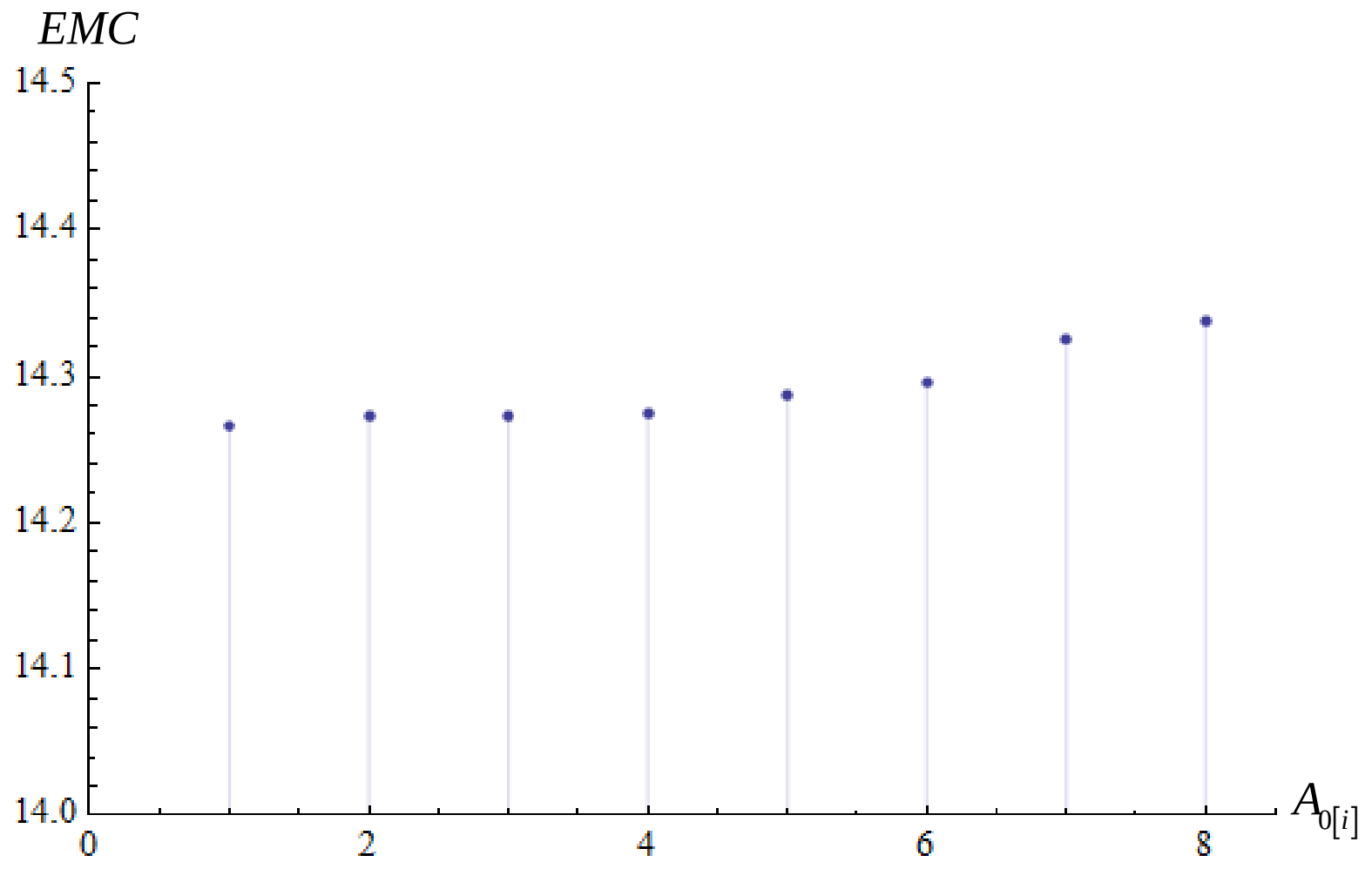

**Figure 17**

*Figure 17*. Effective measure complexity for the investigated eight distinct project organization designs under the constraint that the expected total work is kept on constant level, i.e. $x_{tot}(A_{01}) = \ldots = x_{tot}(A_{08}) = 701.939$. To keep the expected total work constant, the original coupling parameter $f_3 = 0.005$ was adjusted. The additional independent parameters were not changed.

Selected performance results of the Monte Carlo experiments are shown in Figures 18 and 19. Figure 18 shows the histogram of the simulated project duration for organization designs with maximum and minimum diversity of autonomous productivity rates within teams. In Figure 19, the histograms for the total work are given. To simplify the interpretation of the results, the figures also show the means and standard deviations.

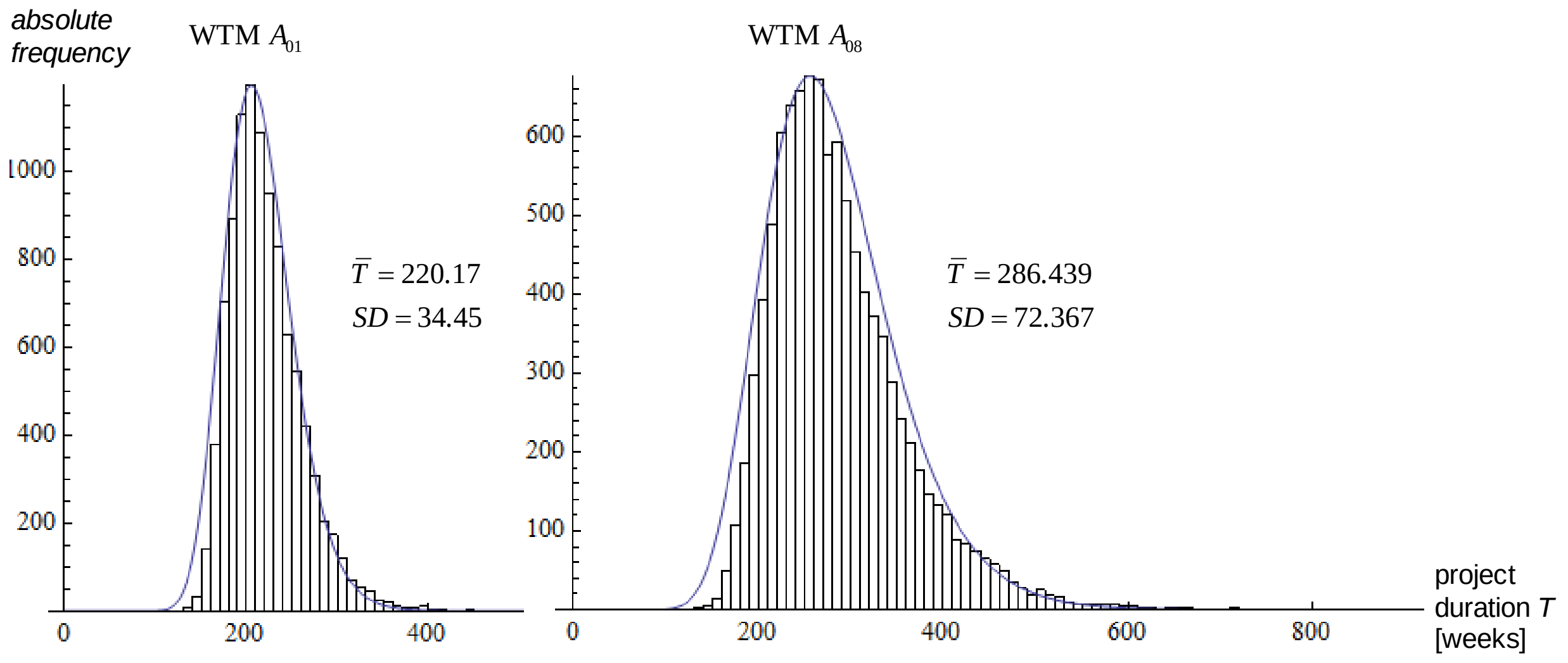

**Figure 18**

*Figure 18*. Histograms of the project duration obtained for organization designs with maximum diversity (encoded by WTM $A_{01}$) and minimum diversity (encoded by WTM $A_{08}$) of autonomous task processing rates within teams. We computed 10,000 independent runs. The stopping criterion for the simulated projects was that a maximum of 5% of work remained for all tasks. The Monte Carlo experiments were based on state eq. 4. The base parameter setting was $a = 0.9$, $f_1$ =0.04, $f_2 = 0.01$, and $\Delta a = 0.01$. The coupling strength $f_3 = 0.005$ was adjusted according to Figure 16 to satisfy the constraint on the total work.

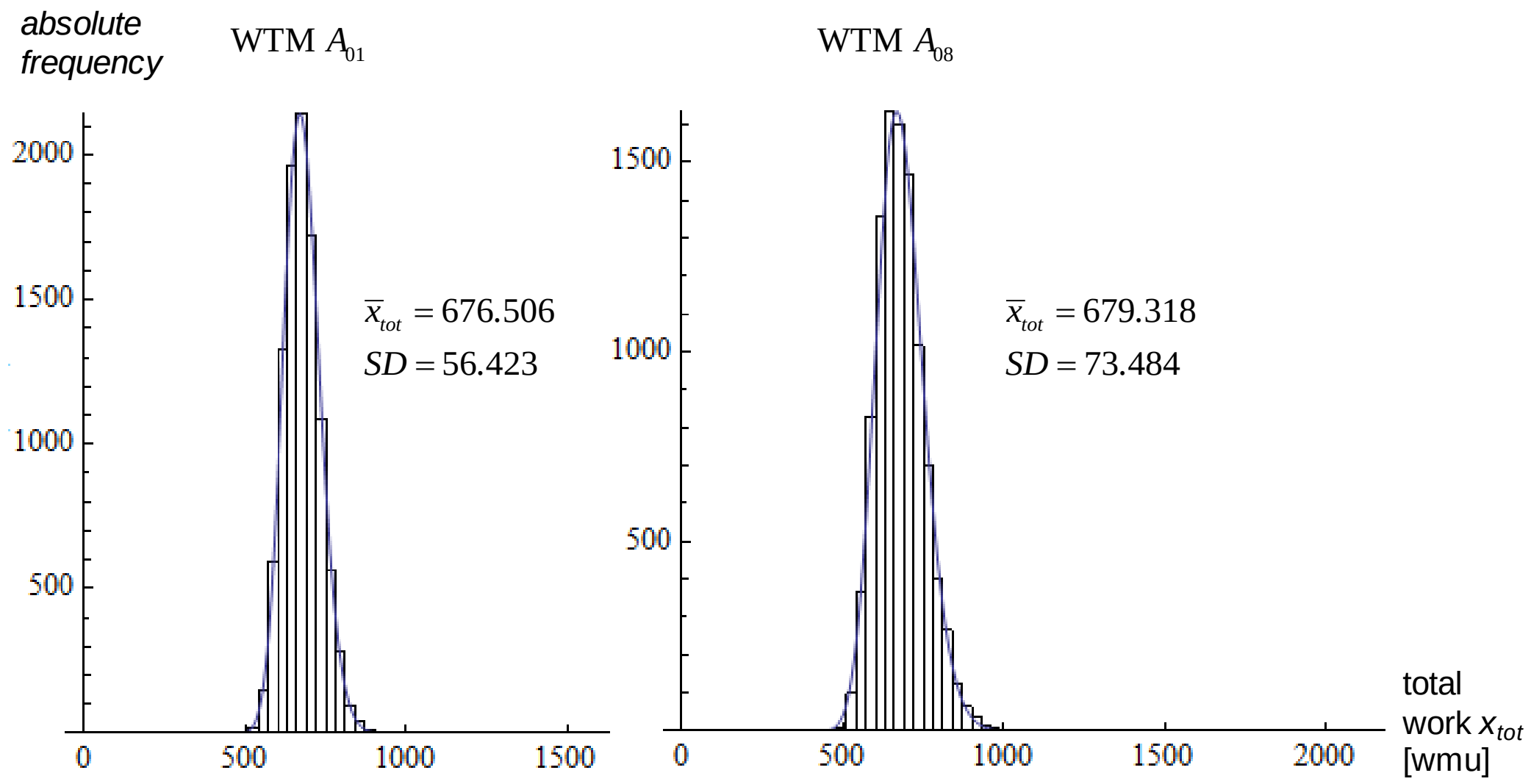

**Figure 19**

*Figure 19*. Histograms of the total work obtained for organization designs with maximum diversity (encoded by WTM $A_{01}$) and minimum diversity (encoded by WTM $A_{08}$) of autonomous task processing rates within teams. We computed 10,000 independent replications. The stopping criterion for the simulated projects is that a maximum of 5% of work remained for all tasks. The Monte Carlo experiments were based on state eq. 4. The base parameter setting was $a = 0.9$, $f_1$ =0.04, $f_2 = 0.01$, and $\Delta a = 0.01$. The coupling strength $f_3 = 0.005$ was adjusted according to Figure 16 to satisfy the constraint on the total work.

According to Figure 18, the organization design with maximum diversity of autonomous task processing rates within teams that is encoded by WTM $A_{01}$ leads an average project duration of $\bar{T}(A_{01}) = 220.17$ [weeks]. If the team members are assigned so that diversity is zero, the average project duration is extended to $\bar{T}(A_{01}) = 286.439$ [weeks]. Furthermore, the standard deviation is more than twice as large. In contrast to these findings, the difference in the mean total work for both organizational conditions is less than 1% (Figure 19). The constraint optimization based on the analytically obtained expected total work therefore leads to very similar mean efforts in the Monte Carlo experiments.

For 100 additional independent runs, the Kruskal-Wallis test on the project duration data shows that the location differences between both project organization designs are significant ($K_T = 68.25$, $p = 8.35 \cdot 10^{-20}$). Hence, the null hypothesis $H_{0,T}$ that the true location parameters of the samples are equal can be rejected on the significance level of $\alpha = 0.05$. The Kruskal-Wallis test on the total work data comes to a different result. It shows that the locations between both organization designs are not significantly different ($K_{x_{tot}} = 1.96$, $p = 0.16$). The null hypothesis $H_{0,x_{tot}}$ that the true location parameters of the samples are equal cannot be rejected on a significance level of $\alpha = 0.05$. Hence, the constraint imposed on the objective function is also effective in the Monte Carlo experiments and leads to very small and insignificant differences in total effort. The goodness-of-fit hypothesis test between both organization designs also indicates that the differences in the distributions from which the total work data were drawn are not significant. The Kolmogorov-Smirnov test statistic is $D_T = 0.0688$. The associated $p$-value is $p = 0.705$. The slight differences in the complexity metric according to Figure 17 therefore do not lead to significant differences in probability distributions of the total work if the work is systematically "leveled out" to satisfy the constraint. As expected, the additional distribution fit test of the project duration data shows significant differences among the organizational conditions. The corresponding test statistic is $D_{x_{tot}} = 0.374$ ($p = 7.25 \cdot 10^{-13}$).

The combined theoretical and computational analyses provide some evidence that the information-theory complexity metric EMC is not only a theoretically very satisfactory quantity but that under certain circumstances it is also a good predictor of the mean and standard deviation of the project duration. Moreover, the results show that the novel management concepts of productivity leveling and design for diversity in productivity are promising in terms of optimizing of NPD projects that use concurrent engineering.

## 5.2 Optimization of Release Period of Design Information Between System-Level and Component-Level Teams

The objective of the second study is to optimize the period for minimal emergent complexity, where information about integration and tests of geometric/topological entities is deliberately withheld by system-level teams and not released to component-level teams. According to section 2.6, new information is kept secret between the releases and work in the subordinate component-level teams is based on the old state of knowledge. This kind of noncooperative behavior is justified by the aim to improve solution maturity and reduce coordination efforts. Optimizing the release period by using a formal complexity metric in conjunction with a mathematical model of periodically correlated work processes is an especially interesting application area because cross-hierarchical teamwork in large-scale NPD projects can be designed systematically and unnecessary coordination efforts can be avoided. Similar to the previous section, it is also very theoretically interesting to analyze whether, in addition to being valid for steady-state processes, EMC can also apply to evaluating the preasymptotic behavior of project dynamics. We start by formulating the unconstrained optimization problem and the presentation of its solution based on a complete enumeration of the release period. We then formulate a constrained optimization problem and solve it by applying the same principle. The constraint is that the expected total work $x_{tot}$ according to eq. 61 is constant among the different release periods.

### 5.2.1 Unconstrained Optimization

#### 5.2.1.1 Methods

The objective function developed in the second study quantifies the complexity of periodically correlated work processes in NPD with predefined release period $s \geq 2$ under the dynamic regime of the state eq. eq. 59. We seek to minimize complexity. The release period was varied systematically in the range [2 ; 20] by increments of one week. The analytical calculations and Monte Carlo experiments consider different correlation lengths and simulate the performance fluctuations accordingly. The time scale is [weeks]. Using state eq. 59 in conjunction with the closed-form solution from eq. 175, we can express the EMC of the generated process in the original state space coordinates as

$$\mathrm{EMC_{PVAR}} = \frac{1}{2}\log_2\left(\frac{\mathrm{Det}\left[\sum_{k=0}^{\infty}((\Phi_0^*)^{-1}\cdot\Phi_1^*)^k \cdot \left((\Phi_0^*)^{-1}\cdot C^*\cdot(\Phi_0^*)^{-\mathrm{T}}\right)\cdot\left((\Phi_0^*)^{-\mathrm{T}}\cdot(\Phi_1^*)^{\mathrm{T}}\right)^k\right]}{\mathrm{Det}[(\Phi_0^*)^{-1}\cdot C^*\cdot(\Phi_0^*)^{-\mathrm{T}}]}\right). \quad (225)$$

The transformed matrix $(\Phi_0^*)^{-1}$ is defined in eq. 55. Its inverse is given by the representation in eq. 60. The autoregressive coefficient matrix $\Phi_1^*$ was defined in eq. 56. The above definition of the complexity metric is an implicit function of the release period $s$ as the dimension of both matrices $\Phi_0^*$ and $\Phi_1^*$ scales linearly with the period (cf. eq. 56).

We defined an example project for the Monte Carlo experiments that includes two component-level tasks and two system-level tasks. Different teams process the tasks simultaneously. Individual task processing is not considered. The component-level tasks involves designing an instrument panel and a center console for a vehicle (cf. McDaniel 1996). The system-level tasks deal with integration testing of these (and other) components and the coordination with the whole vehicle interior design project. Similar to the previous study, the vector components of the state variable $X_{ns+v}$ that are related to processing the system-level and component-level tasks represent the relative number of open issues that need to be resolved before final design release. We assume that both component-level teams work at the same autonomous task processing rate $a_{11}^{C} = a_{22}^{C}$=0.90. The tasks are coupled with symmetric strength and we have $a_{12}^{C} = a_{21}^{C} = 0.05$. Similarly, the teams responsible for system integration and coordination both work at the same (but slightly lower) autonomous task processing rate $a_{11}^{S} = a_{22}^{S}$=0.85. The tasks are coupled at the same (but slightly higher) strength $a_{12}^{S} = a_{21}^{S} = 0.07$. Both system level tasks generate 3% of finished work at each short iteration that is put in hold state until it is released at time step $ns$ ($n \in \mathbb{N}$). Hence, $a_{11}^{SH}$=$a_{22}^{SH}$=0.03. Furthermore, the first component-level task generates 6% of finished work at each iteration for the first system-level task and vice versa. Hence, we have $a_{11}^{CS} = a_{11}^{SC} = 0.06$. The accumulated development issues of the system-level teams are released to component-level teams at the end of the period ($a_{11}^{HC} = a_{22}^{HC}$=1). Additional dynamical dependencies were not considered and therefore all other matrix entries were defined to be zero. The complete representation is as follows:

Combined dynamical operator $A_0^* = (\Phi_0^*)^{-1} \cdot \Phi_1^*$:

$$(\Phi_0^*)^{-1} \cdot \Phi_1^* = \begin{pmatrix} \Phi_1(s)(\Phi_1(1))^{s-1} & 0 & 0 & \cdots & 0 \\ (\Phi_1(1))^{s-1} & 0 & 0 & \cdots & 0 \\ (\Phi_1(1))^{s-2} & 0 & 0 & \cdots & 0 \\ \vdots & \vdots & 0 & \ddots & \vdots \\ \Phi_1(1) & 0 & 0 & \cdots & 0 \end{pmatrix}$$

Transformation matrices:

$$\Phi_1(1) = \begin{pmatrix} A_0^{\mathrm{C}} & A_0^{\mathrm{SC}} & 0 \\ A_0^{\mathrm{CS}} & A_0^{\mathrm{S}} & 0 \\ 0 & A_0^{\mathrm{SH}} & \{1-\varepsilon\}I_2 \end{pmatrix}$$

$$\Phi_1(s) = \begin{pmatrix} A_0^{\mathrm{C}} & A_0^{\mathrm{SC}} & A_0^{\mathrm{HC}} \\ A_0^{\mathrm{CS}} & A_0^{\mathrm{S}} & 0 \\ 0 & 0 & \{\varepsilon\}I_2 \end{pmatrix}$$

Work transformation sub-matrices:

$$A_0^{\mathrm{C}} = \begin{pmatrix} 0.90 & 0.05 \\ 0.05 & 0.90 \end{pmatrix} \tag{226}$$

$$A_0^{\mathrm{S}} = \begin{pmatrix} 0.85 & 0.07 \\ 0.07 & 0.85 \end{pmatrix} \tag{227}$$

$$A_0^{\mathrm{CS}} = \begin{pmatrix} 0.06 & 0 \\ 0 & 0 \end{pmatrix} \tag{228}$$

$$A_0^{\mathrm{SC}} = \begin{pmatrix} 0.06 & 0 \\ 0 & 0 \end{pmatrix} \tag{229}$$

$$A_0^{\mathrm{SH}} = \begin{pmatrix} 0.03 & 0 \\ 0 & 0.03 \end{pmatrix} \tag{230}$$

$$A_0^{\mathrm{HC}} = \begin{pmatrix} 1 & 0 \\ 0 & 1 \end{pmatrix}. \tag{231}$$

As explained in section 2.6, the variable $\varepsilon$ is necessary for an explicit complexity evaluation. EMC simply scales linearly with $\varepsilon$. We calculated with $\varepsilon = 10^{-3}$. By doing so, the finished work after release is set back to a nonzero but negligible amount in terms of productivity.

The initial state $x_0^*$ was defined based on the assumption that all parallel tasks are initially to be fully completed and that no work is in hold state. Hence, for the minimum release period $s_{min}$ =2, we have:

$$x_0^* = \begin{pmatrix} 1 \\ 1 \\ 1 \\ 1 \\ 0 \\ 0 \\ 0 \\ 0 \\ 0 \\ 0 \\ 0 \\ 0 \end{pmatrix}, \tag{232}$$

For larger release periods, additional zeros were appended to the initial state.

Following the procedure of the first study (section 5.1.1), we assumed that the standard deviation $c_{ii}$ of performance fluctuations (eq. 6) influencing task $i$ in the project is proportional to the task processing rate. The proportionality constant is $r = 0.02$. Other correlations among vector components were not considered. Furthermore, it is assumed that the variance of the fluctuations related to the finished work put in hold state is reduced by the factor $10^{-3}$. Hence, we have the covariance matrix $C = E[\varepsilon_{ns+v}\, \varepsilon_{ns+v}{}^{\mathrm{T}}]$:

$$C^* = \begin{pmatrix} C_s & 0 & 0 & 0 \\ 0 & C_1 & 0 & 0 \\ 0 & 0 & \ddots & 0 \\ 0 & 0 & 0 & C_1 \end{pmatrix},$$

where the submatrices are given by

$$C_1 = \{r^2\} \begin{pmatrix} \left(a_{11}^{\mathrm{C}}\right)^2 & 0 & 0 & 0 & 0 & 0 \\ 0 & \left(a_{22}^{\mathrm{C}}\right)^2 & 0 & 0 & 0 & 0 \\ 0 & 0 & \left(a_{11}^{\mathrm{S}}\right)^2 & 0 & 0 & 0 \\ 0 & 0 & 0 & \left(a_{22}^{\mathrm{S}}\right)^2 & 0 & 0 \\ 0 & 0 & 0 & 0 & 10^{-3}(1-\varepsilon)^2 & 0 \\ 0 & 0 & 0 & 0 & 0 & 10^{-3}(1-\varepsilon)^2 \end{pmatrix}$$

$$= \{0.02^2\} \begin{pmatrix} 0.9^2 & 0 & 0 & 0 & 0 & 0 \\ 0 & 0.9^2 & 0 & 0 & 0 & 0 \\ 0 & 0 & 0.85^2 & 0 & 0 & 0 \\ 0 & 0 & 0 & 0.85^2 & 0 & 0 \\ 0 & 0 & 0 & 0 & 10^{-3}(1-10^{-3})^2 & 0 \\ 0 & 0 & 0 & 0 & 0 & 10^{-3}(1-10^{-3})^2 \end{pmatrix} \tag{233}$$

and

$$C_s = \{r^2\} \begin{pmatrix} (a_{11}^C)^2 & 0 & 0 & 0 & 0 & 0 \\ 0 & (a_{22}^C)^2 & 0 & 0 & 0 & 0 \\ 0 & 0 & (a_{11}^S)^2 & 0 & 0 & 0 \\ 0 & 0 & 0 & (a_{22}^S)^2 & 0 & 0 \\ 0 & 0 & 0 & 0 & 10^{-3}\varepsilon^2 & 0 \\ 0 & 0 & 0 & 0 & 0 & 10^{-3}\varepsilon^2 \end{pmatrix}$$

$$= \{0.02^2\} \begin{pmatrix} 0.9^2 & 0 & 0 & 0 & 0 & 0 \\ 0 & 0.9^2 & 0 & 0 & 0 & 0 \\ 0 & 0 & 0.85^2 & 0 & 0 & 0 \\ 0 & 0 & 0 & 0.85^2 & 0 & 0 \\ 0 & 0 & 0 & 0 & 10^{-9} & 0 \\ 0 & 0 & 0 & 0 & 0 & 10^{-9} \end{pmatrix}. \quad (234)$$

The Mathematica software package from Wolfram Research was used to carry out the analytical calculations and the Monte Carlo experiments. The stopping criterion for the simulated projects was that a maximum of 5% of work remained for all tasks. The classic KPIs “project duration” and “total work" were used in addition to EMC. To calculate these KPIs, we generated samples of 10000 independent runs for each release period. We also calculated the expected total work $x_{tot}$ analytically according to eq. 61.

#### 5.2.1.2 Results and Discussion

Figure 20 shows typical traces of work remaining for the parameterized project model assuming a minimal release period of $s_{min} = 2$ [weeks]. The finished work that was put in hold state by both system level tasks at each short iteration is also shown in the list plot around the abcissa. The plots for extended release periods with $s = 10$ and $s_{max} = 2$ [weeks] are shown in Figures 21 and 22, respectively.

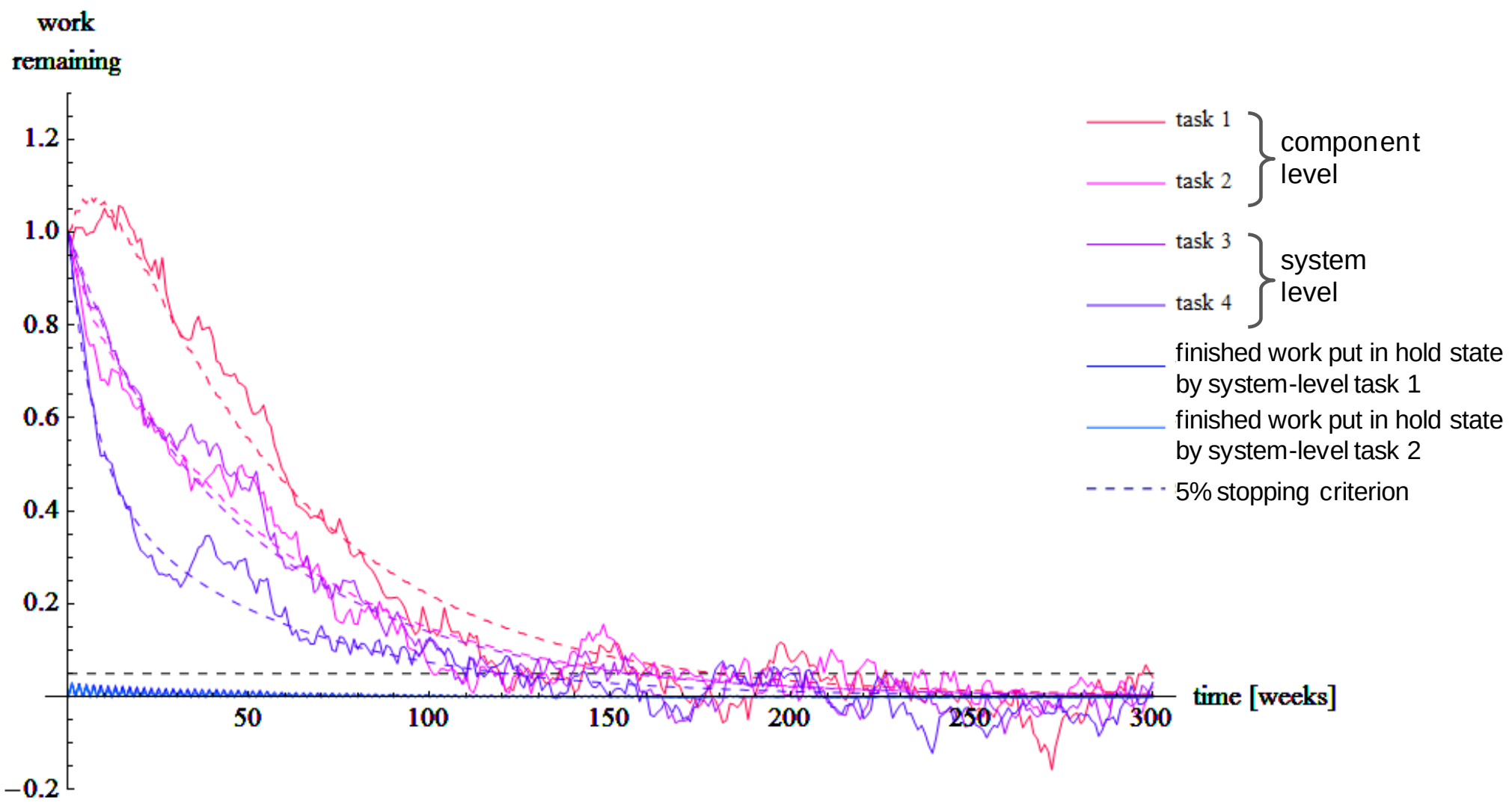

**Figure 20**

*Figure 20*. List plot of work remaining in a simulated NPD project with correlated work processes. It shows the simultaneous processing of all four development tasks. The release period is $s = 2$ [weeks]. The data is based on a single run of the Monte Carlo experiment. The plot also shows the means of simulated time series of task processing as dashed curves. The Monte Carlo experiment was based on state eq. 59. The parameters are given by eqs. 226 to 231. The stopping criterion of 5% is marked by a dashed line at the bottom of the plot.

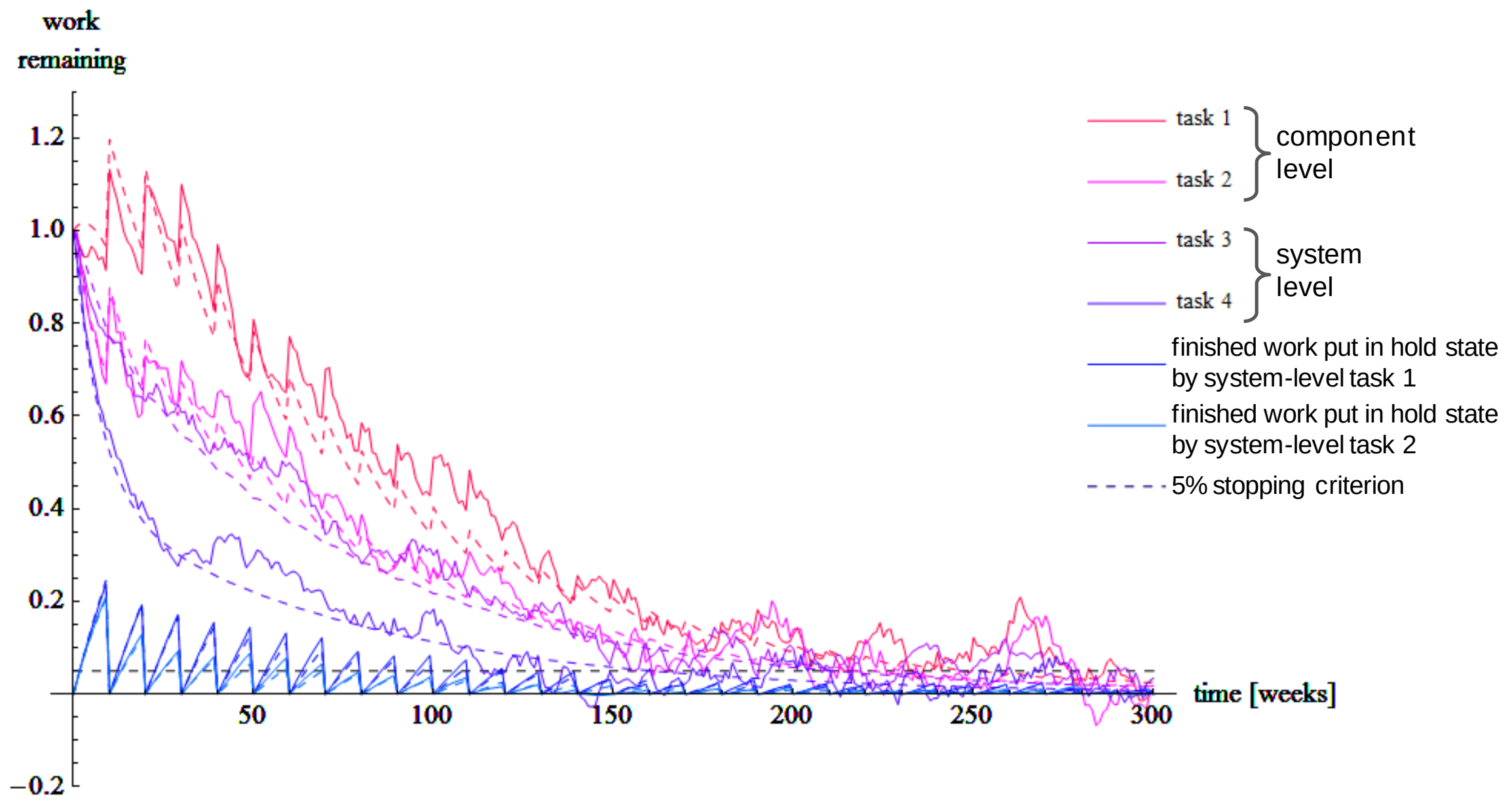

**Figure 21**

*Figure 21*. List plot of work remaining in a simulated NPD project with correlated work processes. The release period is $s = 10$ [weeks]. The other simulation conditions and parameters are the same as in fig. 16.

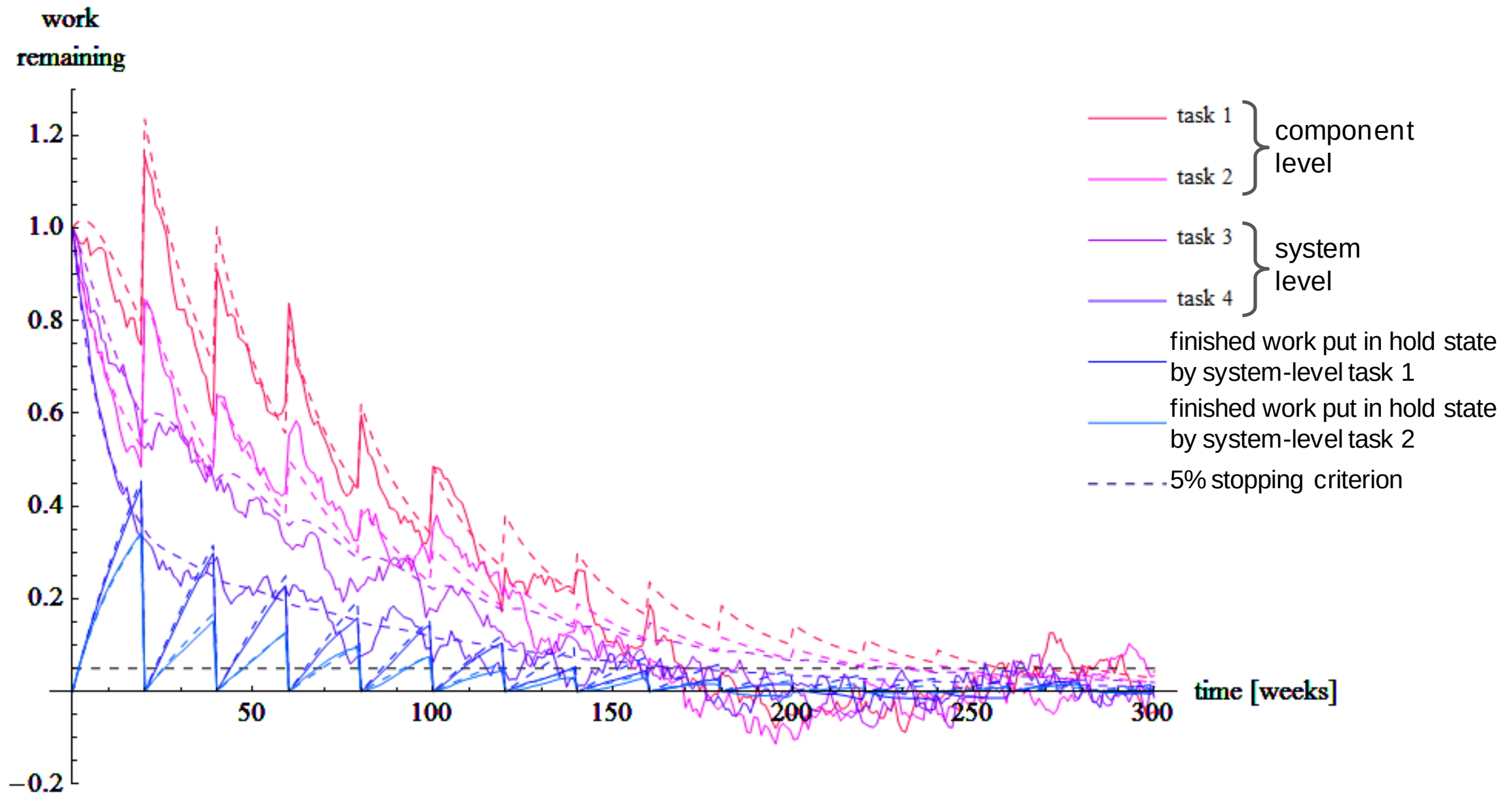

**Figure 22**

*Figure 22*. List plot of work remaining in a simulated NPD project with correlated work processes. The release period is $s = 20$ [weeks]. The data are based on a single run of the Monte Carlo experiment. The other simulation conditions and parameters are the same as in fig. 17.

The comparison of Figures 20, 21 and 22 shows that an extension of the release period from two weeks to ten or 20 weeks increases the average work remaining before the stopping criterion of 5% is met. Furthermore, it is not difficult to see that the intensity of the performance fluctuations is inverse proportional to the period length: the longer the release period, the more that single instances of task processing deviate from the mean (unperturbed) work remaining. For all three release periods significant deviations from the means occur as early as in week 20 and proceed until the project is finished. Figures 21 and 22 also clearly show that the “sawtooth” behavior of finished work that is put in hold state by the teams processing the system-level tasks completely spills over to the component-level tasks and is exacerbated by the performance fluctuations. Conversely, under the given boundary conditions, the processing of the system-level tasks is relatively fast and smooth.

The 10000 runs that were computed for each release period show that the minimum release period of two weeks leads to a mean project duration of $\bar{T}(s=2) = 158.553$ [weeks]. The standard deviation is $SD(s=2) = 34.234$ [weeks]. Both the mean project duration and the standard deviation are minimal within the sample. If the release period is extended to ten weeks, the mean project duration increases to $\bar{T}(s=10) = 212.828$ [weeks] and the standard deviation to $SD(s=10) = 51.181$ [weeks]. An additional extension of the release period to the maximum of 20 weeks further increases the mean project duration and standard deviation, and we have $\bar{T}(s=20) = 226.333$ [weeks] and $SD(s=20) = 54.000$ [weeks]. Figure 23 shows the histograms of the calculated project duration for the three considered release periods. It is interesting that, independent of the release period the distribution of the probability mass follows a long-tailed log-normal form and that durations far from the average are quite likely. The significant deviation from normality is because a 5% stopping criterion was used in the Monte Carlo experiments. Furthermore, for the longest release period of 20 weeks the histogram shows oscillations of the distribution of the probability mass for project durations longer than 175 weeks. This effect is due to the long time span between average release points. The oscillation periods follows the release period.

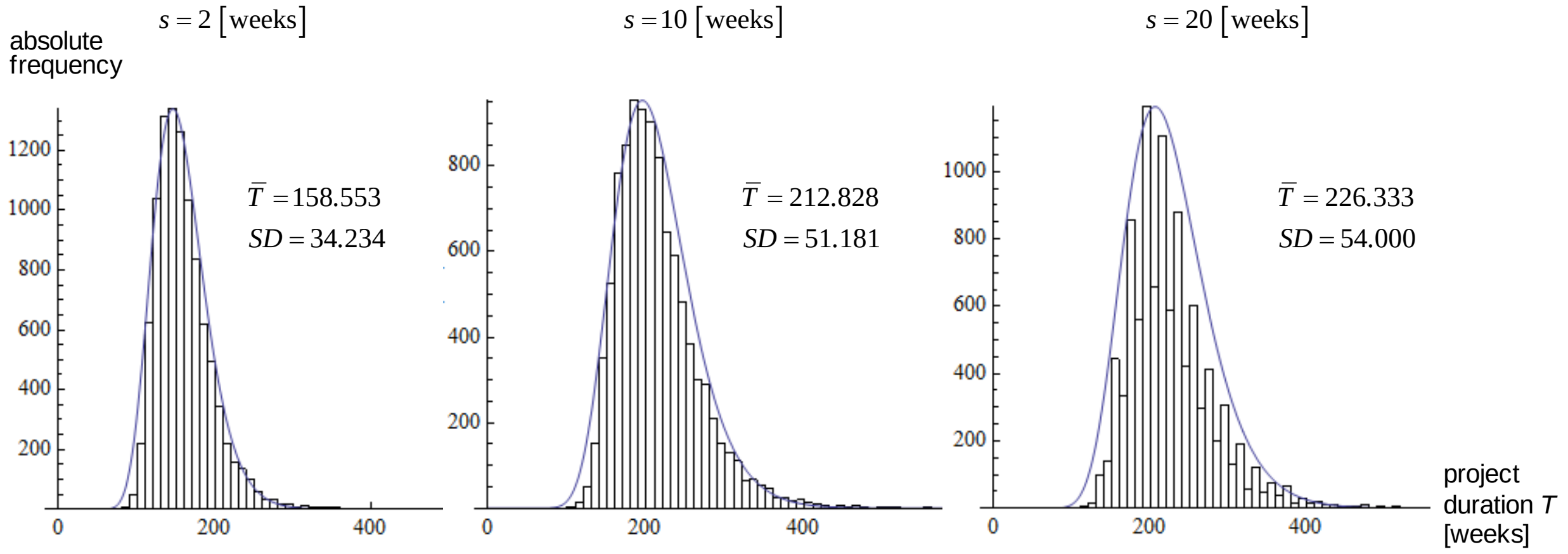

**Figure 23**

*Figure 23*. Histograms of the project duration obtained for three different release periods $s = 2, 10$ and 20 [weeks]. We computed 10,000 independent runs. The stopping criterion for the simulated projects was that a maximum of 5% of work remained for all tasks. The Monte Carlo experiments were based on state eq. 59. The parameters are given by eqs. 226 to 231.

The means and standard deviations of the total work $x_{tot}$ (see section 2.6) in the simulated projects follows a similar grow pattern (Figure 24). However, the accumulation of work reduces the inherent periodic correlations, and an oscillation of the distribution of the probability mass does not occur for the release period of 20 weeks.

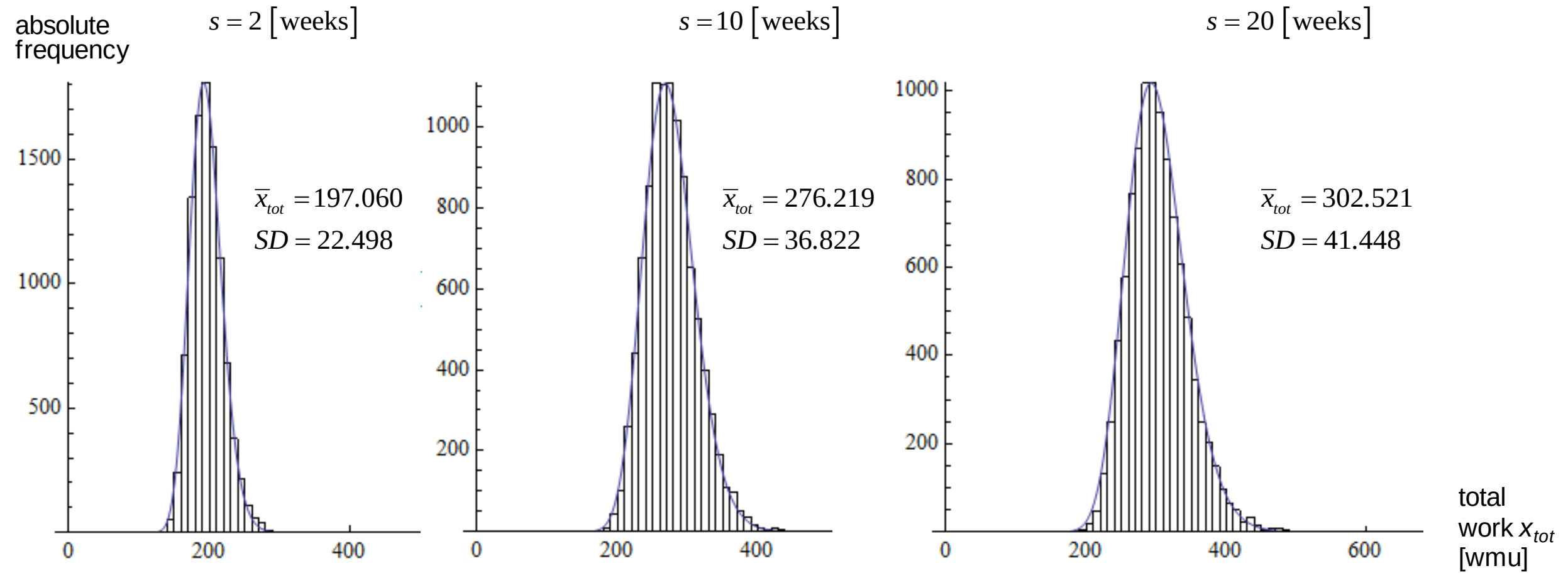

**Figure 24**

*Figure 24*. Histograms of the total work obtained for three different release periods $s = 2, 10$ and $20$ [weeks]. We computed 10,000 independent runs. The stopping criterion for the simulated projects was that a maximum of 5% of work remained for all tasks. The Monte Carlo experiments were based on state eq. 59. The parameters are given by eqs. 226 to 231.

We also calculated analytically the expected total work $x_{tot}$ in the modeled project for different release periods. The expected total work is by definition accumulated over an infinite past history and therefore does not take the stopping criterion of the Monte Carlo experiments into account. The results are presented in Figure 25, which shows the total work is minimal for the shortest period length $s_{min} = 2$ and grows evenly as the period increases.

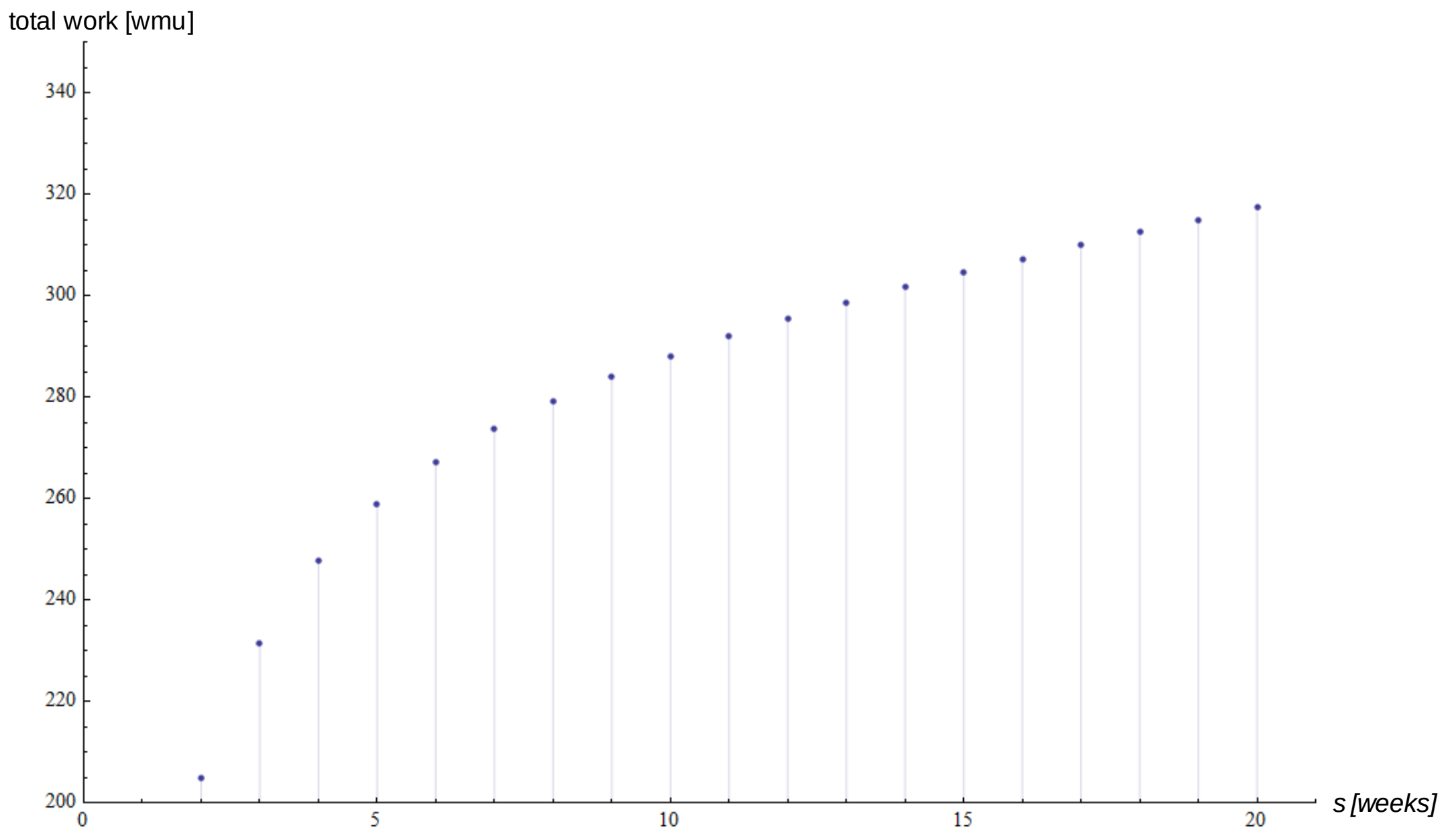

**Figure 25**

*Figure 25*. Expected total work $x_{tot}$ in the modeled project according to eq. 61. The units are work measurement units [wmu] that refer to the definition of the state of work remaining in the project. The parameters are given by eqs. 226 to 231.

The complexity values $\mathrm{EMC}_{\mathrm{PVAR}}$ (eq. 225) that were obtained for different period lengths are shown in Figure 26.

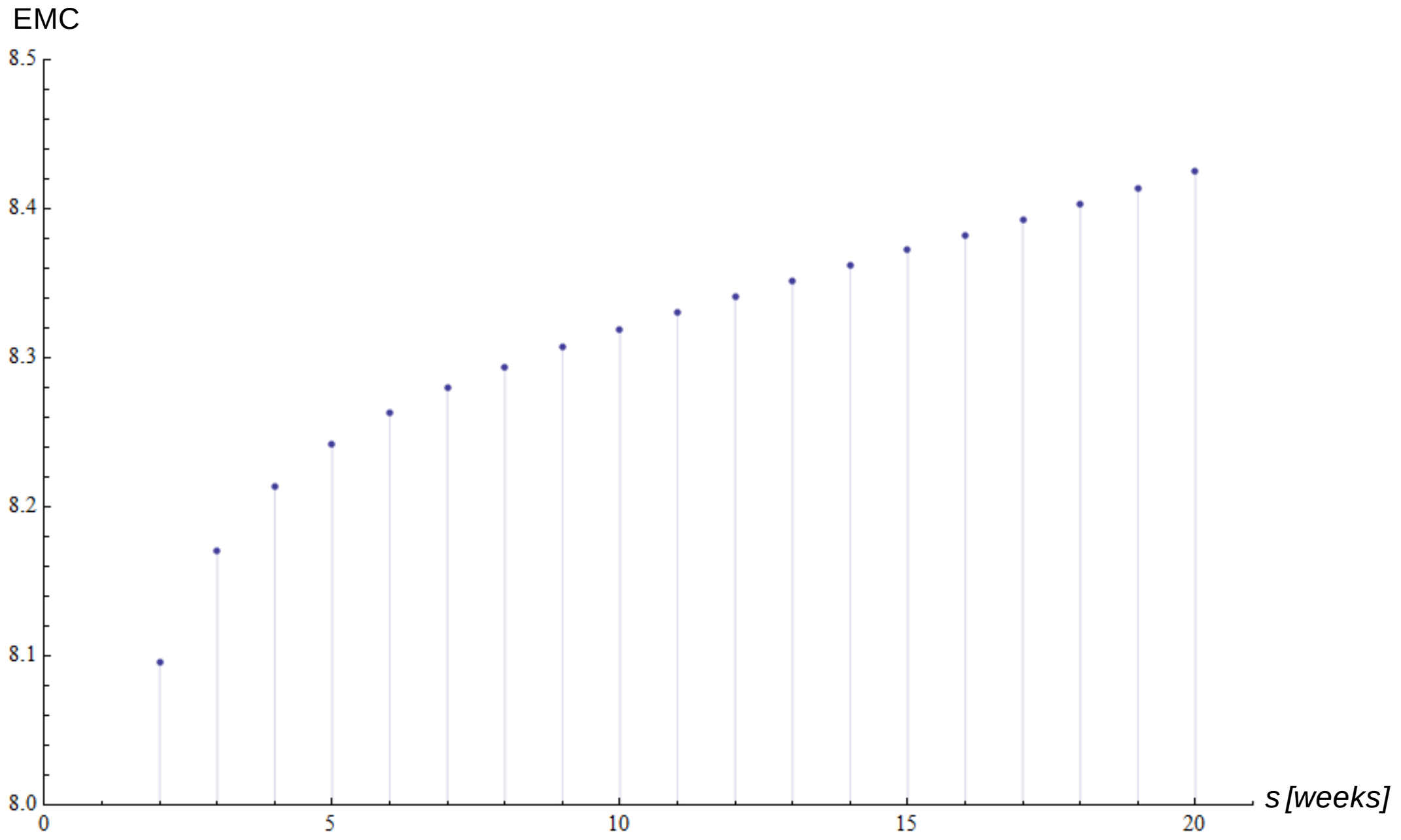

**Figure 26**

*Figure 26*. Effective measure complexity $EMC_{PVAR}$ in the modeled project according to eq. 225. The parameters are given by eqs. 226 to 231.

The comparison of Figures 25 and 26 shows that the complexity metric $\mathrm{EMC}_{\mathrm{PVAR}}$ closely resembles the functional behavior of the expected total work $x_{tot}$ in the modeled project over different periods. Moreover – and most important in view of the objective of the study – the smallest complexity values are assigned to periodically correlated work processes with minimum period $s_{opt} = 2$ [weeks]. In other words, for the given boundary conditions it makes sense to minimize the period in which information about integration and tests of geometric/topological entities is deliberately withheld by system-level teams and not released to component-level teams. By minimizing the period length, the emergent complexity can be kept to a minimum and the total effort involved in the project can be reduced as far as possible.

### 5.2.2 Constrained Optimization

After presenting and discussing the results of the basic unconstrained optimization problem, we move on to formulating and solving a corresponding constrained optimization problem. This is done in a similar manner as in section 5.1.2. The constraint is that the expected total work $x_{tot}$ remains on a constant level among the different release periods by systematic intervention and does not vary with the period as in the previous section. We start by presenting analytical complexity results and go on to present results of the Monte Carlo experiments.

#### 5.2.2.1 Methods

The objective function for the constrained optimization quantifies the complexity of periodically correlated work processes in NPD in the same manner as in the previous section, and is given by eq. 225.

We seek to minimize complexity under the constraint that that the expected total work $x_{tot}$ according to eq. 61 is equal to 204.897 [wmu] for different release periods. This expected total work corresponds to the minimum value that was identified in the previous study. Please recall that this minimum expected total work is obtained for the minimum release period $s_{min} = 2$ [weeks] and for a parameter vector according to eqs. 226 to 234. Starting with the minimum period, the release period was extended by increments of one week until the maximum release period $s_{max} = 20$ [weeks] was reached. The analytical and Monte Carlo experiments must therefore not only consider the different correlation lengths, they must also adjust the independent parameters by systematic algorithmic intervention of the experimenter so that the total expected effort in the project does not change under the different release conditions. To keep $x_{tot}$ constant, the independent parameters $a_{11}^{SH}$ and $a_{22}^{SH}$ were adjusted (see eq. 230). These parameters represent the fraction of work that is put in hold state by the system-integration teams at each short iteration before it is released at the end of period $s$. The parameter adjustment was done by a self-developed iterative method in which the reference values $a_{11}^{SH} = 0.03$ and $a_{22}^{SH} = 0.03$ were reduced incrementally until the expected value $x_{tot}(3 \leq s < 20)$ did not deviate more than $10^{-6}$ [wmu] from the correct value $x_{tot}(s = 2) = 204.897$. Both reference values were reduced by the same amount and it always held $a_{11}^{SH} = a_{22}^{SH}$. The time scale was not modified.

Following the previous procedures, we assumed that the standard deviation $c_{ii}$ of performance fluctuations (eq. 6) influencing task $i$ in the project is proportional to the task processing rate with proportionality constant $r = 0.02$. Other correlations between vector components were not considered. The variance of the fluctuations related to the finished work that is put in hold state is again reduced by the factor $10^{-3}$.

The Mathematica software package from Wolfram Research was used to carry out the analytical calculations and the Monte Carlo experiments. The stopping criterion for the Monte Carlo experiments was that a maximum of 5% of work remained for all tasks in the simulated projects. It was assumed that all development tasks on component- and system-levels were initially fully incomplete. In addition to EMC, the KPIs “project duration” and “total work" were used to evaluate performance. To calculate these KPIs, 10,000 independent replications were considered for each release period.

The results of the Monte Carlo experiments were analyzed by the same inferential statistical methods that were used in section 5.1.2. For the analysis, we drew additional samples based on 100 independent runs. To simplify the interpretation and discussion of the data, we only considered three release periods, namely $s_{min} = 2$, $s_{med} = 10$ and $s_{max} = 20$, in order to guarantee a sufficient coverage of the complete interval. We hypothesized that lower values of the complexity metric $\mathrm{EMC}_{\mathrm{PVAR}}$ lead to a significantly lower expected project duration. We also hypothesized that different levels of the complexity metric $\mathrm{EMC}_{\mathrm{PVAR}}$ do not correspond to significantly different means of the total work in the Monte Carlo experiments. The null hypotheses $H_{0,T}$ and $H_{0,x_{tot}}$ were formulated accordingly. To evaluate these hypotheses the Kruskal-Wallis location equivalence test was used. The level of significance was set to $\alpha = 0.05$.

Following the procedure from section 5.1.2, additional goodness-of-fit hypothesis tests were carried to evaluate the differences between the distributions of performance data for the three release periods. The focus was on paired comparisons between the minimum release period $s_{min} = 2$ and the other periods. The null hypothesis $H_{0,gof}$ was always that performance data drawn from a sample with release period $s_{med} = 10$ or $s_{max} = 20$ does not come from a different distribution than the data obtained for the minimum release period $s_{min} = 2$. The alternative hypothesis $H_{a,gof}$ is that the data comes from a different distribution. The Kolmogorov-Smirnov test was used to evaluate the hypotheses. The level of significance was also set to $\alpha = 0.05$.

#### 5.2.2.2 Results and Discussion

In order to satisfy the constraint imposed for the total work, the fractions $a_{11}^{SH}$=0.03 and $a_{22}^{SH} = 0.03$ of work that are put in hold state by the system-integration teams at each short iteration had to be reduced by a minimum value of $0.00769$ for release period $s = 3$ and a maximum value of $0.01646$ for release period $s_{max} = 20$. A list plot of the necessary reductions of the fractions of work is shown in Figure 27. It shows that longer release period means that the fraction of work put in hold state at each short iteration must be lower in order to satisfy the constraint.

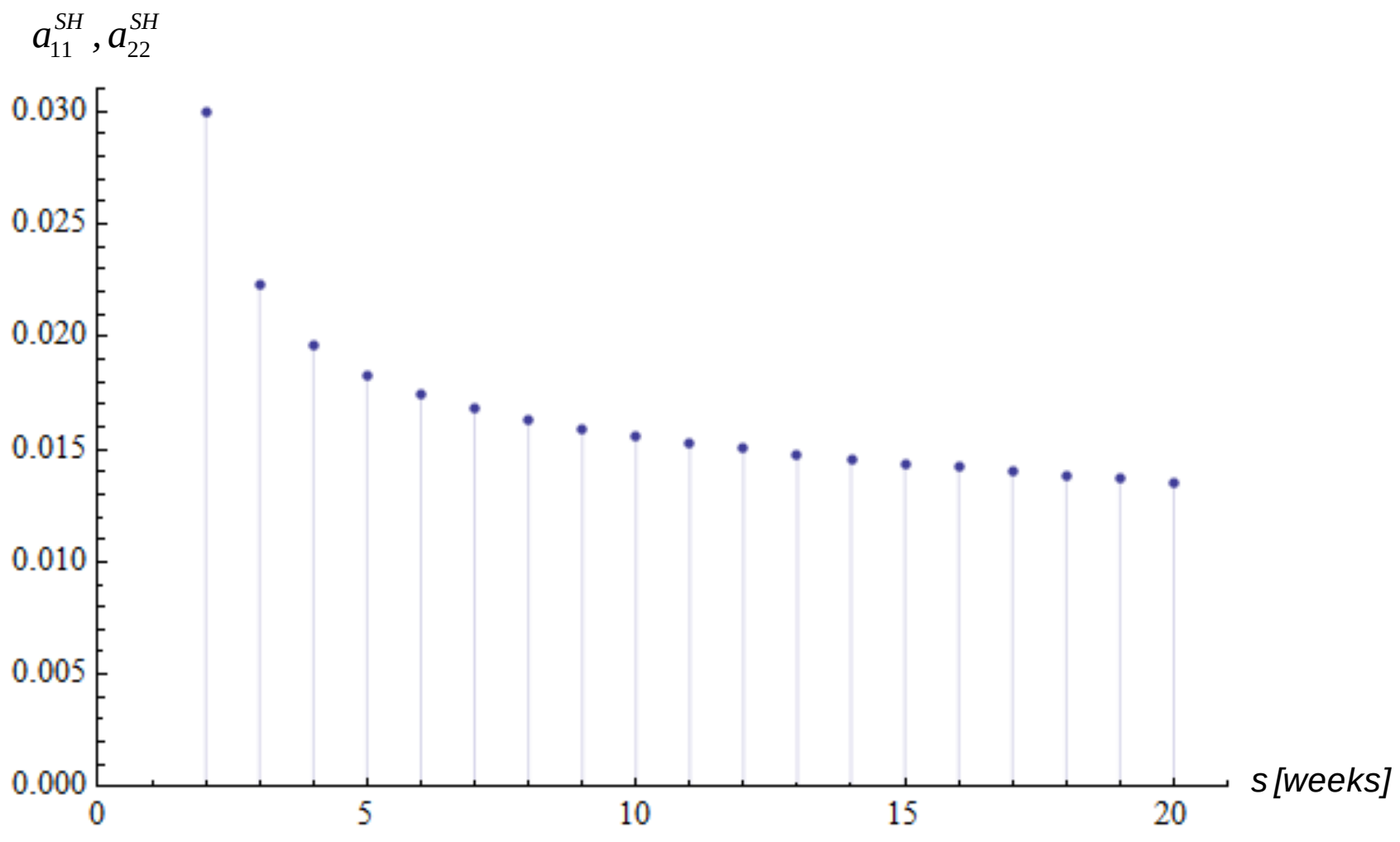

**Figure 27**

*Figure 27*. Adjustments of original parameters $a_{11}^{SH} = 0.03$ and $a_{22}^{SH} = 0.03$ (eq. 230) that were made to satisfy the constraint on the expected total work $x_{tot}(2) = ... = x_{tot}(20) = 204.897$ in the modeled projects with periodically correlated work processes. The additional parameters were not changed and are given by eqs. 226 to 231.

The corresponding values of the complexity metric $\text{EMC}_{\text{PVAR}}$ are shown in Figure 28. Interestingly, the constrained optimization of the release period leads to complexity values that decrease as the period increases. Hence, the release period minimizing emergent complexity in the sense of the complexity metric under the constraint $x_{tot}(2) = ... = x_{tot}(20) = 204.897$ is the maximum period $s_{opt}^{c} = 20$ [weeks]. This result is in stark contrast to the solution of the unconstrained optimization problem, in which minimum complexity values were assigned to the minimum period of $s_{opt} = 2$ [weeks]. The results concerning the project duration for minimum and maximum release periods are shown in Figure 29. The results for a release period of ten weeks are were also included (cf. Figure 23). The associated histograms of the total work are shown in Figure 30.

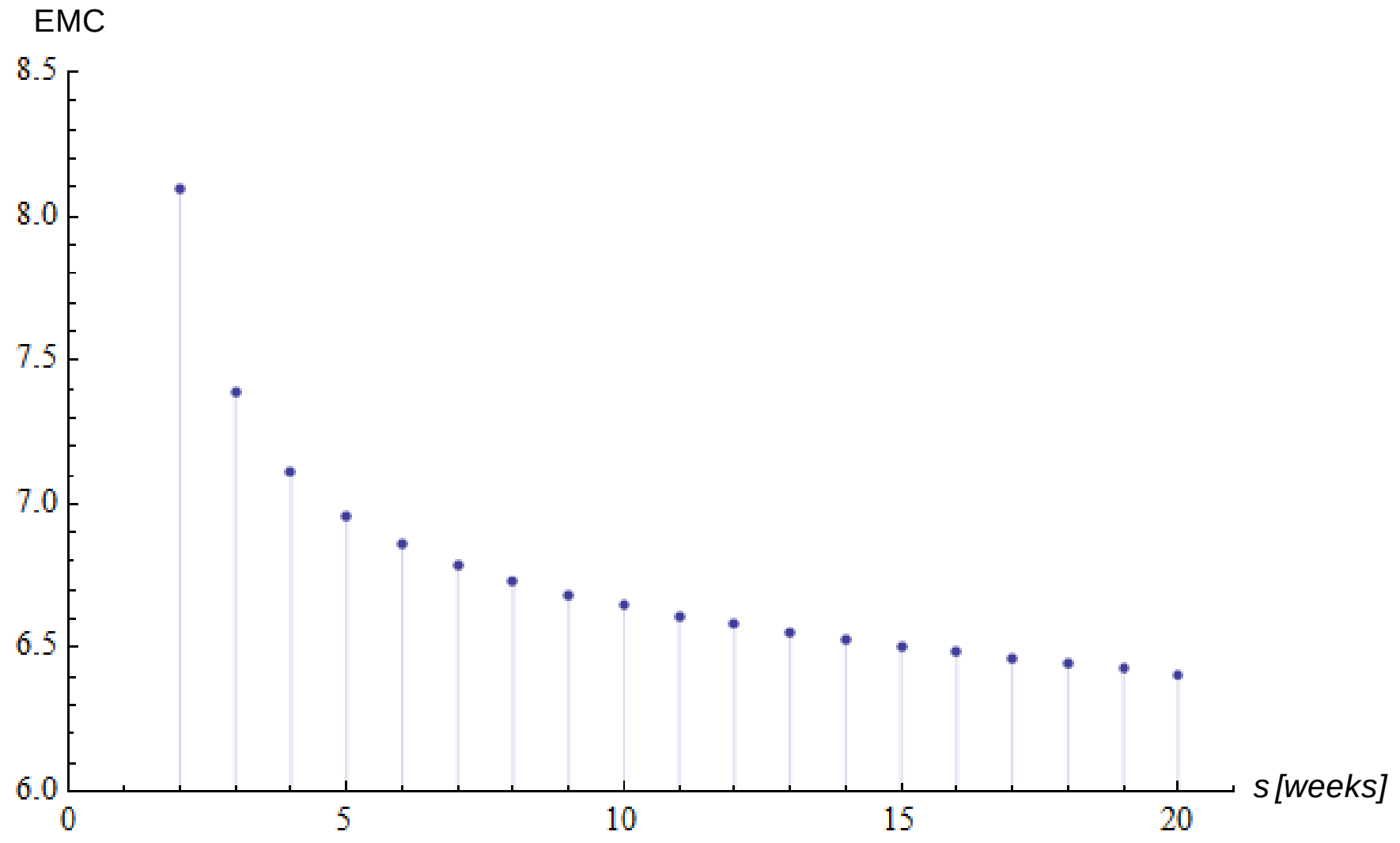

**Figure 28**

*Figure 28*. Effective measure complexity $EMC_{PVAR}$ in the modeled project according to eq. 225 under the constraint that the expected total work is kept on a constant level, i.e. $x_{tot}(2) = ... = x_{tot}(20) = 204.897$. To keep the expected total work constant, the original parameters $a_{11}^{SH} = 0.03$ and $a_{22}^{SH} = 0.03$ (eq. 230) were adjusted by the same value. The additional parameters were not changed and are given by eqs. 226 to 231.

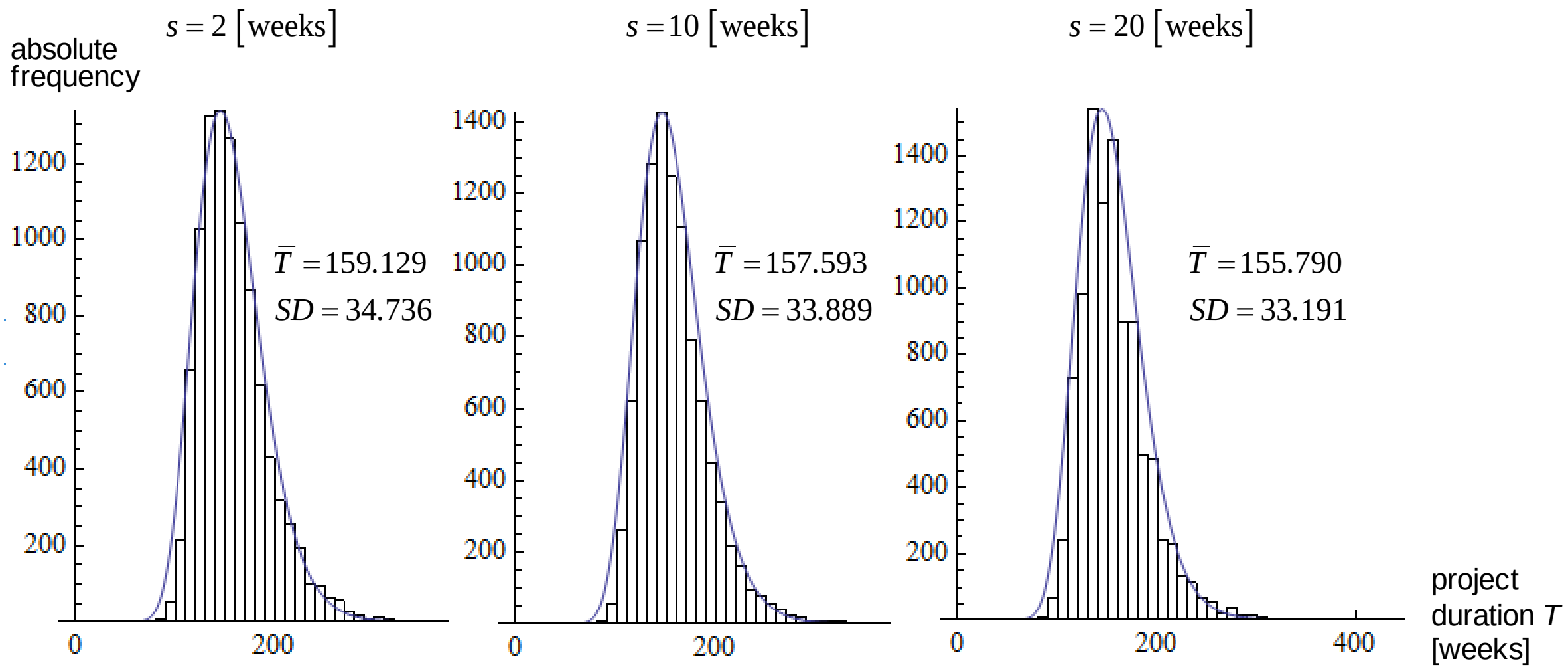

**Figure 29**

*Figure 29*. Histograms of the project duration obtained for the three release periods $s = 2, 10$ and 20 [weeks] under the constraint $x_{tot}(2) = x_{tot}(10) = x_{tot}(20) = 204.897$. We computed 10,000 independent replications. The stopping criterion for the simulated projects is that a maximum of 5% of work remained for all tasks. The Monte Carlo experiments are based on state eq. 59. The parameters are given by eqs. 226 to 231. The adjustments of parameters $a_{11}^{SH}$ and $a_{22}^{SH}$ follows the list plot from Figure 26.

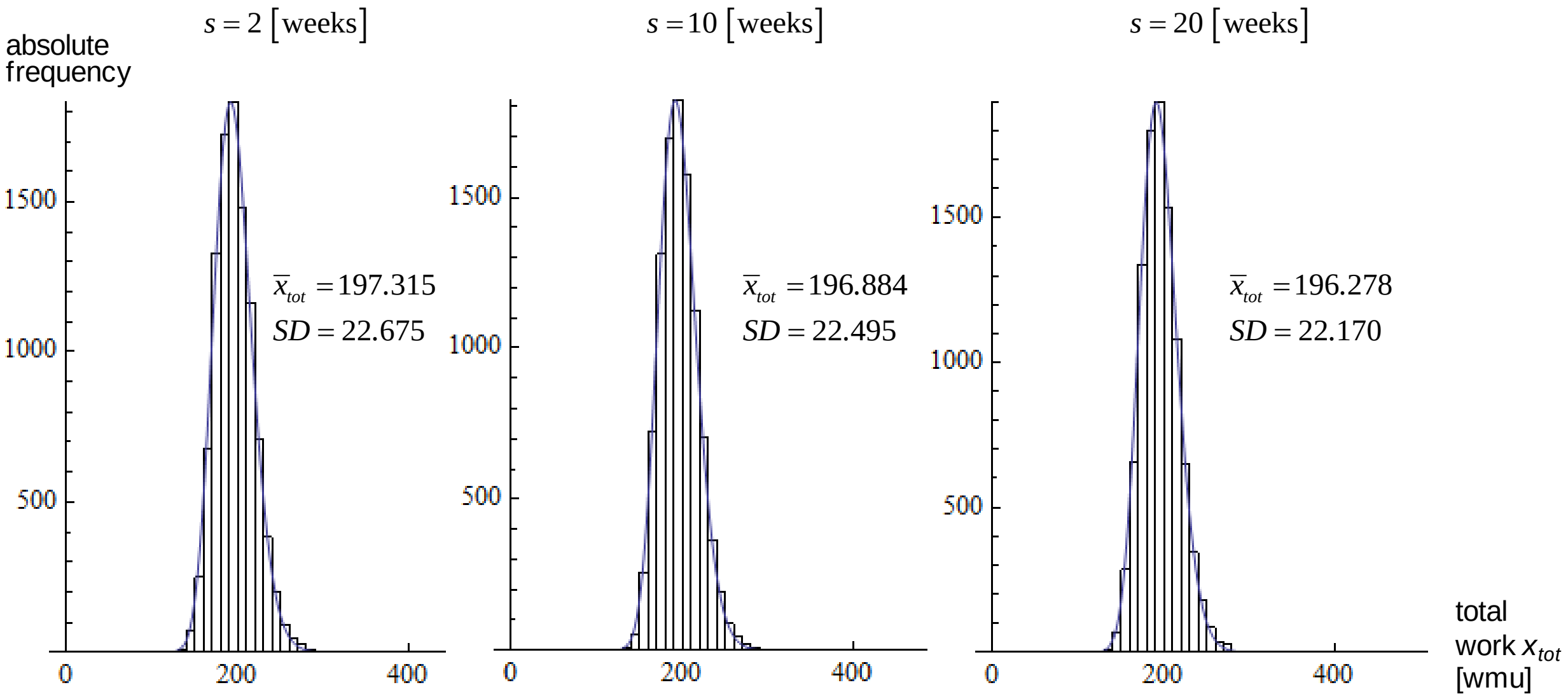

**Figure 30**

*Figure 30*. Histograms of the total work obtained for the three release periods $s = 2, 10$ and 20 [weeks] under the constraint $x_{tot}(2) = x_{tot}(10) = x_{tot}(20) = 204.897$. We computed 10,000 independent replications. The stopping criterion for the simulated projects is that a maximum of 5% of work remained for all tasks. The Monte Carlo experiments are based on state eq. 59. The parameters are given by eqs. 226 to 231. The adjustments of parameters $a_{11}^{SH}$ and $a_{22}^{SH}$ follows the list plot from Figure 26.

For 100 additional independent runs, the Kruskal-Wallis test on the project duration data shows that the location differences between the three release periods are not significant ($K_T = 2.5157$, $p = 0.2852$). Hence, the null hypothesis $H_{0,T}$ that the true location parameters of the samples are equal cannot be rejected on the significance level of $\alpha = 0.05$. The Kruskal-Wallis test on the total work data comes to a similar conclusion and shows that the locations between the three release periods are not significant ($K_{x_{tot}} = 1.2870$, $p = 0.5270$). The null hypothesis $H_{0,x_{tot}}$ that the true location parameters of the samples are equal also cannot be rejected on the significance level of $\alpha = 0.05$. In spite of these insignificant differences in mean performance, the goodness-of-fit hypothesis test between minimum release period $s_{min} = 2$ and maximum period $s_{max} = 20$ indicates significant differences in the distributions of the project duration from which data were drawn. In this case, the Kolmogorov-Smirnov test statistic is $D_{T,2-20} = 0.138$. The associated $p$-value is $p = 0.0401$. Hence, the differences in the complexity metric $\mathrm{EMC}_{\mathrm{PVAR}}$ according to Figure 23 also lead to significant differences in probability distributions. The additional distribution fit tests of the project duration and total work data do not show significant differences among the experimental conditions. The test statistic related to the project duration data for release periods $\mathrm{s_{min}} = 2$ and $\mathrm{s_{med}} = 10$ is $D_{T,2-10} = 0.096$ (p = 0.2961). For the simulated total work the test statistic for release periods $s_{min} = 2$ and $s_{med} = 10$ is $D_{x_{tot},2-10} = 0.0945$ ($p = 0.3135$) and for periods $s_{min} = 2$ and $s_{max} = 20$ it is $D_{x_{tot},2-20} = 0.1038$ (p = 0.2159).

## 6 CONCLUSIONS AND OUTLOOK

In this paper, stochastic models of cooperative task processing in new product development (NPD) projects that are subjected to concurrent engineering (CE) were introduced. The project models are based on the fundamental work of Smith and Eppinger (1997) and Yassine et al. (2003) on deterministic project dynamics and also consider the important developments by Huberman and Wilkinson (2005) toward the theory of stochastic processes. In fact, the new models are VAR model of order 1 that are able to account for unpredictable fluctuations of human performance in cooperative task processing. These fluctuations are inherent in open organizational systems of different kinds. The developed model was validated on the basis of an empirical study in a small industrial company in Germany, where two overlapping development tasks were considered. The company develops sensor technologies for the automotive industry. The models can also capture project dynamics beyond the rather smooth task processing in the validation study: they also, for instance, explain "problem-solving oscillations" (Mihm et al. 2003) with few assumptions about the cooperative problem-solving processes. According to the deterministic and stochastic parts of the state equations, the irregular oscillations between being on, ahead of, and behind schedule can be interpreted as excited performance fluctuations (Schlick et al. 2008) The excitation can occur because of the intimate relationship between the structure of the work transformation matrix (WTM) $A_0$ and the forcing matrix $K$. These mechanisms were uncovered explicitly in the spectral basis (see for instance eqs. 28 and 32 in conjunction with eqs. 186 and 190).

Moreover, and most importantly from a scientific point of view, an information-theory complexity measure termed effective measure complexity (EMC) was introduced, and several closed-form solutions under the premise of the developed stochastic model of cooperative task processing were derived. These solutions are beneficial for evaluating strong emergence in terms of mutual information communicated from the past to the future and for optimizing project organization design. The measure and the associated complexity theory go back to the theoretical physicist Grassberger (1986), whose seminal work has been completely overlooked in organization theory and engineering management literature. His theory allowed us to derive the EMC of the specified class of models of project dynamics from first principles and to calculate closed-form solutions with different numbers of independent parameters and levels of expressiveness. The results are given in eqs. 175, 181, 186, 190 and 193. Especially interesting for complexity analysis and optimization is the closed-form solution from eq. 190, because it significantly reduces the number of independent parameters without blurring the essential spatiotemporal structures that shape emergent complexity in NPD projects. Furthermore, we were able to put upper and lower bounds on the EMC (see eqs. 198, 199 and 200). These bounds are also useful for deriving additional numerical approximations going beyond the scope of this paper. It is important to point out that Grassberger's complexity theory is not limited to a specific class of stochastic project models. If the data are generated by a project in a specific class but with unknown parameters, we can calculate the EMC explicitly, as we did. It is also possible, however, to quantify the complexity of projects that fall outside the conventional models.

In addition to its conceptual validity because of the underlying complexity theory of Grassberger (1986) and other research, the EMC has several favorable properties in the application domain of project management, as follows. 1) It is small for projects in which tasks can be processed independently without cooperation and it assigns larger complexity values to intuitively more complex projects with the same dominant eigenvalue of the work transformation matrix but a stronger task coupling. The importance of the nature, quantity and magnitude of organizational subtasks and subtask interactions is also pointed out in the theoretical and empirical analyses of Tatikonda and Rosenthal (2000). Interestingly, the empirical studies of Hölttä-Otto and Magee (2006) show that estimation of effort in NPD projects is primarily based on the scale and stretch of the project and not on interactions. This is due to the fact that the balancing or reinforcing effects of concurrent interactions in open organizational systems are very difficult to anticipate for project managers. In that sense, the measure can contribute to more reliable effort estimation. The dependencies between tasks were also mentioned as complexity contributing elements in 4 from 6 cases in the empirical analysis of Bosch-Rekveldt et al. (2011). Summers and Shah (2010) consider "complexity as coupling" as one of three main aspects of design complexity. 2) The measure tends to assign larger complexity values to projects with more tasks if the intensity of cooperative relationships is similar, and therefore it is sensitive to the dimensionality of the state space of the project. This property follows, for instance, from the lower bound in eq. 42. The complexity-reinforcing effects of the "size" of a project are also stressed in Mihm et al. (2003), Mihm and Loch (2006), Huberman and Wilkinson (2005), Suh (2005), Hölttä-Otto and Magee (2006), Summers and Shah (2010), and Bosch-Rekveldt et al. (2011). Alternatively, one can divide EMC by the dimension $p$ of the state space and compare projects with different dimensionality. 3) The measure can evaluate both weak and strong emergence in an uncertain NPD environment. According to Chalmers (2002) weak emergence means that there is in principle no choice of outcome. It can be anticipated without detailed inspection of particular instances of task processing. Given the state equation, there are entirely reproducible features of its subsequent evolution that inevitably emerge over time, such as reaching a steady state. In light of our approach, a good technique for the evaluation of weak emergence is the eigenvalue analysis of the WTM. It is obvious that the EMC indicates the same bound of asymptotic stability as does a classic eigenvalue analysis by assigning infinite complexity values: if the dominant eigenvalue has modulus less than 1, the infinite sum in eq. 34 converges, and the project will converge toward the asymptote of "no remaining work"; on the other hand, if the dominant eigenvalue has modulus greater than 1, the sum diverges, and the work remaining grows over all given limits. The emergence of complexity is termed strong if the patterns of project dynamics can only be reliably forecast from the observation of the past of each particular instance of task processing and with relevant knowledge of prior history (Chalmers 2002). In the management literature this phenomenon is also known as "path dependence" (Maylor et al. 2008). Relevant information about the prior history is extracted through the predictive information according to eq. 41. This formula—in conjunction with the formulation of the normalized work transformation and fluctuation matrices—allows a holistic excitation analysis of the design modes under uncertainty. The importance of the factor "uncertainty" in the scope and methods of a project in conjunction with "stability of project environment" is also pointed out in the TOE framework of Bosch-Rekveldt et al. (2011). The information axiom of Suh (2005) addresses both size and uncertainty. The simulation study of Lebcir (2011) shows that development time significantly increases when project uncertainty is changed from low to reference level. 4) The measure is independent of the basis in which the state vectors are represented. It is invariant under arbitrary reparameterizations based on smooth and uniquely invertible maps (Kraskov et al. 2004) and therefore is independent of the subjective choice of the measurement instrument of the project manager. To the best of our knowledge, this fundamental objectiveness is a unique property that other metrics do not possess.

Finally, the manifold theoretical complexity analyses were extended toward the solution of more-practical problems in project management through an applied example of optimizing project organization design. The objective was to minimize project complexity by systematically choosing the "best" team design from within an allowed group of cooperating individuals with different productivities in a simulated CE project. The simulation results show that the EMC is a theoretically highly satisfactory and conceptually valid quantity that also leads to useful results in organizational optimization regarding key performance indicators, such as the duration of the entire project and its empirical variance.

The information-theory approach to evaluating emergent complexity in NPD in conjunction with the state-space representation of project dynamics still will need to be worked out in more detail in the future. A first step in that direction would be to compute the EMC of the most advanced stochastic project model developed by Huberman and Wilkinson (2005), which incorporates multiplicative instead of additive noise (cf. eq. 4). Their model is interesting not only because it can reproduce critical effects of both large groups and long delays with very few assumptions about the cooperative problem-solving processes, but also because of its reasonable assumption that the autonomous task-processing rates and the intensity of cooperative relationships are subject to random influences. However, to the best of our knowledge, the Huberman–Wilkinson model has not been supported by any empirical evidence, and it is an open question whether it is more predictive than our VAR(1) approach. It may be possible to calculate the EMC of the Huberman–Wilkinson model by the following three main steps. First, calculate the covariance matrix in the steady state. Second, show that performance variability in the steady state is governed by a log-normal distribution. The simulation results indicate that the distribution in the steady state may be log-normal (Huberman and Wilkinson 2005). Third, calculate the dynamic entropies of the state variables and derive the corresponding EMC. Unfortunately, the three calculation steps are mathematically very involved. A natural extension on the basis of linear state-space models would be to incorporate both multiplicative and additive noise (Arnold and Wihstutz 1982). However, under this premise the parameter estimation from a small sample size typical for organizational modeling is difficult and can be unreliable. On the other hand, a theoretically and practically promising extension of the dynamic project model was presented in section 2.6, where a so-called periodic vector autoregressive stochastic process was formulated (see e.g. Lütkepohl 2005, Ursu and Duchesne 2009). A periodic vector autoregressive process can capture not only the dynamic processing of the development tasks with short iteration length within CE teams but also a long-scale “seasonal” effect. A seasonal effect is common in large-scale CE projects (e.g. in the automotive or aerospace industry) and is caused by the periodic information release policy of system-level design information across teams. A deterministic model able to simulate this kind of task processing was developed by Yassine et al. (2003). A corresponding stochastic model has been formulated recently (see Schlick et al. 2011). The parameters of periodic vector autoregression model can be calculated quite easily based on maximum likelihood estimation (Lütkepohl 2005) or least square estimation (Ursu and Duchesne 2009). Moreover, the model offers a compact representation as a VAR process, as we showed in eq. 147. Hence, the introduced closed-form solutions and bounds do not have to be reworked but rather can be applied directly to analyze and evaluate emergent complexity.

In the long run, we aim to conduct an external validation study of the complexity theory with experienced project managers in industry. It is hypothesized that the EMC is a conceptually valid complexity variable that can be used for simulation-based optimization of project organization design (see section 4) and that it also has the potential to capture the implicit knowledge of project managers based on the nature, quantity and magnitude of concurrent tasks and their interactions. In general, this complexity measure provides valuable information enabling the project manager and CE teams to better organize their work and to improve coordination. Furthermore, we believe that the results of the optimization study of project organization design can have significant theoretical implications. This is because of the fact that we found certain experimental evidence that the popular lean management technique of production leveling is also effective for the minimization of both the lead time and its empirical variance in NPD projects. The developed theoretical and conceptual framework is a stable foundation for analyzing this interesting phenomenon in detail in the future and for formulating a generalized solution.

ACKNOWLEDGMENT

The research presented in this report was supported by grants from the Deutsche Forschungsgemeinschaft (DFG Normalverfahren SCHL 1805/3-1 and SCHL 1805/3-3). The author would like to thank DFG for their gracious support. Furthermore, Andreas Kräußling, Eric Beutner and Pantaley Dimitrov deserve special acknowledgment for reviewing and correcting the mathematical formulas for the vector autoregression models. My understanding of the relationships between complexity and entropy for hidden Markov processes has significantly benefited from the discussions with Carsten Winkelholz.